\documentclass[final,times,onecolumn,square,numbers]{elsarticle}

\usepackage[T1]{fontenc}
\usepackage[utf8]{inputenc}
\usepackage{textcomp}
\usepackage[english]{babel}
\usepackage{url}
\usepackage{amsmath,amsfonts}
\usepackage{graphicx}
\usepackage{siunitx}
\usepackage{booktabs,caption}
\usepackage{epsfig}
\usepackage{wrapfig}
\usepackage{subcaption}
\usepackage{hyperref}
\usepackage{amsthm}
\usepackage{amssymb}
\usepackage[dvipsnames]{xcolor}
\usepackage{amsmath,color, amssymb}
\usepackage{hyperref}
\usepackage{bm}
\usepackage{mathrsfs}
\usepackage{bbm}
\usepackage{csquotes}
\usepackage[normalem]{ulem}
\usepackage{parskip}

\DeclareRobustCommand{\vect}[1]{
  \ifcat#1\relax
    \boldsymbol{#1}
  \else
    \mathbf{#1}
}

\DeclareRobustCommand{\psv}[1]{
  \ifcat#1\relax
    \boldsymbol{#1}
  \else
    \mathbf{#1}
}

\newcommand{\new}[1]{#1}

\newcommand{\ket}[1]{\vert#1\rangle}
\newcommand{\bra}[1]{\langle#1\vert}
\newcommand{\Hilbert}{\mathscr{H}}
\newcommand{\WW}{\Omega}
\newcommand{\vare}{\varepsilon}
\newcommand{\dd}{\WW}
\newcommand{\D}{\WW_E}
\newcommand{\Tr}{\text{Tr}}
\newcommand{\uu}{\mathsf{P}}
\newcommand{\ee}{\mathsf{E}}
\newcommand{\bbs}{\mathbb{S}(\Hilbert_E)}
\newcommand{\scp}[2]{\langle #1| #2 \rangle}

\newcommand{\MMO}{\hat{M}}
\newcommand{\MMC}{\MMO_1, \ldots, \MMO_K}

\newcommand{\OO}{\mathcal{A}}
\newcommand{\decomp}{\mathscr{D}}

\begin{document}

\title{On the foundations of statistical mechanics}

\author[a]{Marco Baldovin}
\affiliation[a]{organization={CNR, Istituto dei Sistemi Complessi, Universit\`{a} Sapienza},
            addressline={P.le A. Moro 5}, 
            city={Roma},
            postcode={00185},
            country={Italy}}
\author[b]{Giacomo Gradenigo}
\affiliation[b]{organization={Gran Sasso Science Institute},
            addressline={Viale F. Crispi 7}, 
            city={L'Aquila},
            postcode={67100},
            country={Italy}}
\author[a,c]{Angelo Vulpiani}
\affiliation[c]{organization={Department of Physics, Universit\`{a} Sapienza},
            addressline={P.le A. Moro 5}, 
            city={Roma},
            postcode={00185},
            country={Italy}}
\author[d]{Nino Zangh\`i}
\affiliation[d]{organization={INFN sezione di Genova and Dipartimento di Fisica, Università di Genova},
            addressline={via Dodecaneso 33}, 
            city={Genova},
            postcode={16146},
            country={Italy}}

\begin{abstract}
 Although not as wide, and popular, as that of quantum mechanics, the investigation of fundamental aspects of statistical mechanics constitutes an important research field in the building of modern physics. Besides the interest for itself, both for physicists  and philosophers, and the obvious pedagogical motivations, there is a further, compelling reason for a thorough understanding of the subject. The fast development of models and methods  at the edge of the established domain of the field requires indeed a deep reflection on  the essential aspects of the theory, which are at the basis of its success. These elements should never be disregarded when trying to expand the domain of statistical mechanics to systems with novel, little known features.

It is thus important to (re)consider in a careful way the main ingredients involved in the foundations of statistical mechanics. Among those, a primary role is covered by the dynamical aspects (e.g. presence of chaos), the emergence of collective features for large systems, and the use of probability in the building of a consistent statistical description of physical systems. 

With this goal in mind, in the present review we aim at providing a consistent picture of the state of the art of the subject,  both in the classical and in the quantum realm.  In particular, we will highlight the similarities of the key technical and conceptual steps with emphasis on the  relevance of the many degrees of freedom,  to justify the use of statistical ensembles in the two domains.


\end{abstract}

\setlength\parindent{0pt}

  \newcommand{\av}[1]{\left\langle#1\right\rangle}
  \newcommand{\cbr}[1]{\left(#1\right)}
  \newcommand{\sbr}[1]{\left[#1\right]}
  \newcommand{\bbr}[1]{\left\{#1\right\}}
\newcommand{\gcol}[1]{\textcolor{blue}{#1}}
\newcommand{\gcanc}[1]{\gcol{\sout{\textit{#1}}}}
\newcommand{\gcorr}[1]{\gcol{\textit{#1}}}
\newcommand{\gcomm}[1]{\gcol{{\bf #1}}}
\newcommand{\mbremark}[1]{ \textcolor{PineGreen}{\textbf{[[ #1 ]]}}}

\newcommand{\mbnew}[1]{\textcolor{PineGreen}{#1}}
\newcommand{\rn}[1]{\textcolor{Magenta}{#1}}
\newcommand{\rns}[1]{\textcolor{Magenta}{\sout{#1}}}

\maketitle
\vspace*{3cm} 

\vspace*{3cm} 

{\it To the glorious memory\\ of James Clerk Maxwell, Ludwig Eduard Boltzmann, Andrei Nikolaevich Kolmogorov 
and Lev Davidovich  Landau}
  
  \newpage
\tableofcontents

\newpage

\section*{Introduction}
\addcontentsline{toc}{section}{Introduction}
Statistical mechanics (SM) finds its motivation in the study of the behaviour and properties of macroscopic bodies, i.e. of systems composed of a very large number of particles (atoms or molecules). These elementary constituents are supposed to follow mechanical laws, either classical or quantum depending on the case~\cite{ma85}. The description of the whole system's evolution at this level of detail is often referred to as \textit{microscopic dynamics}. Its investigation is not only extremely challenging, but also unnecessary for most purposes: in everyday life, as well as in practical applications, one is rather interested in the \textit{macroscopic dynamics},  i.e. the evolution of a reduced number of observables that are accessible at the macroscopic scale (e.g., temperature, pressure and so on). These  quantities follow  much simpler, deterministic laws, which define the thermodynamics of the system. From an historical point of view, most of them were discovered empirically, without an insightful theoretical basis: as Gibbs once said~\cite{gi02,la69}, 
\begin{quote}
 Thermodynamics is only a blind guide.
\end{quote}
The primary purpose of SM is therefore to give theoretical grounds to thermodynamics,
exploiting the principles
of probability theory and the knowledge of the microscopic dynamical laws.

It is worth stressing that the simplicity of thermodynamical laws, in spite of the expected complexity
of the total microscopic evolution, is due to the fact
that measuring a macroscopic observable is equivalent to averaging
over many degrees of freedom.  The lower level of
complexity at the macroscopic scale does not descend in a
mechanicistic way from the combinations of microscopic interactions:
it is a qualitatively different phenomenon. In this sense SM
is not, as one may think, an archetypal example of rigid reductionism: it is rather
a set of methods and techniques, founded mainly on probability theory,
to explore and analyze the macroscopic behaviour of a mechanical system,
in the limit of a large number of degrees of freedom.

\new{From a foundational standpoint, we take the view that the correct description of microscopic states is quantum mechanical. The universe, as far as current physical understanding goes, is governed by quantum laws, and any classical description should be regarded as an approximation valid under suitable limits. Nevertheless, there is a deep structural analogy between the classical and quantum formulations of statistical mechanics, which allows many classical insights to carry over into the quantum framework. In both cases, microstates are specified with maximal precision: as points in phase space for classical systems and as pure states (or vectors in a Hilbert space) in the quantum setting. Macrostates, on the other hand, correspond to regions in phase space or to subspaces of the Hilbert space characterized by certain macroscopic properties. This parallelism has been emphasized, for example, in the works of J. L. Lebowitz, who has long argued that the Boltzmann approach to entropy and the evolution from less to more probable macrostates retains its core validity in the quantum domain. As a result, much of the classical intuition—particularly in the analysis of the Boltzmann entropy and the role of typicality—can be adapted to the quantum case with appropriate care.}

\new{Roughly speaking, we can divide SM in two branches: equilibrium SM, devoted to the study of properties which do not change in time (e.g., the state equations), and non-equilibrium SM, where the time evolution has an important role (as, for instance, in the study of irreversibility).}
The aim of the present review is to provide a (non-exhaustive)
picture of the main conceptual aspects at the basis of SM for classical and quantum systems\new{, both in and out of equilibrium}.
To this end, we collected those contributions, both from classical
works and from recent literature, that tried to answer the question:
``Why does SM work?''.

Given that the
problem of SM foundations is one and a half century
old, one may wonder why to discuss it today.  In our opinion there are several good reasons for that:
\begin{enumerate}
\item The subject is of interest by itself, both for scientists
  and philosophers,  because  it exemplifies how a radically new phenomenon can
  emerge from the typical behaviour of a lower-level one.   In  particular, the relation between microscopic and macroscopic laws allows to understand why, in  many branches of science, the reductionist approach to complex phenomena is prone to failure~\cite{chibbaro2014reductionism,berry1995asymptotics,ba02}.
\item The subject is pedagogically relevant:  far from being a pure intellectual exercise, it allows instead for a concrete discussion of the main conceptual issues concerning the link between different levels of description of a given reality. Statistical mechanics has proven to be, so far, the only
  methodological and conceptual framework in theoretical physics which
  is valid at all scales, from the cosmic ones dominated by classical
  gravitational interactions to the microscopic ones where quantum
  mechanics is at play, provided that the number of involved degrees of freedom
  is large.  In order to stress the similarities of
  SM foundation problems in the classical and
  quantum domain a whole section of this review work will be dedicated to the
  SM of quantum many-body systems.

   \item Although classical and quantum SM share many similarities in their basic structure, there are also important differences related to quantum phenomena having no classical analog, such as quantum entanglement and the interpretative difficulties associated with macroscopic superpositions. Therefore, the study of the foundations of statistical mechanics may be relevant to the foundations of quantum mechanics and vice versa.
   
\item The subject is important in the development of current
  technology: for instance, challenging frontiers for the applications
  of statistical physics are provided by systems with a small number
  of degrees of freedom, far from the thermodynamic limit, such as
  those of interest in bio- and nano-technologies.  Therefore the
  foundations and applicability of the theory have to be scrutinised~ \cite{ma85},
  in the light of a presumably even higher relevance of
  probabilities than in the original framework of SM.

\end{enumerate}




 Before introducing the content of the single sections of this review, it is useful to retrace quickly the main steps that brought to the formulation of the theory of SM as it is known nowadays. A comprehensive historical digression exceeds instead the scope of this review paper: the interested reader can refer to~\cite{me01,mu07,br03,da18,me01}.

As far as we know, the first primitive form of SM is due to
Daniel Bernoulli. His derivation of Boyle's law relies
on the intuition that the pressure  on the walls of a container
comes from the average effect of the impacts of many
molecules. Bernoulli’s theory (1738) was then almost forgotten
for more than a century, until Rudolf Clausius built
his own kinetic theory of gases (1857), where the fundamental
concept of mean free path (i.e., the free space crossed by a molecule
in the time interval between two collisions with other molecules)
was introduced for the first time.  After Clausius, Maxwell came to
the idea of a probability distribution of the velocities of molecules
in a gas, which is the first important systematic application of
probabilistic concepts to physics~\cite{br03}. 

The progress of SM was strictly related to the development of the
 theory of atomism. The kinetic theory of gases provided empirical
 and theoretical support to the atomistic point of view, predicting
 experimental laws as, e.g., the celebrated result (due to Maxwell)
 that the viscosity of gases is independent of the density.
 The most important contributions in this sense are due to Ludwig
 Eduard Boltzmann. His celebrated H-theorem allowed to reconcile
 the familiar notion of macroscopic irreversibility with the deterministic nature 
 of the micorscopic motion~\cite{ce98}. This success of kinetic theory greatly
 contributed to the building of SM.
 
The third founding father of SM was Gibbs, who, with his celebrated
booklet~\cite{gi02}, contributed to build a systematic theory.
In the modern textbooks we still use Gibbs's terminology for the ensembles
(i.e.  microcanonical, canonical and grand canonical).
Sometimes Boltzmann and Gibbs are viewed as the champions of
different points of view about SM.  According to
this vulgata, Gibbs would be the founder of the ensemble approach, in
contrast with Boltzmann's view based on dynamical theory and ergodicity.
As a matter of fact both ergodicity and ensembles (originally called \textit{monodes}) are inventions of Boltzmann,
although he is not always credited of them~\cite{da18}.  
In Boltzmann's works it is also shown that, in the thermodynamic limit, the {\it
  ergode} and the {\it holode} (respectively the microcanonical and
canonical ensemble) are equivalent. This is also the reason for the name of the
\textit{ergodic hypothesis}, where all states belonging to the same ergode
are supposed to be equiprobable, and of the consequent \textit{ergodic theory}.

  It is worth recalling that kinetic theory was alternative to the traditional understanding of
 thermodynamics employed until the beginning of 19th century
to explain heat flow, which involved the use of an
imponderable material called ``caloric''.  Still in the late 19th century, 
in spite of the important results of kinetic theory, the influential
``energetistic'' school was rejecting  atomism, based
on the success of phenomenological thermodynamics~\cite{me01, ce98}.  This
skepticism, which may now seem surprising, at that time was supported by
important scientists and philosophers such as, e.g., P. Duhem, E. Mach, and
W. Ostwald. In their positivistic point of view, the champions of the
energetistic school believed that macroscopic phenomena (such as chemical
reactions) should be treated solely in terms of phenomenological
descriptions. Their approach was based on the conservation of energy and the spontaneous
increase of entropy, and atoms were only regarded as a useful
mathematical tool, with no realistic physical grounds.
Since atoms and molecules were not visible, a decisive evidence of the atomic
structure was considered to be impossible, and the atomic theory
a physically unverifiable hypothesis.  The
followers of Energeticism also considered the phenomenological thermodynamics
better than atomism in explaining the second law of
thermodynamics. One of the reasons for such an opinion is the fact
that Newtonian mechanics, owing to its time reversal property, cannot
be exploited to distinguish between past and future~\cite{ce98,da18}.

At the beginning of 20th century, a new stimulus to SM came from the study of
Brownian motion by Langevin, Einstein and Smoluchovsky. The stochastic dynamics
of a colloidal particle suspended in a fluid required new mathematical tools based
on probability theory, and set the ground to the development of stochastic processes.
It is interesting that Einstein did not care about the Brownian motion in
itself: his real purpose was to show that microscopic bodies
suspended in a liquid possess an observable fluctuating movement,
which can only be justified in the light of the \textit{molecular} theory of heat.
In other words, he aimed at finding an experimental evidence of the existence of atoms,
in line with Boltzmann's point of view. Using Einstein’s formula on Brownian Motion,
Perrin was even able to ``count'' the number of atoms in a volume of fluid,
by measuring the diffusion of a suspended
colloidal particle.   After Perrin's experimental validation,
even the last followers of Energetisms capitulated to the increasingly compelling
evidence in favour of atomism. 

 Even  after this striking success, a
consistent fraction of the scientific community did not appreciate SM:
beyond the philosophical perplexities of the energetistic school, some
fundamental facts were still hard to explain, e.g., the
temperature dependence of the specific heat in polyatomic molecules
and in the diamond.  After the introduction of Quantum Mechanics (QM),
a suitable generalization of SM taking into account the quantum
effects was able to solve most of these problems.  While the
understanding of specific phenomena was improved,
the impact of QM on SM had mainly a technical nature, leaving
untouched the basic aspects of the theory.

The last century has seen a wealth of contributions in every branch of SM.
In the 1930's, the studies by Birkhoff and von Neumann on the ergodic theory,
and the seminal work of Krylov on mixing gave important contributions to the foundations of SM.
In the 1940's, Onsager's exact solution for the $2D$ Ising model proved that
the formalism of SM can describe phase transitions and critical phenomena.
The Fermi-Pasta-Ulam-Tsingiou (FPUT) numerical experiment on nonlinear
Hamiltonian systems (1955) started a fruitful research line on the 
problem of ergodicity in dynamical systems~\cite{ga07}, which 
culminated with the KAM theorem for the regular behavior of
almost integrable Hamiltonians, proposed by
Kolmogorov and subsequently completed by Arnold and Moser~\cite{du14} (1954-1963).
Between 1960 and 1980 the research in SM was dominated by the study of universality for critical phenomena~\cite{KGHHLPRSAK67}, renormalization group~\cite{PhysRevB.4.3174} and the relation
with limit theorems of probability~\cite{bleher1973, ka57, J75}, with reciprocal influence between SM and 
quantum field theoretic ideas and methods. At the same time, the mathematical
  physics community was able to prove rigorous mathematical results for the Boltzmann
equation~\cite{Cercignani1994}. In more recent times,  the study of ergodicity-breaking transitions
without symmetry breaking~\cite{Kosterlitz_1973}, and the ensemble
formalism for ergodicity-breaking transition in disordered systems by Parisi~\cite{parisi79, mezard1987spin} are among the main achievements.

The above timeline should clarify that the foundation of SM was the result of the collective effort of several generations of scientists, spanned over more than 150 years. The present paper is a (necessarily partial) review of the works that mostly contributed to that fascinating conceptual quest.

In Section~\ref{sec:ensembles} we discuss the two great intuition of Boltzmann that mainly contributed to the birth of Statistical Mechanics, namely the existence of a ``bridge law'' connecting microscopic and macroscopic description, and the ergodic hypothesis. Section~\ref{sec:ergodicitytime} is devoted to the subsequent effort to define ergodicity in a proper physical sense, relating the concept to the time-scales of the dynamics. In Section~\ref{sec:chaos} we analyze the role of chaos in the foundation of SM; in particular we discuss Khinchin's point of view, which does not require the presence of chaos to observe ergodic behaviour of physically meaningful onservables. The problem of irreversibility is reviewed in Section~\ref{sec:irreversibility}, where we report the main contributions stemming from Boltzmann's celebrated $H$-theorem.
In Section~\ref{sec:quantumboltzmann} we show that Boltzmann's approach can be safely extended to the domain of quantum mechanics, introducing the concept of (quantum) thermal equilibrium dominance. Section~\ref{sec:quantumtypicality} is devoted to the review of the notion of typicality in this context: here we also discuss important differences with the classical case, due to  the intrinsically quantum phenomenon of entanglement. Section~\ref{sec:ATTE}  focuses on thermalization, reviewing contributions on the Eigenstate Thermalization Hypothesis and quantum chaos.
In Section~\ref{sec:remarks} we explicitly discuss some misconceptions and possible pitfalls on the conceptual aspects of SM. Finally in Section~\ref{sec:conclusions} we outline some conclusions.

\section{From microscopic dynamics to statistical ensembles}
\label{sec:ensembles}

From a merely practical point of view, the main purpose of SM consists in the 
evaluation of a few suitable quantities, such as free energies or correlation 
functions, in terms of the microscopic features of the considered system.  This is, for instance, how SM is employed in the study of condensed 
matter or physical chemistry. After the discovery of deterministic chaos it is now well established that even 
in systems with few degrees of freedom statistical approaches can be useful, and 
often unavoidable.  At the same time, however, the problem of justifying the use of statistical 
ensembles, for a given microscopic dynamics, is usually 
ignored. The (so-called) Gibbs description is widely accepted due to its 
ability to describe the physical reality, but the reasons of its success are
seldom considered. This pragmatic point of view is of course not satisfactory for 
those who  believe that  understanding the foundations of a theory is as 
important as using it to make calculations. The wide spectrum of viewpoints ranges from 
the Khinchin/Landau perspective, according to which the most relevant ingredient 
is the presence of a large number of degrees of freedom (irrespectively to the 
features of the microscopic dynamics), to the opinion of those (like Prigogine 
and his followers) who credit the key role  to the chaotic nature of the dynamics~\cite{chibbaro2014reductionism}.


Macroscopic systems contain, by definition, a very large number $N$ of particles, where $N$ is of the order of the Avogadro
number. Denoting by $ {\bf q}_i $ and $ {\bf p}_i$ the position and momentum vectors
of the $i$-th particle, the state
of a macroscopic system at time $t$ is described by the vector
$\psv{X}(t) \equiv ({\bf q}_1 (t) , \dots , {\bf q}_N (t), {\bf p}_1
(t),\dots , {\bf p}_N (t) )$ in a $6\, N$-dimensional phase space.
The evolution law is given by the Hamilton's equations: denoting by $V(\lbrace
{\bf q}_j \rbrace)$  the interaction potential, the Hamiltonian is
\begin{equation}
\label{eq:ham}
H = \sum _{i=1}^N {|{\bf p}_i|   ^2 \over 2 m }
 + V(\lbrace {\bf q}_j \rbrace) \, ,
\end{equation}
and the evolution equations are
\begin{equation}
\label{ham1-2}
\displaystyle  { {{\rm d} {\bf q}_i} \over  {{\rm d} t}    } = 
 \displaystyle  { {\partial  H} \over { \partial {\bf p}_i} }= 
\displaystyle { {\bf p}_i \over m}   \,\,\, , \, \,\,
\displaystyle  { {{\rm d} {\bf p}_i} \over  {{\rm d} t}   } =
- \displaystyle  { {\partial  H} \over { \partial {\bf q}_i} }  =
-  \displaystyle { {\partial  V} \over { \partial {\bf q}_i} }  
\end{equation}                                                                
with $i=1, \dots , N$.

Due to their intrinsic phenomenological character, neither temperature nor 
pressure can be trivially associated to the microscopic dynamics above. It is 
therefore mandatory an attempt to find a link between the mechanics of 
microscopic interactions, on the one hand, and the macroscopic thermodynamic 
properties of materials, on the other. The main conceptual  goal of SM is to
find this connection between a dynamics in a very high-dimensional space, 
Eq.~\eqref{ham1-2}, and thermodynamics, which
involves just few global quantities as temperature and pressure.

\subsection{ The bridge between micro and macro}
\label{sec:bridgelaw}

The first attempt to find a bridge between dynamics and thermodynamics
was pursued for the ideal gas, considered as a model of point-like,
non-interacting particles of mass $m$ confined by the walls of a
container (by means of elastic collisions). Clausius
was able to show that in this model the temperature, $T$, is proportional
to the mean kinetic energy:
\begin{align}
m \langle v^2 \rangle \new{= \frac{R}{N_A}T} =k_B T
\end{align}
being $v$ one component of the velocity vector, \new{$R$ the gas constant, $N_A$ the Avogadro number}  and $k_B$ the Boltzmann's
constant. This result was the first step toward the kinetic theory
of gases and the starting point for the building of a consistent
bridge between mechanics and thermodynamics. The most relevant step
forward along this path was then done by Boltzmann, in the form of two
main contributions~\cite{chibbaro2014reductionism,da18,me01}: (i) with a microscopic definition of entropy he proposed a
  precise mathematical relation linking the macroscopic world
  (thermodynamics) to the microscopic one (dynamics);
(ii) he introduced probabilistic ideas and their use for the interpretation of
  physical observables.

Here we are concerned with the conceptual aspects of point (i). Point (ii) is rather subtle, and even nowadays  it is the object of
intense study. Boltzmann's idea was to replace time averages with
averages coming from a suitable probability density (the ergodic hypothesis). We will discuss these aspects in Section~\ref{sec:ergohyp}.

\subsubsection{The definition of entropy}
\label{sec:defentr}

The relation connecting thermodynamics
to the microscopic world  is given by the celebrated
equation (engraved on Boltzmann's tombstone in Vienna.\footnote{Boltzmann never actually wrote down the formula in this explicit form. It is worth noting that among the early pioneers, as suggested by Einstein in the quote below, Planck was one of the first to provide an explicit expression for it.}):
\begin{equation}
S= k_B \log W
\label{SeqlogW}
\end{equation}
where $S$ denotes the entropy of the macroscopic body (a thermodynamical quantity) and 
$W$ is the number of microscopic states (a mechanical-like quantity) realising the macroscopic configuration:
\begin{align}
  W(E,N,V)= \int \delta[E-H(\psv{X},N,V)] \, d \psv{X}\,.
  \label{eq:defWdirac}
\end{align}

Sometimes a different definition of $W$ is also used:
\begin{align}
  W(E,N,V)= \int_{H(\psv{X})< E}  \, d \psv{X} = \int \theta [E-H(\psv{X},N,V)] \, ,
  \label{eq:defWheavy}
\end{align}
which happens to be more convenient to prove analytical results \new{(see e.g.~\cite{campisi2015construction})}.
For systems described by the Hamiltonian dynamics~\eqref{ham1-2}, they are equivalent in
the limit of large $N$. The definition~\eqref{eq:defWdirac} must be adopted instead when
systems allowing negative absolute temperature are considered~\cite{ramsey56, baldovin2021, GILM21, GILM21b}.

If  $W$ can be expressed as a function of the energy
$E$ and of the volume $V$, from the following thermodynamic relations 
\begin{align}
{1 \over T}&= {\partial S \over \partial E} \\
{P \over T}&= {\partial S \over \partial V} \, ,
\end{align}
a mechanical definition of the temperature,
$T$, and of the pressure, $P$, can be obtained.

Equation~\eqref{eq:defWdirac} paves the way for a precise definition of
\textit{macrostate} and \textit{microstate}: while a microstate is
represented by an individual point $\psv{X}$ of the phase space, a
macrostate is represented by the set of all microstates
corresponding to the same value of some macroscopic
constraints. Clearly, within Boltzmann's
perspective, the information on \textit{how many}
microstates concour to the same macrostate is pivotal to our
predictions of macroscopic behaviours in terms of few control
parameters (the macroscopic constraints characterizing the
macrostate).

In 1949 Einstein summarized very well Bolzmann's approach to statistical mechanics~\cite{einstein1949}:

\begin{quote}
On the basis of the kinetic theory of gases, Boltzmann discovered that, aside from a constant factor, entropy is equivalent to the logarithm of the ``probability'' of the [macro] state under consideration. Through this insight, he recognized the nature of the course of events which, in the sense of thermodynamics, are ``irreversible''. However, seen from the molecular-mechanical point of view, all events are reversible. If one calls a molecular-theoretically defined state a microscopically described one, or, more briefly, a microstate, then an immensely large number (\(Z\)) of states belong to a macroscopic condition. \(Z\) is then a measure of the probability of a chosen macrostate. This idea also appears to be of outstanding importance because its usefulness is not limited to microscopic description on the basis of mechanics. Planck recognized this and applied the Boltzmann principle to a system that consists of very many resonators of the same frequency.
\end{quote}

Boltzmann's profound insight, as clarified by Einstein, resides in connecting the entropy  \(S\) of a  system in a microstate $\psv{X}$ with the logarithm of the size $W$  (denoted as $Z$ by Einstein) of the complete set of microstates resulting in the same macrostate. Since $W$ represents the size of the macrostate specified by the microstate $\psv{X}$, Boltzmann's entropy is a property of the individual system.

It is worth to clarify already at this stage the reason why
entropy comes into play within the attempt to relate the world of
microscopic interactions with macroscopic phenomena. The need for
this quantity does not come, as taught sometimes even in university courses, from the necessity to
quantify our ``ignorance'' about the microscopic world, due to the fact
that microscopic details are ``out of focus'' from our perspective.
From Eq.~\eqref{eq:defWdirac} it is clear that there is a huge amount of microscopic configurations corresponding to the same macroscopic properties (which in the case of Eq.~\eqref{eq:defWdirac} are energy, $E$, volume, $V$, and particles number, $N$): entropy just takes into account this degeneracy. The resulting coarse-grained, thermodynamic description would be still valid even if we had access to all the details of the microscopic configuration. The reason why it is adopted has to be searched in its usefulness and practicality, rather than in our lack of knowledge.

Those who, at this point, are still skeptical about the
fundamental role of entropy in theoretical physics,
may raise the objection that it is not a truly ``universal'' quantity.
By this it is meant that it depends, as also the definition in Eq.~\eqref{eq:defWdirac} suggests, on the specific choice of macroscopic observables (in our case, energy and volume). While this
remark may have some philosophical interest, at a practical level it is quite irrelevant: in realistic situations the few 
observables that matter at the macroscopic scale are chosen according to their physical interest 
(e.g. the quantities that can be used as control
parameters for interesting behaviours, like phase transitions). 

A simple, yet important, remark is in order here.
Often, even in good textbooks (see, e.g., Ref.~\cite{nagel1979structure}),
one can read
that the bridge law between thermodynamics and mechanics is given by
the equation
\begin{align}
  \av{ \frac{|\mathbf{p}_i|^2}{2 m}} = {3 \over 2} k_B T
  \label{eq:bridge}
\end{align}
relating the average kinetic energy of a particle with the absolute
temperature $T$.  The relation in Eq.~\eqref{eq:bridge} holds for the
most common Hamiltonian systems of classical particles in three spatial
dimensions, more precisely for all the systems with quadratic kinetic energies
where the equivalence between the ensemble at fixed energy
(microcanonical) and the one at fixed temperature (canonical) works. 
On the other hand it cannot be valid for a system with a generic
Hamiltonian: an example is provided by point-vortex systems in
two-dimensional inviscid fluids~\cite{Onsager1949}.

\subsubsection{Boltzmann's  approach to the Dilute Gas}
\label{eq:Bodiluttoq}

A direct application of the bridge law is represented by Boltzmann's derivation of the  Maxwell distribution of velocities of a dilute gas.
Consider a closed dilute monoatomic gas confined within a volume $V$, comprising \(N\) particles where the total energy \(E\) solely stems from the kinetic energy of these particles. Following the insights of Boltzmann \cite{Bol1898}, a microstate $\psv{X}$, a point on the constant energy  hypersurface $\Gamma_E$, can be conceived as a collection of \(N\) points within the six-dimensional 1-particle phase space, referred to by Boltzmann as \({\mu}\)-space. As the number of particles \(N\) grows significantly large, a more appropriate and coarse-grained description of the system can be given in terms of its macrostate, specified in the new space $\mu$.

This description involves partitioning the \(\mu\)-space into \(K\) cells \(C_j\) (where \(K\ll N\)), each centered on \((\mathbf{q}_{{j}}, \mathbf{p}_{{j}})\) and having volumes \(G_j\). Let \(N_j\) denote the number of particles residing in \(C_j\). Thus, a macroscopic description of the gas state is obtained by specifying these occupation numbers, denoted as \(M=  (N_1, \ldots, N_K)\), up to a given macroscopic resolution.\footnote{Meaning that two microstates may correspond to the same macrostate if their occupation numbers differ by less than a given $\delta N$, small on the macroscopic scale, but large on the microscopic one (e.g. of the order $O(\sqrt{N})$). }. For every microstate, there exists a corresponding macrostate \(M\), and conversely, given \(M\), there exists a substantial subset \(\Gamma_M\) of phase points associated with \(M\). The ensemble of such subsets furnishes the partition of phase space:
\begin{equation}\label{bigcup classical}
\Gamma_E = \bigcup_{M} \Gamma_{M}.
\end{equation}

Within these subsets, one occupies almost the entirety of phase space \(\Gamma_E\). To demonstrate this, Boltzmann computed the volume \(\Omega_M\) of \(\Gamma_M\) using the multinomial formula, yielding\footnote{We have divided the standard formula for the volume by $N!$ which is the right way to count unlabelled particles, also in classical physics; see, e.g., Lanford's \cite{Lan73}.}
\begin{equation}
\label{eq:combbolts2}
W_M =  \prod_{j=1}^K \frac{G_j^{N_j}}{N_j!}.
\end{equation}

For sufficiently large \(N\) and a suitable choice of the cells, one may employ Stirling's formula to approximate \(W_M\) in the limit of large \(N\) and \(N_j\) for a large but fixed number of cells $K$ in the space $\mu$, yielding
\begin{equation}
\label{eq:approxHfun}
    \log W_M \simeq  -\sum_{j=1}^K  f_j\log  f_j \equiv\frac{1}{k_B} S_\text{B} (M),
\end{equation}
where $f_j= N_j/G_j$ is the mean occupation number per cell, neglecting a constant proportional to \(N\). 
We highlight that the approximate expression for $\log W_M $ is given by the Boltzmann entropy $S_\text{B} (M)$, a function of $M$ and thus of the actual microstate belonging to $\Gamma_M$, equivalently it is the negative of the famous Boltzmann's H-function; here $k_B$ is the Boltzmann constant.

Then, one can use a coarse-grained density \(f \) in \({\mu}\)-space such that $N_j = \int_{C_j} f (\mathbf{q,p}) \,d^3\mathbf{q}  d^3\mathbf{p}$. Consequently,
\begin{equation}\label{sec:abmeq1}
  S_\text{B} (M)= k_B   \log W_M  \simeq - k_B\int f(\mathbf{q},\mathbf{p}) \log f (\mathbf{q},\mathbf{p}) \,d^3\mathbf{q}~d^3\mathbf{p} .
\end{equation}
\new{Since we are dealing with dilute systems, the total entropy is close to $NS_\text{B}$.} To find the macrostate of the largest volume, Boltzmann maximized  $ S_\text{B} (M)$ under the constraints:
\begin{align}
    &\int f(\mathbf{q},\mathbf{p}) \, d^3\mathbf{q}~d^3\mathbf{p} = N \label{eq:constraint1}\\
    &\int \frac{\mathbf{|p| }^2}{2m}\, f(\mathbf{q},\mathbf{p})  \, d^3\mathbf{q}~d^3\mathbf{p} = E \label{eq:constraint2}
\end{align}
and found that the equilibrium distribution achieves the  maximum:
\begin{equation}
\label{eq:lln33}
f_\mathrm{MB}(\mathbf{q}, \mathbf{p}) = \frac{N}{V}\left(\frac{1}{2\pi m k_B T}\right)^{3/2} \exp\left(-\frac{|\mathbf{p}|^2}{2m k_B T}\right),
\end{equation}
where \(kT= (2/3)(E/N)\), and MB stands for Maxwell-Boltzmann.

\new{The above probability distribution was obtained by maximizing the
quantity~\eqref{sec:abmeq1} under the constraints~\eqref{eq:constraint1} and~\eqref{eq:constraint2}. The procedure used there may resemble a maximum entropy principle (see Section~\ref{sec:maxent}), but it should not be interpreted as based on a principle of maximum ignorance. Instead, in the limit of large $N$, Eq.~\eqref{eq:lln33} follows directly from the microcanonical distribution. The computation of $f_{MB}$ via the maximization of $S_\text{B}(M)$ subject to the constraints~\eqref{eq:constraint1} and~\eqref{eq:constraint2} is a convenient mathematical shortcut, justified by the physical principle that thermal equilibrium corresponds to the macrostate occupying the largest volume in phase space—i.e., the one with maximal Boltzmann entropy. This reflects the fact that, for a system at equilibrium, the overwhelming majority of microstates lie within this equilibrium macrostate.
}

In modern terms, Boltzmann's derivation of the Maxwellian distribution \eqref{eq:lln33} constitutes the proof of a non-trivial weak Law of Large Numbers for the (weakly dependent) random variables \(N_j/ N\), which represent the fraction of particles in the \(j\)-th cell. These random variables are functions of the microstate \(X \in\Gamma_E\) (the ``sample space''), and their statistics are governed by the probability measure \(\mathsf{P}\), given by the normalized volume on \(\Gamma_E\).

Thus, Boltzmann's finding can be expressed as follows:
\begin{equation}
\label{eq:equidomini}
 \frac{W_{\text{eq}}}{W_E} = 1 - \epsilon,
\end{equation}
where \(W_\text{eq} \equiv W_{M_\text{eq}}\) represents the phase space volume of the equilibrium macrostate described by the Maxwell-Boltzmann distribution \eqref{eq:lln33},\footnote{More precisely, $\Omega_\text{eq}$ is the set of phase points that exhibit negligible deviations from the Maxwell-Boltzmann distribution, as determined by a specified upper tolerance $\delta$. Therefore, Eq. \eqref{eq:equidomini} embodies the weak Law of Large Numbers, signifying that for any given $\delta > 0$, the likelihood of deviations from the Maxwell-Boltzmann distribution diminishes (exponentially) as the system size $N$ approaches infinity.} \(\Omega_E\) is the volume of the entire energy shell, and \(\epsilon\) tends to zero as \(N\) approaches infinity. Indeed, in modern language, implicit in Boltzmann's derivation is the demonstration of a large deviation principle \cite{ellis}, indicating that \(\epsilon\) is exponentially small in \(N\).

Eq. \eqref{eq:equidomini}  expresses the prevalence of Maxwellian microstates within \(\Gamma_E\). Simply put, the overwhelming majority of microstates are ``Maxwellian'', as eloquently expressed by Boltzmann himself~\cite{Bolt96}:
\begin{quote}
[The Maxwell distribution] is characterized by the fact that by far the largest number of possible velocity distributions have the characteristic properties of the Maxwell distribution, and compared to these there are only a relatively small number of possible distributions that deviate significantly from Maxwell’s. Whereas Zermelo says that the number of states that finally lead to the Maxwellian state is small compared to all possible states, I assert on the contrary that by far the largest number of possible states are “Maxwellian” and that the number that deviate from the Maxwellian state is vanishingly small.
\end{quote}

\subsection{The ergodic hypothesis}
\label{sec:ergohyp}

In Section~\ref{sec:bridgelaw} we anticipated that one of the main results of Boltzmann was to introduce probabilistic ideas to interpret physical observables. To understand his point of view, let us consider  the constant-energy hypersurface in the
phase-space of a mechanical system. This hypersurface can be
partitioned in a very large number of small (but finite) cells of the same size, which
can be, in principle, labeled and counted.
The original ``ergodic hypothesis'' by Boltzmann  states that during its time evolution, a
trajectory of the mechanical system eventually passes through all these cells, spending the same amount of time in each of them. This opens the possibility
of replacing a time average by a  phase average, which can be computed by using the tools of probability theory. We will come back to this point in the next paragraph~\cite{da18}.

A later reformulation of the hypothesis was proposed by Paul and Tatyana Ehrenfest~\cite{eh56} in the context of
  the classical mechanics of point particles in the continuum. Their version states that
   the system trajectory on the energy surface visits all its points:
   this is equivalent to say that at infinite time the dynamics has
   taken the system across all microstates compatible with the
   macroscopic constraint on energy~\cite{da18}.
It is well documented that Boltzmann had a resolutely finitist point
of view: in his opinion any concept without discrete
representation was purely metaphysical. Almost certainly this
version of the ergodic hypothesis cannot be attributed to Boltzmann~\cite{ce98}.

The impossibility for a single trajectory to visit every point of the
energy surface was pointed out by M. Plancherel and, independently,
by A. Rosenthal. The Ehrenfests then proposed the
so-called ``quasi-ergodic'' hypothesis, stating that the dynamical
evolution of each initial condition taken
on the energy surface (apart from a zero-measure set of initial
points) covers densely the surface itself~\cite{emch2013logic}.

The modern ergodic theory can be viewed as a branch of the abstract
theory of measure and integration. The aim of this field goes far
beyond its original problem as formulated by Boltzmann in the
SM context~\cite{ga99}.
Consider a dynamical system, i.e., a deterministic
evolution law $U^t$ for the initial datum $\psv{X}(0)$ in the phase
space $\Omega$:

\begin{equation}
\psv{X}(0) \to \psv{X}(t)= U^t \psv{X}(0)
\end{equation}

and a measure $d\mu(\psv{X})$ invariant under the evolution given by
$U^t$, i.e., $ d\mu(\psv{X})=d\mu(U^{-t}\psv{X})$.  The dynamical
system $(\Omega, U^t, d\mu(\psv{X}) )$ is called ergodic, with
respect the measure $d\mu(\psv{X})$, if for every integrable function
$A(\psv{X})$ and for almost all initial conditions $\psv{X}(t_0)$,
with respect to $\mu$, one has:
\begin{equation}
\label{ergo}
\overline{A} \equiv \lim_{{\cal T} \to \infty} {1 \over {\cal T}} 
\int_{t_0}^{t_0+{\cal T}} 
A(\psv{X}(t)) {\rm d} t = \int A(\psv{X}) d \mu(\psv{X}) 
\equiv \langle A \rangle \, ,
\end{equation}
where $ \psv{X}(t)=U^{t-t_0}\psv{X}(t_0)$.
In the context of the SM the phase space $\Omega$ is the hypersurface $H(\psv{X})=E$,
the evolution law $U^t$ is given by the Hamilton equation and as invariant measure
we can use  the microcanical distribution.

\subsubsection{The relation with the problem of the foundations}
It is important to notice that the macroscopic time scale (the one
at which we observe the system) is much larger than the microscopic
dynamics time scale (at which the molecular changes take place).
This means that an experimental measurement is actually the result of
a single observation, during which the system goes through a very large
number of microscopic states.  Calling $A(\psv{X})$ the considered
observable, its measure can be regarded as an average
performed over a very long time (from the microscopic point of view):
\begin{equation}
\overline{A}^{\cal T} = {1 \over {\cal T}} \int_{t_0}^{t_0+{\cal T}} 
A(\psv{X}(t)) {\rm d} t \, .
\end{equation}
The calculation of the time average $\overline{A}^{\cal T}$ requires, in
principle, both the full knowledge of the microstate at a
given time and the determination of its trajectory.  These conditions are
patently impossible to fulfill. Beside the
difficulty to find the detailed time evolution
of the system (\ref{ham1-2}), if the dynamical average
$\overline{A}^{\cal T}$ strongly depends on the initial condition
it is not even possible to make statistical predictions. The ergodic
hypothesis allows to overcome this obstacle, relieving us from the
need to compute dynamical averages, which are replaced by averages
over a suitable probability distribution.

The most natural
  candidate for the invariant probability measure $d\mu(\psv{X})$ is
  the microcanonical measure on the constant energy surface $H(\psv{X})=E$:
\begin{align}
  d\mu _{mc}(\psv{X}|E) = \frac{1}{W(E)} \delta\left( E - H(\psv{X})\right)~d \psv{X},
  \label{eq:micro-1}
\end{align}
where $W(E) = \int d \psv{X} \delta\left( E - H(\psv{X})\right)$.
Some authors~\cite{J67,U95} justify the privileged status of this
distribution as follows: since we cannot access the state of a complex system
with a large number of components, it is fair to assume that its
probability distribution is uniform in the allowed region. It can be
easily checked that such a ``maximum ingnorance principle'', which
regards the uniform distribution as the most
reasonable one, is not so robust. For instance, the choice of a
uniform distribution is related to a specific set of
coordinates used to represent the degrees of freedom of the
system: under the action of any non-linear change of
variables the measure of Eq.~\eqref{eq:micro-1} transforms into a
probability density which is not uniform with respect to the new
coordinates. The statement that the most general and
agnostic probability density is the ``uniform'' one leaves therefore a
question open: uniform with respect to which variables?

As a matter of fact, the microscopic dynamics
puts some stronger constraints on the choice of the invariant
probability distributions. Let us consider for instance a Hamiltonian
system, described by Eq.~\eqref{ham1-2}.
Any probability density $\rho$ which depends only on the Hamiltonian,
$\rho(\psv{X})=\rho[H(\psv{X})]$, is preserved by the
Hamiltonian flow, i.e., it necessarily satisfies a Liouville equation:
\begin{align}
  \frac{\partial \rho}{ \partial t} + \sum_{i=1}^{2N} \dot{X}_i \frac{\partial\rho}{\partial X_i}  = 0.
  \label{eq:liouv}
\end{align}
In the sense specified by the Liouville equation the microcanonical
measure is therefore the \textit{natural} measure for Hamiltonian
systems. The Hamiltonian nature of the dynamics is a sufficient (but not necessary)
condition to satisfy the conservation law expressed by
Eq.~\eqref{eq:liouv}.

The ergodic hypothesis is satisfied if for sufficiently large ${\cal
  T}$ the average $\overline{A}^{\cal T}$ depends only on the energy,
and hence it has the same value for (almost) all the trajectories on
the same constant energy surface. If this is the case, one has
\begin{equation}
  \label{ergo-4}
  \overline{A} \equiv \lim_{{\cal T} \to \infty}\overline{A}^{\cal T}
 = \int A(\psv{X})  {\rm d} \mu _{mc}(\psv{X})
  \equiv \langle A \rangle \, .
\end{equation}
The validity of this equality eliminates both the necessity of
determining a detailed initial state of the system and of solving the
Hamilton's equations. Whether Eq.~\eqref{ergo-4} is valid or not, it
constitutes the main question of the ergodic problem.
 The crucial
role of this issue for SM lies also in the
following consideration. If the statistical properties of a large isolated
system in equilibrium can be properly described in terms of the
microcanonical ensemble, then it is not difficult to show that the
equilibrium properties of a small subsystem (still large at
microscopic level) is correctly described by the canonical
ensemble (in the absence of long-range interactions)~\cite{peliti2011statistical}.
Therefore \textit{a proof of the validity of Eq.~\eqref{ergo-4}
would provide the dynamical justification of the SM description}.  Equation~\eqref{ergo-4}
clarifies why statistical ensembles are a very useful mathematical tool, even if
in the \textit{real} physical world
only single physical systems are typically considered.

The idea of statistical ensembles may nonetheless lead to misunderstandings. Following Gibbs, ensembles
are often described as fictitious collections of macroscopically
identical copies of the object of interest, whose microstates differ
from each other. While in certain cases this may be a convenient
perspective, one should not forget that the purpose of a statistical
ensemble is to describe the properties of a \textit{single} system.
We can say, following Boltzmann's definition of \textit{ergode} and
\textit{monode}~\cite{ga99}, that a statistical ensemble is nothing but the
probability density of a dynamical system governed by mechanicistic
laws. We will come back to this point later.

One is then left with the problem of determining under which conditions a dynamical
system is ergodic.  At an abstract level this question was tackled 
by Birkhoff. Consider a dynamical system given by a phase space, a dynamical evolution and a measure $(\Omega, U^t, d\mu(\psv{X})\,)$.
The two following theorems hold~\cite{emch2013logic}:  

{\bf I }{\it For almost every initial condition
  $\psv{X}_0$ the infinite time average
\begin{equation}
 \overline{A} (\psv{X}_0) \equiv \lim_{{\cal T} \to \infty} 
{1 \over {\cal T}} \int_0^{\cal T} A(U^t\psv{X}_0) {\rm d} t
\end{equation}
 exists.}  

{\bf II }{\it A necessary and sufficient condition for the system to
be ergodic, i.e. for the time average $\overline{A}(\psv{X}_0)$ to be independent of the initial condition (for almost all $\psv{X}_0$), is that 
the phase space $\Omega$ must be metrically indecomposable. The latter 
property means that $\Omega$ cannot be splitted into 
two invariant parts (under the dynamics $U^t$), both of positive measure.}  

Sometimes instead of ``metrically indecomposable'' it is used the
equivalent term ``metrically transitive''.  The part {\bf I} implies that the time average can depend, in principle, on the initial condition. This, from a physical point of view, is not very
stringent. The result in the part {\bf II} is more
interesting, although almost useless from the point of view of
SM. It is in fact generally not possible to decide
whether a given system satisfies the condition of metrical
indecomposability.

\subsubsection{Ergodicity: a phase-space geometry perspective}

The problem of determining whether a
Hamiltonian system is ergodic or not is directly related to the existence
of non-trivial integrals (i.e. conserved quantities).  Consider a Hamiltonian $ H( {\bf q}, \, {\bf p}) $,
with $ {\bf q}, \, {\bf p} \in \mathbb{R}^N $. If there
exists a canonical transformation (i.e. a change of variables
which does not change the Hamiltonian structure of the system) from
the variables $ ( {\bf q}, \, {\bf p}) $ into the action-angle
variables $ ( {\bf I}, \, \vect{ \phi}) $, such that the new Hamiltonian
depends only on the actions $ {\bf I} $:
\begin{equation}
\label{h0}
H=H_0 ({\bf I})\, ,
\end{equation} 
then the system is called integrable.
In this case the time evolution of the system is
\begin{equation}
\label{evoluzione}
\left\{\begin{array}{ll}
I_i (t)  = I_i (0) \\   
\phi _i (t)   = \phi _i (0) + \omega _i ({\bf I}(0) ) \, t  \, ,
\end{array} \right.
\end{equation}
where $\omega_i = \partial H_0 / \partial I_i$ and $i=1, \dots, N$. There are therefore $N$ independent first integrals, since all the actions $I_i$
are conserved and the motion evolves on $N$-dimensional torus. An
integrable system cannot therefore be ergodic, since for any initial
condition the trajectories are confined on a zero-measure subset of
the constant energy hypersurface.

It is then fairly natural to wonder about the effect of
perturbations on (\ref{h0}), i.e. to study the Hamiltonian
\begin{equation}
\label{hpert}
H ( {\bf I}, \, \vect{ \phi})  = H_0 ({\bf I}) + \varepsilon 
H_1 ( {\bf I}, \, \vect{ \phi})\, ,
\end{equation} 
also called a near-integrable Hamiltonian.
Do the perturbed system (\ref{hpert}) trajectories result ``close'' to
those of the integrable system (\ref{h0})?  Does the introduction of
the perturbation term $\varepsilon H_1 ( {\bf I}, \, \vect{ \phi})$ still
allow for the existence of integrals of the motion besides the energy?
These questions are of obvious interest to 
 the ergodic problem.  Of course, if there are first
integrals, beyond the energy, the system cannot be ergodic: choosing
one of the first integrals as the observable $A$, one has
$\overline{A}=A(\psv{X}(0))$, which depends on the initial condition
$\psv{X}(0)$ and therefore, in general, it cannot coincide with the
phase average $\langle A \rangle$.

The importance of $H_1({\bf I},\boldsymbol{\phi})$ for the validity of
the ergodic hypothesis and the consequential possibility to apply
statistical mechanics, at least according to the standard viewpoint, is
transparent: while the argument sketched above made clear that with
$\varepsilon=0$ the dynamics is not ergodic, one expects (or at least
hopes) that, even for $\varepsilon \ll 1$, the non-integrable part of the
Hamiltonian is sufficient to produce and ergodic behaviour, so that
$\langle A \rangle_{\varepsilon}\simeq \langle A \rangle_0$.

\subsubsection{Poincar\'e theorem: a hint towards ergodicity?}

In his seminal work on the three body problem Poincar\'e showed that, under quite general assumptions, a system like the one in Eq.~\eqref{hpert}, with $\epsilon
\neq 0$, does not allow analytic first integrals, besides
energy~\cite{po92}.  The existence of first integrals is equivalent to
the possibility of finding a change of variables $({\bf I}, \, \vect{
  \phi}) \to ( {\bf I'}, \, \vect{ \phi'})$ preserving the Hamiltonian
nature of the system (canonical transformation) such that the
Hamiltonian is a function only of the (new) actions ${\bf I'}$.
Practically one has to find a generating function $S({\bf I'}, \, \vect{
  \phi})$, which links $({\bf I}, \, \vect{ \phi})$ to $( {\bf I'}, \,
\vect{ \phi'})$:
\begin{equation}
\label{PO1}
I_n={ {\partial S( {\bf I'}, \, \vect{ \phi})} \over
{\partial \phi_n} } \,\,\,\,\,\,\, 
\phi'_n={ {\partial S( {\bf I'}, \, \vect{ \phi})} \over
{\partial I'_n} } \,\,\, , 
\end{equation} 
in such a way that the perturbed Hamiltonian
\begin{equation}
\label{PO2}
H ( {\bf I}, \, \vect{ \phi})  = 
H_0 \cbr{{ {\partial S} \over {\partial \phi}}} 
+ \epsilon  H_1 \cbr{{ {\partial S} \over {\partial \phi}}
, \, \vect{ \phi}}\, 
\end{equation} 
is only function of $I'$. Notice that Eq.~\eqref{PO2} depends upon $I'$ because $I$ is a function of $I'$ and $\phi$, see~\eqref{PO1}.  One approach is to look for a
solution in the form of power series in $\epsilon$:
\begin{equation}
\label{PO3}
S=S_0+ \epsilon S_1 + \epsilon^2 S_2 + ...
\end{equation} 
The zero order $S_0= {\bf I'} \cdot \vect{ \phi}$  corresponds
to the identity transformation, obtained for $\epsilon=0$. 
 Substituting the
series (\ref{PO3}) for $S$ in (\ref{PO2}) one gets an equation for
$S_1$:
\begin{equation}
\label{PO4}
{  {\partial H_0({\bf I'})} \over
 {\partial {\bf I'} } } \cdot
{ {\partial S_1( {\bf I'}, \, \vect{ \phi})} \over
{\partial \vect{ \phi} } }=-H_1( {\bf I'}, \, \vect{ \phi}) \,\,\, .
\end{equation} 
If one expresses $H_1$ and $S_1$ as Fourier series in the angle vector
$\vect{ \phi}$:
\begin{equation}
\label{PO5}
H_1=\sum_{\bf m} h^{(1)}_{\bf m}({\bf I'}) e^{ i {\bf m} \cdot \vect{ \phi}}
\,\,\,\,\,\, , \,\,\,\,\,\,
S_1=\sum_{\bf m} s^{(1)}_{\bf m}({\bf I'}) e^{ i {\bf m} \cdot \vect{ \phi}}
\,\,\,\,\,\, , 
\end{equation} 
where ${\bf m}$  is an $N$-component vector of integers, one obtains
\begin{equation}
\label{PO6}
S_1= i \sum_{\bf m}
{  {h^{(1)}_{\bf m}({\bf I'})}
 \over 
{{\bf m} \cdot \vect{ \omega}_0({\bf I'})} }
 e^{ i {\bf m} \cdot \vect{ \phi}}
\,\,\,\,\,\, , 
\end{equation} 
where $\vect{ \omega}_0({\bf I'})= \partial H_0({\bf I'}) /\partial{\bf
  I'}$ is the unperturbed $N$-dimensional frequency vector for the
torus corresponding to action ${\bf I'}$.     From the above result it
is clear the origin of the nonexistence of first integrals: this is
the celebrated {\it problem of small denominators}.  Of course
(\ref{PO6}) is not acceptable for the values of ${\bf I'}$ such that
${\bf m} \cdot \vect{ \omega}_0({\bf I'})=0$ for some value of ${\bf
  m}$.  Also in the case where the unperturbed frequencies $\vect{
  \omega}_0({\bf I'})$ are rationally independent the denominator ${\bf
  m} \cdot \vect{ \omega}_0({\bf I'})$ can be arbitrarily small,
therefore one has to conclude that first integrals (beyond energy)
cannot exist.

Poincar\'{e}'s result sounds rather positive for
SM: from the non existence of conservation laws one
can hope that a generic Hamiltonian system is ergodic.  On this line
there is an important result by Fermi, who generalized Poincar\'e's
work. Fermi showed that in a Hamiltonian system obtained as a generic perturbation of an integrable system, if $N > 2$, it is impossible to find a surface of dimension $2N-2$
embedded in the $2N-1$ dimensional constant energy surface that (i) is
analytical in the variable $({\bf I}, \vect{ \phi})$ and $\epsilon$ and (ii) contains all the trajectories starting on it~\cite{F23}.  This
(correct) result induced Fermi to argue that Hamiltonian systems
(apart from the integrable ones, which must be considered atypical) are, in
general, ergodic as soon as $\epsilon \neq 0$. As we will discuss in the following Section, the ergodic problem was still far from being solved.

\subsubsection{An interlude on monocicles and ergodicity}

Usually the convolute path followed by Boltzmann to reach a
consistent formulation of SM, starting from
mechanics, is summarised in the two main contributions that we have 
briefly analyzed above, namely the bridge law relating entropy to mechanics
and the ergodic hypothesis~\cite{da18,ce98}.
While these two concepts are usually regarded as independent, there
is an interesting way to introduce ergodicity using the second
law of thermodynamics and formula~\eqref{SeqlogW}, thus establishing a link
between the two. Boltzmann himself was aware of this connection,
which had an important role in the
development of the ideas of the founder of SM. For a clear discussion on
this topic, which is almost unknown even to scholars interested in the
history of physics, the reader is referred to the nice paper by Campisi
and Kobe~\cite{CK10}.

The argument is based on a result due to Helmholtz, concerning
one-dimensional Hamiltonian systems. 
Consider the separable Hamiltonian
\begin{align}
 H(q,p,V)={p^2 \over 2m} + \phi(q,V)
\label{eq:H1d-separable}
\end{align}
 where  $V$  is a control parameter. For instance, if $H$ describes the dynamics of a pendulum, $V$ may be the length of the string, and we assume to be able to vary it in time. If, for any $V$, the potential
 $\phi(q,V)$ (i) has a unique minimun and (ii) diverges as  $|q| \to \infty$, then, for any value of $E$,
the motion is surely periodic. The period 
  $\tau(E,V)$ depends on $E$ and $V$.
 It is easy to see that   the motion is ergodic, i.e. the time averages coincide with the averages 
 computed with the microcanonical
 distribution:
\begin{equation}
d\mu(p,q)=
{ \delta(H(q,p,V)-E) dq dp \over  \int \int \delta(H(q,p,V)-E) dq dp}  \,\, .
\label{5.9}
\end{equation}

Let us define the temperature $T$ and the pressure $P$ in terms of time averages $ \overline{( ....)}$
computed on the period $\tau(E,V)$:
\begin{equation}
T= {2 \over k_B} \overline{\frac{p^2}{2m}} \,\, , \,\,
P= -{1 \over k_B} \overline{\partial \frac{\phi(q,V)}{\partial V} } \,,
\label{5.10}
\end{equation}
 the entropy function is
\begin{equation}
S(E,V)=k_B \ln 2 \int_{q_{-}{(E,V)}}^{q_{+}(E,V)} \sqrt{2m[E-\phi(q,V)]} \, dq \,,
\label{5.11}
\end{equation}
 with  $q_{-}(E,V)$  and $q_{+}(E,V)$ 
 the minimum and maximal value of $q$ respectively.
Helmholtz's theorem states that it is possible to  prove the  following relations:
\begin{equation}
{\partial S \over \partial E} ={1 \over T} \,\, , \,\,
{\partial S \over \partial V} ={P \over T} \,\, .
\label{5.12}
\end{equation}
Let us note that $S(E,V)$ can be written in the familiar form
$$
S(E,V)=k _B\ln \int_{H(q,p,V)<E} dp\, dq \,\, ,
$$
which is nothing but Eq.~\eqref{eq:defWheavy}.
The above  results imply a rather interesting consequence, namely
the {\it existence of  a mechanical analogue for the entropy}:
the quantity
$$
{dE + PdV \over T}
$$
where $T$ and $P$ are expressed via time averages of mechanical observables, is an exact differential.

Boltzmann's  purpose was to generalise the above result, which is
exactly proven for $1d$ Hamiltonians, to systems with many particles, i.e. to find a function
$S(E,V)$  such that the relations  (\ref{5.10})  and (\ref{5.12}) are
 valid in spaces of arbitrary dimension.
For a system composed of $N$ particles
and described by the Hamiltonian~\eqref{eq:ham}, assuming ergodicity it is indeed possible to show
a {\it Generalised   Helmholtz's theorem} for the function
\begin{align}
S(E,V)=k_B \ln \int_{H({\bf q}, {\bf p},V)<E} d{\bf q} d{\bf p} \,\, .
\end{align}

The proof is quite similar to the $1d$ case,
see~\cite{CK10}. With respect to the $1d$ case, in systems with many particles one lacks the periodicity of the trajectory; this property has to be replaced by ergodicity, hence the need for the ergodic hypothesis.  From the
Generalised Helmholtz's theorem one has the second law of
thermodynamics, i.e. the existence of a function (the entropy $S$)
which can be expressed in mechanical terms, such that $dS/T$
is an exact differential.

The above approach has severe
limitations\new{, since} it works only for specific choices of the
Hamiltonian,  such as the one in Eq.~\eqref{eq:H1d-separable}. Indeed, one is forced to introduce the temperature in terms of the average of the kinetic energy: the above method is therefore not suitable for a generic Hamiltonian, as discussed already in Section~\ref{sec:defentr} about the case of negative temperature systems~\cite{baldovin2021,GILM21}. \new{Still, the class of physical Hamiltonians for which this strategy is allowed is quite broad, so that the method is of undeniable interest.}

\section{``Physical'' ergodicity: a matter of time-scales}
\label{sec:ergodicitytime}

After Fermi's 1923 work, even in the absence of a rigorous
proof, the ergodicity problem appeared solved to
the physics community.  There was a vast consensus that the non-existence
theorems for regular first integrals implied ergodicity.  In the
1930's the ergodic problem thus became a subject for
mathematicians, who tackled it in a rather abstract way,
without particular emphasis on its connections with statistical
mechanics~\cite{cornfeld2012ergodic}. The interest of the Physics community around this
subject was woken up again in the '60s, when the newly introduced tools of numerical simulations allowed to
investigate ergodicity from a completely different perspective.

\subsection{The FPUT ``experiment'': emerging of slow time-scales}
\label{sec:fput}

An important step towards the  understanding of the link between dynamics and
SM is represented by the so-called ``FPUT'' problem~\cite{FPU55}
(from the names of the
authors: Fermi, Pasta, Ulam and Tsingou\footnote{ The original list
of authors only included Fermi, Pasta and Ulam, with an acknowledgement to
Mary Tsingou for her work in programming the simulations on the MANIAC computer. 
For many years the programmer's contributions were largely ignored by
the community, so the problem is typically referred to as ``FPU'' in the
literature. More recent publications introduced the acronym ``FPUT''
to give Mary Tsingou proper
credit for her contribution.}). The work was completed in 1955
(Fermi died before the report was written), and appeared for the first
time in 1965 as a contribution to an anthology of Fermi's
papers~\cite{fe65}.  For a general discussion on the FPUT
problem see~\cite{ga07,BI05,BCP13,CGG05}.

This work played a unique role in the development of different fields
of research as dynamical chaos and numerical simulations. Fermi and
collaborators studied the time evolution of $N$ particles of mass $m$,
interacting with weakly non-linear springs.  The Hamiltonian of the
system is

 \begin{equation}
\label{fpu}
H= \sum_{i=0}^N \left[  { p_i^2 \over 2 m } + {K \over 2} 
    \bigl( q_{i+1} - q_i \bigr)^2 + {\epsilon \over r} 
 \bigl( q_{i+1} - q_i \bigr)^r
         \right]
\end{equation}
where $q_0=q_{N+1}=0=p_0$ and the cases $r=3$ and $r=4$ are considered.

If $\epsilon=0$ the system is integrable. Indeed, introducing  the normal
modes:
\begin{equation}
\label{normali}
a_k = \sqrt{ {2 \over N+1} } \, \sum_{i} q_i 
\sin \cbr{ i\, k\, \pi  \over N+1 } \qquad   (k=1,\dots,N) \, ,  
\end{equation}
the system reduces to $N$ non-interacting harmonic oscillators whose
angular frequencies are
\begin{equation}
\label{omega}
\omega_k = 2\, \sqrt{ {K \over m} } \,  
\sin \cbr{ k\, \pi  \over 2 (N+1) } 
\nonumber
\end{equation}
and the energies  of the modes are
\begin{equation}
\label{energia}
E_k = {1 \over 2} \, \left(  
\dot{a}_k ^2 + \omega _k ^2 a_k^2 
\right)\, .
\nonumber
\end{equation}
The $E_k$'s, in the case $\epsilon=0$, are constant during the time
evolution.  For small values of $\epsilon$ it is not difficult to
compute perturbatively all the thermodynamically relevant
quantities in the framework of equilibrium SM,
i.e. in terms of averages over a statistical ensemble
(\textit{e.g.} the canonical or microcanonical ones).  In particular,
it can be shown that
\begin{equation}
\label{equiparti}
\langle E_k \rangle  \simeq {E_{tot}  \over N}  \,,
\end{equation}
where by $E_{tot}$ we denote the total energy of the system.
We note that equipartition, assuming the validity of equilibrium SM,
can be valid either for $\epsilon =0$ or $\epsilon\ll 1$; however
$\overline{E_k}$, the time average computed along a trajectory, can
coincide with $ \langle E_k \rangle $ only if $\epsilon \neq 0 $, so
that the normal modes interact, and can lose the memory of their
initial conditions.

What happens if an initial condition is chosen in such a way
that all the energy is concentrated in a few normal modes (e.g.
$E_1 (0) \neq 0$, while $E_k (0) = 0 $ for $k=2,\dots, N$)?
Before the FPUT experiment, the general expectation was (based on the
results by Poincar\'e and Fermi) that the first normal mode would 
progressively transfer energy to the others and that, after some
relaxation time, every $E_k(t)$ would fluctuate around the common
value given by (\ref{equiparti}). According to the introduction of the
paper written by Ulam in~\cite{fe65}, it is very likely that Fermi had
no doubt about the thermalization of the system: he was
not seeking a numerical test of his own and Poincar\'e's results. Most probably
he was interested in a numerical simulation for the only purpose of
investigating the thermalization times, i.e., the times
necessary for the system to go from a non-equilibrium state (energy
concentrated in only one mode) to the equipartition state predicted by
SM. In the FPUT work a numerical simulation was
performed with $N=16, 32, 64, \, \epsilon \neq 0$ and all the energy
concentrated initially in the first normal mode.  Unexpectedly, no
tendency toward equipartition was observed, even for long times.  In
other words, a violation of ergodicity and mixing was found.

\begin{figure}[t!]
\centering
\includegraphics[clip=true,width=0.8\columnwidth, keepaspectratio]{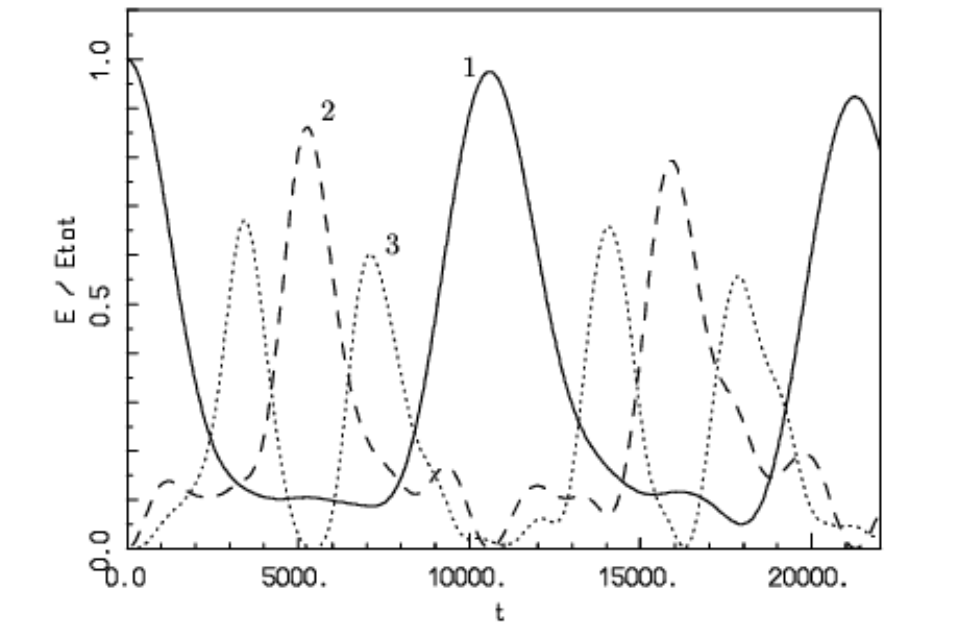}
\caption{$E_1(t)/E_{tot}$, $E_2(t)/E_{tot}$, $E_3(t)/E_{tot}$ 
for the FPUT system, with $N=32$, $r=3$, $\epsilon= 0.1$ and 
energy density ${\cal E} = E_{tot}/N = 0.07$. The figure is 
a courtesy of G.~Benettin.}
\label{fig1}
\end{figure}

\begin{figure}[t!]
\centering
\includegraphics[clip=true,width=0.8\columnwidth, keepaspectratio]{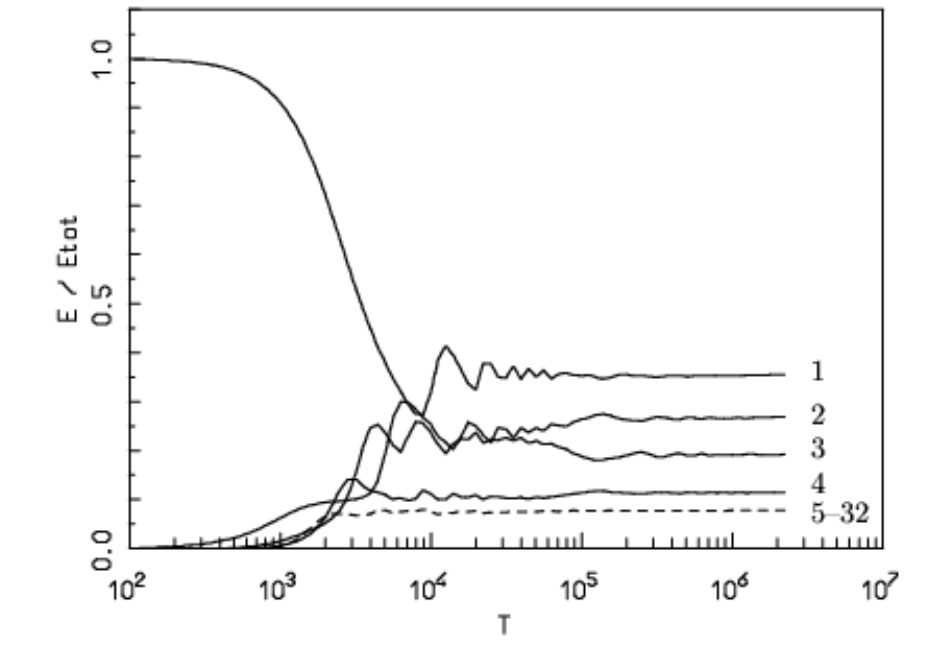}
\caption{Time averaged fraction of energy, in modes $k=1,2,3,4$ (bold
  lines, from top to below) and $ \sum _{k=5}^{32} E_{(av)k}
  (T)/E_{tot}$ (dashed line).  The parameters of the system are the
  same as in Fig.~\ref{fig2}.  Courtesy of G.~Benettin.}
\label{fig2}
\end{figure}

In Fig.~\ref{fig1} the time behaviour of the quantities
$E_k/E_{tot}$, for several $k$'s, $N=32$ and $r=3$ is shown.  Instead of a
loss of memory of the initial condition, an almost periodic
mode is observed: after a long time, $E_1$ reverts practically back to its initial
value. Fig.~\ref{fig2} shows that\new{, when considering time averages,} this periodic motion
corresponds to a breaking of equipartition between Fourier modes.

The numerical computations of FPUT strongly contrasted with
expectations, and, according to Ulam in his preface to the report
which appeared on {\it Note e Memorie}, Fermi considered the result an
important discovery which unambiguously showed how the prevalent
opinion (at that time) on the generality of mixing and thermalization
properties of non-linear systems, might not always be justified.

\subsection{KAM theorem: a scenario for slow time-scales}

Actually Fermi was not the first one to realize that the conjecture about
the validity of the ergodic hypothesis of a generic non integrable
system does not hold.  One year before the numerical computations of
FPUT, Kolmogorov proposed (without a detailed proof, but clearly
expressing the basic idea) an important theorem~\cite{K54}, which was
subsequently completed by Arnold~\cite{A63} and Moser~\cite{M62}. The
theorem, now known as KAM, reads as follows. 

Given a Hamiltonian
$$
H ( {\bf I}, \, \vect{ \phi}) = H_0 ({\bf
  I}) + \epsilon H_1 ( {\bf I}, \, \vect{ \phi})\,,
  $$
  with $H_0 ({\bf I})$
  sufficiently regular and
\begin{equation}
    \label{eq:condkam}
  \det \vert \partial^2 H_0 ({\bf I})/
  \partial I_i \partial I_j \vert \neq 0\,,
\end{equation}
  if $\epsilon$ is small
  enough, then on the constant-energy surface, invariant tori survive
  in a region whose measure tends to $1$ as $\epsilon \to 0$.  These
  tori, called KAM tori, result from a small deformation of those
  present in the integrable system ($\epsilon =0$).

The FPUT results can be seen (a posteriori) as a non-trivial check of
the KAM theorem and of its physical relevance,
i.e. of the fact that the KAM tori survive for physically
significant values of the non linear parameter $\epsilon$. Let us notice that in the FPUT system Eq.~\eqref{eq:condkam} does not hold, so \new{additional effort is needed in order to obtain results similar to those provided by KAM theory}~\cite{henrici2008results}.

Taking into account the ``small denominators'' problem discussed in the previous Section,
the existence
of the KAM tori is a rather subtle and strongly counterintuitive fact.
Indeed, for every value (even very small) of $\epsilon$, some tori of
the perturbed system, the so-called resonant ones, are destroyed, and
this forbids analytic first integrals.  In spite of that, for small
$\epsilon$ most tori survive, even if slightly deformed; thus the
perturbed system (at least for ``non-pathological'' initial
conditions) behaves similarly to the integrable one. 

In a nutshell the idea of KAM theorem is the following: Poincar\'e's
result shows that, because of the small denominators, it is not
possible to find a canonical transformation such that in the new
variables the system is integrable. On the other hand one can try to
obtain a weaker result: i.e. not in the whole phase space but in a
region of non zero measure.  This is possible if the Fourier
coefficients of $S_1$ in (\ref{PO6}) are small. Assuming that $H_1$ is
an analytic function, the $h^{(1)}_{\bf m}$ decrease exponentially
with $m=|m_1|+|m_2|+ \dots +|m_N|$.  On the other hand there exist
tori, with frequencies $\vect{ \omega}_0({\bf I})$, such that the
denominator is {\it not too small}, i.e.
\begin{equation}
\label{LIU}
\vert {\bf m} \cdot \vect{ \omega}_0({\bf I'})\vert > 
K(\vect{ \omega}_0) m^{-(N+1)} \,\,\,\,\,\, ,
\end{equation} 
for all integer vectors ${\bf m}$ (except the zero vector).  The set
of the $\vect{ \omega}_0$ for which Eq. (\ref{LIU}) holds has a non zero
measure in the $\vect{ \omega}_0$-space, and thus one can build the
$S_1$ in a suitable non zero measure region (around the nonresonant
tori). Then one has to repeat the procedure for the $S_2, S_3, ...$
and to control the convergence. For a non technical discussion on the
KAM theorem and its relevance in physics see~\cite{du14}.

\subsection{Thermal equilibrium: a choice of time-scale and observables}

As seen in the previous Section, understanding whether a system
will asymptotically reach thermal equilibrium or not is not the only important
problem concerning the thermalization dynamics.
For any practical purpose, predicting how long the thermalization process will take
is at least as relevant. In particular, in physical applications one
needs to know whether the thermalization time-scale is long or short
compared to the characteristic time of the measurement process, or
the observational time-scale.  Clearly, if the relaxation time (the
time needed to thermalize) is much larger than the observational time,
any reasoning about the system in terms of thermal equilibrium becomes
dangerous~\cite{ma85}.

In order to clarify the role of the observational time let us
take the example of a cold cup filled with boiling water.  After a few minutes
the water and the cup reach the same temperature and we can
consider the two to be in equilibrium.  After a few hours
the temperature of the water, and that of the cup, become equal to the room
temperature: another equilibrium state. But water molecules
also evaporate, so if the observation time is of some days, we will be able to observe
another equilibrium state.  What we refer to as ``equilibrium'' turns out to be in most cases a concept which depends on the choice of time scales. 

A striking example of the relevance of relaxational time-scales
is represented by glassy systems, which are ubiquitous in
condensed matter. At low temperature, these systems 
are typically found in disordered, out-of-equilibrium states
characterized by the presence of competing
interactions between the molecules (also referred to as \textit{frustrated} interactions).
This phenomenon may arise as a consequence of intrinsic randomness in the internal forces, as it happens in all the prototypical models of spin glasses~\cite{parisi79,mezard1987spin}, or by their competitive nature, as it happens in liquids and colloids~\cite{mezard1999thermodynamics,cavagna2009supercooled,parisi2020theory,berthier2011theoretical}. The out-of-equilibrium nature of glassy systems is evident when considering phenomena like the so-called \textit{aging}, which is represented by the dependence of macroscopic observables on the time elapsed since the beginning of the experiment. An equilibrium stationary state, characterized by properties such as time-translational invariance, cannot be reached on experimentally accessible times. In a sense, glassy systems ``never'' lose memory of the initial conditions.

\begin{figure}
  \centering
  \includegraphics[clip=true,width=0.6\columnwidth]{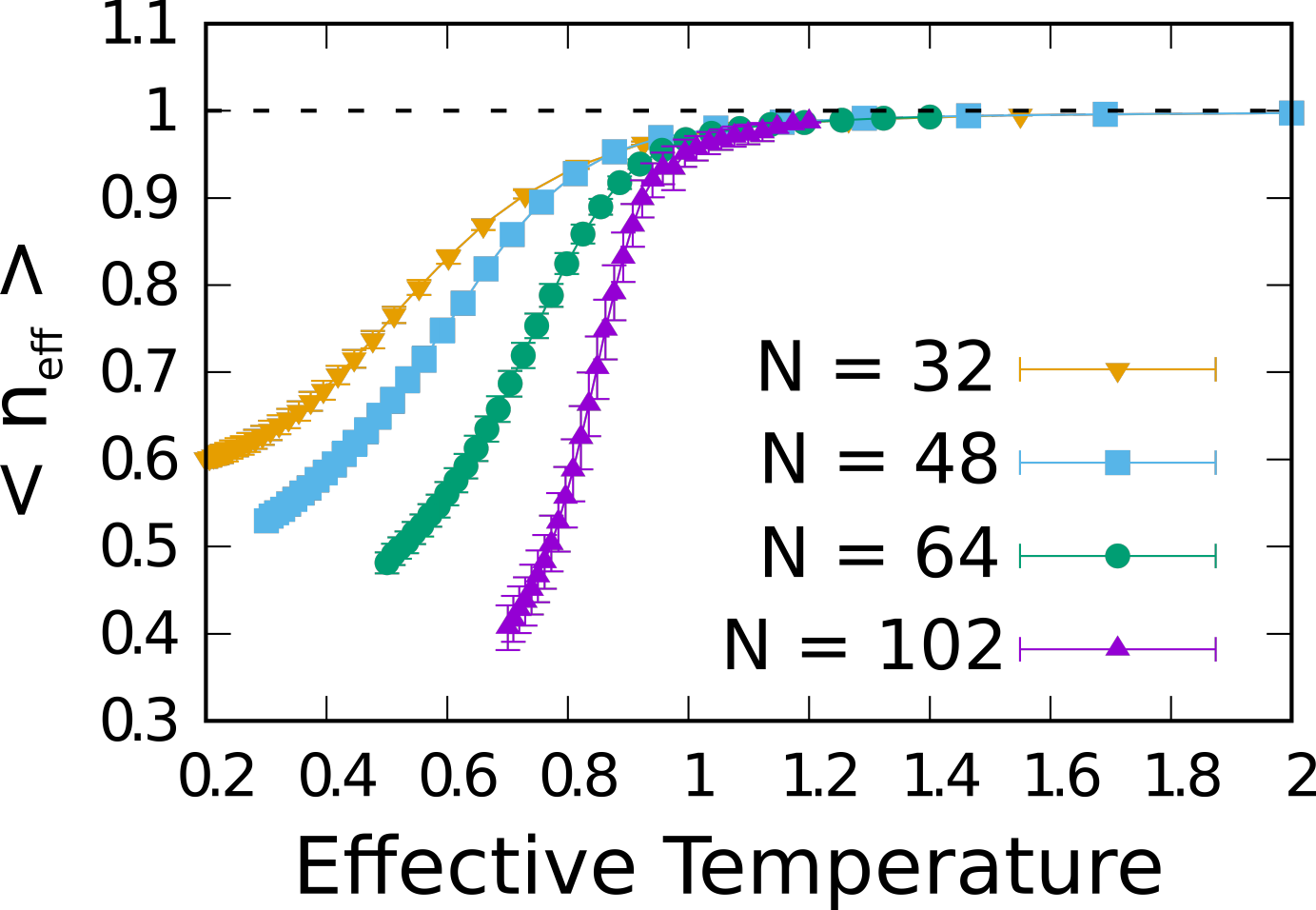}
  \caption{From Ref.~\cite{gradenigo2020glassiness},
    effective number of degrees of freedom $\av{n_{\text{eff}}}$ as an
    indicator of amplitude sharing between light modes; notice the
    decay moving from right to left from an equipartited phase
    $n_{\text{eff}}=1$ to an heterogeneous phase and the sharpening of
    the behaviour for increasing number of modes in the spectrum.}
  \label{fig:laser}      
\end{figure}
  Recent investigations have shown signature of the glass transition also for the distribution of amplitudes and phases of light modes in random lasers~\cite{ghofraniha2015experimental,antenucci2015glassy}. In particular, the numerical investigations of~\cite{gradenigo2020glassiness} presented an attempt to establish a connection between, on the one hand, the ergodicity breaking transition for light in random systems, analyized with the standard thermodynamic approach of spin glasses and, on the other hand, the dynamical
approach used to study the lack of thermalization in the FPUT and other non-linear models (such as the Discrete Nonlinear Schr\"odinger Equation).  The
numerical analyses of~\cite{gradenigo2020glassiness,niedda2023universality,niedda2023intensity} show that standard indicators for
thermodynamic ergodicity breaking, borrowed from the literature on
glasses, signal a glass transition at the same value of the parameters
where an equipartition breaking transition takes place, like  it happens for ergodicity breaking in non-linear models. In this case the breaking of equipartition is
observed with respect to the sharing of amplitude on different light
modes. In Fig.~\ref{fig:laser}, taken from Ref.~\cite{gradenigo2020glassiness}, it is
reported the behaviour of light spectrum, showing the
unequal sharing of light amplitude between field modes, with an
increasing heterogeneity as the system moves inside the glassy phase.

 Quite clearly, whether we can talk
about equilibrium or not does not depend on the ratio between the
relaxational time-scale and the observational one, but it also
crucially depends on which physical process we are interested in.  This
observation anticipates an issue that we will discuss thoroughly
across all next sections: the ``proper choice of observables'' will
turn out to be a crucial point in establishing the robustness of the
SM description of many-body systems. 

\subsection{A big trouble: choosing the wrong observable}
\label{subsec:wrong-obs}

The mathematical infinite time limit $\overline{A}$ needs
a physical interpretation: specifically, one should know how large the averaging time ${\cal T}$ has to be taken in order to have $\overline{A}^{\cal T} \simeq \langle A
\rangle$.  It is easy to realize that the answer to this question 
depends both on the observable $A$ and on the number of particles $N$.
As a simple example let us consider a region $G$ of the phase space
$\Omega$, and the observable

\begin{equation}
\label{A}
A(\psv{X})=    \begin{cases} \begin{array}{ll}
1 \quad  \textrm{if} \quad  \psv{X} \in G \\
0 \quad \textrm{otherwise}   \, .
\end{array} 
\end{cases}
\end{equation}

Let $G$ be   a $6\,N$
dimensional hypercube of side $\epsilon$.  Of course $\langle A
\rangle \propto \epsilon^{6N}$ is nothing but the probability $\mu(G)$  to stay
in $G$, and $\overline{A}^{\cal T}$ is the fraction of time the
trajectory $\psv{X}(t)$ spends in $G$ during the interval $[0,{\cal T}]$.  
An estimation of the equilibration time ${\cal T}_{eq}$, i.e.  the time necessary to have a fair agreement
between $\overline{A}^{\cal T}$ and $\langle A \rangle$, is  given by the
Poincar\'e  recurrence time ${\cal T}_{rec}$, which can be easily estimated using 
Kac's lemma~\cite{ka57}:
\begin{align}
{\cal T}_{rec} \simeq {1 \over \mu (G)}\,.
\end{align}
  
By  assuming ${\cal T}_{eq}\simeq{\cal T}_{rec} $, one  obtains then
\begin{align}
{\cal T}_{eq} \simeq \epsilon^{-6N} \,.
\end{align}
The above result was basically obtained by Boltzmann in his answer to
the criticisms of Zermelo about the validity of his $H$
theorem~\cite{ce98,da18} (see also Section~\ref{sec:irreversibility}).  Because of the very huge number of
particles in a macroscopic system the time of recurrence of a non
equilibrium state is astonishingly large. For instance, as estimated
by Boltzmann, in a sphere of air of radius $1$ cm at temperature $300
K$ and standard pressure, for a state in which the
molecules concentration will differ from the average value by $1\%$
one has to wait $10^{{10}^{14}}$ seconds, which is much larger than
the age of the universe.
 On the other hand, one can note that the
above considered observable has a rather poor relevance for the SM,
while the important observables for the thermodynamics, the ones by
which equilibrium states are characterized, are not generic functions. Therefore the choice of $A$ as a reference observable to determine the equilibration time-scales is in some sense pathological, and the actual equilibration times are much shorter for physically meaningful quantities. This aspect will be discussed in the next Section.

\section{Is chaos necessary for thermal equilibrium?}
\label{sec:chaos}

Before moving further, let us summarize the conclusions that
can be drawn from the previous Sections. First, we have formulated the ergodic
problem as a question on the equivalence between statistical and dynamical
averages for generic initial data and asymptotic times. Then, we
realized through several examples that any ``real world
experience'' such as a measurement or observation process has a
characteristic time scale: therefore, although it has some interest in
itself as a mathematical-physics problem, determining the asymptotic properties of
a certain microscopic dynamics may be not so relevant after all. What
we really need to know is on which time scale dynamical
averages are \textit{almost the same thing} as statistical averages.
The FPUT numerical experiments discussed in Section~\ref{sec:fput} taught us that in nearly-integrable
systems there are peculiar initial conditions for which this
time scale grows enormously. So, initial conditions can be important. The choice of the observables turns out to be crucial as well: from the counter-example discussed in Section~\ref{subsec:wrong-obs}, conceived by Boltzmann to argue against the recurrence paradox, we learned that a choice of untypical
observables can lead to the wrong estimation of equilibration time scales.  We will get back to this aspect in the
last part of this Section.

 Once realized that the possibility to exploit the SM
toolkit is a matter of relaxation times, one needs to
understand what are the factors that concur to define them. In this way it is possible, in principle, to determine on which time scales SM is expected to work for a certain system, given its microscopic dynamics. A widespread opinion is that
relaxation times are controlled by the ``amount of
chaoticity'' of the dynamics, which would imply that the validity of
SM strictly relies on the presence of chaos. \new{With this word we indicate the sensitive dependence on initial conditions}~\cite{LPRV87} \new{which can be formalized in terms of Lyapunov exponents, see the following Section}~\ref{sec:lyap}. In what follows we illustrate this point of
view, and then provide some counter-examples showing its limits. The Section ends with the discussion of
Khinchin's approach, which is able to provide some ergodicity statements
on the basis of hypothesis that totally ignore whether the
microscopic dynamics is chaotic or not. We will see in Section~\ref{sec:vonneumantyp} that the situation is quite similar in the context of Quantum Mechanics: independence from microscopic dynamic details is at the core of the ``quantum ergodic theorem'' formulated by
Von Neumann~\cite{cite66}.

\subsection{Chaos and KAM tori: the threshold problem}

After the discovery of the KAM theory and the FPUT numerical
experiment, it became clear that in weakly non-integrable systems
the phase space is characterized by the presence of regular behaviours, which are surely non ergodic, that 
might crucially have a role for their statistical features.

When dealing with high-dimensional systems it is challenging to grasp
the relevance of the KAM theorem on analytical grounds.  Numerical
studies by mean of computers are practically unavoidable.  The great
merit of numerical computations is to allow for specific tests, and
thus to serve as a guide towards future theories. This has been for
instance the case for the simulations on the FPUT. Izrailev and
Chirikov~\cite{IC66} were the first to notice that for high values of the nonlinear coefficient $\epsilon$, when KAM theorem does not hold, a good statistical
behaviour is observed. The energy, initially concentrated in the
lowest frequency normal modes, can be seen to spread equally on all
normal modes. As a consequence, the time averages are in agreement
with those of the equilibrium statistical mechanics.

\begin{figure}[t!]
\centering
\includegraphics[clip=true,width=0.8\columnwidth, keepaspectratio]{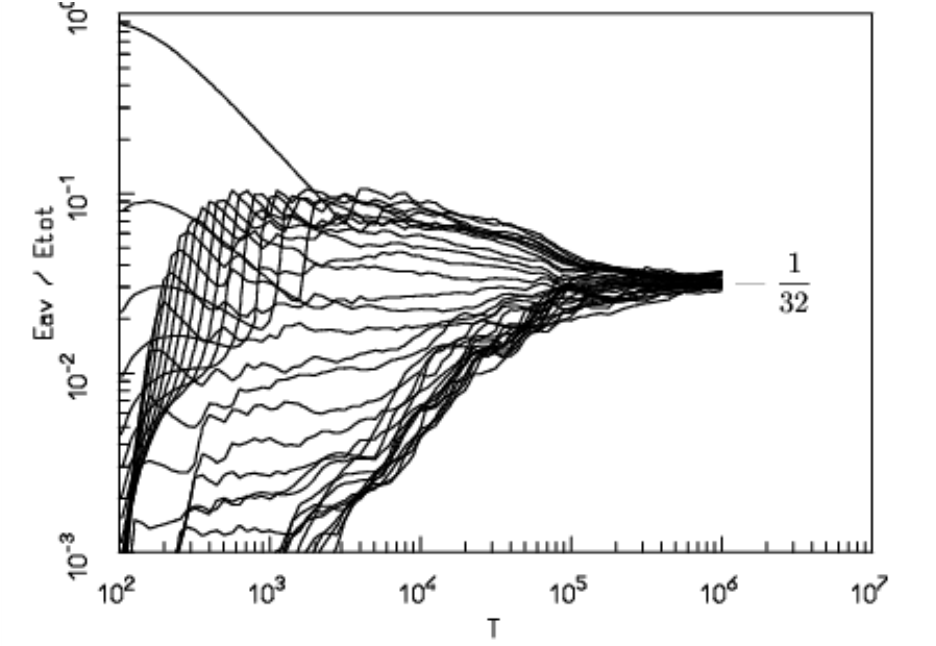}
\caption{Time averaged fraction of energy, in all the modes
  $k=1,\dots,32$. The parameters of the system are $N=32$, $r=3$,
  $\epsilon= 0.1$ and energy density $\varepsilon = E/N = 1.2$.
  Courtesy of G.~Benettin.}
\label{fig-4.3}
\end{figure}

Let us recall that, even keeping fixed the non-linearity coefficient
$\epsilon$, the relative importance of nonlinear terms is controlled
by the specific energy $\varepsilon=E/N$, where $E$ is the total
energy and $N$ the number of particles. For instance, for the FPUT
system with quartic non-linearity, the relative importance of
non-linear terms is proportional to $\varepsilon$. In what follows we
will always consider the coefficient $\epsilon$ fixed and the specific
energy $\varepsilon$ as the control parameter of the non-linearity, as
is customary in the literature on the subject\cite{CGG05}. Numerical
simulations provide a clear evidence that at finite $N$ it exists a
threshold value $\varepsilon_c$ for the specific energy such
that~\cite{LPRV87,R01,CGG05}:

\begin{itemize}
\item if $\varepsilon < \varepsilon_c $ the KAM tori play a major role
and the system does not follow equipartition, even after a very long time;
\item if $\varepsilon > \varepsilon_c $ the KAM tori have a minor effect: 
the system follows equipartition and 
standard SM holds, at least for large (but finite) times.
\end{itemize}

On the one hand, from numerics there is some evidence that a threshold value
$\varepsilon_c$ of the energy density separates regular from irregular
behaviour. On the other hand, since in Hamiltonian systems there is no
sharp transition between regular and chaotic behaviour, meaning that
regular and chaotic motions can coexist at every value of
$\varepsilon$, there are some physical questions which ``naturally''
arise:

\begin{enumerate}
\item \label{item:chaos1} Are there other systems that, just like the FPUT Hamiltonian, behave regularly for small non-linearities, and irregularly for large ones?
\item \label{item:chaos2} Is there any dependence of the threshold
  $\varepsilon_c$ on $N$? In other words, is what we are really after
  a size-dependent parameter $\varepsilon_c(N)$? In this case, which
  is the beahviour of $\varepsilon_c(N)$ in the thermodynamic limit?
\item \label{item:chaos3} What are the characteristic times of the equipartition
process?
\end{enumerate}

About point~\ref{item:chaos1}, even in the absence of a rigorous
application of the KAM theorem, we expect the mechanism of the
transition to chaos for increasing $\varepsilon$ to be generic for all
systems that are obtained as a perturbation of an harmonic system (as
FPUT).  As for the points~\ref{item:chaos2} and~\ref{item:chaos3}, the
answers are less clear. Whether $\varepsilon_c$ depends on $N$ is
clearly a crucial and still debated point: in fact, if one was able to
establish on the basis of numerical analysis that
$\lim_{N\rightarrow\infty}\varepsilon_c(N) = 0$, then the traditional
point of view relating the presence of non-linear terms to ergodicity
would be re-established. On the contrary, if $\varepsilon_c$ was not
dependent on $N$, there would be a serious discrepancy with the
results expected by equilibrium SM. Detailed numerical simulations and
analytic computations have been performed on points~\ref{item:chaos2}
and~\ref{item:chaos3}. In particular, the results of the very careful
numerical analysis presented in the work of Benettin and
Ponno~\cite{BP11} provided strong evidence that the existence of an
\textit{ergodicity threshold} $\varepsilon _c$ separating ``fast'' from
``slow'' relaxation is a finite-size effect.  More precisely, the
analysis of~\cite{BP11} focused on the time $\tau_R(\varepsilon)$
needed by the system to reach the equipartition starting from a
far-from-equilibrium initial condition (for instance all the energy is
concentrated in a few normal modes). It is shown that, even if
$\tau_R(\varepsilon)$ is increasing for decreasing values of
$\varepsilon$, this dependence is a power-law kind, $\tau_R \sim
\varepsilon^{-\gamma}$ (for some value of the positive constant
$\gamma$) in the $N\rightarrow\infty$ limit. This means far for
increasingly large values of $N$ there is no signature of an {\it
  ergodicity threshold} energy $\varepsilon_c$. \\
 
Let us finally notice that in certain systems, i.e., the system of
coupled rotators to be discussed in Sec.~\ref{sec:rotators},
which have been studied numerically at finite values of $N$ where the
presence of a stochasticity threshold $\varepsilon_c$ can be
established, even the choice of initial conditions at energies {\it
  above} the threshold, $\varepsilon > \varepsilon_c$, does not
guarantee the validity of SM~\cite{LPRV87}.\\


Let us conclude this brief look at the
FPUT problem noting that system~\eqref{fpu} is close to an integrable dynamics, but the proper $H_0$ is not the
harmonic chain: the choice that fulfills~\eqref{eq:condkam} has been
found to be the Toda Hamiltonian~\cite{BCP13}.

\subsection{Lyapunov exponents and entropy}
\label{sec:lyap}

In order to fully understand the role of chaos for the
foundations of SM, it is useful to recall how it is defined in quantitative terms. The observables that are universally acknowledged as the estimators of chaos are the Characteristic Lyapunov Exponents (CLEs).  They are a set of parameters
associated to each trajectory $\psv{X}(t)$ in phase space, providing a characterization of its instability: they quantify the mean
rate of divergence of trajectories that start infinitesimally close
to a reference one~\cite{vulpiani2009chaos, benettin1980lyapunov}. To introduce CLEs we consider the
discrete-time dynamics of a $d$-dimensional map:
\begin{align}
  \psv{X}(t+1) = U_1\left[\psv{X}(t)\right],
\end{align}
where $U_1$ is an operator that maps the phase space into itself. The
generalization to continuous-time dynamics is straightforward. The
stability of a single trajectory $\psv{X}(t)$ can be studied by
looking at the evolution of its nearby trajectories $\tilde{\mathbf{X}}(t)$, obtained from initial conditions $\tilde{\mathbf{X}}(0)$ that are
displaced from $\psv{X}(0)$ by an infinitesimal vector: $\tilde{\mathbf{X}}(0)= \psv{X}(0)+ \delta\psv{X}(0)$ with $\Delta(0) \equiv
|\delta\psv{X}(0)| \ll 1$. In non-chaotic systems, the distance
$\Delta(t)$ between the reference trajectory and perturbed ones either
remains bounded or increases algebraically. On the contrary,  in chaotic systems $\Delta(t)$ grows exponentially with time
(provided that the phase-space is unbounded):
\begin{align}
  \Delta(t) = e^{\gamma t}~\Delta(0).
\end{align}
A rigorous characterization of trajectory instability is obtained by introducing the parameter:
\begin{align}
\label{eq:mle}
  \lambda_{\text{max}} = \lim_{t\rightarrow\infty}\lim_{\Delta(0)\rightarrow 0} \frac{1}{t} \log\left( \frac{\Delta(t)}{\Delta(0)} \right),
\end{align}
which is the mean exponential rate of divergence and is called the
\textit{maximum Lyapunov exponent}. When the limit
$\lambda_{\text{max}}$ exists and is positive, the trajectory shows
sensitivity to initial conditions and thus the system is chaotic.  One should notice that the two limits appearing in its definition cannot be
exchanged: by doing so, we would get that $\lambda_{\text{max}}$ trivially vanishes in the presence of bounded dynamics, and that chaos is not possible in those cases (which is of course not true).

The maximum Lyapunov exponent alone does not fully characterize the
instability of a $d$-dimensional dynamical system. Actually, there are $d$ Lyapunov exponents (defining the \textit{Lyapunov spectrum}), which can
be computed by studying the time-growth of $d$ orthogonal
infinitesimal perturbations  with respect to a reference
trajectory: $\lbrace \delta{\psv{X}}_i(0)\rbrace_{i=1,\ldots,d}$, where
$$
  \delta\psv{X}_i(0) \perp \delta\psv{X}_j(0)~~~~\forall~~~~i\neq j\,.
$$

If we now consider a Hamiltonian system (which is the relevant case for
equilibrium SM), we have that, due to the
symplectic structure of the dynamics, also the Lyapunov 
spectrum has a specific structure:

\begin{align}
  \lambda_k+\lambda_{d-k+1}=0 \quad \forall k=1,...,d
\end{align}
hence $\sum_{n=1}^d \lambda_n =0$ .
The definition of the CLEs for a
deterministic dynamical systems should be compared with estimators, analogous in spirit, that are defined in quantum mechanics, where the role of points in phase space is played by wavefunctions in the infinite dimensional Hilbert space (see Section~\ref{sec:chaosquantum}).

When looking for possible links
between CLEs and thermodynamics, it is important to recall
the relation between the Lyapunov spectrum and the
Kolmogorov-Sinai entropy $h_{\text{KS}}$. The latter is an information theoretic quantity expressing
how efficient  the dynamics is in loosing track of initial conditions, when moving across the phase space. We will not enter into details about its precise definition here: the interested reader can find it in~\cite{boffetta2002predictability}.  What we want to stress is the result, due to Pesin~\cite{MR0410804}, stating that
\begin{align}
h_{\text{KS}} = \sum_{k=1}^{d} \lambda_k\,.
\label{eq:KS-entropy}
\end{align}
Whether Eq.~\eqref{eq:KS-entropy} has a deep relevance for SM is a crucial question. To evaluate it in macroscopic systems composed of $N$ particles, one has to consider $d = N$ and take the
large-$N$ limit of the Lyapunov spectrum. There is clear numerical
evidence~\cite{KK87,LPR86}, supported by few analytical
results~\cite{BS93}, that in this limit the Lyapunov
spectrum can be characterized in terms of a certain scaling function
$f(x)$ such that:
\begin{align}
\lambda_n = \lambda_{\text{max}}^*~f(\lambda_n/N),
\end{align}
where $\lambda_{\text{max}}^* = \lim_{N\rightarrow\infty}
\lambda_{\text{max}}$. This result implies that the Kolmogorov-Sinai
entropy is proportional to the number of degrees of freedom,
\begin{align}
h_{\text{KS}} \sim N,
\end{align}
so that it is an extensive quantity, as expected from any sensible
definition of entropy.

The above discussion shows that in the limit of large $N$ the Lyapunov exponents can have a good
asymptotic behaviour. Although this result is surely interesting,
in general there is no biunivocal relation between chaos and SM:
$$
  \text{chaos} \quad \nLeftrightarrow \quad \text{SM}.
$$
As we are going to discuss in the next Section, chaos is
neither a necessary nor sufficient condition to guarantee the validity
of SM~\cite{castiglione2008chaos,LPR86,FMV91,BVG21,BGV21,Cocciaglia2022}.

\subsection{Chaos without equilibrium (and equilibrium without chaos)}

Some of the results discussed above suggest that the presence of chaos (i.e.,
positive Lyapunov exponents of the dynamics) is a necessary ingredient for
the validity of SM. Unfortunately the scenario
seems to be much more intricate.

Chaos is a concept that involves, at an abstract mathematical level,
both the limit $t \to \infty$ and an arbitrary observational resolution on the initial
state.  In realistic situations one has to deal instead
with finite accuracy and finite time, and it is important to take into
account these limitations. For instance the limit of
arbitrary high accuracy can be relaxed by introducing suitable tools,
such as the Finite
Size Lyapunov Exponent and the $\epsilon$-entropy, which is a finite-size generalization of the Kolmogorov-Sinai entropy~\cite{boffetta2002predictability}.
In addition there are situations where the system is, strictly speaking, non chaotic
(i.e. the Lyapunov exponent is zero) but its features are rather
irregular.  Such a property (denoted by the term pseudochaos by
Chirikov and Ford)~\cite{F83,VUF93,M00} is basically due to the
presence of long transient effects and can have a key role in the
dynamics of quantum systems.

The points discussed above can be understood, for instance, by considering the following examples.

\subsubsection{Symplectic maps}

The present example is to show that, even in chaotic systems where most
KAM tori are destroyed, the validity of SM is not guaranteed, at least
over long (but finite) times~\cite{FMV91,HGO94}.
This behaviour is not restricted to systems
similar to FPUT, i.e. weak perturbations of the harmonic chain.

Let us consider the high-dimensional system
of symplectic coupled maps of the form
\begin{equation}
\label{map-4}
\left\{\begin{array}{ll}
 \phi_n(t+1)= \phi_n(t)+ I_n(t) \quad & \textrm{mod} \; 2\pi       \\
  I_n(t+1)= I_n(t)+ \epsilon \nabla F({\bf \phi}(t+1)) 
&  \textrm{mod} \; 2\pi
\end{array} \right.
\end{equation}
where $n=1, ..., N$ and $\nabla=(\partial/\partial \phi_1,....,
\partial/\partial \phi_N)$. The above symplectic dynamics is nothing but a
canonical transformation from the ``old'' variables $({\bf I},\vect{
  \phi})$ at time $t$ to the ``new'' variables
$({\bf I'},\vect{ \phi'})$ at time $t+1$, via the generating function:
\begin{equation}
S({\bf I},\vect{ \phi'})=\sum_n   \phi'_n I_n 
-{1 \over 2}\sum_n I_n^2  + 
 \epsilon  F(\vect{ \phi'}) \,\,\, .
\end{equation}
The map in Eq.~\eqref{map-4} can be seen as the Poincar\'e section of
an Hamiltonian system with $N+1$ degrees of freedom.  When $\epsilon$
vanishes the system is integrable, and the term $\epsilon F(\vect{
  \phi})$ plays the role of the non-integrable perturbation of the
Hamiltonian. In~\cite{FMV91,HGO94} clear numerical evidence is found that
the irregular behaviour becomes dominant as $N$ is large.
Specifically, it is observed that the volume of the phase space occupied
by the KAM tori decreases exponentially as $N$ increases. Despite this, which may sound positive for the foundations of SM, let us remark that the probability distribution of Lyapunov exponents in some systems shows that long time scales are important at all $N$'s, in such a way even in the large-$N$ limit there is a persistent memory of initial conditions.
Furthermore, in very large systems one finds that the so-called Arnold diffusion~\cite{arnold2009instability},
i.e. the wandering of trajectories in a phase space characterized
by the coexistence of chaotic regions and regular
ones~\cite{FMV91,HGO94}, can be very weak, and different trajectories,
although with a high value of the Lyapunov exponent, maintain some of
their own features for a very long time. For instance, the correlation functions decay with a characteristic time much larger than $1/\lambda_{\text{max}}$.

\subsubsection{Thermodynamic anomalies in chaotic coupled rotators}
\label{sec:rotators}

In~\cite{LPRV87} the relevance of chaos in nonlinear
Hamiltonian systems is studied in the case of a FPUT chain,
using as a benchmark the prediction of SM
obtained from the canonical ensemble.

The canonical ensemble
can be viewed as the distribution describing the statistical features of
a small part of a larger, conservative system, provided that the interactions are short ranged. This remark suggests
a natural way to simulate it, avoiding the need of any noise
source. The numerical simulation of a
canonical ensemble for the FPUT system can be performed subdividing a chain
of $N$ particles into $N_1$ subsystems of $N_2=N/N_1$ oscillators each,
with $N_1 \gg 1$ and $N_2 \gg 1$. In this way the time averages
of observables defined in the subsystems can be computed, and they can be compared
to the canonical ensemble predictions.
For instance, one can define the internal energy $U$ as the mean value
(over time) of the energy $E_j$ in the $j$-th subsystem:
$U=\overline{E_j}/N_2$. Similarly, for the specific heat $C_V$
one has:
\begin{equation}
C_V={ {\overline{E_j^2} - \overline{E_j}^2} 
\over {N_2 T^2}  }
\end{equation}
where the temperature is defined as $T=\overline{p^2}$.
 Detailed
numerical computations show that both the internal energy $U$ and the
specific heat $C_V$, as function of the temperature, are close
to the predictions of the canonical ensemble.  Remarkably, this agreement holds also in the region at small energy
(i.e., small temperature) where the system behaves regularly
(the KAM tori are dominant).
 These results are in agreement with the
perspective of the ergodic approach of Khinchin, to be
discussed in the next section, which suggests a rather marginal role of
dynamics.

In~\cite{LPRV87} it was also pointed out that in certain
regimes (namely high energies, at least for the system of coupled
rotators studied therein) the typical behaviour of the system does not follow
the prediction of SM. Consider the rotators ruled
by the Hamiltonian:
\begin{equation}
\label{rota}
H= \sum_{i=0}^N \left[  { p_i^2 \over 2 m } 
        + {\gamma}(1- \cos( q_{i+1} - q_i )) \right] \,.
\end{equation}
For any fixed $\gamma$ this system has two different integrable limits: (i) for
small values of the total energy it is a perturbation of a harmonic chain
of oscillators;  (ii) for large values of the energy, since the potential is
bounded, it is a perturbation of a system of independent rotators. Numerical simulations show that $U$ is always very close to the canonical
prediction, while the agreement for $C_V$ fails at large temperature, despite the positivity of $\lambda_{\text{max}}$, which implies the presence of dynamical chaos.
 The different
behaviour of $C_V$ in the two near-integrable regimes of low and high
temperature can be understood as follows.  For the FPUT system and for
the low temperature rotators the ``natural'' variables are the normal
modes, which are able to put in evidence, even in a statistical
analysis, the regularity of the dynamics. When
observing the energy of a subsystem, non-negligible fluctuations of the
``local'' energy can be seen even in the presence of almost decoupled
normal modes.  For the chain of
rotators at large energy, instead, the carriers
of the energy are the ``local'' variables $\{p_j,q_j\}$ themselves,
and therefore the fluctuations of the local energy are strongly
depressed, as well as the exchange of energy among the subsystems.

\subsubsection{Thermalization of integrable systems}

Let us now discuss an integrable system that, in spite of its
integrability, shows good statistical features in agreement with
the equilibrium SM.  Consider the Toda lattice,
\begin{align}
  H(q,p) = \sum_{i=1}^N \frac{p_i^2}{2} + \sum_{i=0}^N V(q_{i+1}-q_i),
  \label{eq:HToda}
\end{align}
where $V(x)$ is 
\begin{align}
  V(x) = \exp(-x)+x-1\,.
  \label{eq:Vx}
\end{align}

A complete set of independent integrals of motion is known~\cite{henon1974integrals}, whose explicit form is rather involved. Their physical meaning is not transparent. Remarkably, Spohn was recently able to find a rigorous statistical description of this model in terms of its Generalized Gibbs Ensemble, which takes into account the extensive number of conservation laws of the system~\cite{spohn2020generalized}.

In the spirit of the FPUT
one may wonder whether the system is able to reach thermalization  with respect to different canonical variables, for instance  the Fourier modes. A possible test consists in evolving the system starting from an initial condition in which only the first few Fourier modes are
excited.  With initial conditions at specific energies of order $O(1)$ or larger one
observes that, in finite times, the harmonic energy reaches
equipartition.  This can appear to be not particularly surprising:
basically, it is a sort of virialization of the global kinetic energy
starting from initial conditions which are not at equilibrium.

A less obvious result is provided by a test for the specific heat, which
involves the behavior of the energy fluctuations of a subsystem by considering the Fourier modes as the reference variables: in
the Toda model the numerical result for the specific heat are in
good agreement with the analytical computation obtained with the
canonical ensemble. This is quite analogous to the one obtained in the
FPUT system~\cite{ga07}, where chaos is present; therefore we have
evidence that even the dynamics of an integrable system such as the
Toda chain has good ``equilibrium'' properties, provided one looks at
``generic'' observables.

Numerical investigations of trivially integrable systems, e.g. harmonic chains, provide additional evidence of the poor role of chaos for the SM of large systems. Studying relaxation from a given initial condition, atypical with respect to the random distribution of energy among the modes, one observes relaxation to a thermal equilibrium state. Such a result is practically independent from the orthonormal base used to represent the chain, provided it is selected in a random way~\cite{Cocciaglia2022}. A similar result can be proven analytically for the single-particle momentum distribution of a harmonic chain~\cite{baldovin2023}.


\subsection{Khinchin's approach: ergodicity without chaos}

In his celebrated book {\it Mathematical
  Foundation of the Statistical Mechanics}~\cite{kh49}, Khinchin  presents some important
results on the ergodic problem, which do not need the metrical
transitivity of the Birkhoff theorem.  The general idea of his
approach is based on the following facts:

\begin{enumerate}
\item In  systems which are of interest to SM the
number of degrees of freedom is very large.
\item In SM the relevant macroscopic observables are not generic
 functions of the phase space (in the mathematical sense): they are always defined as averages over many degrees of freedom. The
validity of Eq.~(\ref{ergo-4}) only needs to be shown for this kind of quantities.
\end{enumerate}

\begin{figure}
\centering
\includegraphics[width=.5\linewidth]{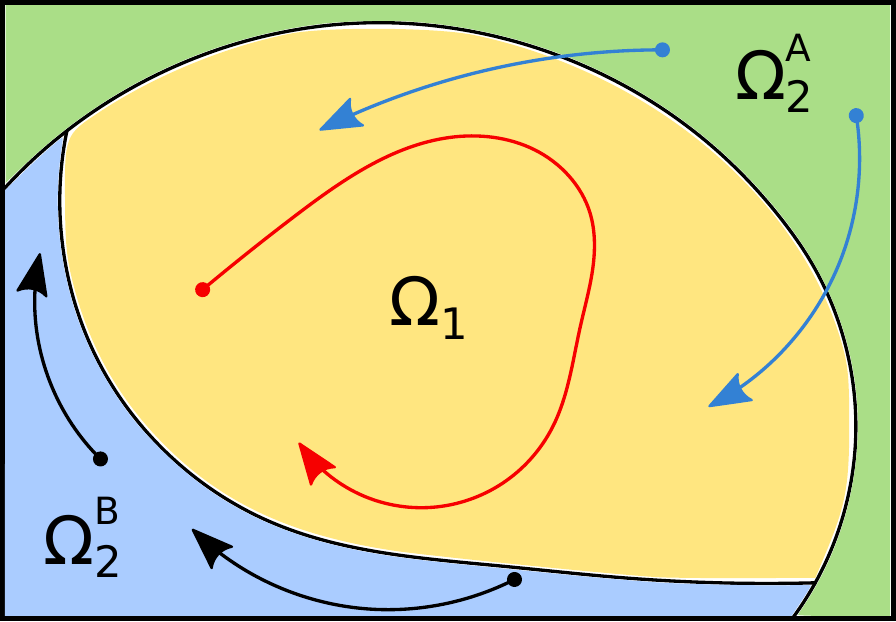}
\caption{   Schematic representation of the partition of  $\Omega $ for $N \gg 1$;
the volume of the subregions are  not drawn to scale, actually $\mu (\Omega_1)$ is close to $1$ and
$\mu (\Omega_2) \simeq 0$.}
 \label{fig4}      
\end{figure}

To summarise Khinchin's point of view it is useful to look at Fig.~\ref{fig4}, which schematically represents the phase space $\Omega$ of a system with $N\gg 1$ particles and a certain energy. Given a macroscopic observable $f(\mathbf{X})$, the total phase space can be divided in two regions: $\Omega_1$, whose points are ``typical'' states with respect to the considered observable, in a sense to be better specified later; and
$\Omega_2$, containing configurations  such that the value of $f(\mathbf{X})$ is ``far'' from its equilibrium average.
  Everyday experience and numerical simulations suggest that $\Omega_2$ includes a
subregion $\Omega^A_2$ whose points, after a certain thermalization time, evolve into states belonging to $\Omega_1$. This is for instance the fate of the out-of-equilibrium states considered in the FPUT numerical experiment, as discussed before.
The remaining
region $\Omega^B_2$ contains the initial states that never
enter into $\Omega_1$ (at least, not in an experimentally accessible time).

 For SM to hold true, the validity of Eq.~(\ref{ergo-4})
is allowed to fail for a set of initial conditions $\psv{X}(0)$ of small  measure (i.e., going to zero as $N \to \infty$). Therefore, if one is able to show that the volume of $\Omega_2^B$ vanishes in the $N\gg1$ limit, an ``effective ergodicity'' is automatically proven: the system spends most of its time in the region $\Omega_1$ where the value of $f(\mathbf{X})$ is close to its equilibrium value, but for a set of initial conditions with vanishing measure.

Keeping these points in mind, it is possible to go beyond Birkhoff's theorems,
which starts from stronger assumptions: generic dynamical systems (i.e., also low-dimensional ones), non-specific observables, no dependence on the initial conditions.
Kinchin considers a separable Hamiltonian
system, i.e.:
\begin{equation}
\label{k1}
H=\sum_{n=1}^N H_n({\bf q}_n,{\bf p}_n)
\end{equation}
and a special class of observables (called \textit{sum functions}) of
the form
\begin{equation}
\label{k2}
f(\psv{X})=\sum_{n=1}^Nf_n({\bf q}_n, {\bf p}_n)
\end{equation}
where $f_n=O(1)$. Typical examples of sum functions are
the pressure, the kinetic energy, the total energy and the
single-particle distribution function. 

Using the fact that the Hamiltonian is separable one has:
\begin{equation}
\label{k3}
\langle f \rangle =O(N) \quad \textrm{and} \quad \sigma^2=\langle
(f-\langle f\rangle)^2 \rangle=O(N)\, ,
\end{equation}
where  $\langle \, \, \rangle$ indicates the microcanonical
ensemble average.  Consider the time average $\overline{f}(\psv{X})$
of the observable $f$ along a trajectory starting from $\psv{X}$.
Under quite general hypotheses (without invoking the metrical
transitivity) one has:
\begin{equation}
\langle \overline{f}(\psv{X})\rangle =\langle f \rangle \quad
\textrm{and}  \,\,
\langle (\overline{f} - \langle f \rangle)^2\rangle \le \langle (f -
\langle f \rangle)^2 \rangle=O(N) \,\, .
\end{equation}
The above result is rather intuitive: the uncertainty for the time average of a sum function $\overline f (X)$ cannot be larger than the uncertainty of $f(X)$.

Now we can use the Markov inequality
\begin{equation}
\textrm{Prob} \left( { {|\overline{f} - \langle f \rangle|} \over
 |\langle f \rangle|} \ge a \right) \le { \sigma \over {a |\langle f
 \rangle|} } = {1 \over a } O(N^{-1/2})
\end{equation}
to obtain, taking $a=O(N^{-1/4})$,
\begin{equation}
\label{eq:khin}
\textrm{Prob} \left( { {|\overline{f} - \langle f \rangle|} \over
 |\langle f \rangle|} \ge K_1 N^{-1/4} \right) \le K_2 N^{-1/4}
\end{equation}
where $K_1$ and $K_2$ are $O(1)$.  Thisproves that for the
class of sum functions the set of points for which time and phase
averages differ more than a given, small amount (which goes to zero as $N \to
\infty$), has a measure which vanishes in the limit $N \to \infty$. 

The price to pay in order to avoid the metrical transitivity, besides the reasonable assumptions discussed before, is the separability of the Hamiltonian, Eq.~\eqref{k1}, i.e. the absence of interactions in the dynamics.  This undesirable
restriction to the separable structure of the Hamiltonian\new{, i.e. the fact that particles do not interact,} was removed by Mazur and Van der Linden~\cite{ML63}.  They extended the result~\eqref{eq:khin}\new{, valid for sum functions}~\eqref{k2}, to systems of particles
interacting through a short range potential, showing that the
intuition of Khinchin on the marginal relevance of the interaction among
the particles, was basically correct. Letting apart the technical
contents of the work, the physical interpretation of the result is
that, due to the short range of the interactions, a many-particle
system behaves as if it consists of a large number of noninteracting
components. As Mazur and van der Linden write:
\begin{quotation}
One might think of
  subsystems consisting of large numbers of particles; the interaction
  between these subsystems is then a surface effect and very small
  compared to the energy content of the subsystems themselves. 
\end{quotation}
 Their
calculations then imply that the energies of these subsystems
  behave as almost independent random variables, so that a central
  limit theorem still applies. It is interesting to note that the
obtained result is valid for all but a finite number of values of the
temperature, where the system may undergo a phase transition, at
variance with the noninteracting case considered by Khinchin.  

Let us stress again that in Khinchin's result, as well as in the
generalization of Mazur and van der Linden, basically the dynamics has
no role and the existence of good statistical properties is due to the
fact that $N \gg 1$.  In other words, the physically relevant
observables are self-averaging, i.e. they are practically
constant (except in a region of small measure) on a constant-energy
surface. This means that the time average operation is not so
important to assign a physical meaning to the quantity $\langle f
\rangle$, that results to be equal to the near constant value of the
observable. As a consequence -- and this can be viewed as the
essence of the Khinchin's results --  the ensemble-based
SM works independently of the validity of
ergodicity (in the mathematical sense). This was also (and
already) the point of view of Boltzmann himself~\cite{ga99}.

\subsection{Importance-sampling dynamics: irrelevance of ergodicity?}

Let us spend few words about the reasons of the success of numerical simulations in equilibrium SM.  The two main
methods used in  numerical computations are  molecular dynamics
(MD) and local Monte Carlo dynamics (MC). The former consists in the
numerical integration of the Newton equation for all the microscopic
constituents. The latter is an importance sampling
dynamics~\cite{J75,J01}. Although the two approaches are different in many respects (one is deterministic, the other one is stochastic), they are rather similar from the point of view of the practical
implementation. In both cases one generates a long trajectory
\begin{align}
\psv{X}(\Delta t)\,,\,\, \psv{X}(2\Delta t)\,,\,\,...,\,\,  \psv{X}(M\Delta t)
\end{align}
where $\Delta t$ is the integration step and $M\gg1$. Once this series of data has been
obtained, a time average can be computed as
\begin{align}
\overline A ^ M= {1 \over M} \sum_{k=1}^M A( \psv{X}(k\Delta t)) \,.
\end{align}
If the dynamics is ergodic, then one expects to get a fair estimate of the  value $\langle A
\rangle$,  for large values of $M$.  In the MD the trajectory is obtained via a numerical
integration of the Hamilton equations, while in the MC one has a
stochastic rule (a Markov chain) that provides $\psv{X}((k+1) \Delta t)$, given $\psv{X}(k\Delta t)$.

 As discussed in
Sect.~\ref{subsec:wrong-obs}, a consequence of the Kac
lemma is that some variables need to be averaged over a time span
which grows exponentially with $N$. In addition, bearing in mind the KAM
theorem we know that, strictly speaking, any realistic physical system
described by deterministic mechanical rules is not ergodic. We have
seen that both these remarks are circumvented by the ergodic
theorem of Khinchin, which proves a \textit{softer} version of
ergodicity, emphasizing, among the other aspects, that only certain observables are relevant for thermodynamic
purposes. Khinchin therefore achieves the goal of proving ergodicity by
putting some (very reasonable) constraints on the class of initial
conditions and observables. And one could be happy enough with this
solution: it is a matter of taste. But let us forget Khinchin's result for a moment, and insist on having a dynamics that can be assumed
to be ergodic for \textit{any} choice of the initial conditions and
\textit{any} choice of the observables.

The main feature of MC is that it allows to easily select a Markov chain in such a way that the corresponding stochastic dynamics is ergodic. This said, it is not difficult to figure out that such ergodicity cannot be the true reason of the
method's success. This can be easily argued by means of the following
example. Let us consider a ``small'' system, e.g., a two-dimensional
Ising model on a $100 \times 100$ square lattice. The number of
possible microstates is gigantic: there are $2^{10^4} \simeq
10^{3000}$ microscopic configurations. By no mean a numerical
simulation of this model can sample a finite fraction (even very small) of all the microscopic configurations. This goal cannot be achieved in a human life-time, not even using the fastest
supercomputer available on earth.  Nevertheless, it is matter of fact
that, at least in the cases not too close to a critical point, few
thousands iterations MC are enough to compute the main thermodynamic
quantities.
From the previous example we can conclude
that the reason of the success of  numerical computations is that, in
numerical experiments, we are only interested in a few observables (e.g.
potential energy) involving a large number of degrees freedom. In this sense there is a strong analogy with the approach to ergodicity of
Khinchin, Mazur and van der Lynden.

 Let us briefly discuss how the above reasoning applies to
the Monte Carlo computation in a system with discrete states
$k=1, 2, ...,  {\cal N}$, with probability $p_k$. As an example one can think of the Ising model, where
each state corresponds to a certain configuration of the $N$ spins and
${\cal N}=2^N$.  The simulation performs a random walk among the states, according
to a Markov chain that is ergodic and has $\{ p_k \}$ as
invariant probabilities (e.g., using the Metropolis algorithm). A trajectory $(i_1, i_2, .... , i_M)$ is generated as the sequence of states explored by the random walk.  Indicating by $f_k(M)$ the frequency of the state $k$ in the trajectory $(i_1,
i_2, .... , i_M)$, we can write the time average as
\begin{align}
 {1 \over M} \sum_{t=1}^M A_{i_t} = \sum_k A_k f_k(M) \,.
\end{align}
If $A$ is a variable {\it \`a la} Khinchin, i.e. many spins are
involved, for $N \gg 1$ one expects that $\langle A
\rangle$ is given by two terms
\begin{align}
  \langle A \rangle= \sum_{k \in S_1}  A_k p_k + \sum_{k \in S_2}  A_k p_k
\end{align}
where by $S_1$ we denote the set of states with non-negligible probability, while the contribution to
$\langle A \rangle$ of all the states in $S_2$ is negligible for $N \to
\infty$.

The time average computed along a trajectory $(i_1, i_2, .... , i_M)$,
is able to estimate the first term with $M\gg 1$, but much smaller
that ${\cal N}$, i.e. the time necessary to span the whole phase space.
For instance in the study of the statistical features far from the
critical points, even with $M=O(10^5)$ one can obtain very accurate results
for the main thermodynamic quantities.

\subsection{Some conclusions on the ergodic problem}

Ergodicity is, at the same time, extremely restrictive
(i.e., time and phase averages must be equal for almost all the
initial conditions) and not really conclusive at a physical level
(because of the average on an infinite time interval). 
The available analytical results (e.g. the KAM
theorem) give only qualitative indications, even in
the quasi-integrable limit; detailed numerical investigations
are unavoidable to understand the validity of the hypothesis
at a practical level.

Extended simulations on high dimensional Hamiltonian systems show in a
clear way that chaos is not necessarily a fundamental ingredient for
the validity of the equilibrium SM, and the naive idea
that chaos implies good statistical properties is
not correct~\cite{LPRV87,BVG21}. Even in the
absence of chaos it is possible to observe
(in agreement with Khinchin's ideas) a good agreement
between the time averages and the predictions of
equilibrium SM.

About the question of whether the
systems described by SM must have a large
number of degrees of freedom,
opposite opinions can be found in the literature. 
Gibbs~\cite{gi02} believed that
``{\it the laws of SM apply to conservative systems
  of any number of degrees of freedom, and are exact}''.  Instead Grad~\cite{G67} explicitely writes
``{\it the single feature which distinguishes SM is
  the large number of degrees of freedom}''.  One can read rather
similar sentences in the well known textbook of Landau and
Lifshitz~\cite{la69}. 
Perhaps the best synthesis, that fairly summarizes the meaning of the results by Khinchin, Mazur and van der
Lynden, is due to Truesdell~\cite{T66}: 
\begin{quotation}
 I should like to be able to say that
  statistics is unnecessary, that the name ``SM'' is
  a misnomer for ``asymptotic mechanics'', as far as equilibrium is
  concerned. This is almost true, but not quite so.
\end{quotation}

\section{Irreversibility and macroscopic dynamics}
\label{sec:irreversibility}

Many processes that are experienced and observed in everyday life are irreversible. A broken glass cannot be brought back into its original intact state, a cake cannot be ``reversely cooked'' to recover the raw ingredients, and a hen will never spontaneously turn into an egg. Despite our instinctive familiarity with the idea of irreversibility, this concept stands among the most intriguing challenges that modern physics had to face, in order to provide a consistent description of nature. Until the second half of 19th century, there was indeed no reliable physical theory explaining the reason why macroscopic objects behave irreversibly, despite the governing laws of nature being reversible (Newton's laws in the classical framework, as well as Schr\"odinger's equation in quantum physics). Among the achievements of SM, one of the most important is thus to have clarified the emergence of irreversibility in macroscopic systems, conciliating the empirical observations of thermodynamics with the established microscopical mechanical descriptions. As it will be discussed along this section, the main contribution to this understanding is due again to Boltzmann, who was able to identify the large number of degrees of freedom as the origin of non-reversible behaviors in thermodynamics. 
Despite many alternative attempts focusing, e.g., on the role of chaos or quantum effects, Boltzmann's interpretation remains, in our opinion, the most convincing one. Part of the wide corpus of work that has been developed around his ideas is reviewed in the present section. For further analysis of Boltzmann's contribution and the discussion of some misconceptions often encountered among physicists and philosophers of science, the reader is  also referred to Chapter 5 of Ref.~\cite{ce98}, Chapters 3 to 7 of Ref.~\cite{bricmont1995chaos} and Chapter 4 of Ref.~\cite{chibbaro2014reductionism}.

\subsection{The problem}

The Second Law of Thermodynamics, in its different formulations, clearly expresses that some physical processes do not admit a time-reversed counterpart. Heat can flow from hotter to colder bodies, but the reverse process will never be observed to happen spontaneously, i.e. without an external action on the system.  This statement, which may look pretty obvious at a first glance, leads to an apparent contradiction when related with the fundamental equations at the basis of our understanding of the physical world, either in the classical (Newton) or in the quantum (Schr\"odinger) description.

To grasp the idea of the problem let us consider the motion of a system of classical particles, whose canonical coordinates (positions and momenta) will be identified by $(\mathbf{q},\mathbf{p})\in \mathbb{R}^{2N}$, where $N$ is the number of degrees of freedom. Given the Hamiltonian 
\begin{equation}
    H(\mathbf{q},\mathbf{p})=\sum_{i=1}^{N} \frac{p_i^2}{2m_i} +U(\mathbf{q}) \,,
\end{equation}
where $U(\mathbf{q})$ is the interaction potential, the Hamilton equations describing the dynamics read
\begin{equation}
\begin{cases}
    \dot{q}_i=\frac{p_i}{m}\\
    \dot{p}_i=-\frac{\partial \mathcal{H}}{\partial q_i}
\end{cases}
\quad \quad i=1,...,N\,.
\end{equation}
It can be easily checked that if the above equations are satisfied by a trajectory, they are also verified by its time-reversed counterpart, i.e. the trajectory obtained by operating the time-reversal transformation:
\begin{equation}
    t \to -t \quad \quad\mathbf{q} \to \mathbf{q} \quad \quad \mathbf{p} \to -\mathbf{p}\,.
\end{equation}
In other words, any evolution admits a reversed trajectory that is consistent with the laws of mechanics, and there is no preferential direction for the ``arrow of time''.
By extending this line of reasoning to macroscopic objects, it is not immediately clear why some reversed dynamics, which are \textit{a priori} not forbidden by the laws of mechanics, are never observed in practice. 

\subsubsection{The many points of view}

The apparent conflict between the Second Law of thermodynamics and the fundamental equations of mechanics has been justified in different ways over the years~\cite{chibbaro2014reductionism}.

A first possible explanation is that the considered mechanical theory is not accurate enough to account for irreversible processes, and more detailed descriptions are needed to catch this aspect of reality. The Standard Model of particle physics, for instance, allows time reversal symmetry to be violated. The well known ``CPT Theorem'' assures that symmetry is preserved under the simultaneous occurrence of (i) charge conjugation, (ii) inversion of spatial coordinates (parity transformation) and (iii) time reversal; this means however that if a phenomenon is not symmetric under the simultaneous application of the first two transformations (CP violation), then it also violates time-reversal symmetry. Not only are these phenomena possible in principle, but they have also been observed in celebrated experiments. The first CP violation has been measured for the decay of $K$ mesons in 1964~\cite{christenson1964evidence,cronin1981cp}, leading to a Nobel Prize for this discovery. $B$ mesons (2001)~\cite{abe2001observation, aubert2001measurement}, strange $B$ mesons (2013)~\cite{aaij2013first} and $D^0$ (2019)~\cite{aaij2019observation} have been then proven to violate the symmetry as well.

A similar point of view, under some respects, is the one expressed by those who look for the origin of irreversibility in the quantum mechanism of the collapse of the wave function during a measurement. Also in this case the emergence of irreversibility is ascribed to a fundamental aspect of the dynamics that is not encoded in the mechanical equations, the collapse of the wave function. This line of reasoning would also imply that the degree of irreversibility of a process somehow depends on the presence, or not, of an external observer.

Others claim that the modelization of physical objects as isolated mechanical systems described by the fundamental equations is unrealistic. In real-world situations no system is isolated: all objects can exchange energy with the environment, meaning that stochastic baths always have to be included in their mathematical description. \new{For a discussion on this topic in the specific case of quasi-isolated systems, see~\cite{YUKALOV2003313}.}

All the three interpretations above propose to overcome the apparent paradox by modifying the equations ruling the dynamics, for different conceptual reasons. According to this line of thought, the asymmetry under time reversal should be inserted, somehow, ``by hand'' in the fundamental description~\cite{prigogine1979nouvelle}. The underlying idea is that, in the words of Prigogine
\cite{prigogine1984irreversibility}, ``Irreversibility is either true on all levels or on none''.



A different school of thought identifies in the role of chaos the origin of the contradiction. It is known that in chaotic systems the uncertainty on the initial conditions exponentially increases, so that after a while it is impossible to get back to the original state. According to this line of reasoning, irreversibility is not present at the level of single trajectories, but it is instead an intrinsic property of complex systems and must be described in terms of ensembles of trajectories~\cite{nicolis1977self}. This interpretation does not require to change the fundamental equations ruling the dynamics, but it restricts the emergence of irreversibility to a specific (although very wide) class of systems, i.e. the chaotic ones~\cite{barnum1994boltzmann}.

An explanation of the apparent paradox that does not require any change in the equations of motion and is valid for every macroscopic system can be drawn by following the lines of Boltzmann's work on $H$-theorem and subsequent contributions~\cite{lebowitz1993boltzmann}. The fundamental idea is that the origin of irreversibility lies in the choice of out-of-equilibrium initial conditions and in the large number of degrees of freedom of the system. As it will be discussed later, these two elements are common to all systems to which the Second Law of Thermodynamics applies, including non-chaotic ones.

\subsubsection{About chaos and irreversibility}

Let us  open a short digression about  chaos   and its relation with macroscopic irreversibility, a topic that led to a  certain amount of confusion in the past.

In mathematical terms a dynamical system described by the variables $\psv{X}(t)$ is said to be mixing if it relaxes to the invariant measure, i.e. if, 
independently of the initial density distribution 
$\rho({\bf x},0)$,
\begin{equation}
\label{a1}
\rho({\bf x},t) \to 
\rho^{inv}({\bf x}) \,
\end{equation}
in the large $t$ limit.
The convergence must be interpreted 
in a proper mathematical sense: for every  $\varepsilon >0$ and ${\bf x}\in \mathbb{R}^N$ the limit
$$
\int_{|{\bf x}- {\bf y}| < \epsilon}  \rho({\bf y},t) \, d {\bf y} \to
\int_{|{\bf x}- {\bf y}| < \epsilon}  \rho^{inv} ({\bf y}) \, d {\bf y} \, 
$$
must hold. Although  only in few cases it is possible to show in rigorous way that 
a system is mixing, there are clear numerical evidences for
the validity of  Eq.~\eqref{a1}  in a large class of chaotic systems~\cite{vulpiani2009chaos, gaspard_1998, Dorfman1999}. 

At a first glance the above described mixing property may appear to be equivalent to 
irreversibility, and this identification has been even claimed to be exact by some authors. In the following we briefly recall the essential difference between the two concepts. A more detailed discussion on this controversy can be found in Ref \cite{bricmont1995chaos}.

First, let us notice  that
the relaxation to an invariant measure is a property of a whole ensemble of initial conditions, and
it just reflects the fact that different points in the support of $\rho({\bf x},0)$, even if they are very close each other at $t=0$,
will be separated as a consequence of chaos.
However  this relaxation of $\rho({\bf x},t)$ to $ \rho^{inv} ({\bf x})$,
has no relation with  the irreversibility observed in everyday life,  which concerns the
behaviour of  a unique  (macroscopic) system. We do not need to consider the whole ensemble of initial conditions of a cup of hot coffee in order to conclude that it gets cold in an irreversible way: we just need one observation \new{of some macroscopic variables. This point will be discussed in detail in Section~\ref{sec:typicality}}.

The second point, which will turn out to be very relevant also in the following discussion, is that Eq. \eqref{a1} refers  to generic  chaotic dynamical systems, also low dimensional ones; in such systems the distinction between microscopic and macroscopic variables is meaningless. This would imply that irreversibility is a property that has to be expected also for microscopic systems, as suggested by the claim present in the letter by  Driebe reported in \cite{barnum1994boltzmann}:

 {\it Irreversible   processes are well observed in systems with few degrees of freedom,
  such as the baker or multibaker transformation.} 
  
It is however possible to find microscopic systems where the  relation \eqref{a1} holds, without implying any irreversible feature.
Consider for instance a $2D$ symplectic chaotic map, e.g. the celebrated \textit{Arnold's cat}
$$
\begin{aligned}
x_{t+1}&=x_t+y_t \,\,\, \text{mod} \, 1 \,,\\
y_{t+1}&=y_t+x_{t+1} \,\,\, \text{mod} \, 1\,.    
\end{aligned}
$$
This system is chaotic and  mixing, i.e. \eqref{a1}  holds \cite{vulpiani2009chaos}.
On the other hand,   comparing  the direct trajectory 
${ {\bf x}_1, {\bf x}_2, .... , {\bf x}_{T-1} ,{\bf x}_T \ }$
 with the reverse one  ${ {\bf x}_T, {\bf x}_{T-1}, .... , {\bf x}_2 ,{\bf x}_2 \ } $, 
it is impossible to  observe 
any irreversibile feature.

\subsection{Boltzmann's contribution}

In his fundamental work~\cite{boltzmann1872}, Boltzmann provided at the same time a powerful result to understand the evolution of the probability density function of a molecule in a gas, the celebrated Boltzmann equation, and the first microscopic explanation of the emergence of irreversibility, the $H$-theorem. These ground-breaking results have attracted, over the years, the critics of many physicists and philosophers, before being eventually accepted by the scientific community. In this Section we try to review the main points of the debate arisen around Boltzmann's contribution, and the corpus of results supporting his view.

\subsubsection{The Boltzmann equation}
 Let us consider a gas of $N$ identical spherical particles with radius $\sigma$. We are interested in the evolution of the probability density function $f(\psv{X}, \vect{v}, t)$ that any particle is found at position $\psv{X}$ with velocity  $\vect{v}$. This law is known to be given by the Boltzmann equation
\begin{equation}
\label{eq:evf}
    \partial_t f+ \vect{v}\cdot \partial_{\psv{X}} f+\frac{\vect{F}(\psv{X})}{m}\cdot \partial_{\vect{v}} f  =N \sigma^2Q(f,f)\,,
\end{equation}
where $\vect{F}(\psv{X})$ represents the external force, $m$ is the mass of the particles, and the right-hand side accounts instead for the role of the collisions between the considered particle and the other molecules of the gas. This interaction introduces a nonlinear term into the evolution equation for the single-particle pdf, as it is clear from the explicit evaluation of $Q$ (see below). When this term is negligible, the left-hand side of the equation is equal to zero, and the Liouvulle equation for the probability density function of free non-interacting particles is recovered.

A complete derivation of the Boltzmann equation, as well as a thorough discussion of its implications, can be found in~\cite{ce98}. Here we will only recall the main points of the derivation, stressing the fundamental assumption made by Boltzmann, the so-called ``molecular chaos hypothesis'' (\textit{Stosszahlansatz}). As we will see in a the following, in order to obtain his famous result Boltzmann had to assume the velocities of the colliding particles to be \textit{independent} before the collision. This assumption is intuitively expected to hold true in a very diluted gas, and it has been rigorously proven in some limit cases, as discussed in the next subsection.

Let us restrict ourselves, for the sake of simplicity, to the case of homogeneous gas in the absence of external forces. This will allow us to consider only the behaviour of the velocities of the particles, assuming that their positions are always homogeneously distributed. The collision term is given by the sum of two contributions: a loss and a gain term. 
The loss term is due to the possibility that a particle changes its velocity from $\vect{v}$ to a different value during a collision; the probability rate of this event is equal to the probability rate that a particle with initial velocity $\vect{v}$ collides with any other particle (whose initial velocity will be denoted by $\vect{v}_{\star}$), hence:
$$
 Q_{loss}[f,f]= -\int_{\mathbb{R}^3} d\vect{v_{\star}} \int_{S^{+}}d \vect{n} \,f_2(\vect{v}, \vect{v}_{\star}, t) | (\vect{v}-\vect{v}_{\star})\cdot \vect{n}|\,,
$$
where $f_2(\vect{v}, \vect{v}_{\star}, t)$ is the probability density function that two particles have velocities $\vect{v}$ and $\vect{v}_{\star}$. Here $\vect{n}$ is the unit vector along the direction joining the centers of the two particles at the moment of the collision. The second integral is taken over the values of $\vect{n}$ on the half-sphere $S^{+}$ such that the collision can actually take place, given the initial velocities of the particles, i.e. 
$$
S^{+} = \{\quad \vect{n}\in S \quad |\quad (\vect{v}-\vect{v}_{\star})\cdot \vect{n}>0\quad \}\,,
$$
where $S$ is the unitary sphere. 
The second contribution is a gain term, which accounts for the particles that attain velocity $\vect{v}$ after a collision. Denoting by $\vect{v}_{\star}$ the final velocity of the other particle involved, it is known from classical mechanics that the initial velocities before the collision are given by 
$$
\vect{v}'=\vect{v}-\vect{n}\cbr{\vect{n}\cdot(\vect{v}-\vect{v}_{\star})}\,,\quad \quad \vect{v}_{\star}'=\vect{v}_{\star}+\vect{n}\cbr{\vect{n}\cdot(\vect{v}-\vect{v}_{\star})}\,.
$$
We have therefore
$$
 Q_{gain}[f,f]= \int_{\mathbb{R}^3} d\vect{v}_{\star} \int_{S^{+}}d \vect{n} \,f_2(\vect{v}',\vect{v}'_{\star}, t) |(\vect{v}'-\vect{v}'_{\star})\cdot \vect{n}|\,.
$$
By gathering the two contributions and taking into account the \textit{Stosszahlansatz}, one has
\begin{equation}
   Q[f,f]= \int_{\mathbb{R}^3} d\vect{v} \int_{S^{+}}d \vect{n} \,\sbr{f(\vect{v}', t)f(\vect{v}'_{\star}, t)-f(\vect{v}, t)f(\vect{v}_{\star}, t) } |(\vect{v}-\vect{v}_{\star})\cdot \vect{n}|\,,
\end{equation}
where the probability of the velocities of two particles \textit{before the collision} has been expressed as the product of single-particle probabilities (it has been ``factorised''), i.e. $f_2(\vect{v}_1,\vect{v}_2,t)=f(\vect{v}_1,t)f(\vect{v}_2,t)$.

Without the molecular chaos assumption, it would not be possible to close the evolution equation for $f(\vect{v})$: the r.h.s. would depend indeed on the two-particle pdf, whose estimation would require an additional equation, where joint pdf of larger number of particles would be present in turn. The resulting hierarchy of equations is known as  BBGKY (Bogoliubov–Born–Green–Kirkwood–Yvon). A careful discussion of its links with the Boltzmann equation can be found in~\cite{ce98}.

\subsubsection{The H-theorem}
\label{sec:htheorem}

The Boltzmann equation has a fundamental relevance \textit{per se}. It is worth mentioning that, historically, it is the first equation for the evolution of a probability density, a decisive step toward the foundation of statistical physics, with relevant conceptual implications. But, apart from that, the Boltzmann equation is also the starting point to derive one of the most important results for the understanding of statistical irreversibility from microscopical arguments, i.e. the celebrated $H$-theorem.

One may notice that, because of the symmetries of the operator $Q$,
\begin{equation}
\begin{aligned}
   \int d \vect{v}_j \, Q(f,f) g(\vect{v})=\frac{1}{4}\int d \vect{v} &\, d \vect{v}_{\star} \, \sbr{f(\vect{v}', t)f(\vect{v}'_{\star}, t)-f(\vect{v}, t)f(\vect{v}_{\star}, t) }\times\\
   &\times |(\vect{v}-\vect{v}_{\star})\cdot \vect{n}|\sbr{g(\vect{v})+g(\vect{v}_{\star})-g(\vect{v}')-g(\vect{v}'_{\star})}\,.
\end{aligned}
\end{equation}
If one chooses
\begin{equation}
    g(\vect{v})=\log[f(\vect{v})]
\end{equation}
it is found
\begin{equation}
    \int d \vect{v} \, Q(f,f) \log[f(\vect{v})]=-\frac{1}{4}\int d \vect{v} d \vect{v}_{\star} \, \sbr{f(\vect{v}')f(\vect{v}'_{\star})-f(\vect{v})f(\vect{v}_{\star}) } |(\vect{v}-\vect{v}_{\star})\cdot \vect{n}| \log\sbr{\frac{f(\vect{v}')f( \vect{v}'_{\star})}{f(\vect{v})f(\vect{v}_{\star})}}\,,
\end{equation}
where we have dropped the time dependence to simplify the notation. The right-hand side can be shown to be non positive, by recalling the general inequality
\begin{equation}
    \cbr{x-y}\log\cbr{\frac{x}{y}}\ge 0\,.
\end{equation}
By defining the quantity
\begin{equation}
    H(t)=\int d\vect{v}\, f(\vect{v}) \log[f(\vect{v})]\,,
    \end{equation}
it is immediately noticed that
\begin{equation}
\label{eq:htheorem}
    \frac{d H}{d t}\le 0\,,
\end{equation}
where use has been made of Eq.~\eqref{eq:evf} and of the  inequality just proven. It can be shown that the equal sign holds only if the distribution of the single-particle velocity is Maxwell-Boltzmann
$$
f_{MB}(\vect{v}) \propto \exp\cbr{-\frac{\beta v^2}{2}}\,,
$$
i.e. if the system is at equilibrium.
This result is known as $H$-theorem.

The physical interpretation of Eq.~\eqref{eq:htheorem} is clear: it is possible to exhibit a macroscopic observable whose evolution, under suitable assumptions, is irreversible, even if the underlying microscopic dynamics is not. The quantity $H(t)$ defined before reduces to the single-particle contribution to the entropy (with a minus sign) when the system is at equilibrium. By identifying $H(t)$ with the thermodynamic entropy one obtains a formualtion of the Second Law for diluted gases.

The \textit{Stosszahlansatz} plays a crucial role in the argument. Assuming that the hypothesis holds for the whole dynamics is equivalent to asking that the correlations between different particles, induced by the collisions, are always negligible. This assumption sounds reasonable for diluted gases. As it will be discussed in the following section, rigorous results have been proven for short times (fractions of the typical relaxation time of the system). To fully understand the importance of the assumption, let us observe that a surprising and apparently paradoxical result can be obtained by imposing the \textit{Stosszahlansatz} at time $t=t_0$ and evolving the system forward and backward in time. Due to the reversibility of the dynamics, by repeating the above argument for $t<0$ one finds, in a small interval of time around $t=0$,
$$
\begin{cases}
\frac{dH}{dt}\le0\quad \quad t>0\\
\frac{dH}{dt}\ge0\quad \quad t<0\,,
\end{cases}
$$
i.e. the $H$ function increases until the time when the molecular chaos hypothesis is satisfied, then it decreases again. This simple observation clearly shows that the emergence of macroscopic irreversibility is related to the initial conditions that are taken into account. In particular, typical out-of-equilibrium initial conditions fulfill the \textit{Stosszahlansatz}. One may think, for instance, of two gases at different temperatures that are merged together; initially the velocities are not correlated, as the two gases have been prepared independently, and $H$ will then decrease. The reverse situation, i.e. a gas prepared in such a way that, after a while, the particles will divide into two uncorrelated groups, is instead almost impossible to observe, at least when the number of degrees of freedom is large.

\subsubsection{Classical objections and ``paradoxes''}

Boltzmann's work raised objections from other physicists of his time. Most of them can be summarised into two famous criticisms by Loschmidt~\cite{loschmidt1876, Darrigol2021BoltzmannsRT} and  Zermelo~\cite{zermelo1896, steckline83}.

Loschmidt's remark concerns the apparent contradiction between the reversibility of Newton's mechanics and the irreversibility of Boltzmann's result. It is known as ``irreversibility paradox'' (\textit{Umkehreinwand}). Loschmidt imagines an out-of-equilibrium gas where all the energy is given to a single particle, while all the other molecules are at rest. After a chain of collisions, the energy is shared among all particles and thermal equilibrium is reached. If now the velocities of the particles are suddenly reversed, Newton's dynamics will eventually bring the system to the original state where all the energy is given to a single particle, thus violating the Second Law of thermodynamics. Loschmidt uses this example to show that the Second Law cannot be derived by mean of mechanical arguments only.
In his answer to Loschmidt~\cite{boltzmann1877}, Boltzmann shows to appreciate the remark. He agrees about the fact that the dynamics by itself is not enough to derive the Second Law: a prominent role is played by the initial conditions, and by the fact that those giving rise to a measurable violation of the Second Law are extremely rare for macroscopic objects. In particular, the chance that the system achieves a microstate in which the velocities of the particles are those required by the \textit{Gedankenexpetiment} of Loschmidt is negligible in the thermodynamic limit. Quantitative studies of Loschmidt's paradox have been executed by mean of numerical simulations~\cite{Orban1967-gq,vulpiani2009chaos}, showing that a small inaccuracy in the velocity-inversion procedure forbids the system from coming back to its original macrostate. 

Zermelo's objection is based instead on the so-called ``recurrence paradox'' (\textit{Wiederkehreinwand}). Poincar\'{e}'s recurrence theorem states that, given an isolated mechanical system starting in any initial condition, the dynamics will eventually bring the system to a configuration which is arbitrarily close to the starting one in the phase space. This theorem appears to be in open contradiction with Boltzmann's result, as it is incompatible with the existence of a quantity whose evolution is monotonic in time. Boltzmann's answer~\cite{boltzmann1896} clarifies that his result does not only stem from Newton's Laws of Mechanics, but additional hypotheses are made (namely, the \textit{Stosszahlansatz}). This hypothesis is a reasonable approximation if the number of particles of the system is large; if, instead, the number of degrees of freedom is small (Zermelo only considered 3 particles in his argument), the collisions will quickly correlate the distribution of the particles' velocities, invalidating the assumptions. The occurrence of local violations of the monotonicity of $H(t)$, or bumps, is thus always possible with finite sistems, and it is even possible, in principle, that $H(t)$ reaches larger values than the initial one. However, as Boltzmann notices, the typical time that one has to wait to observe one of these violations on macroscopic systems is extremely long (he explicitly computes the recurrence time for 1 cm$^3$ of gas at room temperature, finding a result much larger than the age of the universe). This idea was then formalised by M. Kac in 1947~\cite{kac47}: he was able to show that in a space equipped with a measure, the recurrence time of a set of initial points is inversely proportional to the measure of the set, a result of ergodic theory known as Kac's lemma (as already discussed in Section~\ref{subsec:wrong-obs}).

The so-called ``paradoxes'' have the important merit to show that the origin of Second Law and macroscopic irreversibility has not to be searched for in the laws of the dynamics, but instead in the atypicality of initial conditions and in the large number of degrees of freedom. In Boltzamnn's world: ``The second Law can be explained mechanically by the unprovable assumption that... the universe is in an improbable state''~\cite{boltzmann1897, steckline83}. As a consequence, systems that are isolated out from this out-of-equilibirum universe start from out-of-equilibirum conditions as well. Among the possible explanations to justify the out-of-equilibrium nature of the universe, Boltzmann mentions the possibility of a fluctuation, local in time, whose incredible duration is due to the large number of degrees of freedom that are involved~\cite{Darrigol2021BoltzmannsRT}.

\subsubsection{Lanford’s result and other mathematical contributions}

These ground-breaking results, obtained by mean of Boltzmann's deep physical intuition about the kinetic theory of gases, are an extremely difficult challenge for mathematicians. The community had to wait nearly a century after the formulation of the $H$-theorem before a rigorous proof was available~\cite{ce98}. The problem at the time was to prove that, in the so-called \textit{Boltzmann-Grad limit}
$$
\sigma \to 0 \quad \text{while} \quad N\sigma^2 = const.\,,
$$
the evolution of the one-particle distribution is close to the one obtained from the Boltzmann equation (for most microscopic initial conditions). The $H$-theorem would have followed as a direct consequence, as recalled in Section \ref{sec:htheorem}. 

First important results were found by Carleman, Morgenstern and Arkeryd~\cite{Carleman1933,morgenstern54, Arkeryd1972}, who were able to prove the existence of Boltzmann's dynamics in the spatially-homogeneous case, thus not taking into account the role of correlations arising from collisions.

In 1972 Gallavotti proved the validity of the Boltzmann equation for a particle moving in an environment of fixed obstacles distributed according to the equilibrium distribution (Lorenz model)~\cite{gallavotti1972nota}.
In the same year Cercignani was able to show~\cite{cercignani72} the validity in the non-homogeneous case, provided that, in the Boltzmann-Grad limit, the involved mathematical objects exist, and the solution of the evolution equation is unique. Although not a proof, this argument clarified what was the roadmap to the proof of the $H$-theorem.

The main result in this direction is due to Lanford~\cite{Lan75}, who proved the missing properties of existence and uniqueness for the solution, for a time that is rather small when compared to the macroscopic dynamics, but still non-negligible. The existence and uniqueness properties are indeed shown to hold for one fifth of the average time interval between two collisions of a given particle with any other particle. During this time, about 20\% of the molecules has the chance to collide, and the result is thus far from trivial. More precisely, what Lanford was able to prove is that, given a system in an initial microstate that is failry described by the single particle distribution $f(\vect{v},0)$, the evolution of the observable $f(\vect{v},t)$ is described with high accuracy by Boltzmann equation, in the considered dilute gas and short-time limits, for most of the microstates (i.e., with probability arbitrarily close to 1 in the $N\to \infty$ limit)~\cite{LANFORD198170}.
Rigorous estimates on the errors of Lanford's results are provided in~\cite{Pulvirenti2017}. Results inspired to Lanford's work were recently found for interaction potentials different from the hard-sphere one~\cite{Gallagher2012FromNT, pulvirenti2014, Ayi2017}.

Further advancements were attained later, about the case of a rarefied gas expanding in vacuum~\cite{Illner1986,Illner1989}. In this case it is possible to justify the Boltzmann equation for arbitrary long times.
The same result has not yet been derived for the case of a gas in a box, even if the existence of the solutions of the corresponding Cauchy problem has been shown in  general~\cite{diperna1989}. Results in this direction are available only in the special case of small perturbations of the homogeneous case~\cite{Arkeryd1987}.

\subsection{Irreversibility, typicality and number of degrees of freedom}
\label{sec:typicality}

One of the main messages of Boltzmann's results is that macroscopic irreversibility is \textit{almost always} observed because macroscopic observables behave \textit{almost always} in the same way. This property is usually called ``typicality'' ~\cite{goldstein2012typicality,lazarovici2015typicality, zanghi2005fondamenti}. A typical time evolution of a system may be defined as a trajectory in the macroscopic phase space that stays ``close'' to the average one (in a sense that has to be carefully specified) for the whole considered time interval. For most physical systems, when a large number of degrees of freedom is present, most trajectories are typical: this means that macroscopic properties that can be proven to hold for the average trajectory (e.g., macroscopic irreversibility) are also valid for the overwhelming majority of the single observed realizations. The average trajectory is somehow a mathematical abstraction, whose actual observation is practically impossible, as it would require an enormous number of independent experiments prepared in the same macroscopic initial conditions, but different microscopic state (see discussion in Section~\ref{sec:errors}). Single realizations can be instead actually observed, and typicality allows to expect for them the same macroscopic properties that hold for the average trajectory. In this sense typicality is a fundamental aspect of the dynamics of statistical systems with a large number of degrees of freedom. In this section we will review some explicit examples of the emergence of typicality (and macroscopic irreversibility) in the limit of large number of degrees of freedom. \new{Let us notice that from a mathematical point of view typicality is somehow related to the theory of large deviations, namely to the law of large numbers: see the following Section~\ref{sec:ehrenfest} for the discussion of a simplified model. }

\subsubsection{The Ehrenfest model and other analytical results}
\label{sec:ehrenfest}

Boltzmann's results about irreversibility are quite subtle, and they may appear paradoxical even after a careful analysis of the argument. In order to clarify some of the most important conceptual points, especially at a pedagogical level,  simpler analytical models have been introduced that share the conceptual features of the diluted gas studied by Boltzmann.

In this context, an interesting model that qualitatively reproduces macroscopic irreversibility was however introduced by Paul and Tatiana Ehrenfest, and then also investigated by Kac~\cite{ehrenfest1907, kac1947}. 
Let us consider two urns containing, at time $t$, a number of spheres equal to $n_t$ and $N-n_t$, respectively. The spheres are numbered, and at each time step one of the $N$ total spheres is randomly extracted and moved to the other urn. The evolution of $n_t$ is thus given by 
\begin{equation}
\label{eq:evehrenfest}
n_{t+1}=
\begin{cases}
    &n_t+1\quad\text{with probability }\quad P_{n_t \to n_{t+1}}=\frac{N-n_t}{N}\\
    &n_t-1\quad\text{with probability }\quad P_{n_t \to n_{t+1}}=\frac{n_t}{N}\,.
\end{cases}
\end{equation}
In this sense the Ehrenfest model, also known as \textit{Urn model} or \textit{Dog-fleas model}, can be seen as a random walk for the discrete variable $n_t$. It is however instructive to think of $n_t$ as a collective variable describing the macroscopic state of the system, where the microscopic state is given by the position of each particle (urn $A$ or urn $B$) at time $t$. As in the case of Newton equations for the microscopical motion of the molecules of a hard-spheres gas, the evolution equations for the microstate are reversible: if particle $n$ has been just moved from urn $A$ to urn $B$, it can now brought back to $A$ with the same probability.  Nonetheless, the evolution of the macrostate, which follows Eqs.~\eqref{eq:evehrenfest}, is irreversible.

\begin{figure}
    \centering
    \includegraphics[width=.7\linewidth]{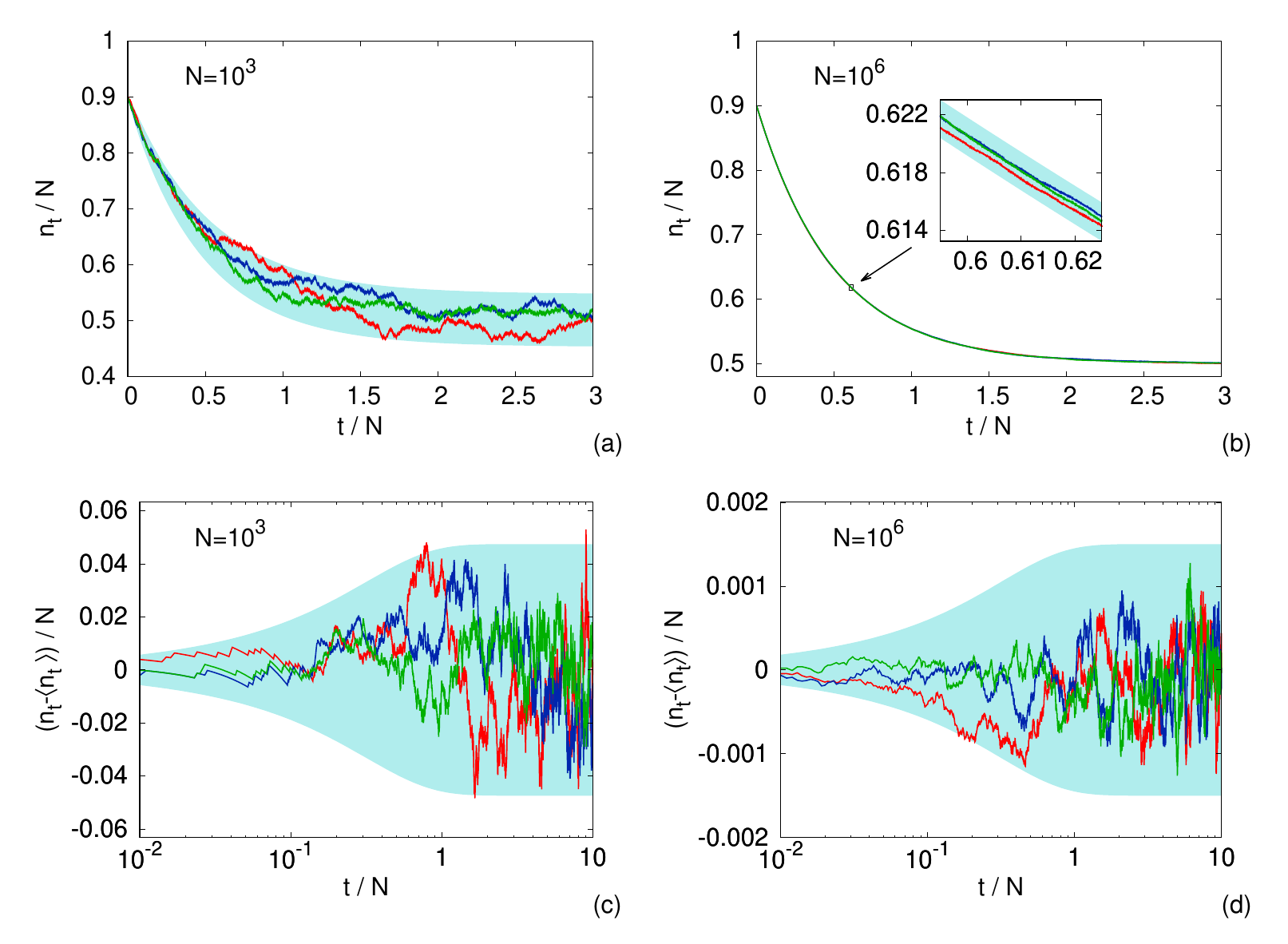}
    \caption{For two values of $N$, typical trajectories of the macroscopic system, initially prepared in the macroscopic state $n_t=0.9N$. Absolute value [panels (a) and (b)] and relative distance from the average [panels (c) and (d)] are shown. The spreading of $\sigma$ is also reported (light-blue area). Figure from Ref.~\cite{baldovin2019irreversibility}.}
    \label{fig:ehrenfest}
\end{figure}

To appreciate this point, let us briefly discuss the typical dynamics of $n_t$, assuming that it starts from a value $n_0=N$. It can be shown~\cite{baldovin2019irreversibility} that:
\begin{flalign}
\label{eq:conditionaln}
\langle n_t\rangle &= {N \over 2}\sbr{1 + \left(1- {2 \over N} \right)^t} \,\, , \\
\label{eq:conditionalvar}
\sigma_t^2 &={N \over 4}\sbr{ 1 + \left(N-1\right)\left(1- {4 \over N} \right)^t-
N\left(1- {2 \over N} \right)^{2t} } \,,
\end{flalign}
where $\sigma_t^2$ is the variance of the process.  The value of $n_t$ in the stationary state is given by $N/2$, which is obtained in the long-time limit, when the spheres are equally divided into the two urns. 
From Eq.\eqref{eq:conditionaln} is clear that the relaxation is 
monotonic, with an exponential 
decay ruled by the characteristic time $\tau= - 1/\ln(1- 2/N)\simeq N/2$. Similarly, the conditional standard deviation $\sigma_t$ tends to its 
equilibrium value $\sqrt{N}/2$ with a characteristic time $O(N)$. See Fig.~\ref{fig:ehrenfest} for typical examples of the dynamics.

The stationary distribution for the state of the system is given by
\begin{equation}
    P(n)=\frac{1}{2^N}\frac{N!}{n!(N-n)!}\,.
\end{equation}
The model is reversible, i.e. in the stationary state the detailed balance holds: $P(n)P_{n\to n'}=P(n')P_{n'\to n}$. Let us notice that $n_t$, at equilibrium, has an extremely small probability $2^{-N}$ to attain the value $n_t=N$. Therefore the ``inverse trajectory'' (a trajectory starting from $N/2$ and reaching $n_t=N$), to all practical purposes, never appears if $N$ is large enough.  

The origin of the apparent mismatch between microscopic reversibility and macroscopic irreversibility can be explained by recognising that the macrostate in which the system is initially prepared is extremely improbable according to the stationary distribution selected by the dynamics; in this sense, it is equivalent to a ``far-from-equilibrium'' state for a thermodynamic system, and this is the reason why a ``reverse trajectory'' reaching it is almost never observed. In other words, irreversibility does not emerge as a consequence of the dynamics (which is perfectly reversible), but of our peculiar choice of the initial condition. Needless to say, trajectories starting from a state close to equilibrium would be reversed with much larger probability. An equivalent way to look at the problem  is to consider the probability of recurrences. It was proven by  Kac~\cite{kac1947} that if we start from the initial condition where all the particles are contained in the first urn, then the average recurrence time to get to the initial condition is $2^N$.

It is important to remark the role of the number of degrees of freedom in the emergence of irreversibility. The fact that a macrostate corresponding to $n_t=N$ is so ``unlikely'' mostly depend on the fact that $N\gg 1$. If we had considered just a few spheres, let's say $N=3$ as in Zermelo's objection against Boltzmann, the initial condition with all spheres in the first urn would have been observed to occur, in the stationary state, roughly once in eight time steps.

Finally, let us stress that irreversibility is observed for \textit{typical} macroscopic trajectories, once the initial condition is chosen. It can be shown~\cite{baldovin2019irreversibility} that the 
probability:
\begin{equation}
\mathcal{P}_{N} = \text{Prob}\cbr{\frac{ \left|n_t - \langle n_t \rangle \right|}{N} < N^{-\beta} \quad \forall t\,:\, 0 < t \le cN}
\end{equation}
where $0<\beta<1/3$ and $c$ is an arbitrary positive constant, tends to 1 when $N$ tends to infinity. In other words, a single realization is almost surely 
contained in a stripe $\av{n_t}\pm N^{1-\beta}$, if $N$ is large enough~\cite{cerino2016role}.
The proof is based on repeated applications of Chebichev's inequality and on the bounds to the variation of $n_t$ imposed by the dynamics. The important aspect to stress here is that, once the system is prepared in an out-of-equilibrium initial condition, the macroscopic state will typically follow a relaxation to equilibrium quite close to the average one, which we know to be irreversible. It is \textit{not} necessary to observe a set of copies of the same system prepared in the same (macroscopic) initial condition in order to observe irreversibility. At a qualitative level, this behavior is the same of the diluted gas studied by Boltzmann, where the macroscopic observable $H$ behaves irreversibly \textit{almost always}, provided that the system is prepared in an out-of-equilibrium condition. Also in this case, typicality is observed when the number of degrees of freedom is large~\cite{lazarovici2015typicality}. \new{Let us notice that irreversibility can be detected also by looking at quantities other than $H$, as in the present case: the $H$ function does not play a fundamental role from a conceptual point of view.}

The Ehrenfest model is not the only one for which analytical calculations can be carried out. Among the others, one should certainly mention the so-called Kac's ring~\cite{gottwald09}, a toy model whose dynamics is completely deterministic and microscopically reversible, while irreversible at a macroscopical level. The system under consideration is a periodic 1-dimensional lattice composed of $N$ sites, each of them occupied by a binary variable $\sigma_n=\pm1$, with $n=1, ..., N$. The (discrete) dynamics is given by
\begin{equation}
    \sigma_n(t+1)=J_n\sigma_{n-1}(t)\quad 1 \le n \le N\,,
\end{equation}
where $\sigma_0(t)\equiv \sigma_N(t)$ and the  $\{J_n\}$ are binary constants whose value ($1$ or $-1$) is assigned randomly at the beginning of the dynamics. We can think of the $\{\sigma_n\}$ as points that move clockwise along a circle, and that change their color from black to white, or vice-versa, when they cross some marked angles. 

The microscopic evolution is clearly reversible, as we just need to invert the direction of the dynamics (from clockwise to counterclockwise) to recover the previous configuration. Macroscopical observables, on the other hand, behave irreversibly: for instance, the fraction of positive $\sigma$ variables approaches, in a few steps, a stationary value which only depends on the fraction of negative $J$ edges. Then, for macroscopic times (finite fraction of $N$) its value will only experience small fluctuations.
Interestingly, the evolution shows exact recurrencies, on a time scale that diverges with the size of the system. Indeed, after a time equal to $2N$ each moving point has encountered every marked angle twice, so it is found again in its initial state. In this sense there is a strong analogy with the deterministic gas described by Boltzmann, and criticized by Zermelo; also in this case, the answer to the objection relies on the scale separation between the typical times of the macroscopic dynamics, which quickly reaches stationarity, and the long recurrence time of the microscopic configuration.

\subsubsection{Numerical computations}

A different line of research has focused on the numerical investigation of models able to reproduce macroscopic irreversibility.

We have already mentioned the early numerical experiments reported in Ref~\cite{Orban1967-gq}: there a diluted system of  hard colliding disks was considered, with an istantaneous inversion of the velocities of the particles enforced at a given time $t_{inv}$. Since the dynamics is deterministic, the system is brought back to its original state at time $2t_{inv}$. However, if an error is introduced in the inversion process, the chaotic nature of the model ``spoils'' the behaviour of the reversed macroscopic observable, which is not able to reach the initial value.

In Ref.~\cite{cerino2016role} a pipe was considered, containing $N$ non-interacting
particles of mass $m$ described by canonical coordinates $\{x_i, p_i\}$ , $1 \le i \le N$. The particles are horizontally confined, on the left, by a
fixed wall and, on the right, by a wall free to move without friction
(a piston). The piston has mass $M \gg m$, and its position $X$ changes due to collisions
with the gas particles and under the action of a
constant force $F$. The total Hamiltonian reads
\begin{equation}
\label{eq:HP_NI}
\mathcal{H}_0=\sum_{i=1}^{N}\frac{\vect{ p_i}^2}{2m}+ \frac{P^2}{2M} + FX + Q(X,P,\{x_i\},\{p_i\})\,,
\end{equation}
where the term $Q(X,P,\{x_i\},\{p_i\})$ accounts for the interactions with the walls against
which the particles collide elastically.

The equilibrium statistical properties of the 
system can be easily computed using the microcanonical
ensemble~\cite{FALCIONI11,CERINO14}.  At equilibrium, the gas
particles are uniformly distributed within the available volume, in
particular the horizontal coordinate $x_i$ is uniform in $[0,X]$, with
velocities following the Maxwell-Boltzmann distribution at the
equilibrium temperature, $T_{eq}$.  In one dimension the
equilibrium values read 
\begin{equation}
T_{eq}=\frac{2NT_0+2FX_0}{3N+3}\,,\quad X_{eq}=(N+1) \frac{T_{eq}}{F}\,,\quad
\sigma^{eq}_X=\sqrt{N/3}\frac{T_{eq}}{F} + o(\sqrt{N})\,.
\label{eq:eqtermomento}
\end{equation}

Eqs.~(\ref{eq:eqtermomento}) show that the piston position provides a 
measure of the temperature, once $F$, $N$, the initial position $X_0$ of the piston and the initial temperature $T_0$ of the gas are given.
We notice that the average becomes more and more sharp, 
$\sigma^{eq}_X/X_{eq}=O(N^{-1/2})$, as $N$
increases

\begin{figure}
\centering
\includegraphics[width=.5\linewidth]{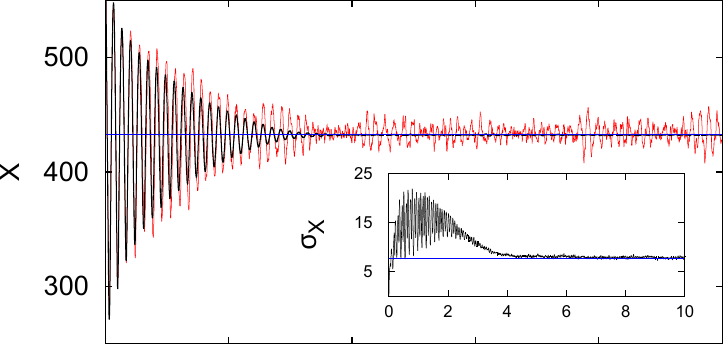}
\caption{Irreversibility in the piston model. Black and red curves denote ensemble averaged and
single-realization trajectory, respectively, for the piston position $X$. Blue
horizontal lines denote the equilibrium values in the one-dimensional
non-interacting case.  Inset: standard deviation, $\sigma_X(t)$.  Figure from ~\cite{cerino2016role}, further details therein. \label{fig:irr}}
\end{figure}

Macroscopic irreversibility can be appreciated in this system by following the time evolution of
the piston position.  At
time $t=0$, the piston is fixed at position $X_0$, and the initial microscopic state of the gas is at equilibrium, with a given temperature $T_0$. Then the piston is released, meaning that it can slide horizontally. Numerical simulations show that when the initial state is
sufficiently far from from equilibrium, i.e.
$|X_0-X_{eq}| \gg \sigma_X^{eq}$, its evolution $X(t)$ exhibits an
irreversible behavior.
Figure~\protect\ref{fig:irr}a reports a single trajectory, $X(t)$, and
the behavior of the ensemble average, $\langle X(t)\rangle$, obtained
by repeating the simulation from the same macroscopic initial
condition (the same $X_0$ and $T_0$) but different microscopic
initializations of the gas particles. Far from equilibrium, the
fluctuations of the trajectories with respect to the average are relatively small: for almost every initial configuration of the system
compatible with the macroscopic state, the time evolution of the
piston position is close to the average
one. The standard deviation of the position, $\sigma_X(t)$,
 as shown in the inset of Fig.~\ref{fig:irr}, evolves from the initial value  $0$ (by construction) and reaches the equilibrium
value at long time.  In particular, we notice the
non-monotonic behavior of $\sigma_X(t)$ in the short-time oscillatory
phase.  Similar behaviors are not uncommon for systems starting from
an unstable state~\cite{TOMBESI81}. Still, $\sigma_X(t)$ remains small with respect to the
average value, and macroscopic irreversibility can be observed in a
single trajectory of the macroscopic system, when initialized in a
non-equilibrium state.

\begin{figure}
    \centering
    \includegraphics[width=.39\linewidth]{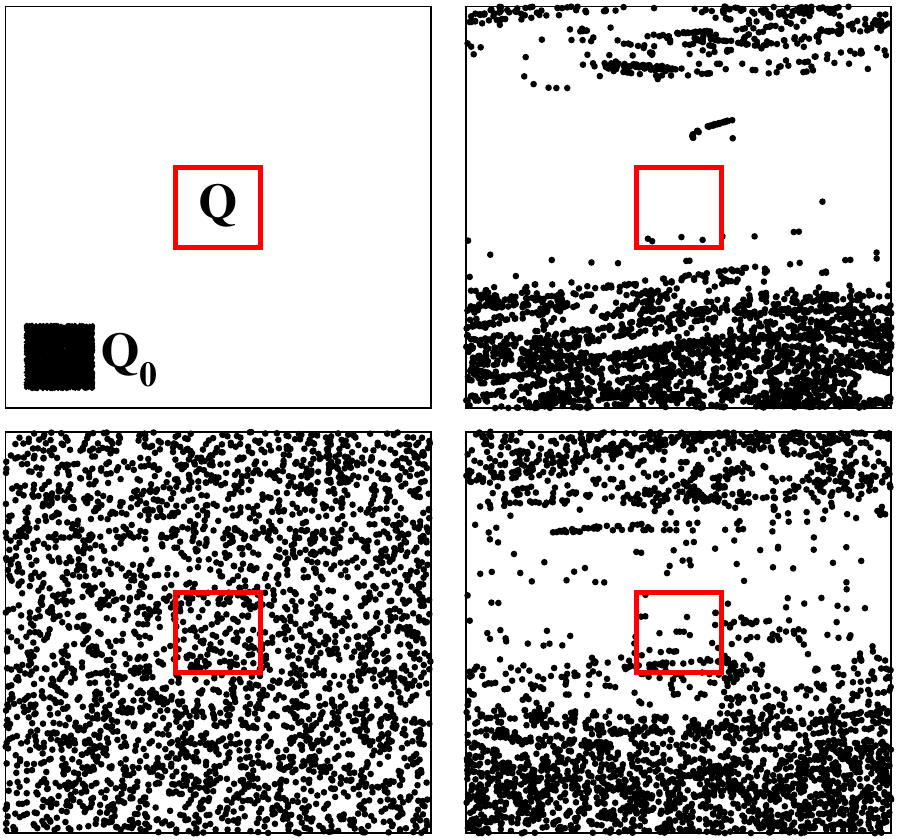}
    \includegraphics[width=.39\linewidth]{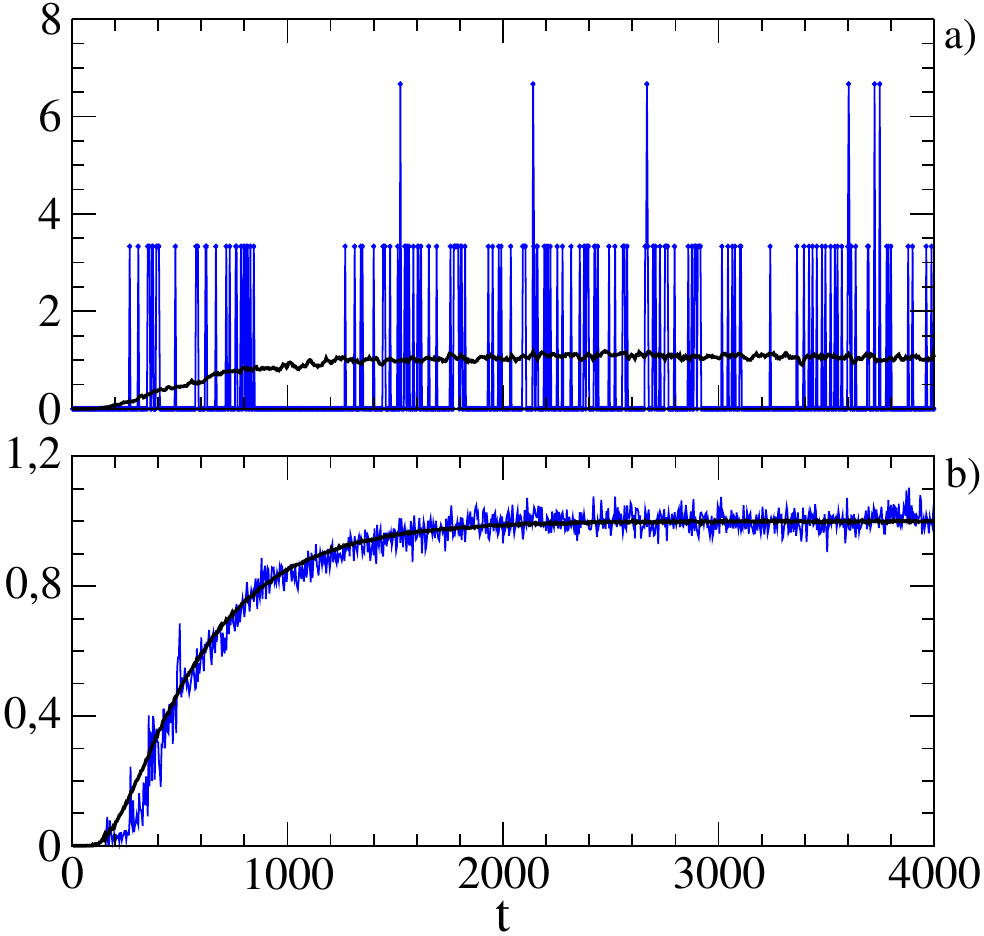}
    \caption{ Irreversible spreading of an ink drop, as modeled by Eq~\eqref{mappette}. Left panel: evolution in the $x-p$ space for the ink particles (in clockwise order
from top left). The ink particles start uniformly distributed in
$Q_0 \equiv [0.3:1.3]\times[0.3:1.3]$, while the solvent
ones have been thermalized in a previous time integration.  The
instantaneous occupation $n(t)$ is monitored in the (red) box $Q$
centered in $(\pi,\pi)$ with side $\pi/5$. Right panel: instantaneous occupation $n(t)/n_{eq}$ of the set $Q$ (blue,
fluctuating curve) and its average $\langle n(t) \rangle/n_{eq}$
(black, smooth curve) over 500 independent initial conditions starting
from $Q_0$: (a) $n_{eq}=0.3$ (drop with very few particles) and (b) $n_{eq} = 10^3$ (drop with many particles). Figure from ~\cite{cerino2016role}, further details therein.  }
    \label{fig:ink}
\end{figure}

It is worth noticing that the whole system here is not chaotic, i.e. it has vanishing Lyapunov exponents. Repeating the computations for interacting particles one has a chaotic system. However the results for $x(t)$ vs $t$ are rather similar to those obtained in the non interacting case~\cite{cerino2016role}. From the above numerical computations it is clear  that chaos is not an essential feature of irreversibility~\cite{vulpiani2009chaos}.

Let us discuss another simple model introduced in Ref.~\cite{cerino2016role}, which  mimics the diffusion of an ink drop in a fluid. The dynamics is time discrete and has a symplectic structure, i.e. it preserves the phase-space volume during the evolution. It represents a special case of a system previously proposed in Ref.~~\cite{Boffetta03}, namely
\begin{equation}
\label{mappette}
\left\{
\begin{array}{lll}
p_i(t+1) &=& p_i(t) + \epsilon\cos[x_i(t)- \theta(t)]\\ x_i(t+1) &=&
x_i(t) + p_i(t+1) \\ J(t+1) &=& J(t)
- \epsilon \sum_{j=1}^N\cos[x_j(t)- \theta(t)],\\
\theta(t+1) &=& \theta(t) + J(t+1) \,.
\end{array}
\right. 
\end{equation}
Each pair $(x_i,p_i)$ identifies an ink ``particle'' ($i=1,\ldots,
N$) in the phase space, and periodic boundary conditions on the
two-dimensional torus $\mathbb{T}_2 = [0,2\pi]\times[0,2\pi]$ are
assumed. The parameter $\epsilon$ tunes the intensity of the
mean-field-like interaction, mediated by the auxiliary
variables $\theta$ and $J$.   In the presence of interactions the system
is found to exhibit complex trajectories, as realistic gases or liquids in molecular
dynamics systems.  System (\ref{mappette}) can be shown to be 
time-reversible~\cite{quispel92}.

In~\cite{cerino2016role} a subset of the particles is labelled as ``ink'', the rest of them as ``fluid'', and the system is initialized in a state where the fluid particles are at equilibrium, while the ink particles are contained in a small region of the phase space. The number of ink particles in a different region of the phase space are monitored in time, see Fig.~\ref{fig:ink}. The irreversible behavior of this macroscopic dynamics is clear for the average trajectory, but it is also typical only when a large number of ink particles is considered: in the other case, the single realization does not show clear irreversibility, as the considered observable is not macroscopic.

\subsubsection{Macroscopic behavior}

\begin{figure}
\centering
\includegraphics[width=.9\linewidth]{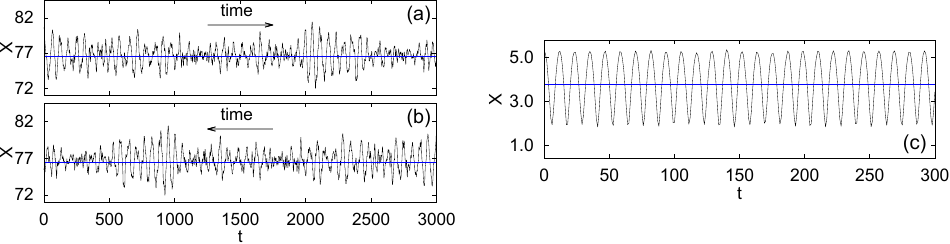}
\vspace{-0.3truecm}
\caption{ Evolution from close-to equilibrium initial conditions, 
(a) and (b), or for small systems (c), for the piston model in Ref.~\cite{cerino2016role}. Piston position $X$ vs time.  Horizontal (blue)
lines denote the microcanonical ensemble average position of the
piston $X_{eq}$. Figure from ~\cite{cerino2016role}, further details therein.}
\label{fig:noirr}
\end{figure}

To understand better the relevance of the initial condition and the many degrees of freedom, it is useful to come back to the piston model~\eqref{eq:HP_NI} discussed in Ref.~\cite{cerino2016role}. In Figure~\ref{fig:noirr}a the typical evolution from a (close-to)
equilibrium initial condition,
i.e. $|X_0-X_{eq}| \approx \sigma_X^{eq}$ is shown.  Irreversibility does not manifest in this case: the time reversed
trajectory is basically indistinguishable from the forward trajectory (compare Figs.~\ref{fig:noirr}a
and b).  Similarly, irreversibility cannot be observed when the number of degrees of freedom ($N$) is small. In the last case no notion of typicality can be defined: it is even
meaningless to speak of far-from-equilibrium initial conditions, as
fluctuations are of the same magnitude of mean
values.
Though the evolution is statistically
stationary, we cannot define a (thermodynamic) equilibrium state when
$N$ is small. A large number of degrees of freedom, as well as
starting from a very non-typical initial conditions, are essential requirements for observing
macroscopic irreversibility.

As a final remark, let us comment about the double role played by the large size of the system when dealing with irreversibility. On the one hand, having a large system allows to distinguish between a microscopic and a macroscopic description, and to define thermodynamic equilibrium; macroscopic irreversibility comes from the relaxation to equilibrium of systems prepared in out-of-equilibrium initial conditions, and it cannot be observed if no thermodynamics can be defined. On the other hand, the large number of degrees of freedom tend to insure, as a consequence of the Central Limit theorems, that most of the observed evolutions of the system are typical at a macroscopic level, so that irreversibility does not only concern the average trajectory, but (almost always) also the single realizations.

\subsection{Some mistakes about entropies and ensembles}
\label{sec:errors}

Boltzmann's results about macroscopic irreversibility have been often misinterpreted. Many excellent contributions clearly explain the meaning of his work~\cite{ce98,Gol99}, in a much more accessible way then the original papers; nevertheless, some confusion is often encountered about the conclusions that  can be drawn from the $H$-theorem. \new{Often, mainly for historical reasons, irreversibility is discussed in terms of the monotonic behaviour of entropy (or entropy-like functions). However, there is no actual necessity for this. In the following we discuss the irreversible behaviour of macroscopic observables of the Ehrenfest model and in some numerical examples.} We conclude this section with some comments about the most common misinterpretations concerning macroscopic irreversibility.

First,  it is important to remark that macroscopic irreversibility is not a property of entropy only: it can be observed in the behaviour of any observable that is macroscopic in the Khinchin sense. Examples of this have been provided in Section~\ref{sec:typicality}. Considering the evolution of the single-particle entropy, i.e., the $H$ function defined by Boltzmann (with opposite sign), is however very convenient as it allows to write down the Boltzmann inequality~\eqref{eq:htheorem}.

Secondly, the entropy that is considered in the $H$-theorem is defined from a single-particle distribution of a given macroscopic system. In order to compute it, one should measure the velocities of all the particles of the considered system at a given time, make an histogram to infer the distribution of the single-particle velocities and then apply the definition of $H$. It is important to notice that $H$ is defined for a single system, composed by many particles. This is \textit{not} equivalent to the so-called ``Gibbs entropy''
\begin{equation}
    S_G=-k_B\int dv_1\,dq_1 ... \int dv_N\,dq_N \mathcal{P}(v_1,q_1,...,v_N, q_N) \log  \mathcal{P}(v_1,q_1,...,v_N, q_N)\,,
\end{equation}
which is instead defined starting from the ensemble distribution $\mathcal{P}(v_1,v_2,...,v_N)$, i.e. the probability for the microstate \textit{of the whole system}. Its physical meaning is completely different from the Boltzmann one, as $S_G$ involves many copies of the same system, prepared in different conditions (a requirement quite hard to realize experimentally)~\cite{caprara2018law,zurek2018eliminating}. This quantity does not fulfill any $H$-theorem and, as a matter of fact, it can be shown that $S_G$ is exactly conserved (as a direct consequence of Liouville theorem). 

A possibility to observe an increase of $S_G$ in time is to coarse-grain the microscopic phase space by introducing a partition in cells with typical size $\varepsilon$. One may define the probability $\tilde{\mathcal{P}}_j(t,\epsilon)$ that the point of the phase space identifying the microscopic state at time $t$ is found in the $j$th cell. The corresponding coarse-grained Gibbs entropy
\begin{equation}
\tilde{S}_G(t,\epsilon) = -k_B\sum_j \tilde{\mathcal{P}}_j(t,\epsilon) \ln
\tilde{\mathcal{P}}_j(t,\epsilon) \,\,
\end{equation}
increases in time, after a transient time in which it stays constant. The duration of this transient depends on $\varepsilon$, and diverges as $-\log(\varepsilon)$ in the limit of small $\varepsilon$: this would imply that macroscopic irreversibility is valid only on a time scale that depends on the chosen value of $\varepsilon$~\cite{cerino2016role}. This topic, as long as an introduction to the use of $\tilde{S}_G(t,\epsilon)$ in chaos theory and dynamical systems, is discussed in~\cite{castiglione2008chaos}.

It is also important to stress that $f(\vect{v},t)$ can be measured: for instance from the velocities of the particles at time $t$ one can estimate $f(\vect{v},t)$ by filling an hystogram. Therefore, also Boltzmann's $H$ function is a measurable object. The ensemble probability  $\mathcal{P}(v_1,q_1,...,v_N, q_N)$, instead, is not: this means that Gibbs entropy is not experimentally accessible.


\section{Quantum Statistical Mechanics: A Boltzmannian Approach}
\label{sec:quantumboltzmann}

At the core of statistical mechanics lies the fundamental question: What does it mean, from a microscopic perspective, for a system to be in thermal equilibrium?  
Boltzmann's response to this question seamlessly extends into the quantum domain.  Central to this extension is the notion that a system in a pure quantum state can achieve thermal equilibrium, even when its density matrix significantly diverges from the microcanonical and canonical density matrices. 

This line of investigation, actively pursued by researchers over several years, delves into understanding how a closed quantum system in a pure state can exhibit behavior akin to thermal equilibrium. Notable studies in this field encompass references such as \cite{cite11, PSW06, cite23, cite60, cite46, cite48, cite52, Lin2009, GLMTZ10a, cite49, cite53, cite50, cite12, GHLT2015, GHLT17, Tas2016}, building upon pioneering contributions predating these endeavors \cite{cite66, cite29, cite9, cite57}. These are the kind of researches we will examine here and in the two following sections.

Several comprehensive reviews have already explored the implications of Boltzmann's concept of thermal equilibrium and the phenomenon of thermalization in quantum mechanics. Notably, works by Mori, Ikeda, Kaminishi, and Ueda \cite{Mori2018}, as well as Tumulka \cite{Tumulka2019}, are distinguished for their meticulous examination of these subjects. However, this review adopts a distinct approach, emphasizing the conceptual foundations while refraining from extensive exploration of technical intricacies.

Our goal is to ensure that readers grasp the fundamental concepts surrounding thermal equilibrium and thermalization in quantum systems. While we aim to avoid overly technical details, we strive to improve readability and accessibility for a wider audience. We will outline the key points covered in this review below to guide readers through our discussion.
\bigskip 

Boltzmann's approach to statistical mechanics is distinguished by its focus on bridging the microscopic description offered by mechanics with the macroscopic perspective provided by thermodynamics. At its core, this approach hinges on establishing a connection between the individual behavior of a system's constituents at the microscopic level---governed by the laws of mechanics---and the collective, observable properties at the macroscopic level as described by thermodynamics. The distinction between the microscopic and macroscopic levels is crucial for addressing the question of what it means, from a microscopic standpoint, for a system to be in thermal equilibrium.

Transitioning from classical to quantum mechanics introduces the microscopic description---the microstate---of a closed quantum system as provided by a pure quantum state and the laws governing its evolution generated by the system's Hamiltonian operator. Here, the foundations of statistical mechanics meet the foundations of quantum mechanics in the following sense. 

It is widely acknowledged that a pure quantum state may not provide a complete description of a physical system, as exemplified by Schr\"odinger's cat paradox, leading us into the broader territory of the foundations of quantum mechanics—a topic beyond the scope of this review. For the purposes of quantum statistical mechanics, however, one can often proceed {\em as if} the quantum state provides a complete description of a body's microscopic state, acknowledging that exceptions exist. With this caveat, unless otherwise stated, we will consider the notion of a microstate as provided by the concept of a pure quantum state.

In both classical and quantum scenarios, a macrostate comprises a collection of microstates sharing identical macroscopic characteristics. These microstates are determined by coarse-grained values of key observables, which vary depending on the level of description employed.

For classical systems, macroscopic observables often include coarse-grained values of molecular positions and velocities within the single-particle phase space (kinetic description), as well as quantities like particle number, energy, momentum, and magnetization within each cell of a spatial volume division occupied by the system (hydrodynamic description).

In the quantum realm, a macroscopic state is similarly characterized by the eigenvalues associated with coarse-grained operators representing relevant macroscopic observables. These operators can capture various aspects, such as molecular positions and velocities within the single-particle Hilbert space (kinetic description), or properties like particle number, energy, momentum, and magnetization within each cell (hydrodynamic description). However, it's important to note that quantum operators may not commute, meaning joint eigenvalues need not exist in general. Nevertheless, by carefully selecting cell sizes and applying appropriate techniques, it is possible to effectively utilize commuting operators to represent the macrostate in the quantum realm, akin to classical systems.

A system is deemed to occupy a specific macrostate if its pure state, which encapsulates the microstate, exhibits negligible components in subspaces orthogonal to the eigenspace representing the designated macrostate. All such microstates share identical macroscopic attributes. Thus, while the quantum perspective introduces certain complexities compared to its classical counterpart, it fundamentally encapsulates the same underlying concept: in the quantum scenario as well, it is meaningful to assert that the microstate of the system, at any given moment, typically corresponds to some macrostate.

Boltzmann's approach to statistical mechanics unveiled a profound aspect of thermal equilibrium: 
the size of the subset encompassing the pure states meeting the conditions of thermal equilibrium---in both classical and quantum scenarios---is nearly identical to that of the entire energy shell. This principle, termed here\footnote{It aligns with what Tasaki calls ``thermodynamic typicality" \cite{Tas2016}.} ``thermal equilibrium dominance," stands as the cornerstone of statistical mechanics. It elucidates the effectiveness of equilibrium thermodynamics and the process of thermalization. Notably, this principle underpins both classical and quantum statistical mechanics, representing a lasting achievement of Boltzmann's work. Furthermore, this principle takes on a new dimension in the quantum realm, extending to the microscopic scale in a manner devoid of classical counterparts, thanks to the exquisite quantum property of entanglement.

\bigskip

The central theme of this section revolves around thermal equilibrium dominance from a quantum perspective.  Its implications will be further analyzed in the subsequent sections. 

%

\label{Sec:q-intro}


Thermal equilibrium is of course the core business of thermodynamics.
In his book on thermodynamics \cite{Callen}, Callen notes that thermodynamics, instead of being rooted in novel and specific laws of nature, reflects a universal feature inherent in all laws. Essentially, thermodynamics explores the limitations on the possible properties of matter arising from the symmetry properties of fundamental physics laws. Central to thermodynamics is the concept of thermal equilibrium~\cite{Callen}:
\begin{quote}
in all systems there is a tendency to evolve toward states in which the properties are determined by intrinsic factors and not by previously applied external influences. Such simple terminal states are, by definition, time-independent. They are called equilibrium states.   
\end{quote}

Regarding the symmetry properties inherent in fundamental physical laws, it's crucial to recognize the deep connection between the time-translation symmetry and the existence of a conserved macroscopic energy function within thermodynamics. This principle serves as the foundation for our forthcoming exploration of closed quantum systems.

Our analysis, primarily centered around abstract notions of quantum state and observables, could potentially encompass relativistic quantum field Hamiltonians. However, for the sake of clarity and focus, we shall refrain from venturing into that realm and instead confine our discussion within the framework of nonrelativistic quantum mechanics.

\subsection{Setting the Stage}
\label{Sec:q-ensembles}
Here, we offer a summary of standard material concerning closed quantum systems, providing a brief overview of quantum gases and microscopic hydrodynamic fields, along with a summary of the role of density matrices in quantum mechanics, with particular emphasis on their significance in describing equilibrium ensembles. This recap lays the foundation for the forthcoming discussion, and Table \ref{tab:my_label2} illustrates the parallels between classical and quantum statistical mechanics.

\begin{table}\fontsize{10}{10}
    \centering
    \begin{tabular}{llll}
        & & Classical  & Quantum\\
       1.&  {\small phase-space/Hilbert space}& $\Gamma$  & $ \mathcal{H}\;$ $\; \left(\text{unit sphere }\; \mathbb{S}(\mathcal{H})  \right)$  \\
      2. & {\small microstate}  & $\psv{X}=(\{\vect{q}_i\},\{\vect{p}_i\})\in \Gamma$  &  $\Phi\in \mathcal{H}$ $\left(\text{unit  vector }\Phi\in \mathbb{S}(\mathcal{H})  \right)$\\
      3.&  {\small dynamics (symplectic/unitary)} & $\psv{X}\mapsto  \psv{X}_t$   & $\Phi\mapsto \Phi_t$\\
      4.&  {\small energy shell/ Bloch sphere} &  $\Gamma_{E}$ 
   & $\bbs$ \\
   5. &{\small partition into macrostates} &  $\Gamma =   \bigcup_M \Gamma_M $ 
   & $\mathcal{H} = \bigoplus_M\mathcal{H}_M$ \\
   6. & {\small size $\WW_M$ of the macrostate} & volume $(\Gamma_M)$  &  dimension ($\Hilbert_M)$\\
   7. & {\small macroscopic observables} & {\small functions
   $\{\MMO_j\}$ on $\Gamma$}  &  {\small operators
   $\{\MMO_j\}$  on $\mathcal{H}$} \\
   8. & {\small typicality measure} & {\small volume } & {\small surface area}  
    \end{tabular}
    \caption{Analogies between classical and quantum statistical mechanics.}
    \label{tab:my_label2}
\end{table}

\subsubsection{Closed Quantum Systems}
\label{SEC:QNMICRO}
In quantum mechanics, the microscopic state of a closed system, its microstate, is represented by a pure state $\Phi\in \mathcal{H}$, where $\mathcal{H}$ is the Hilbert space of the system.  The time evolution of the state is  governed by Schr\"odinger's equation
\begin{align}
\label{eq:screq}
i \hbar \frac{d \Phi_t}{d t}  = {H} \Phi_t\,,
\end{align}
with time-independent Hamiltonian $H$, which define  a unitary evolution in $\mathcal{H}$:
\begin{equation}
    \Phi_t=U_t \Phi\,,\quad  U_t= e^{-\frac{i}{\hbar}Ht }\,.
\end{equation}
Specifically, we will focus on quantum non-relativistic systems composed of a large number $N$ of particles, such as atoms or molecules, confined in a bounded region of physical space.

Let $\{\phi_\alpha\}$ be an orthonormal basis of $\mathcal{H}$ consisting of eigenvectors of $H$ with eigenvalues $E_\alpha$. Consider an energy interval $[E,E+\delta E]$, where $\delta E$ is small on the macroscopic scale but large enough for the interval $[E,E+\delta E]$ to contain very many eigenvalues, say $E\sim N$ and $\delta E \sim \sqrt{N}$. Let $\mathcal{H}_{E}\subseteq \mathcal{H}$ be the corresponding subspace,
\begin{equation}
\mathcal{H}_{E} = \text{span}\bigl\{\phi_\alpha: E_\alpha\in[E,E+\delta E]\bigr\} \,.
\end{equation}
This  subspace is commonly referred to as the
 \emph{micro-canonical energy shell}.  Denote by $\WW_{E}$ the dimension of $\mathcal{H}_{E}$, which is finite,  and is commonly referred to as the ``number of states'' in the shell.  Asymptotically, in the number of particles $N$ (assuming negligible degeneracy, as is the case for realistic systems), $\WW_{E}$ grows exponentially, i.e.,
\(\log \WW_{E} \sim N\).

Since a quantum state $\Phi$ is normalized to 1, 
\[
\| \Phi \|^2 =  \sum_\alpha |c_\alpha|^2 = 1,
\]
microstates are best represented as points on the unit sphere 
\begin{equation}
\label{eq:unitsphere}
\mathbb{S}(\mathcal{H}) = \{\Phi \in \mathcal{H}: \|\Phi\| = 1\}.
\end{equation}

The state space associated with the microcanonical energy shell is then the unit sphere $\bbs$ in $\mathbb{C}^n$, where $n =\WW_{E}$ and the system's temporal evolution is then represented by a curve on this sphere.

Felix Bloch referred to $\bbs$ as the \emph{quantum phase space} \cite{Walecka89} and we shall refer to it also as the  \emph{Bloch sphere}\footnote{While the term ``Bloch sphere'' is typically reserved for the geometrical representation of the pure state space of a two-level quantum mechanical system (qubit) in quantum mechanics and computing, our use of the terminology, as explicitly referred to in Bloch's notes on statistical mechanics \cite{Walecka89}, is particularly fitting. Here, $\bbs$ is recognized as the proper phase space of a quantum system.}. Notably, $\bbs$ can be treated as a classical phase space—the unit sphere in $2n$ real dimensions with the complex structure endowing it with a symplectic structure. Consequently, Schr\"odinger's evolution of the state can be viewed as a Hamiltonian evolution (in the classical sense) of a finite-dimensional linear system. Thus, in contrast to the classical case, the quantum dynamics of microstates concerns exclusively a completely integrable linear system executing quasi-periodic motion with characteristic frequencies $\omega_1 = E/\hbar, \ldots, \omega_n = (E+\delta E)/\hbar$.

This notwithstanding, the inherent simplicity of quantum dynamics should not obscure the peculiarities of quantum mechanics. In classical mechanics, observables (or properties) are aptly described by functions on the phase space. In the quantum realm, the analog might be functions on Hilbert space or its unit sphere. However, it's essential to note that we do not measure the wave function, making functions on Hilbert space not physically measurable and so not serving as observables in quantum theory. In quantum mechanics, observables are associated with self-adjoint operators or, more broadly, with positive-operator-valued measures (POVMs).\footnote{A POVM, or Positive-Operator-Valued Measure, is a generalization of the concept of observable in quantum mechanics. In traditional quantum mechanics, observables are associated with collections of orthogonal projectors $E_i$, each corresponding to a possible outcome of a measurement. The family of such projectors is called a projection-valued measure (PVM) and, via the spectral theorem, is in correspondence with the usual notion of self-adjoint-operator-as-observable. However, a POVM allows for more flexibility by allowing $E_i$ to be positive operators that are not necessarily projectors, but are such that $\sum_i E_i = I$, where $I$ is the identity operator. Then the standard probabilistic rules of quantum mechanics extend to these more general quantum observables: for a system in the state $\Phi$ and measurement associated with the POVM $E=\{E_i\}$, the probability of outcome $i$ is given by $\langle\Phi E_i|\Phi\rangle$. Indeed, it can be shown that the probabilities of \emph{any} quantum experiment can be expressed in terms of a POVM. What has just been described for observables with discrete values can be easily extended to observables with continuous values, replacing $E_i$ by $E(dx)$.}

\subsubsection{Generalities on Macroscopic Quantum Systems}
\label{sec:qugas}
An illustrative example of a closed system involves a gas consisting of $N$ particles confined within volume $V$. In this scenario, the microstate is described by the wave function $\Phi = \Phi(q)$, where $q = (\mathbf{r}_1, \ldots, \mathbf{r}_N) \in V^N$ represents the configuration of the system. Each $\mathbf{r}_i$, $i=1, \ldots, N$, denotes the triple of position coordinates for the $i$-th particle.

More precisely, for a system of $N$ spinless bosons (or fermions) with mass $m$ confined within a cubic box $V = [0,L]^3 $, the Hilbert space $\mathcal{H}$ consists of square-integrable (anti-)symmetric functions defined on $V^N$. The state $\Phi$ of such a system evolves 
according to by Schr\"odinger's equation 
\eqref{eq:screq} with 
Hamiltonian   
\begin{equation}
\label{eq:standham}
{H} = -\frac{\hbar^2}{2m} \sum_{i=1}^N \boldsymbol{\nabla}_i^2 + \sum_{i<j} v\left(|\mathbf{r}_i-\mathbf{r}_j|\right).
\end{equation}
Here, $\boldsymbol{\nabla}_i^2$ denotes the Laplacian satisfying Dirichlet (or periodic) boundary conditions at the boundary of $V$, and $v(r)$ represents a specified pair potential. The inclusion of spin is straightforward by considering $\Phi$ to take values in the tensor product of the single-particle spin spaces.

In analyzing the thermal properties of a dilute gas, the pair potential in \eqref{eq:standham} is often disregarded, resulting in the model of the quantum ideal gas. This simplification is justified by assuming that the gas is sufficiently dilute, enabling molecules to move mostly freely with occasional collisions..\footnote{While interactions between molecules are considered negligible compared to individual kinetic energies, collisions remain crucial in the gas's approach to equilibrium, especially in the Boltzmann analysis.}
For a quantum ideal gas, \eqref{eq:standham} is conveniently rewritten as
\begin{equation}\label{eq:frehamsq}
H = \sum_\alpha \varepsilon_\alpha \hat{n}_\alpha,
\end{equation}
where $\varepsilon_\alpha$ are the eigenvalues of the single-particle Hamiltonian $H^{(1)} = -\frac{1}{2m}\boldsymbol{\nabla}_i^2$ (assuming that the single-particle energies are ordered: $\varepsilon_1 \leq \varepsilon_2 \leq \ldots \leq \varepsilon_N \leq \ldots$). Here, $\hat{n}_\alpha = a_\alpha^* a_\alpha$ is the occupation number operator of the $\alpha$-th eigenstate $\ket{\alpha}$ of $H^{(1)}$, and $a_\alpha^*$ and $a_\alpha$ denote bosonic or fermionic creation and annihilation operators of the state $\ket{\alpha}$.\footnote{That is, $a_\alpha^*$ and $a_\alpha$  satisfy the commutation relations $[a_\alpha, a_\beta^*]_\pm = \delta_{\alpha\beta}$, $[a_\alpha, a_\beta]_\pm = [a_\alpha^*, a_\beta^*]_\pm = 0$, where the minus sign stands for the commutator (bosons) and the plus sign for the anticommutator (fermions). } In general, this Hamiltonian has a very large number of constants of motion. This is because $[H, \hat{n}_{\alpha}] = 0$, implying that the number of particles in any single-particle state $\psi_\alpha$ is conserved.

The structure of the microstates of the quantum ideal  gas is conveniently expressed as the action of creation operators on the ``vacuum'' $\ket{0}$, the unique state containing no particles (any annihilation operator acting on it is defined to give zero). The state
\begin{equation}\label{eq:qumicrostate}
    \Phi_{n_1, n_2, \ldots} = \prod_{\alpha} \left[ \frac{(a_\alpha^*)^{n_\alpha}}{\sqrt{n_\alpha}} \right] \ket{0}
\end{equation}
contains $n_1$ particles with energy $\varepsilon_1$, $n_2$ particles with energy $\varepsilon_2$, and so on, with $\sum_\alpha n_\alpha = N$. Any state of the system can be constructed as a linear combination of such states. The ground state for bosons corresponds to placing all $N$ particles in the lowest state, given by
\[\Phi_G^B = \frac{1}{\sqrt{N!}} (a_1^*)^N \ket{0},\]
while for fermions, the ground state fills the lowest $N$ states:
\[ \Phi_G^F = a_1^* a_2^* \ldots a_N^* \ket{0}.\]

The equilibrium properties of the quantum ideal gas will be considered in Sec \ref{sec:qeba}, while some recent findings on  its  entropy increase will be reviewed in Sec. \ref{sec:engrqidgas}.


\label{sec:michydfie}
To analyze the dynamics of Hamiltonians given by \eqref{eq:standham} or \eqref{eq:frehamsq}, it is often convenient to introduce bosonic or fermionic field operators\footnote{Usually introduced as follows: The single-particle eigenstates of the free case are labeled by the momentum $\mathbf{k}$. So we can use the non-interacting Hamiltonian above, but with the following changes in notation: $\alpha \rightarrow \mathbf{k}$, $a_\alpha \rightarrow \psi(\mathbf{k})$, $a^*_\alpha \rightarrow \psi^*(\mathbf{k})$. Then $\psi (\mathbf{r})$ and $\psi^* (\mathbf{r})$ are introduced as the Fourier transforms of $\psi (\mathbf{k})$ and $\psi^* (\mathbf{k})$, whence their commutation relation: $[\psi(\bf r), \psi^*(r') ]_\pm = \delta( r-r')$.} $\psi(\mathbf{r})$, and rewrite \eqref{eq:standham} as 
\begin{equation}
H = \int  \vare (\mathbf{r})  d^3\mathbf{r}
\end{equation}
where
\begin{equation}\label{eq:hyden1}
\vare (\mathbf{r})=   \psi^{*}(\mathbf{r}) \left( - \frac{\hbar^2}{2m}\boldsymbol{\nabla}^2 \right) \psi(\mathbf{r}) + \frac{1}{2} \int  d^3 \mathbf{r}' v(|\mathbf{r} - \mathbf{r}'|) \psi^{*}(\mathbf{r}) \psi(\mathbf{r}) \psi^{*}(\mathbf{r}') \psi(\mathbf{r}') .
\end{equation}
is the energy density operator\footnote{It is essential to ensure proper symmetrization of the operators to guarantee self-adjointness. Despite this,  the choice of $\varepsilon (\bf r)$ is not unique.}.  The particle density operator $n(\mathbf{r})$ and the momentum density operator $\mathbf{g}(\mathbf{r})$ are similarly defined:
\begin{equation} \label{eq:hyden2}
n(\mathbf{r}) = \psi^{*}(\mathbf{r})\psi(\mathbf{r}), \quad \mathbf{g}(\mathbf{r}) = \frac{\hbar}{2mi} \left[\psi^{*}(\mathbf{r}) \boldsymbol{\nabla}\psi(\mathbf{r}) - (\boldsymbol{\nabla}\psi^{*}(\mathbf{r}))\psi(\mathbf{r})\right].
\end{equation}

\new{Similarly, more complex expressions hold for the stress tensor $\mathbb{T}$ and the energy flux density $\mathbf{J}$. Operators representing local magnetization and other relevant quantities are also introduced if spin is considered. These operators define the microscopic hydrodynamic fields, which are crucial for understanding the macroscopic behavior of a large quantum system. When integrated over sufficiently large spatial regions, these fields yield extensive observables whose fluctuations are governed by a law of large numbers, akin to the classical Khinchin-type functions arising from sums of weakly dependent variables. In this coarse-grained sense, they play a role analogous to that of classical macroscopic observables, forming the cornerstone of the system's thermodynamic and hydrodynamic description.
}
In this context, it is crucial to highlight the presence of \emph{local conservation laws}, which are most clearly expressed in the Heisenberg representation:
\begin{align}\label{eq:hydrofield}
\begin{aligned}
  &\frac{\partial n}{\partial t} + \boldsymbol{\nabla} \cdot \mathbf{g}   = 0, \\
  &\frac{\partial \mathbf{g}}{\partial t} + \boldsymbol{\nabla}\cdot \mathbb{T}  = 0, \\
  &\frac{\partial \varepsilon}{\partial t} + \boldsymbol{\nabla} \cdot \mathbf{J}   = 0.
\end{aligned}
\end{align}

The significance of hydrodynamic fields on a macroscopic scale and their relevance in statistical mechanics, particularly in the very definition of thermal equilibrium as emphasized in Sec. \ref{sec: TVs}, arises from their local conservation properties. This sets them apart from arbitrary operators in the quantum realm or, in classical physics, arbitrary functions across phase space. Notably, an excess of a locally conserved quantity within a specific region cannot swiftly dissipate locally, as might occur otherwise. Instead, it gradually diffuses throughout the entire system, contributing to extended relaxation times \cite{Forster}. Consequently, the transport processes governing such quantities are extremely long at the microscopic scale, rendering them macroscopically relevant. These processes represent the observable, large-scale manifestations of the intricate microscopic dynamics occurring within the system.   This characteristic is crucial in evaluating the effectiveness of certain models of thermalization, a topic that will be examined in detail in Section \ref{sec:ctts}.

\subsubsection{Density Matrices in Quantum Mechanics}
\label{sec:denmatrqm}
A density matrix, denoted by $\rho$, is a positive trace-class operator with a unit trace on the Hilbert space $\mathcal{H}$ and is widely acknowledged as offering the most general characterization of a quantum state for a given system. For a closed system with Hamiltonian $H$, its density matrix evolves according to
\begin{equation}\label{eq:evoldenmat}
  \rho_t  = e^{-\frac{i}{\hbar}Ht } \rho \,e^{\frac{i}{\hbar}Ht }  \,.
\end{equation}

When a system is represented by the density matrix $\rho$, the expected value of an observable $A$ is given by $\text{Tr}  \rho A$. A density matrix that is a one-dimensional projection, i.e., of the form $|\Phi\rangle\langle\Phi|$ where $\Phi$ is a unit vector in the Hilbert space of the system, describes a \emph{pure state} (namely, $\Phi$), and a general density matrix can be decomposed into a \emph{mixture} of pure states $\Phi_{k}$,
\begin{equation}
\rho =\sum_k p_k |\Phi_k\rangle\langle\Phi_k| \,,\quad\mbox{where} \quad \sum_{k} p_{k} =1\,,
\label{eq:dmsd}
\end{equation}
and is called a {\em mixed state}.

Naively, one might regard $p_{k}$ as the probability that the system \emph{is} in the state $\Phi_{k}$. However, this interpretation is untenable for various reasons. First of all, the decomposition \eqref{eq:dmsd} is not unique. A density matrix $\rho$ that does not describe a pure state can be decomposed into pure states in various ways.

It is always possible to decompose a density matrix $\rho$ in such a way that its components $\Phi_k$ are orthonormal. Such a decomposition will be unique except when $\rho$ is degenerate, i.e., when some $p_k$'s coincide. For example, if $p_1=p_2$, we may replace $\Phi_{1}$ and $\Phi_{2}$ by any other orthonormal pair of vectors in the subspace spanned by $\Phi_{1}$ and $\Phi_{2}$. Even if $\rho$ were nondegenerate, it need not be the case that the system is in one of the states $\Phi_k$ with probability $p_k$. This is because, for any decomposition \eqref{eq:dmsd}, regardless of whether the $\Phi_k$ are orthogonal, if the pure quantum states of the system were $\Phi_k$ with probability $p_k$, this situation would be described by the density matrix $\rho$.

However, whenever  the actual state of the system is pure, but uncertainty exists, expressed by a probability measure ${\mu}$ on the Bloch sphere $\mathbb{S}$, one may describe the system with the density matrix
\begin{equation}\label{covmatrix}
\rho_{\mu}=\int_{ \mathbb{S}}  \ket{\Phi}\bra{\Phi}\, {\mu}(d\Phi),
\end{equation}
which also serves as the covariance matrix of ${\mu}$ provided ${\mu}$ has a mean zero. Nevertheless, all the additional statistical information embodied in ${\mu}$ is lost in $\rho_{\mu}$. As clarified earlier, the correspondence ${\mu}\mapsto \rho_{\mu}$ is not one-to-one.

Moreover, a density matrix can represent a system that lacks a pure state altogether. This occurs when the system, with Hilbert space $\mathcal{H}_S$, functions as a subsystem of a larger system with Hilbert space $\mathcal{H}_S\otimes \mathcal{H}_B$. In such cases, the density matrix $\rho_S$ of the system is obtained from the (perhaps even pure) state $\rho_{S+B}$ of the larger system through a partial trace, by demanding that the expected value of any system observable $A$ pertaining only to the system (and thus represented by the operator $A\otimes I$ for the larger system) be given by:
\begin{equation}
\label{eq:partrace2}
    \text{Tr}_S  \rho_S  A  = \text{Tr}_{S+B}  A \otimes I_B \rho_{S+B} \,.
\end{equation}
This defines $\rho_S$ as the  \emph{reduced state} of the system $S$, denoted as 
\begin{equation}
\label{eq:partrace}
\rho_S = \text{Tr}_B \rho_{S+B} \,,
\end{equation}
with $\text{Tr}_B$ representing the operation of ``tracing out" the degrees of freedom of the system $B$. Simply put, $\rho_S$ is the ``marginal" of $\rho_{S+B}$.

It is important to observe that the reduced state of a pure state $\Phi \in \mathcal{H}_S\otimes \mathcal{H}_B$ can be mixed. Specifically, $\rho_S$ is pure if and only if $\Phi$ factors, i.e., $\Phi = \Phi_S\otimes \Phi_B$. Otherwise, $\rho_S$ is mixed, indicating entanglement between the system $S$ and its environment $B$.

\subsubsection{Thermal Equilibrium Ensembles in Quantum Mechanics}
\label{sec:densitymatrices}
In classical statistical mechanics, the defining characteristics of a system in thermal equilibrium find succinct expression through appropriate—for a closed system, time-invariant—probability distributions, commonly referred to as ``classical ensembles." Similarly, in the realm of quantum mechanics, a parallel presumption is made to encapsulate macroscopic features in thermal equilibrium. Departing from the reliance on probability distributions, quantum mechanics employs density matrices for this purpose. These matrices, representing the ``quantum ensembles," play a crucial role in characterizing a system in thermal equilibrium.

The density matrices representing the quantum ensembles arise straightforwardly by analogy with their classical counterparts.  Below, we summarize:
\begin{description}
 \item[{\it Microcanonical ensemble}]
 \begin{equation}
 \label{eq:microens}
   \rho_{E}  =\frac{1}{\WW_{E} }\mathbbm{1}_{[E, E +\delta E]} ({H})\,,
 \end{equation}
 where $\mathbbm{1}_S(x)$ is the set characteristic function ($=1$ if $x\in S$, and $=0$ otherwise), $\mathbbm{1}_{[E, E +\delta E]} ({H})$ is the orthogonal projector $P_{\mathcal{H}_{E}}$ onto the microcanonical energy shell subspace $\mathcal{H}_{E}$,\footnote{\label{fn:spth} Recall that the application of a function $f$ to a self-adjoint operator with pure point spectrum $A$ is defined to be $f(A)=\sum f(a_\alpha)|\varphi_\alpha\rangle\langle\varphi_\alpha|$ if the spectral decomposition of $A$ reads $A=\sum a_\alpha |\varphi_\alpha\rangle \langle\varphi_\alpha|$.} and 
 $$ \WW_{E} = \text{Tr}  \left[\mathbbm{1}_{[E, E +\delta E]} ({H})  \right] = \text{Tr} \left[ P_{\mathcal{H}_{E}} \right] $$
 is the dimension of  $\mathcal{H}_{E}$, i.e., the number of states in the energy shell. This density matrix is diagonal in the energy representation and gives equal weight to all energy eigenstates in the interval $[E, E +\delta E]$.
 \item[{\it Canonical ensemble}]
 \begin{equation}
 \label{eq:qcan}
   \rho_{\beta}  =\frac{1}{Z_{\beta} } e^{-\beta {H}} \,,
 \end{equation}
 where  
 \begin{equation}Z_{\beta} = \text{Tr} \left[ e^{-\beta {H}} \right] 
\end{equation}
 is the canonical partition function.
 \item[{\it Grand canonical ensemble}]
 \begin{equation}
 \label{eq:qgcan}
   \rho_{\beta, {\mu}}  =\frac{1}{\mathcal{Z}_{\beta, {\mu}}}  e^{-\beta ({H}-{\mu} N)} \,,
 \end{equation}
 where  {\em here} $N$ is the number operator, ${\mu} $ is the chemical potential,   and 
 \begin{equation}\mathcal{Z}_{\beta, {\mu}}   = \text{Tr} \left[ e^{-\beta ({H}-{\mu} N)} \right] 
\end{equation}
 is the grand canonical partition function.
 \end{description}

Parallel to the classical case, one expects \textit{equivalence of the ensembles} for large quantum systems. This refers to the idea that, under certain conditions, different ensembles provide equivalent descriptions of a system in the limit of an infinitely large system---the thermodynamic limit. In other words, as the size of the system becomes large, the predictions of different ensembles concerning the values of the macroscopic observables converge to the same results. It corresponds to the equivalence, under Legendre transformations, of the different representations of the fundamental thermodynamic equation of a system; see, e.g., \cite{Callen}. This equivalence allows for the interchangeability of ensembles in the study of quantum statistical mechanics, although some care is needed in performing the limit, see, e.g,  \cite{roeck}. Moreover, there are well known particular cases where the equivalence of ensembles is lost~\cite{BMR01,CDR09,CDFR14}.

 \new{The equivalence of ensembles plays a central role in simplifying the analysis of statistical mechanics, both classical and quantum. It allows the freedom to choose the most convenient ensemble—microcanonical, canonical, or grand canonical—for a given problem or calculation, with the assurance that, under suitable conditions, the thermodynamic predictions will coincide in the appropriate limit. Several concrete mathematical techniques support this flexibility.}

\new{In the classical setting, the equivalence is typically established by comparing the entropy function in the microcanonical ensemble with the Legendre transform of the free energy in the canonical ensemble, often using large deviation estimates and convex analysis. In the quantum case, techniques such as canonical typicality, concentration of measure on high-dimensional Hilbert spaces, and explicit estimates on the trace-norm distance between reduced density matrices play a similar role. These approaches rigorously justify the replacement of ensemble averages with expectation values over typical pure states or subsystems, especially in the thermodynamic limit. We will appeal to these results repeatedly throughout the following analysis.}

\subsubsection{From the Microcanonical to the Canonical}
 
\label{sec:fmttce}

In view of the forthcoming discussion in Sec. ~\ref{sec:cantyp} regarding the universality of the canonical ensemble, it is beneficial to revisit the standard textbook derivation of the canonical ensemble from the microcanonical ensemble.

Consider the reduced density matrix of a system weakly coupled to a heat bath, where the composite system $S+B$ is described by the microcanonical density matrix at a suitable total energy $E$. This rationale assumes the negligibility of the interaction between the system and the bath, allowing the total Hamiltonian of the composite system to be expressed as $H = H_{S+B}= H_{S} + H_{B} +\lambda V$, with the energy eigenvalues of the bath much larger than those of the system and interaction term $\lambda V$ negligible. The composite system $S+B$ is then assumed to be represented by a microcanonical ensemble \eqref{eq:microens}.

Let's consider the standard assumption that both $H_{S}$ and $H_{B}$ have pure point spectra and are bounded from below. In the Hilbert spaces $\mathcal{H}_{S}$ and $\mathcal{H}_{B}$, choose eigenbases of $H_{S}$ and $H_{B}$ denoted by $\phi^{(S)}_1, \phi^{(S)}_2, \ldots{}$ and $\phi^{(B)}_1, \phi^{(B)}_2, \ldots{}$ respectively, with corresponding eigenvalues $E^{(S)}_1 \le E^{(S)}_2 \le\ldots{}$ and $E^{(B)}_1 \le E_2 \le \ldots{}$.

Recalling Eq. \eqref{eq:partrace} for $\rho_{S+B}$ given by \eqref{eq:microens}, the reduced density matrix $\rho_{S}$ of the system is expressed as:
\begin{equation} 
\rho_{S} = \frac{1}{\WW^{(S+B)}_{E} }\sum_\alpha \WW^{(B)}_{E_\alpha} |\phi^{(S)}_\alpha \rangle \langle\phi^{(S)}_\alpha|\,, \label {star}
\end{equation}
where $\WW^{(S+B)}_{E}$ is the dimension  of the microcanonical energy shell $[E, E+\delta E]$ of the composite and $\WW^{(B)}_{E_\alpha}$ is the dimension of the microcanonical energy shell $[E- E^{(S)}_\alpha, E- E^{(S)}_\alpha + \delta E]$ of the bath.

It is then evident that when the bath is sufficiently large, 
\begin{equation}
\label{rhosapprox}
    \rho_{S}\approx\rho_\beta \,,
\end{equation}
as given by \eqref{eq:qcan} for $H= H_{S}$,  with $\beta = dS(E)/dE$ where $S(E)$ is the bath's entropy. This follows from the basic fact that $S(E)\approx \log \WW^{(B)}_{E} $, so that:
\begin{equation}
\WW^{(B)}_{E_\alpha}\approx e^{S(E-E^{(S)}_\alpha)} \approx e^{ S(E)-\beta E^{(S)}_\alpha} \sim e^{-\beta E^{(S)}_\alpha }.
\end{equation}
More precisely, one can prove that  in the thermodynamic limit $\rho_{S}\to \rho_{\beta}$, the canonical ensemble of the system $S$ at inverse temperature  $\beta = ds(e)/de$, where $s(e) = \lim [S(E)/N]$ and $e=\lim[ E/N ]$.

\subsection{Boltzmann's  Notion of Thermal Equilibrium}

\label{sec:gvsbne}
\new{The discourse on thermal equilibrium can be broadly characterized by two distinct and often contrasting perspectives that, following the terminology adopted in \cite{GLTZ2020}, are referred to as the ``ensemblist'' and the ``individualist'' views, respectively.
}

In alignment with Gibbs' approach, the ensemblist view suggests that a classical system reaches thermal equilibrium when its phase points adhere to a particular probability distribution associated with a relevant thermodynamic ensemble, be it microcanonical or canonical. Similarly, in the realm of quantum systems, a parallel assumption is made concerning the evolution of its density matrix towards a quantum equilibrium ensemble.

This perspective implies the fundamental need to represent a system through a mixed state, characterized by a probability distribution across phase space in classical mechanics or a density matrix in the quantum context. This requirement arises from the fact that a pure state cannot evolve into mixed states within a closed system.

\new{According to this perspective, a system initially in a non-equilibrium ensemble is expected to evolve toward an equilibrium ensemble over long times. This convergence is typically understood, within the ensemblist framework, to occur when the system---whether classical or quantum---exhibits sufficient chaotic behavior, such as mixing or ergodicity.
}

\new{However, the fundamental significance of these dynamical conditions is debatable and largely context-dependent. While they may be regarded as essential within the Gibbsian conception of thermal equilibrium, they are not necessary from the Boltzmannian or individualist viewpoint. In the latter, chaoticity is not a prerequisite for equilibrium; rather, equilibrium is a typical property of microstates within a suitable macroscopic description, and chaotic dynamics may or may not emerge.
}

Following  Boltzmann's approach, the individualist perspective emphasizes that an individual system is characterized at any given time by a particular pure state—represented by a phase point in classical mechanics or a unit vector in quantum mechanics. According to this view, determining whether a system is in thermal equilibrium involves examining various factors, such as the spatial distribution of energy within the system, and assessing the constancy of local temperature and other intensive variables.

In other words, a pure state is deemed in thermal equilibrium if it manifests properties associated with thermal equilibrium, such as a uniform spatial distribution of energy within the system's volume—effectively residing in a specific equilibrium subset of the totality of accessible states. This subset encompasses the pure states meeting the conditions of thermal equilibrium. \new{Within this conception of thermal equilibrium, chaoticity is not postulated as a fundamental ingredient; instead, it may arise as a secondary feature, but is not required to explain the emergence or stability of equilibrium.
}

\new{Nevertheless, this picture must be reconciled with the Poincaré recurrence theorem, which implies that an isolated system may eventually return arbitrarily close to its initial state. As a result, even a system starting in or near equilibrium is expected to exhibit rare but significant fluctuations far from equilibrium over extremely long timescales. However, these fluctuations are negligible on physically relevant timescales, and equilibrium remains the overwhelmingly typical state.
} 

The concept of approaching equilibrium for an individual system implies that an initial non-equilibrium pure state eventually reaches thermal equilibrium and resides in that state not forever but, for the most part, in the long run. Section \ref{sec:ATTE} explores the mathematical implementation of this idea in the quantum domain, scrutinizing the conditions under which it holds true.

It turns out that an essential condition for thermal equilibrium lies in Boltzmann's discovery of the property of ``thermal equilibrium dominance.'' This property indicates that the size of the subset encompassing the pure states meeting the conditions of thermal equilibrium in both classical and quantum scenarios is nearly identical to that of the entire energy shell.

\subsubsection{Quantum Extension of Boltzmann's Approach}
\label{sec:qeba}
Boltzmann's analysis of the dilute gas, recalled in subsection~\ref{eq:Bodiluttoq}, seamlessly extends to an ideal  quantum gas. The microscopic state of the gas, whether bosonic or fermionic, can be described by a state $  \Phi_{n_1, n_2, \ldots}$, as given  by equation \eqref{eq:qumicrostate}. 
This state specifies that there are $n_1$ particles with energy $\varepsilon_1$, $n_2$ particles with energy $\varepsilon_2$, and so on, with the total number of particles equal to $N$ and $\sum n_\alpha \varepsilon_\alpha = E$, with tolerance $\delta E \ll E$.

Following textbook analysis, a macroscopic description can be formed by grouping particles into energy intervals, denoted as $[\mathcal{E}_{j-1}, \mathcal{E}_j]$. It's important to ensure that the spacing $\delta_j \mathcal{E} = \mathcal{E}_j - \mathcal{E}_{j-1}$ between consecutive energy values within each interval is much larger than the spacing between neighboring energy levels of individual particles. These intervals can be likened to cells in classical $\mu$-space, ensuring each interval covers numerous one-particle energy levels.

Let the total number of such intervals be represented by $K$, satisfying the condition $1 \ll K \ll N$. Additionally, let $G_j$ denote the number of single-particle states within each interval. Then, $M= (N_1,\ldots, N_K)$ represents a macrostate of the gas, where $N_j$ is the occupation number of the $j$-th interval. If $\Hilbert_M$ denotes the Hilbert space associated with such a macrostate, by summing over all possible macrostates, we recover the microcanonical space for bosons or for fermions:
\begin{equation}
\label{eq:bigoplus1}
    \Hilbert_E = \bigoplus_{M} \Hilbert_{M}
\end{equation}

Since the one-particle spectral intervals $[\varepsilon_{j-1}, \varepsilon_j]$ are microscopically large, we can assume that the wave functions of the particles contained in different intervals have disjoint support, and therefore, the effects due to quantum statistics for particles in different intervals are absent. Then,
\begin{equation}
\label{eq:bigoprod1}
    \Hilbert_{M} = \Hilbert_1 \otimes \cdots \otimes \Hilbert_K
\end{equation}

However, for the $N_j$ particles contained in the spectral subspaces $\Hilbert_j$, we cannot neglect the effect due to quantum statistics. So the dimension of  $\Hilbert_{M}$ is
\begin{equation}
\label{eq:omegafermibos}
    \Omega_M = \prod_{j=1}^K \Omega_j
\end{equation}
where $\Omega_j$ are the dimension of the spaces $\Hilbert_j$. 

If there are $G_j$ single-particle states in the $j$-th interval, according to whether they are fermions or bosons, the dimension of $\Hilbert_j$ is given by (see, e.g., \cite{la69})
\begin{equation*}
    \Omega_j = \frac{G_j!}{N_j! (G_j - N_j)!} \quad \text{for fermions,}
\end{equation*}
and
\begin{equation*}
    \Omega_j = \frac{(N_j + G_j - 1)!}{(G_j - 1)! N_j!} \quad \text{for bosons.}
\end{equation*}

Using the Stirling formula\footnote{In the Fermi case, the Pauli exclusion principle dictates that there can be no more than one particle in each quantum state. However, when dealing with macroscopic systems, the numbers $N_j$ and $G_j$ can be large and of the same order of magnitude. While a more detailed analysis is warranted in such cases, for brevity, we note that the final result remains the same as that obtained from a more nuanced analysis.}, one obtains the asymptotic formula for $\log\Omega_M$, indeed two,  one for fermions and one for bosons. These formulas are exactly similar in form for Fermi and Bose statistics, differing only regarding one sign. The upper sign corresponds to Fermi statistics, and the lower sign corresponds to Bose statistics:
\begin{equation}\label{eq:omegamfermibose}
    \log\Omega_M =   \sum_j G_j \left[ -f_j \ln(f_j) \mp (1 \mp f_j) \ln(1 \mp f_j) \right]\equiv\frac{1}{k} S_\text{B} (M)\,,
\end{equation}
where, as in the classical case, $f_j=N_j/G_j$ is the mean occupation number per cell, and, analogously to the classical expression \eqref{eq:approxHfun}, $S_\text{B} (M)$ is the Boltzmann entropy of the macrostate $M$, and thus the entropy of each microstate in $\Hilbert_M$.

Maximizing the entropy  \( S_\text{B} (M)= k\log\Omega_M  \) over the occupation numbers \( N_1, \ldots, N_K \) for a large total number of particles \( N \) and a large number of energy states \( K \), subject to the constraints
of constancy of particle number $N$ and energy $E$,
leads to the equilibrium distributions
\begin{equation}
\label{eq:distrfermibose}
f_j = \frac{1}{e^{ (\mathcal{E}_j - {\mu})/({k} T) } \pm 1}.
\end{equation}
Here, \( {\mu} \) and \( T \), representing the chemical potential and the temperature of the system respectively, are implicitly determined by the constraints. These distributions are the celebrated Fermi-Dirac and Bose-Einstein statistics, for fermions and bosons respectively.

By further refining the analysis, one can derive a distribution \( f(\mathbf{q}, \mathbf{p}) \) in the single-particle phase space. This is achieved by dividing the phase space into cells \(d\tau=  d^3 \mathbf{r} d^3 \mathbf{p} \) that are sufficiently large to suppress quantum fluctuations (each cell having a volume significantly greater than \( h^3 \), where \( h \) is the Planck constant), yet small enough to allow the representation of the particle distribution as a continuous function \( f(\mathbf{q}, \mathbf{p}) \). In the continuum limit, the entropy of the macrostate $M$  becomes
\footnote{For a rough justification, consider a particle confined in a one-dimensional interval \( \delta x \) with periodic boundary conditions which exhibits a certain number of normal modes within a wave number interval \( \delta k \). The level spacing is \( \frac{2\pi}{\delta x} \), leading to \( \frac{\delta x}{2\pi} \delta k = \frac{\delta x \delta p}{h} \) normal modes in \( \delta k \), where \( p = \hbar k \). Consequently, the number of normal modes in a phase space cell \( \delta^3 \mathbf{q} \delta^3 \mathbf{p} \) is given by \( G = \frac{\delta^3 \mathbf{q} \delta^3 \mathbf{p}}{h^3} \).}
\begin{equation}
  S_\text{B}(M) =  k_B \int \left[ -f \ln(f) \mp (1 \mp f) \ln(1 \mp f) \right] \frac{d\tau}{h^3}\,.
\end{equation}
Similarly, one arrives at a continuum version of the Fermi-Dirac and Bose-Einstein distributions, akin to the Maxwell-Boltzmann distribution \eqref{eq:lln33}. However, for now, let's stop here and take note of some key observations. 

The situation closely parallels the classical case: an equilibrium subspace emerges, linked to the Fermi-Dirac or Bose-Einstein distributions, contingent upon the particle's nature, that dominates the available Hilbert space. This subspace, denoted as $\Hilbert_\text{eq} \equiv \Hilbert_{M_\text{eq}}$, corresponds to the occupation numbers of the spaces $\Hilbert_j$, $j=1, \ldots K$, in \eqref{eq:bigoprod1} following the distribution given by \eqref{eq:distrfermibose}. Within the context of such an equilibrium distribution, the dimension of $\Hilbert_\text{eq}$ dramatically surpasses the combined dimensions of all other spaces in the orthogonal sum \eqref{eq:bigoplus1}. In essence, the equality \eqref{eq:lln33}, encapsulating the notion of ``thermal equilibrium dominance,'' remains valid, with Hilbert space dimension replacing phase space volume. Here, $\epsilon$ still represents an exponentially small factor with the number of particles, and \eqref{eq:lln33} embodies the quantum analog of the law of large numbers.

However, there's a crucial distinction to note: unlike \eqref{bigcup classical}, which partitions phase space, \eqref{eq:bigoplus1} offers an orthogonal decomposition of Hilbert space. This prompts important questions regarding quantum states that are superpositions of occupation numbers.  While microscopic superpositions within each $\Hilbert_j$ lead to states with undefined microscopic occupation numbers, this doesn't pose an issue since they share the same macroscopic occupation number, making all states in each $\Hilbert_j$ valid microstates.

However, the situation regarding states arising from superpositions of single-particle states in different $\Hilbert_j$ is more delicate. Given that these states are macroscopically distinct, their superposition introduces the `measurement problem' or `Schr\"odinger's cat' paradox, as alluded to in the introduction. Nevertheless, due to the largeness of the equilibrium subspace as expressed by \eqref{eq:equidomini}, this nuanced aspect of quantum mechanics doesn't challenge a robust quantum notion of thermal equilibrium.

To address this issue, we could stipulate that thermal equilibrium microstates have negligible components in the other subspaces of the decomposition \eqref{eq:bigoplus1}. Instead of mandating an equilibrium microstate to precisely reside within $\Hilbert_\text{eq}$, which would be excessively strict, it's reasonable and practically sufficient to require that a microstate $\Phi$ be sufficiently close to $\Hilbert_\text{eq}$ in the Hilbert space norm to classify it as an equilibrium microstate. However, in nonequilibrium scenarios, this matter warrants further examination.

\subsubsection{The Quantum Notion of Macrostate}

Our previous discussion revealed a connection between quantum and classical statistical mechanics within the context of ideal gases. The central insight is the emergence of a dominant equilibrium subspace, $\Hilbert_\text{eq}$, within the quantum Hilbert space. This finding has broad implications and prompts us to explore its applicability beyond ideal gases. To achieve this, we need a more general notion of a macrostate. This broader concept of a macrostate will also serve to clarify items 5, 6, and 7 in Table \ref{tab:my_label2}, Sec.~\ref{Sec:q-ensembles}, allowing us to compare classical and quantum statistical mechanics.

Let's consider first the classical case. Consider a single macrostate described by a single number, denoted as $M$, which represents the value of a physical quantity defined by a function $\MMO$ on phase space, i.e., $M= \MMO(X)$.

We adopt a coarse-graining approach to address the inherent limitations in measuring macroscopic properties. This involves discretizing the values of the function $\MMO$ into intervals that are large on a microscopic scale but small on a macroscopic scale, such as $\delta M, 2\delta M, \ldots, n \delta M, \ldots$. So, instead of  the function $\MMO$, we use its coarse-grained representation ${^\text{cg}}\MMO$, defined as
\begin{equation}\label{coarsegrain}
g(\MMO) = \left[ \frac{\MMO}{\delta M} \right] \delta M \equiv  {^\text{cg}}\MMO\,,
\end{equation}
where $[x]$ represents the nearest integer to $x \in \mathbb{R}$. Two phase points in phase space are considered macroscopically indistinguishable if and only if the function ${^\text{cg}}\MMO$ takes on the same value of $M$ at these points. This implies that the phase space subset representing the macrostate $M$, denoted as $\Gamma_M$,  is defined as $\Gamma_M\equiv {^\text{cg}}\MMO^{-1} (M)$.

In cases where the macrostate $M$ is defined by several numbers, we have multiple macroscopic coarse-grained variables ${^\text{cg}}\MMO = ({^\text{cg}}\MMO_1, \ldots, {^\text{cg}}\MMO_K)$. As a result, the macrostate $M$ can be characterized by a list of coarse-grained values $M = (M_1, \ldots, M_K)$. Just as before, this macrostate corresponds to a phase space subset $\Gamma_M$, which includes all phase points $X$ with coarse-grained values  $M= {^\text{cg}}\MMO (X) $.

Since coarse-grained energy is one of the macro variables, every $\Gamma_M$ is contained within a micro-canonical energy shell  $\Gamma_E$ of finite thickness $\delta E\ll E$. Consequently, for a system with a fixed constant macroscopic energy, we establish the partition
\begin{equation}
\label{partition}
\Gamma_E = \bigcup_M \Gamma_M
\end{equation}
In this partition, each subset $\Gamma_M$ represents a macrostate, which, in turn, can be viewed as a collection of microstates sharing identical macroscopic properties.\footnote{This generalizes the case of the classical ideal gas, where the functions ${^\text{cg}}\MMO_j$ represent the number of particles in the cell $C_j$ into which the single-particle phase space has been divided, with coarse-graining implicitly assumed in our earlier discussion.}

Transitioning to the quantum realm, a phase function is replaced by a quantum observable, represented by a self-adjoint operator, which we'll continue to denote by $\MMO$ for convenience. By using the same symbol for both a phase function and an operator,  we wish to highlight the structural analogy between the values $M$ of a function $\MMO$ on phase space and the eigenvalues $M$ of an operator $\MMO$.

While the operator has a discrete spectrum for a macroscopic system with constant energy $E$ (and tolerance $\delta E$), its eigenvalues are often densely packed, appearing as a continuum.\footnote{Indeed, energy levels in macroscopic systems, have  level spacings that decrease exponentially as the number of particles in the system increases. This dense packing effectively results in nearly continuous energy levels, making defining distinct stationary states or energy boundaries challenging.}  Analogously to the classical case, we perform coarse-graining by considering the coarse-grained operator ${^\text{cg}}\MMO = g(\MMO)$, with $g$ again defined by \eqref{coarsegrain}.\footnote{See footnote~\ref{fn:spth}.} Instead of values of a function we now have to do with the eigenvalues of an operator, but formally nothing changes: 
the eigenvalues of ${^\text{cg}}\MMO$ become $\delta M, 2\delta M, \ldots, n \delta M, \ldots$. These eigenvalues are separated by gaps $\delta M$, and the size of these gaps reflects the imprecision of macroscopic measurements. 
For example, in the case of the Hamiltonian $H$ for a macroscopic system, the gaps between its eigenvalues are much smaller than the imprecision in macroscopic energy measurements, $\delta E$. Consequently, coarse-graining at a level of $\delta E$, as in $ g(H) = [H/\delta E] \delta E$, results in a high degree of degeneracy for each eigenvalue of $g(H)$.  

The eigenvalues  $M= \delta M, 2\delta M, \ldots, n \delta M, \ldots$ of $\MMO$ exhibit a high degree of degeneracy. We denote the spectral subspace associated with eigenvalue $M$ as $\Hilbert_M$. Finally, we note that coarse-graining naturally extends to a commuting family of self-adjoint operators: ${^\text{cg}}\MMO= ({^\text{cg}}\MMO_1, \ldots, {^\text{cg}}\MMO_K)$. Similar to the classical case, a coarse-grained energy $ g(H)$ is one of the macro observables in the list.

Let $M= (M_1,\ldots,M_K) $ denote the eigenvalues of the commuting operators obtained by restricting $\MMO_1, \ldots, \MMO_K$ to $\Hilbert_E$ and performing coarse-graining.\footnote{We will omit the 'cg' upper index and refer to $\MMC$ as a family of coarse-grained observables, unless specified otherwise, to simplify notation.} Denote $\Hilbert_M$ as the spectral subspace associated with $M $. Then, an orthogonal decomposition of the microcanonical energy shell emerges as follows:
\begin{equation}\label{eq:orthoDecom}
\Hilbert_E = \bigoplus_M \Hilbert_M.
\end{equation}
In this decomposition, each macro-space $\Hilbert_M$ corresponds to a macro-state of the system, and, as in the classical case, all quantum states in $\Hilbert_M$ have the same macroscopic appearance. The dimension of $\Hilbert_M$ is denoted as $\WW_M$, and the sum over all $M$ satisfies $\sum_M \dd_M = \D$. In practice, $\dd_M \gg 1$ since we are dealing with a macroscopic system.

Indeed, in applications to concrete physical systems, the observables $\MMC$ may not necessarily commute. In this case, there will not be a decomposition such as \eqref{eq:orthoDecom}. However, according to the insightful analysis by von Neumann \cite{cite66} and others (cf.~\cite{J69}), macroscopic (coarse-grained) observables in a macroscopic quantum system can be naturally ``rounded'' to constitute a set of commuting operators.\footnote{Von Neumann justified the decomposition \eqref{eq:orthoDecom} by starting with a family of operators corresponding to coarse-grained macroscopic observables. He argued that by 'rounding' these operators, the family can be transformed into a set of operators $\MMC$ that commute with each other, possess pure point spectra, and exhibit substantial degrees of degeneracy. This line of reasoning has spurred investigations into determining whether, for given operators $A_1, \ldots, A_K$, whose commutators are small, one can find approximations  that commute exactly. The answer is negative for $K \ge 3$ and general $A_1, \ldots, A_K$~\cite{Choi}.
However, the operators $A_1, \ldots, A_K$ we focus on are not generic; they are essentially sums or averages of local observables across spatial cells (see Sec. \ref{sec: TVs} below). A result by Ogata \cite{Ogata} suggests that, in practice, macro observables of this type can effectively be made to commute.} In the following, we will proceed by assuming the validity of von Neumann's ``rounding," while acknowledging that there are alternative strategies to address the issue of noncommutativity.\footnote{It is also  feasible to introduce a notion of macro-spaces that doesn't require the commutation of corresponding macro-observables; see, for instance, \cite{J69}. More recently, there have been proposals along the same lines by Tasaki \cite{Tas2016}.}

\subsubsection{A Simple Example of Coarse-Grained Macro-Observable} We present an example directly quoted from \cite{GLMTZ10a} to aid in familiarizing oneself with the concept of quantum macrostates and observables. 

Consider a system composed of two identical subsystems designated 1 and 2,  with Hilbert space $\mathcal{H} = \mathcal{H}_1 \otimes \mathcal{H}_2$. The Hamiltonian of the total system is
\begin{equation}\label{eq:hamiltonian}
H = H_1 + H_2 + \lambda V\, ,
\end{equation}
where $H_1$ and $H_2$ are the Hamiltonians of subsystems 1 and 2
respectively, and $\lambda V$ is a small interaction between the two subsystems. We assume that $H_1$, $H_2$, and $H$ are positive operators. Let $\mathcal{H}_E$ be spanned by the eigenfunctions of $H$ with energies between $E$ and $E+\delta E$. 

In this example, we consider just a single macro-observable $O$, which is a projected and coarse-grained version of $H_1/E$, i.e., of the fraction of the energy that is contained in subsystem 1 alone. We cannot take $O$ to simply equal $H_1/E$ because $H_1$ is defined on $\mathcal{H}$, not $\mathcal{H}_E$, and will generically not map $\mathcal{H}_E$ to itself, while we would like $O$ to be an operator on $\mathcal{H}_E$. To obtain an operator on $\mathcal{H}_E$, let $P$ be the projection $\mathcal{H}\to\mathcal{H}_E$ and set
$$H_1'=PH_1P$$
(more precisely, $H_1'$ is $PH_1$ restricted to $\mathcal{H}_E$). Note that $H_1'$ is a positive operator, but might have eigenvalues greater than $E$. Now define 
$${^\text{cg}}O =g(H_1')$$
with the coarse-graining function \eqref{coarsegrain}.

The $\mathcal{H}_M$ are the eigenspaces of $O$; clearly, $$\bigoplus_M\mathcal{H}_M = \mathcal{H}_E\,. $$ If, as we assume, $\lambda V$ is small, then we expect $\mathcal{H}_{0.5 E}=\mathcal{H}_\text{eq}$ to have the overwhelming majority of dimensions. In a thorough treatment, we would need to prove this claim, as well as that $H_1'$ is not too different from $H_1$, but we do not give such a treatment here.

\subsubsection{Thermodynamics and Local Observables}
\label{sec: TVs}

The relevant macroscopic observables for specific systems like dilute gases include the kinetic variables discussed in Sec. \ref{sec:qeba}. However, the physically relevant variables are the hydrodynamic fields for a general system. As highlighted in Sec. \ref{SEC:QNMICRO}, these fields give rise to observables that exhibit extensive behavior on the macroscopic scale, playing a fundamental role in the system's thermodynamic and hydrodynamic description.

The coarse-grained observables corresponding to the microscopic hydrodynamic fields are defined by dividing the system's volume $V$ into cells, which are small on the macroscopic scale yet large enough to accommodate a significant number of particles.
Within each cell, macroscopic observables are derived by integrating the microscopic fields over the cell's volume, applying coarse-graining, and utilizing von Neumann's ``rounding'' technique (where noncommutativity becomes practically negligible as the cells grow sufficiently large). This procedure enables us to obtain relevant macroscopic observables associated with individual cells, encompassing various quantities such as the number of particles for each type, total energy, momentum, magnetization, and more. These choices harmonize seamlessly with established conventions in thermodynamics and hydrodynamics, particularly in the realms of fluids and magnetic materials.

A similar approach is employed in the context of lattice systems, where quantum particles like electrons, bosons, or fermions occupy specific lattice sites. In this scenario, another grid is introduced, consisting of cells significantly larger than those of the original lattice. Within each of these macroscopic cells, sums or averages of the microscopic observables associated with the lattice points are considered, and coarse-graining and ``rounding'' techniques are applied.

\subsubsection{Thermal Equilibrium Dominance}
\label{sec:ctt}

As explained earlier, Boltzmann originally established that for the dilute gas, there exists a specific macrostate $M_\text{eq}$ that overwhelmingly dominates the others. We refer to this property as ``thermal equilibrium dominance,''  which is expressed by \eqref{eq:equidomini}. 
 The phenomenon of exponential smallness of $\epsilon$ in the number of particles per cell within the available volume of the system is widely accepted and supported by rigorous results for specific systems \cite{Lan73}. The macro sets $\Gamma_{M}$ that are distinct from the equilibrium set exhibit significantly smaller volumes. Indeed, the volume of smaller macro sets in an energy shell is typically much smaller than the next larger macro set.

When it comes to quantum systems,  an analogous notion holds for a specific macro-space $\Hilbert_{\text{eq}}$ in the decomposition \eqref{eq:orthoDecom}. This equilibrium subspace satisfies the formally identical condition
\begin{equation}
\label{eq:thertyp2}
\frac{\dd_{\text{eq}}}{\WW_E} = 1 - \epsilon,
\end{equation}
with $\epsilon\ll 1$ and now $\dd_{\text{eq}}$ and $\WW_E$ representing the dimensions of $\Hilbert_{\text{eq}}$ and $\Hilbert_E$, respectively.\footnote{Exceptions exist, such as in the ferromagnetic Ising model with a vanishing external magnetic field and not-too-high temperature. In this case, two macro states (one with a majority of spins up, the other with a majority of spins down) jointly dominate but have equal volume. However, this exception does not significantly impact the overall discussion.} Moreover, $\epsilon$ is expected to be exponentially small in the number of particles per cell of the partition of the system's available volume, as in the classical case. 

What we refer to as ``thermal equilibrium dominance" has been termed ``thermodynamic typicality" by Tasaki \cite{Tas2016}, who emphasized the utility of employing large deviations methods to establish it (see Eq. \eqref{eq:trentasei} below).

Understanding the typical range of values for $\epsilon$ in \eqref{eq:thertyp2} can provide valuable insights. For macroscopic systems with realistic sizes ($N \geq 10^{20}$), equilibrium values can be precisely defined across a multitude of cells (macroscopically small but microscopically large) into which the available volume of the system can be divided, say $m \sim 10^9$ cells for a liter of air. Moreover, one should consider the inaccuracies, say $\delta M_j \sim 10^{-12}$, in the coarse-graining involved in the construction of the macro observables $M_j$, in this case, the mean number of particles per cell. The ensuing estimate, both simple and enlightening, is as follows \cite{GHLT17}:
\begin{equation}\label{realvalues}
    \epsilon \approx e^{-mN\delta M_j^2} \approx \exp(-10^{-15}N) \approx 10^{-10^5}.
\end{equation}
Adjustments can be flexibly made for smaller systems or different precision requirements. However, as $N$ diminishes to a certain extent, the endeavor to define equilibrium values becomes impractical.

Thermal equilibrium dominance is frequently observed in many-body systems. However, proving this property from a microscopic Hamiltonian perspective can be challenging. Some researchers have established it by demonstrating a property of large deviation in either the microcanonical or canonical ensemble, as summarized below.

Consider a single macroscopic observable $\hat{m}$. This observable represents an intensive local quantity and is defined as $\hat{m}= \frac{1}{N}\sum_{i=1}^{N} \MMO_i$, where each $\MMO_i$ corresponds to translational copies of a local operator $\MMO$. Several rigorous results have been successfully established in the domain of translation-invariant quantum spin systems characterized by short-range interactions. Consider the canonical ensemble given by \eqref{eq:qcan} and introduce the notation $\langle\hat{m}\rangle_{\text{can}} \equiv \text{Tr} \rho_\beta \hat{m}$.

For the case where the dimensionality $d = 1$, it has been rigorously proven that $\hat{m}$ exhibits a significant property known as a large deviation property. Specifically, for any positive values of $\epsilon$ and $\delta$, there exist $\eta > 0$ such that the following inequality holds for sufficiently large values of $N$ \cite{ogata2}:
For the case where the dimensionality $d = 1$, it has been rigorously proven that $\hat{m}$ exhibits a significant property known as a large deviation property. Specifically, for any positive values of $\epsilon$ and $\delta$, there exist $\eta > 0$ such that the following inequality holds for sufficiently large values of $N$ \cite{ogata2}:
\begin{equation} 
\label{eq:trentasei}\left\langle 
P \left[\, \left|\hat{m}  - \langle\hat{m}\rangle_{\text{can}}\right|  \geq \delta\, \right] \right\rangle_{\text{can}} \leq e^{-N\eta},
\end{equation}
where $\langle \dots \rangle_{\text{can}}$ denotes a canonical average. Here, for a self-adjoint operator $A=\sum_n a_n \ket{a_n}\bra{a_n}$, $P{\left[A \geq a\right]}$ denotes the projection operator onto the subspace spanned by the eigenstates of $A$ that have eigenvalues greater than or equal to $a$, i.e., 
\[
P{\left[A  \geq a\right]} = \sum_{n : a_n \geq a} |a_n\rangle\langle a_n|.
\]

Eq. \eqref{eq:trentasei} signifies that the likelihood of observing significant deviations from the equilibrium value decreases exponentially as the system size increases. Furthermore, due to the principle of equivalence of ensembles, we can replace the canonical ensemble in equation \eqref{eq:trentasei} with the microcanonical ensemble, where the energy $E$ corresponds to the inverse temperature $\beta$, and this is denoted as $E = \langle{H}\rangle_{\text{can}}$ \cite{Tas2016}. In this context, equation \eqref{eq:trentasei} represents thermal equilibrium dominance for a single macroscopic observable.

For systems with dimensionality $d \geq 2$, equation \eqref{eq:trentasei} has been established for sufficiently high temperatures \cite{netoc,gallavotti2002large,lenci2}. It is widely believed that the concept of thermal equilibrium dominance generally holds in physically realistic many-body Hamiltonians for any equilibrium ensemble that corresponds to a single thermodynamic phase.

\subsubsection{Macroscopic Thermal Equilibrium}
\label{sec:MATEQu}
While the concept of thermal equilibrium dominance closely mirrors its classical counterpart, the assertion that a microstate $\Phi$ is in thermal equilibrium is not a direct translation of the analogous classical notion. A microstate does not necessarily belong to a single macrostate in quantum mechanics. Therefore, for a state $\Phi$ to be in thermal equilibrium, we require
\begin{align}
\label{Phiinnu1}
     \langle \Phi | P_{\text{eq}} | \Phi \rangle \approx 1
\end{align}
where $P_{\text{eq}}$ denotes the projection operator onto $\mathcal{H}_{\text{eq}}$.

More generally and precisely,  a quantum state $\Phi$ is said to be in the macrostate ${M}$ if it is close (in the Hilbert space norm) to $\mathcal{H}_{{M}}$. If $P_{M}$ denotes the orthogonal projector onto $\mathcal{H}_{{M}}$, the degree of closeness is quantified by the condition:
\begin{align}
\label{Phiinnu}
     \langle \Phi | P_{M} | \Phi \rangle \ge 1 - \delta_{M}  \,,
\end{align}
where $0 < \delta_{M} \ll 1$.
In particular, $\Phi$ is said to be in thermal equilibrium if
\begin{equation}\label{eq:MATE}
\langle \Phi | P_{\text{eq}} | \Phi \rangle \ge 1 - \delta\,, 
\end{equation}
where $0 < \epsilon \ll \delta \ll 1$, $\epsilon$ being defined in Eq.~\eqref{eq:thertyp2}. (We write $\delta$ instead of $\delta_{M_\text{eq}}$ for ease of notation.)

This definition aligns with the approach used in several works \cite{Gri1994, Pen2004, Rig2008, GLMTZ10a, Tas2010, Har2013, Har2014, GLMTZ10a, GLMTZ10}. The set of all quantum states $\Phi$ satisfying the condition \eqref{eq:MATE} has been termed macroscopic thermal equilibrium (MATE) by the authors of \cite{GHLT2015, GHLT17}\footnote{Explanation of the terminology: ``MATE'' was meant by the authors to be contrasted with ``MITE'' (Microscopic Thermal Equilibrium), a notion we shall analyze in a subsequent section.}, more precisely they defined the set
\begin{equation}\label{eq:MATE2}
    \text{MATE} =
    \left\{ \Phi \in \bbs  \, :\,  \bra{\Phi} P_{\text{eq}} \ket{\Phi} \ge  1 - \delta  \right\}
\end{equation}

A corresponding condition for density matrices arises: a system with state $\rho$ is said to be in macroscopic thermal equilibrium if
\begin{equation}
\label{eq:densidelta}
\Tr(P_{\text{eq}} \rho) = 1 - \delta,
\end{equation}
where $\delta$ represents a small deviation from unity. Analogous conditions hold for expressing that a density matrix is in a macrostate ${M}$.

Concerning the choice of \(\delta\) in \eqref{eq:MATE}, it should strike a balance between not being too small (to ensure that  the majority of available states are in thermal equilibrium) and not being too large (to maintain the significance of \(\Phi\) being in thermal equilibrium). Realistic \(\epsilon\) values (\(10^{-10^{5}}\) or smaller, according to \eqref{realvalues}) provide flexibility when selecting \(\delta\). For instance, one may opt for \(\delta < 10^{-200}\), in line with Borel's argument that events with probabilities \(< 10^{-200}\) can be assumed to never occur in our universe \cite{Borel}. Another reasonable choice is \(\delta < \sqrt{\epsilon}\) \cite{GHLT17}.

\subsubsection{Other Notions of Thermal Equilibrium}
\label{sec:onte}

\label{sec:vNNTE}

Von Neumann proposed a definition of thermal equilibrium distinct from macroscopic equilibrium\footnote{As well as from microscopic equilibrium, a notion that we shall discuss later in Sec.  \ref{sec:mte}.}. Similar to macroscopic equilibrium, this definition relies on a family $\MMC$ of commuting macroscopic observables, which realize an orthogonal decomposition \eqref{eq:orthoDecom} of the system's energy shell. However, according to Von Neumann, a state $\Phi\in\mathcal{H}_E$ is considered in thermal equilibrium if and only if the following condition holds:
\begin{equation}\label{vonNcondi}
\| P_M \Phi \|^2 = \langle\Phi| P_M |\Phi\rangle \approx \frac{\Omega_M}{\Omega_E} \quad \text{for all } M
\end{equation}
where, like in Eq. \eqref{Phiinnu}, $P_M$ are the projectors of the subspaces of orthogonal decomposition \eqref{eq:orthoDecom}. If the dominance of thermal equilibrium holds, then von Neumann's equilibrium states correspond to the set MATE, thus highlighting macroscopic equilibrium as the more natural definition in such circumstances.

 As the authors of \cite{GHLT17} note, Von Neumann's consideration involves the case where inaccuracies \(\delta M_j\) in the coarse-graining process are smaller than the typical size of fluctuations in thermal equilibrium. According to the estimate \eqref{realvalues}, i.e., \(\epsilon = e^{-mN\delta M_j^2}\),  the relative error   \(\delta M_j/M_j\) (meaning here by $M_j$ the  eigenvalue of the operator $\MMO_j$ on  \(\Hilbert_{\text{eq}}\)),  is  $3 \times 10^{-6}$ or less for $m = 10^9$ (number of 3-cells) and $N = 10^{20}$ (number of particles). Indeed,  ``that a macroscopic measurement could determine the number of particles in a given cubic millimeter of a macroscopic system (or the amount of energy, or charge, or magnetization in that volume) with an accuracy of 6 digits seems not realistically feasible, so the assumption of such a small $\delta M_j$ is perhaps overly stretching the idea of ``macroscopic' \cite{GHLT17}". 

The approach proposed by Reimann \cite{Reim2015}, which involves a typical observable \(A\), shares close ties with von Neumann's method employing an orthogonal decomposition \eqref{eq:orthoDecom} derived from the eigenspaces of the single observable \(A\).

Another definition,   due to Tasaki \cite{Tas2016} 
in the same direction as \cite{roeck,cite60} closely resembles the concept of macroscopic equilibrium. Unlike von Neumann's approach, it bypasses the need for rounding to ensure commutativity among macro observables, a step that can pose significant practical challenges. Instead, consider \(\MMC\) as the family of macro observables before rounding off and coarse-graining.

\subsubsection{The Effectiveness of Thermodynamics}
\label{sec:eesm}

Classically, the existence of the dominant macrostate ${M}_\text{eq}$ implies that all macroscopic variables $M_j = \hat{M}_j (X)$ are nearly constant functions on $\Gamma_E$, meaning the set where they deviate (beyond tolerances) from their dominant values has a `vanishingly small' volume.
In other words, in thermal equilibrium, the value of a macrovariable stabilizes around its nearly constant equilibrium value $M_j^\text{eq}$. This equilibrium value is \emph{trivially} expressed as the average over the equilibrium subset $\Gamma_{\text{eq}}$. But since $\Gamma_{\text{eq}}\approx \Gamma_E$, this average can be extended to provide an exponentially accurate approximation to the entire energy shell $\Gamma_E$. 
Consequently, the equilibrium value $ M_\text {eq}$ of a macrovariable can be accurately calculated by employing the microcanonical ensemble average $\langle M_j\rangle_\text{mc}$.
The proximity of $\dd_{\text{eq}}/\WW_E$ to 1 signifies for the Boltzmann entropy that 
\begin{equation}
\label{sec:entroforeq}
S_\text{eq} = {k} \log \WW_\text{eq} \approx
{k} \log \WW_E = S(E).
\end{equation}
Thus $S_\text{eq}$ is well aproximated by the equilibrium thermodynamic entropy.

The quantum case follows a similar narrative. For a state $\Phi$ near macroscopic thermal equilibrium, a quantum measurement of the macroscopic observable $\MMO_j$ provides the equilibrium value with a probability of at least $1 - \delta$. Given our estimated values for $\delta$, deviations from the equilibrium value become exceedingly rare. Moreover, a joint measurement of $\MMO_1, \ldots, \MMO_K$ ensures the consistent observation of their respective equilibrium values. Like in the classical case, the Boltzmann entropy in thermal equilibrium is expressed by the same formula \eqref{sec:entroforeq}, and the same conclusions apply.

\section{Typicality in Quantum Statistical Mechanics}
\label{sec:quantumtypicality}

\subsection{Typical Properties of Thermal Equilibrium}
\new{Typicality is a versatile and indispensable concept with applications spanning diverse fields, including mathematics and physics. In the context of statistical mechanics, and in particular in the quantum framework, typicality provides a conceptual bridge between microscopic and macroscopic descriptions of equilibrium. It allows us to understand how macroscopic equilibrium properties emerge from the overwhelming majority of microscopic configurations, without invoking detailed dynamical evolution.}

\new{To clarify the quantum mechanical setting, we note that a microstate is described by a pure state in a high-dimensional Hilbert space, typically within a narrow energy shell. A macrostate, in contrast, corresponds to a subspace of this Hilbert space that reflects the values of certain coarse-grained observables—such as total energy, particle number, or magnetization—within a specified tolerance. The core idea is that, for a sufficiently large quantum system, most pure states within the energy shell are macroscopically indistinguishable: they yield expectation values of macroscopic observables that are extremely close to their microcanonical averages. In this sense, equilibrium becomes a typical property of states in the relevant energy shell.}

\new{This perspective is at the heart of the modern understanding of thermal equilibrium in isolated quantum systems, as well as in studies of the eigenstate thermalization hypothesis (ETH). These approaches formalize the idea that microscopic equilibrium does not require detailed time evolution or external randomness: rather, equilibrium emerges from the geometric structure of the Hilbert space and the concentration of measure phenomenon in high-dimensional spaces.}

\new{Before exploring in detail the typical properties of thermal equilibrium, we shall therefore provide some background on the mathematical notion of typicality and its implications for quantum statistical mechanics.
}

\subsubsection{Typicality}
\label{se.tyvspr}

A feature or behavior is considered typical in a set $\mathscr{X}$ if it occurs for most elements in $\mathscr{X}$.  More precisely,  asserting  that a statement $\mathscr{S}(x)$ is typical or true ``for most $x$" within the set $\mathscr{X}$ signifies that:
\begin{equation}
\label{eq:gmost}
\uu\{x | \mathscr{S}(x)\} \geq 1 - \epsilon\,,
\end{equation} where  $0 < \epsilon \ll 1$, and $\uu$ is a suitable probability measure on  $\mathscr{X}$. \footnote{In mathematics, typicality theorems provide valuable insights into various phenomena, such as the concentration of area near the equator or the localization of volume near the surface of a sphere in high-dimensional space. They also shed light on the prevalence of irrational numbers within the real numbers in the interval~$ [0, 1]$ and the uniform distribution of orthonormal bases in high-dimensional spaces~\cite{GLTZ17}.}

We say  `suitable' to refer to a measure $\uu$ that meets some necessary criteria, such as being non-contrived, uniform, and, in physics,  invariant under the system's temporal evolution---after all, the notion of typicality should not change with time.  

Though $\uu$ functions as a probability measure, its primary role is to formalize the concept of ``most." It's important to note that absolutely continuous measures yield the same notion of ``most," implying that typicality is represented by an equivalence class of measures. Therefore, statements such as $\uu(A) = 1/3$, while meaningful for a genuine probability, lack significance for a measure whose sole purpose is to capture the notion of ``most.''

This interpretation of $\uu$ as a measure of typicality aligns with Andrey Kolmogorov's views  as outlined in his 1933 work ``Grundbegriffe,'' and summarized by his pupil Yuri Vasilyevich Prokhorov. Prokhorov, paradoxically, noted in the Soviet Encyclopedia that only probabilities close to zero or one are meaningful. Arguably,  he was referring to measures of typicality rather than genuine, objective probabilities.

The assertion of thermal equilibrium dominance in the classical case is indeed a statement that follows the same pattern as in equation \eqref{eq:gmost}. More generally, 
an instance of typicality, is provided by any scenario in which  the (weak) Law of Large Numbers is applicable to a sequence of random variables ${Y_i}$ such that, for any $\delta > 0$,
\begin{equation}
\label{eq:lln}
\uu\left\{ \left|\frac{1}{N} \sum_i^N Y_i - {m} \right| < \delta \right\} = 1-\epsilon \,,
\end{equation}
with $\epsilon>0$ that tends to zero as $N$ approaches infinity thereby showing the convergence of the empirical mean to the theoretical mean $m$.  As we have seen, in statistical mechanics, this may lead to probabilities of the order of $1-10^{-200}$  for $N\sim 10^{20}$. In this particular scenario, it does not seem appropriate to say that we are dealing with genuine probabilities.

\subsubsection{The Surface Area as a Typicality Measure}
\label{sec:tsaaatm}
In the context of quantum mechanics, thermal equilibrium dominance is expressed in terms of the notion of ``dimension'' of a vector space: it is the condition that the dimension of the equilibrium subspace $\Hilbert_\text{eq}$ is almost equal to that of the entire energy shell Hilbert space $\Hilbert_E$. 

A probability measure that encapsulates this property is the normalized surface area, $\uu$, on the Bloch sphere $\bbs$ (see Sec. \ref{SEC:QNMICRO}). 

This measure was initially proposed by Schr\"odinger \cite{Sch1927, Sch1952}, and later refined by Bloch \cite{Walecka89}, in conjunction with the microcanonical density matrix $\rho_E$ given by \eqref{eq:microens} through \eqref{covmatrix}. It is readily observed that 
\begin{equation}\label{covmatrix2}
\rho_E=\int_{ \mathbb{S}(\mathcal{H})} \ket{\Phi}\bra{\Phi} {\uu}(d\Phi)\,.
\end{equation}

The surface area measure $\uu$ is stationary, remaining invariant under the unitary time evolution generated by ${H}$. When quantum dynamics is interpreted as classical dynamics (as outlined in Sec.  \ref{Sec:q-ensembles}), this measure corresponds to phase space volume  for the dynamical system. Significantly, it is maximally dispersed over the set $\bbs$ of permissible microstates---it is the uniform measure over the sphere. 

We note that the normalized surface area ${\uu}$ provides insights concerning standard formulas in quantum statistical mechanics. A useful formula arising from \eqref{covmatrix2}  for the micro-canonical average of an observable $A$ on $\Hilbert_E$ is 
\begin{equation}
\label{eq:usefultr}
   \Tr\rho_E A =\frac{\Tr A}{\D} = \int_{ \mathbb{S}(\mathcal{H})} \bra{\Phi} A \ket{\Phi} {\uu}(d\Phi) \equiv \ee   \bra{\Phi} A \ket{\Phi}  
\end{equation}
(just recall \eqref{eq:microens}, namely, recall  that the micro-canonical density matrix $\rho_{E}$ is  $1/\D$ times the identity operator on $\Hilbert_E$). 
Here $\ee$ denotes  the expectation with respect to $\uu$
and $\Phi$ is regarded as  a random vector with uniform distribution $\uu$  over $\bbs$.

Another application concerns the relation between the microcanonical and the canonical ensembles. Consider Eq. \eqref{rhosapprox}, which expresses that the reduced density matrix of a system weakly interacting with a large heat bath is canonical when the combined System + Bath is characterized by the microcanonical density matrix $\rho_E$.
The right-hand side is just the average of the reduced density matrix $\rho_S^\Phi = \Tr_B \ket{\Phi}\bra{\Phi}$ for $\Phi \in \bbs$:
\begin{equation}\label{marginalcanonical}
\int_{\bbs} \Tr_B \ket{\Phi}\bra{\Phi} \uu(d\Phi) = \Tr_B \int_{\bbs} \ket{\Phi}\bra{\Phi} \uu(d\Phi) = \Tr_B \rho_E\,.
\end{equation}
Hence, we infer that the marginal with respect to the normalized surface area of the microcanonical ensemble is canonical---a noteworthy observation.

\subsubsection{Gaussian Adjusted Projected Measures}
As highlighted in Sec.  \ref{sec:densitymatrices}, there exist numerous probability distributions that lead to the same density matrix, consequently yielding identical predictions for experimental outcomes. However, Goldstein et al.  \cite{GAP1} have argued that there are ``natural'' measures associated with quantum ensembles, and such measures can be both useful and physically meaningful.
According to their analysis, the normalized surface area ${\uu}$ is naturally associated with the microcanonical ensemble.

The general problem of associating probability distributions with quantum ensembles has been extensively explored \cite{GAP1,TZ2005,GLMTZ2016}. It has been demonstrated that any density matrix $\rho$ can be linked to a natural measure on the unit sphere $\bbs$, known as GAP$(\rho)$ (where GAP stands for Gaussian Adjusted Projected measure). 

These studies extend the correspondence \eqref{covmatrix2} between the microcanonical ensemble and the normalized surface area to an open quantum system, one that interacts with its environment and is almost always entangled with it. This is challenging because such a system is typically not ascribed to a wave function but only a reduced density matrix. However, in the cited works, it has been established that there is a precise way of assigning to it a pure state, known as its conditional pure state. 

Several universality (or typicality) results have been established, suggesting that if the environment is sufficiently large, a GAP measure can be attributed to the open system, and this measure corresponds to the thermal equilibrium measure associated with the canonical (or grand canonical) quantum ensemble. In the ensuing discussion, our exclusive attention will be on the GAP measure ${\uu}$ linked with the microcanonical density matrix, as it offers a notion of typicality for quantum closed systems, which is our main focus here.

\subsubsection{Useful Notions of `Most' for Quantum Systems}\label{sec:OBNs}
As we shall explain below, the notion of ``most'' expressed by $\uu$ can be naturally extended  from states to orthonormal bases,  unitary operators, Hamiltonians,  density matrices, and orthogonal decompositions \eqref{eq:orthoDecom} of the energy shell. For the sake of simplicity and consistency, we will refer to the uniform measures on these sets also as $\uu$. 

The uniform measure on the set of orthonormal bases  (ONBs) $ \{\phi_\alpha\}$ is constructed as follows: Begin by randomly selecting the first basis vector $\phi_1$ uniformly on $\bbs$ according to the measure $\uu$ on $\bbs$. Then, let $\Hilbert_1$ represent the orthogonal complement of $\phi_1$. Subsequently, choose $\phi_2$ uniformly on $\mathbb{S}(\Hilbert_1)$, and let $\Hilbert_2$ denote the orthogonal complement of the subspace spanned by $\phi_1$ and $\phi_2$, and so forth. In this way the notion of random  ONB  uniformly distributed is well-defined. 
.

The uniform probability distribution over the unitary group, consisting of unitary matrices on $\Hilbert_E$, is known as the Haar measure. It is the unique (up to a constant) normalized measure that remains invariant under multiplication by any fixed unitary matrix, whether from the left or from the right. There is a one-to-one correspondence between orthonormal bases (ONBs) and unitary operators, as each ONB corresponds to a unique unitary operator that transforms the standard basis into that ONB. Therefore, transferring $\uu$ accordingly to the group of unitary operators indeed yields the Haar measure on that group.

We will regard the eigenvalues for Hamiltonians or density matrices as fixed and consider the uniform measure for their eigenbasis, which (assuming non-degeneracy) forms an ONB. Therefore, $\uu$ can be transferred to Hamiltonians or density matrices in this manner.

For subspaces, we will regard the dimension $\dd$ as fixed; the measure over all subspaces of dimension $\dd$ arises from the measure on $\text{ONB}(\Hilbert)$ as follows. If the random orthonormal basis $\phi_1,\ldots,\phi_{\D}$ has uniform distribution, we consider the random subspace spanned by $\phi_1,\ldots,\phi_\dd$ and call its distribution uniform.

For decompositions \eqref{eq:orthoDecom}, denoted now $\decomp=\{\Hilbert_M\}$, of the system's Hilbert space, we will regard the number $K$ of subspaces as fixed, as well as their dimensions $\dd_M$; the measure over decompositions arises from the measure on ONBs as follows. Given the orthonormal basis $\phi_1,\ldots,\phi_{\D}$, we let $\Hilbert_M$ be the subspace spanned by those $\phi_\alpha$ with $\alpha\in J_M$, where the index sets $J_M$ form a partition of $\{1,\ldots,\D\}$; we also regard the index sets $J_M$ as fixed.

\subsubsection{Typicality of Macroscopic Thermal 
Equilibrium}
\label{sec:TMTE}
In quantum systems, macroscopic thermal equilibrium is characterized by several typical properties. These properties help us understand why thermal equilibrium is so prevalent. 

Firstly, a significant portion of the Bloch sphere is occupied by states within the energy shell which are in thermal equilibrium. Secondly, most mixed states in the energy shell display thermal behavior. Lastly, the majority of eigenstates of the Hamiltonian are also in thermal equilibrium. Let's now examine these properties in more detail.

\begin{itemize}  \item[1]{\em Most States in the Energy Shell are in} thermal equilibrium \cite{GLMTZ10a}. This conclusion arises from the application of \eqref{eq:thertyp2} and \eqref{eq:usefultr}:
\begin{equation}\label{eq:mostVectors}
\int \bra{\Phi}P_\text{eq}\ket{\Phi}\, \uu(d\Phi) = \frac{\Tr P_\text{eq}}{\D} = \frac{\dd_\text{eq}}{\D} \approx 1\,.
\end{equation}
Given that the quantity $\bra{\Phi}P_\text{eq}\ket{\Phi}$ is bounded above by 1, this implies that most states $\Phi\in\bbs$ must satisfy \eqref{eq:MATE}, establishing their inclusion in the thermal equilibrium set MATE defined in \eqref{eq:MATE2}. In other words,  the set $\text{MATE}$ occupies most of the normalized surface area of $\bbs$. More precisely the following inequality holds \cite{GHLT17}:
\begin{equation}
    \uu \left( \text{MATE} \right) \ge 1- \frac{\epsilon}{\delta}
\end{equation}
 where $\epsilon \ll \delta$. According to the numerical estimates provided in Sec.  \ref{sec:MATEQu}, almost all of the Bloch sphere is included in  the set MATE, with only a vanishingly small portion not belonging to it.
\end{itemize}
Two other crucial properties of macroscopic thermal equilibrium deserve emphasis \cite{GHLT2015, GHLT17}:

\begin{itemize}
    \item[2] \textit{Most Mixed States within the Energy Shell are in} MATE. ``Most''
    in this context refers to most ONBs in the sense specified in Sec.  \ref{sec:OBNs}, implying that the density matrices are decomposed with respect to a suitable orthonormal basis.

    \item[3] \textit{Most Eigenstates of $H$ are in} MATE. In other words, denoting as usual by $\phi_\alpha$, the eigenstates of $H$, we have that 
    \begin{equation}\label{eq:mosteigen}
        \scp{\phi_\alpha}{{P_\text{eq}}|\phi_\alpha} = 1-\epsilon \quad\text{ for most }  \alpha
    \end{equation}
    This is indeed a direct consequence of thermal equilibrium dominance. In fact, for any OBN $b_1, \ldots, b_{\WW_E}$ of $\Hilbert_E$, a fraction close to 1 (specifically, at least $1-\epsilon/\delta$, where $\epsilon \ll \delta$) of all basis vectors are in MATE, since 
$$  \frac{1}{\WW_E}  \sum_{n=1}^{\WW_E} \bra{b_n} P_\text{eq}\ket{b_n} = \frac{1}{\WW_E} \Tr P_\text{eq} > 1- \epsilon \,.$$
\end{itemize}

Property 3 is significant because it embodies a weak form of the Eigenstate Thermalization Hypothesis (ETH), according to which all eigenstates of $H$ (possibly within a specific range) are in thermal equilibrium, see below.
\bigskip

It should be noted that the concept of macroscopic thermal equilibrium discussed earlier is applicable to scenarios where many-body Anderson localization (MBL) occurs \cite{anderson1958, basko2008, huse2007}. In systems exhibiting MBL, the eigenfunctions of the Hamiltonian exhibit a form of spatial localization, which presents challenges in achieving thermal equilibrium. However, according to Property 3, the majority of the surface area of $\bbs$ is occupied by pure states that exhibit thermal behavior, even in cases where specific assumptions about the Hamiltonian $H$ are not relied upon.

\subsubsection{The Eigenstate Thermalization Hypothesis}
\label{sec:ETHT}
In 1994, Mark Srednicki introduced the eigenstate thermalization hypothesis (ETH) as a theoretical framework to explain the behavior of individual eigenstates within closed quantum systems characterized by a large number of degrees of freedom. In essence, ETH establishes specific conditions on the Hamiltonian $H = \sum_\alpha E_\alpha \ket{\phi_\alpha}\bra{\phi_\alpha}$ to ensure that all eigenstates $\phi_\alpha$ approach thermal equilibrium. 

In the context of macroscopic thermal equilibrium, ETH is defined as the condition that $\phi_\alpha \in \text{\sc MATE}$ for all $\alpha$. Equivalently, it is required that a stronger version of \eqref{eq:mosteigen} holds, namely,
\begin{equation}
\label{ETH_MATE}
\scp{\phi_\alpha}{P_\text{eq} | \phi_\alpha} \ge 1-\epsilon\;\; \text{for all } \alpha,.
\end{equation}

It is important to emphasize, as discussed in Sec.  \ref{sec:MATEQu}, that the fact that most eigenstates of $H$  are in macroscopic equilibrium is solely a consequence of the dominance of equilibrium. So, it is the distinction between ``most'' and ``all'' that makes the difference.

\subsubsection{Typicality of Macroscopic  ETH} 
While some Hamiltonians adhere to \eqref{ETH_MATE}, others diverge from it. However, it is noteworthy that {\em most}  Hamiltonians do indeed satisfy the ETH as expressed by \eqref{ETH_MATE}. This notable theorem has been rigorously proven by Goldstein et al. \cite[Lemma 1]{GLMTZ10a}, and when we say ``most'' Hamiltonians, we are referring to 
the notion of typicality expressed by the uniform measure $\uu$, as discussed in Sec. \ref{sec:OBNs}. 

The proof presented in \cite{GLMTZ10a} leverages methods from probability theory, by considering the eigenstates $\phi_\alpha$ as forming  a ``random'' basis uniformly distributed on the set of orthonormal bases in $\Hilbert_E$.  As an illustration of the tecniques, one step of the proof in \cite{GLMTZ10a} involves utilizing a Gaussian random vector $G = (G_1, \ldots, G_d)$ in $\Hilbert_E$, characterized by a mean of 0 and a covariance of $I$, to represent random points on the sphere $\bbs$. These techniques enable the authors to demonstrate that for a large system, equation \eqref{ETH_MATE} holds with a probability close to 1, thereby establishing the typicality of ETH.

It is important to appreciate that while probability theory is used to derive such results, there's no inherent randomness or genuine probability involved in this context. This is a ``typicality'' result, akin to the findings discussed in Sec. \ref{sec:TMTE}.

\subsubsection{Further Remarks on the Method of Typicality}
\label{sec:otmot}

We conclude our discussion on the typical properties of thermal equilibrium with some remarks, restating and elaborating on what has been discussed in Sec. \ref{se.tyvspr}.

Firstly, we observe that when we refer to most Hamiltonians $H$ or states $\Phi$ with respect to a normalized measure $\uu$, we're describing a scenario where these entities are assumed to be randomly chosen according to the distribution represented by $\uu$. However, this doesn't imply that in real-world situations, $H$ or $\Phi$ are inherently random, nor does it suggest that repeated experiments would yield a distribution close to $\uu$. Instead, this notion emphasizes that the sets of ``bad'' Hamiltonians or states have a very small measure with respect to $\uu$. This doesn't imply a specific probability distribution of Hamiltonians or states in concrete situations.

This method of tackling the study of complex real-world systems in quantum physics traces its roots back to von Neumann's work \cite{cite66}, further developed by Wigner in nuclear physics during the 1950s. According to this method, instead of becoming entangled in the intricacies of specific systems or focusing solely on simplified models, it strikes a balance. It examines the behavior of generic systems, avoiding the need to immerse itself in the intricate details of their unique Hamiltonians. Essentially, the method of typicality offers a way to gain insights into a wide range of systems without becoming bogged down in specific details.

Wigner's description aptly captures  the  essence of the method:
\begin{quote}
One deals with a specific system, complete with its proper (although often unknown) Hamiltonian, yet acts as if dealing with a multitude of systems, each with its own Hamiltonian, and then averages over the properties of these systems. This approach can be meaningful only if it turns out that the properties of interest are consistent across the majority of admissible Hamiltonians.\cite{wigner67}
\end{quote}

Wigner pioneered this method, which has played a pivotal role in generating unique predictions about the intricate properties of complex quantum systems in nuclear physics, particularly within the framework of his work on random matrices. Additionally, this method has proven valuable in exploring the path to thermal equilibrium and shedding light on the dynamics guiding a system toward such equilibrium.

Despite potential objections regarding the applicability of a certain property to a specific system, we maintain that a typicality theorem remains enlightening. At the very least, it helps differentiate between typical and exceptional behavior. Essentially, when a property turns out to be typical, it reasonably suggests that it may also apply to a specific system unless compelling reasons indicate otherwise.

\subsection{Beyond Boltzmann's Thermal  Equilibrium}
The discussion so far has underscored the resemblance between Boltzmann's concept of thermal equilibrium and its quantum counterpart. However, within the quantum domain, another notion of thermal equilibrium emerges, characterized by a distinctly microscopic flavor. This concept is rooted in quantum principles, particularly quantum entanglement, which lack classical counterparts. In the subsequent discussion, we will explore this new perspective on thermal equilibrium.

\subsubsection{Canonical Typicality}
\label{sec:cantyp}

In Sec.  \ref{sec:fmttce}, we revisited the well-established fact that a system, denoted as $S$, weakly coupled to a heat bath, labeled as $B$, conforms to the canonical ensemble characteristics when the composite system $S+B$ is governed by the microcanonical ensemble associated with a suitable energy shell. This principle holds for both classical distributions on the phase space and quantum density matrices. Notably, in the realm of quantum systems, an even more compelling assertion emerges which has no classical analog. 

It turns out that not only is the marginal of the microcanonical ensemble canonical, as expressed by Eq. \eqref{marginalcanonical}, but  a more robust statement holds true, namely that for the majority of $\Phi$ of the composite, the reduced density matrix of the system $\rho_S^\Phi$ is canonical.

To understand how this comes about, consider the framework of Sec.  \ref{sec:fmttce} where $S$ and $B$ are large, with $B$ being much larger than $S$. If $S+B$ is in the state $\Phi$, and an observable pertaining only to $S$ is measured, the statistics of the results are governed by the reduced density matrix $\rho^\Phi_{S} = \text{Tr}_B \ket{\Phi}\bra{\Phi}$ (see Eq. \eqref{eq:partrace}). While the quantum state $\Phi$ of the total system is pure, the state $\rho^\Phi_{S}$ of $S$ is a mixed state unless $\Phi$ factorizes, i.e., $\Phi= \psi\otimes \phi$. This stands in stark contrast to classical systems, where a subsystem is in a single microstate (and not a mixture of microstates) whenever the total system is in a single microstate.

Even if the two quantum states $\Phi$ and $\Phi'$ are significantly different, their respective reduced density matrices $\rho^\Phi_{S}$ and $\rho^{\Phi'}_{S}$ could be similar. A notable finding by Goldstein et al.~\cite{cite23} and Popescu et al.~\cite{PSW06} is that {\em nearly all} quantum states $\Phi\in \bbs$ within the energy shell of $S+B$ result in nearly identical reduced density matrices, denoted as $\rho_S^{\text{\sc Mc}}$. Here, $\rho_S^{\text{mc}}$ represents the reduced density matrix of the microcanonical ensemble $\rho_E$ for the entire system, i.e., $\rho_S^{\text{mc}} \equiv \text{Tr}_B \rho_E$.

This property, called \emph{canonical typicality}, amounts to the statement: for most $\Phi \in \bbs$,
\begin{equation}
\label{eq:cantyp}
 \rho^\Phi_S \approx \rho_S^{\text{mc}}   \,,
\end{equation}
with `most' defined by the uniform measure $\uu$ on $\bbs$. 

To express the sense of the  approximation in \eqref{eq:cantyp}, one needs to introduce a norm, and here for density matrices, it is appropriate to use the trace norm. It is noteworthy that the precise estimate found in \cite{PSW06}, based on Levy's Lemma, allows for a subsystem that is quite large, indeed almost half of the composite system!

The authors of \cite{cite23} focused on a conventional scenario involving a composite system $S+B$, where $S$ and $B$ are both large, but with $B$ significantly surpassing $S$ in size, allowing for the neglect of interactions between the two. Under these conditions, the expression for $\rho_S^{\text{mc}}$ in \eqref{eq:cantyp} transforms into $\rho_{S, \beta}$---the canonical density matrix of the subsystem $S$. Specifically, this corresponds to the canonical density matrix given by \eqref{eq:qcan}, where $H= H_{S}$ and $\beta$ is determined by $E$ in the standard manner, as discussed in Sec.  \ref{sec:fmttce}. This establishes that, for most $\Phi \in \bbs$,
\begin{equation}
\label{eq:cantyp1}
 \rho^\Phi_S\approx \rho_{S, \beta}  \,. 
\end{equation}

In this scenario, the strategy of the proof  is simpler: To establish Eq. \eqref{eq:cantyp1}, one needs to demonstrate that \eqref{star} holds, at least approximately, when $\rho_{S}$ is replaced by $\rho^{\Phi}_S $ for typical $\Phi\in \bbs$ \cite{cite23}.

Undoubtedly, canonical typicality  provides a  very robust justification for the universality of the  quantum canonical ensemble. It's worth noting that canonical typicality, as manifest in the proofs of \eqref{eq:cantyp} or \eqref{eq:cantyp1}, does not rely on any specific properties of the system dynamics, nor does it assume whether the spectrum of the Hamiltonian corresponds to an integrable or chaotic system. Only the largeness of the composite $S+B$ is relevant. 

As already noted, no analog of canonical typicality exists in classical mechanics for microstates. A classical microstate is represented by a point $X$ in phase space, specifying the positions and momenta for all particles. When considering a subsystem $S$, defined by a subset of labeled particles or the set of particles within a specific region $R$ of the available volume, its state is determined by the positions and momenta within $S$ and is represented by $X_S$ in the phase space of $S$, derived from the overarching point $X$.

Re-elaborating on the preceding: In classical mechanics, the reduced state of a pure state is always pure. However, in quantum mechanics, the reduced state of a pure state can be mixed. Specifically, $\rho_S^\Phi$ is pure if the joint state $\Phi$ factors, i.e., $\Phi = \psi \otimes \phi$. On the other hand, $\rho_S^\Phi$ is mixed whenever the system $S$ is entangled with its environment $B$. Quantifying the entanglement in a pure state $\Phi$ usually involves estimating the extent to which $\rho_S^\Phi$ is mixed. Therefore, the non-classical character of canonical typicality is intimately connected with a remarkable quantum phenomenon---namely, \textit{entanglement}.

\bigskip  

We note that elements of the argument supporting canonical typicality can be traced back to Schr\"odinger's 1952 work \cite{Sch1952}, where he discusses the reluctance to regard a system in thermal equilibrium as being in a sharp energy eigenstate. However, Schr\"odinger's exposition lacks an explicit connection between the assumption and typicality or a link to the reduced density matrix $\rho^\Phi$,  which he does not mention. Tasaki \cite{Tasaki1} delved into the reduced density matrix of a system coupled to a heat bath, where $S+B$ is described by a wave function $\Phi$. He demonstrates the canonical typicality of the long-time average of $\rho^{\Phi(t)}$ for a specific coupling Hamiltonian.  Gemmer and Mahler \cite{gm} have established canonical typicality by computing appropriate Hilbert space volumes, assuming very large degeneracy for energy eigenstates.

\subsubsection{Local Observables and Canonical Typicality}
\label{sec:loct}
An alternative perspective on canonical typicality relevant to the subsequent discussion views it as a local property intrinsic to large systems. More precisely, one can regard the subsystem  $S$ as the portion within a small spatial region $S$ of a closed, large system, with $B$ representing the complementary set denoted by $S^c$. In this context, $\rho^\Phi_{S}$ and $\rho_S^{\text{mc}}$ in \eqref{eq:cantyp} denote the reduced density matrices of $S$ obtained by tracing out the complement of $\ket{\Phi}\bra{\Phi}$ and $\rho_E$, respectively. They are expressed as $\rho_S^{\Phi} = \text{Tr}_{S^c} \ket{\Phi}\bra{\Phi}$ and $\rho_S^{\text{mc}} = \text{Tr}_{S^c} \rho_E$, respectively.

Canonical typicality then ensures that for most $\Phi \in \bbs$,
\begin{equation}
\label{eq:localcantyp}
 \langle\Phi| A_S |\Phi\rangle \approx \Tr \rho_E  A_S\,,   
\end{equation}
for every $S$-local observable $A_S =  A \otimes I_{S^c}$ related to $S$. This observation implies that a single pure quantum state can faithfully reproduce the results of the microcanonical ensemble when probing any small part of a large system. Utilizing this insight, one can develop efficient algorithms for simulating equilibrium states in quantum systems, see \cite{Sugiura13} and references therein.

The intriguing connection between canonical typicality and the emergence of thermal-like features at the small scales will be explored further below.

\subsubsection{Microscopic Thermal Equilibrium}
\label{sec:mte}

The notion of ``microscopic thermal equilibrium'' was formalized by Huse, Goldstein, Lebowitz, and Tumulka \cite{GHLT2015, GHLT17}. A similar concept found application in previous works, as exemplified by references \cite{cite48, Lin2009, cite53, Lyc2013, Nand2015}.

This concept relies on canonical typicality, in the form presented in Sec.  \ref{sec:loct}, asserting that for a subsystem within a spatial region $S $  of a closed, large system, the majority of states $\Phi\in \bbs$ satisfies \eqref{eq:cantyp}. For clarity, we restate it here:
\[\rho_S^{\Phi} \approx \rho_S^{\text{mc}}\,, \]
where $\rho^\Phi_{S}$ and $\rho_S^{\text{mc}}$ denote the reduced density matrices of $S$ obtained by tracing out $S^c$ in the pure state $\ket{\Phi}\bra{\Phi}$ and the microcanonical density matrix $\rho_E$, respectively.

We recall that for sufficiently small $S$, neglecting interactions between $S$ and its complement, the reduced density matrix $\rho_S^{\text{mc}}$ assumes a thermal character, expressed by \eqref{eq:cantyp1}, where
\begin{equation}\label{eq:firteqmc}
    \rho_S^{\text{mc}}= \rho_{S, \beta} \quad \text{with} \quad \rho_{S, \beta} = \frac{1}{Z_S} e^{-\beta H_S}.
\end{equation}

However, for sufficiently small $S$, it becomes inconsequential whether one starts from the microcanonical density matrix $\rho_E$ of the large system or its canonical density matrix $\rho_\beta$, given by \eqref{eq:qcan}, where $H$ is the Hamiltonian of the large system. This is a manifestation of the equivalence of ensembles and can be expressed by redefining $ \rho_{S, \beta}$  as 
\begin{equation}\label{eq:cangeneral}
    \rho_{S, \beta}= \Tr_{S^c} Z^{-1} e^{-\beta H}
\end{equation}
for suitable $\beta > 0$. Thus we
 characterize  $S$ to be ``sufficiently small" if 
\begin{equation} \label{eq:firteqmc2}
    \rho_S^{\text{mc}}\approx \rho_{S, \beta}
\end{equation}
with $\rho_{S, \beta}$ given by \eqref{eq:cangeneral}. The term ``canonical" is assigned to the density matrix $\rho_{S, \beta}$ given by \eqref{eq:cangeneral}, whether or not the interaction can be neglected.

If $S$ is sufficiently small according to \eqref{eq:firteqmc2}, then canonical typicality implies 
\begin{equation}\label{eq:cantypreloaded}
    \rho_S^{\Phi} \approx \rho_S^{\text{mc}}\approx \rho_{S, \beta}\,.
\end{equation}
This, in turn,  implies that for most $\Phi$, the $S$-subsystem is governed by a canonical density matrix, behaving ``thermally" with respect to the measurement of $S$-local observables, as expressed by \eqref{eq:localcantyp}. If \eqref{eq:firteqmc2} holds for the $S$-subsystem, the same is true for every smaller subsystem $S'$ contained in $S$, determined by taking another partial trace. There is nothing special about the $S$-subsystem except its linear size, say its diameter $\ell$. If \eqref{eq:firteqmc2} holds for $S$, it also holds simultaneously for all subsystems with diameter $\le \ell$.

Now, we can precisely define microscopic thermal equilibrium with all the essential elements in place. Let $\ell_0$ be the largest $\ell$ small enough to ensure that \eqref{eq:firteqmc2} holds for every subsystem $S$ with diameter $\le \ell_0$. MITE is then defined as the set of states $\Phi\in \bbs$ such that \eqref{eq:firteqmc2} holds at the scale $\ell_0$. In other words, a state $\Phi$ is in  MITE if for every subsystem of diameter $\ell_0$ or smaller, the reduced density matrix is close to the thermal equilibrium reduced density matrix. Explicitly,
\begin{align}\label{eq:subsetMITE}
    \text{MITE} &= \bigcup_{S : \text{diam}(S) \leq \ell_0} \left\{ \Phi \in \bbs : {\rho}^\Phi_S \approx 
    \rho_{S}^\text{mc} \right\}
\end{align}
The careful choice of the scale $\ell_0$ is important for the system in question, according to the strength of the interactions among its distinct components on that specific scale. Essentially, microscopic equilibrium implies being in thermal equilibrium on the relevant microscopic length scale. As a practical value, for example, one may take $\ell_0 \approx 10^{-3} \text{diam}(V)$, where $V \subset \mathbb{R}^3$ is the volume of the whole system \cite{GHLT2015, GHLT17}. 

\subsubsection{Relations between Macro and Micro Thermal Equilibrium}

The implication from microscopic to macroscopic equilibrium for macroscopic systems, $\text{MITE} \subset \text{MATE}$,  can be understood as follows. Considering that macro observables result from sums or averages of local observables over spatial cells (e.g., with a characteristic length $a$), it becomes evident that, once $a \leq \ell_0$, states exhibiting thermal behavior for micro observables (i.e., yielding the same probability distribution over the spectrum as $\rho_E$) also exhibit thermal behavior for macro observables. This category encompasses those in MITE. The condition $a \leq \ell_0$ implies that $\rho_S \approx \rho_E$ holds, at least up to the length scale of the macro observables. This condition is frequently met; for instance, considering a cubic meter of air at room conditions, for  $\ell_0 \approx 10^{-3}$ m, a realistic choice for $a$ could be $a \approx 10^{-4}$ m.


We will now  summarize the differences and similarities between the two notions of thermal equilibrium; for a detailed analysis, see \cite{GHLT2015, GHLT17}.

Whether macroscopic or microscopic, most states within the energy shell are in equilibrium. In macroscopic equilibrium, states align closely with equilibrium values derived from the microcanonical ensemble, with very small quantum fluctuations of macro variables ensuring that a single measurement almost surely yields the equilibrium value.

Conversely, in microscopic equilibrium, the expectation values of local operators closely approximate their corresponding equilibrium values. Although quantum fluctuations may not be negligible, repeated measurements of a microscopic observable result in an average that matches the equilibrium value. All states in microscopic equilibrium are also in macroscopic equilibrium, suggesting that the latter is stronger due to its stricter requirements regarding both expectation values and quantum fluctuations.

The validity of these equilibrium concepts is contingent upon system size. Macroscopic equilibrium is meaningful only for large systems, as thermal equilibrium dominance requires equilibrium fluctuations ($\sim \sqrt{N}$) to be much smaller than the mean value ($\sim N$). In contrast, microscopic equilibrium indicates equilibrium across all observables for any subsystem with a diameter less than or equal to a certain threshold. It is meaningful even for small systems (\(N \sim 10\)) with a weaker condition, as these numerical estimates follow from the basic estimates establishing the validity of canonical typicality \cite{PSW06}. For a thorough explanation, see \cite{GHLT2015, GHLT17}.\footnote{Indeed, Saito, Takesue, and Miyashita \cite{SaitoEtAl} examined the system-size dependence of quantum dynamics in a spin system and found that \(N = 8\) is enough to achieve a good agreement between the long-time average of the expectation value of a local quantity and the prediction by equilibrium statistical mechanics.}

\subsubsection{Unified Framework for Thermal Equilibrium}
\label{sec:uften}
The different notions of thermal equilibrium can be seen as specific instances of the following framework \cite{GHLT17,GLMTZ10a}:

{\em A state $\Phi$ in $\mathcal{H}_E$ is considered to be in thermal equilibrium relative to a given family $\mathcal{A}$ of observables if, for each $A \in \mathcal{A}$, the probability distribution over the spectrum of $A$ as defined by $\Phi$ is approximately equivalent to that defined by the microcanonical density matrix $\rho_E$, i.e.,}
\begin{equation}\label{eq:phiophieq}
\langle \Phi | A | \Phi \rangle \approx \text{Tr}(\rho_E A), \text{ for all } A \in \mathcal{A}.
\end{equation}

Macroscopic equilibrium corresponds to the family $\OO= \OO_\text{MATE}$  of  macroscopic observables $\{ \MMO_1, \ldots , \MMO_K \}$, defining the decomposition \eqref{eq:orthoDecom} of $\mathcal{H}_E$ into macrostates.  Note that verifying the validity of \eqref{eq:phiophieq} is sufficient to check macroscopic equilibrium \eqref{Phiinnu1}, i.e., that $ \langle \Phi | P_{\text{eq}} | \Phi \rangle \approx 1$.

Microscopic equilibrium arises for the family  \begin{equation}\label{familymite}\OO= \OO_\text{MITE} = \bigcup_S \OO_S \, \end{equation}
with the union taken over all spatial regions $S$ of diameter $\leq \ell_0$ and $\OO_S$ is the set of all self-adjoint operators on $\Hilbert_S$, more precisely
\[
\OO_S = \left\{ A \otimes {I}_{S^c}  :  A \text{ self-adjoint on } \Hilbert_S \right\}
\]
(with ${I}_{S^c}$ the identity operator and $S^c$ again the complement of $S$). In that case, \eqref{eq:phiophieq} becomes 
\begin{equation}\label{eq:phiophieq2}
\Tr \rho_S^\Phi A \approx \Tr \rho_S^\text{mc}  A\,.
\end{equation}
In this sense, macroscopic thermal equilibrium is relative to the macro observables, whereas microscopic thermal equilibrium is relative to all local observables concerning any $S$ of diameter $\leq \ell_0$. The latter observables include those of a more microscopic and local nature.

Although only briefly sketched in this review, other characterizations of thermal equilibrium fit within the framework discussed above. In particular, Tasaki's approach \cite{Tas2016} parallels macroscopic equilibrium but accommodates noncommutative algebras of observables. Von Neumann's concept of thermal equilibrium, outlined in Sec \ref{sec:vNNTE}, aligns with macroscopic equilibrium in terms of algebraic structure but requires projectors onto subspaces $\Hilbert_M$ to meet the conditions \eqref{vonNcondi}. Reimann's proposal, outlined in \cite{Reim2015}, involves a typical observable \(A\) and shares affinities with von Neumann's method, employing an orthogonal decomposition \eqref{eq:orthoDecom} derived from the eigenspaces of the single observable \(A\).

\section{Approach to Thermal Equilibrium}
\label{sec:ATTE}

In the preceding section, we arrived at a conclusive response to the central question raised in the introduction: What does it signify, from a microscopic viewpoint, for a system to be in thermal equilibrium? As elucidated, from a Boltzmannian, individualist perspective, it means that the pure state characterizing the system resides within a specific subset of the energy shell. Although various subsets correspond to different interpretations of thermal equilibrium, they are interconnected; for instance, as previously discussed, attaining microscopic equilibrium implies macroscopic equilibrium.

Exploring the path to equilibrium for an individual system, characterized by a specific pure state at any given moment, involves delving into several related aspects: understanding the circumstances under which a system initially outside the equilibrium subset transitions into that subset over time, determining the speed at which it approaches equilibrium, and evaluating the duration for which it remains in that state. Addressing these issues requires a detailed examination of various aspects of the evolution of a large closed system over time. 

Initially, the system's microstate resides within a specific macrostate. As time progresses, the system undergoes transitions among different macrostates, driven by the temporal evolution of microstates. These changes in macroscopic variables are typically governed by autonomous phenomenological equations, which include kinetic and hydrodynamical equations. A paradigmatic example is Boltzmann's equation for the distribution function of a dilute gas. A notable reference for this discussion, covering both classical and quantum cases, is Volume 10 on Physical Kinetics in the renowned course of Theoretical Physics by Landau and Lifshitz \cite{la81}. These irreversible phenomenological equations yield stationary solutions corresponding to thermal equilibrium states and provide information on the rates of thermalization. Deriving these equations from microscopic dynamics provides invaluable insights into a system's progression toward thermal equilibrium.

A complementary perspective on examining the approach to thermal equilibrium---less pragmatic and more foundational in nature---involves exploring the overarching emergence of thermodynamic principles from the temporal evolution of microstates. The aim is to discern the general conditions under which this emergence occurs. This is the perspective that we shall explore in this section.

Before proceeding with the analysis, it's useful to highlight two fundamental factors that elucidate thermalization and irreversible behavior in macroscopic systems: the abundance of degrees of freedom and the prevalence of an equilibrium subset of microstates encompassing almost the entire energy shell. These factors hold relevance in both classical and quantum contexts. Interestingly, the quantum perspective offers an even more robust explanation for macroscopic irreversibility within the Boltzmannian framework.

In classical systems, thermalization is expected to occur for most initial non-equilibrium microstates, while in the quantum realm, our current understanding suggests that it occurs for all initial non-equilibrium microstates.

\subsection{Thermalization in Large Quantum Systems }
\label{sec:Thermalization}
The most fundamental principle of thermodynamics asserts that any closed macroscopic system initially departing from equilibrium will eventually progress to a state of thermal equilibrium and predominantly persist in that state over time. This phenomenon is commonly known as the approach to equilibrium or thermalization. 

\new{In the classical setting,  a classical phase point, under Hamiltonian evolution, spends most of its time within the equilibrium region of phase space, provided this region dominates the measure. In the quantum setting, the situation is similar in spirit, yet more subtle in its mathematical realization. For large systems, the quantum state typically evolves toward a configuration that remains extremely close—though not exactly confined—to the equilibrium subspace in the Hilbert space, in the sense of the $L^2$ norm. That is, the overlap between the evolving state and the equilibrium subspace becomes very large and remains so for the vast majority of times. This reflects the concentration of measure in high-dimensional Hilbert spaces and supports the interpretation of thermal equilibrium as a typical property, much like in the classical case.}

\new{Before delving more deeply into the mechanisms and mathematical structure behind thermalization, we now briefly recall some basic elements of quantum dynamics that will be needed in the sequel.}

\subsubsection{Dynamics of Closed Quantum Systems}
As previously noted, the temporal evolution of the system is depicted by a trajectory $\Phi \mapsto \Phi_t$ on the Bloch sphere $\bbs$, arising from a unitary evolution within the microcanonical shell $\mathcal{H}_{E}$,
\begin{equation}\label{eq:basuniphi}
\Phi_t = U_t \Phi_0, \quad U_t = e^{-iHt}
\end{equation}
for simplicity, we have adopted $\hbar=1$, and $\Phi$ is the initial state at $t=0$. In the ONB $\{\phi_\alpha\}$  diagonalizing $H$, $H\phi_\alpha = E_\alpha \psi_\alpha$ (assuming non degeneracy) \eqref{eq:basuniphi} becomes
\begin{equation}\label{eq:basuniphi2}
\Phi_t = \sum_\alpha c_\alpha  e^{-iE_\alpha t} \phi_\alpha
\end{equation}
where $c_\alpha= \langle\phi_\alpha| \Phi_0\rangle$.
Given that $|\langle\phi_\alpha| \Phi_t\rangle| = |c_\alpha|$, it is evident that these moduli remain unchanged as time progresses. Within the quantum state space $\bbs$, surfaces characterized by fixed moduli $|c_\alpha|$ and arbitrary phases take the form of tori. These tori exhibit invariance under the system's dynamics. Importantly, in cases where the Hamiltonian operator $H$ eigenvalues are rationally independent, which is typically the scenario, the dynamics on each torus display ergodic behavior. This well-established fact, see, e.g., \cite{Arnold}, underscores the ergodic nature of these tori within $\bbs$.
To sum up, the dynamics of a quantum state are similar to those of a classical, completely integrable linear system. 

Moreover, it's worth noting that due to the compact nature of the state space $\bbs$, the Poincaré recurrence theorem holds. Formulated by Henri Poincaré for classical systems, this theorem also has significant implications in the context of our discussion. It states that a closed quantum system will inevitably return arbitrarily close to its initial state after a sufficiently long time---many eons, Boltzmann would say. This recurrence property provides insights into the long-term behavior of quantum states within $\bbs$.

\subsubsection{Unified Framework for Thermalization}
\label{sec:Ufft}

Given the diverse characterizations of thermal equilibrium, spanning macroscopic, microscopic, and von Neumann's perspectives, thermalization manifests in various forms. Building upon the unified framework expounded in Section \ref{sec:uften}, we define thermalization relative to a specified family $\mathcal{A}$ of observables that characterize thermal equilibrium. In this context, we posit that thermalization relative to $\mathcal{A}$ ensues when equation \eqref{eq:phiophieq} holds true for $\Phi = \Phi_t = U_t \Phi_0$ for most times $t$ within the interval $[0, T]$, where $T$ is arbitrarily large and the initial state $\Phi_0$ is not necessarily in equilibrium. In essence, thermalization relative to $\mathcal{A}$ signifies that:
\begin{equation}\label{eq:phiophieq1}
\langle \Phi_t | A | \Phi_t \rangle \approx \text{Tr}(\rho_E A), \text{ for all } A \in \mathcal{A} \text{ and most } t \in [0, T]
\end{equation}

Two related questions immediately arise: for which Hamiltonians $H$, and for which class of initial microstates $\Phi_0$, does thermalization relative to $\mathcal{A}$ occur? In the subsequent sections, we will delve into the state-of-the-art research concerning these questions. However, before delving into specifics, let's elucidate the general strategy employed to tackle this problem.

Note that the term ``most" with respect to $t$ in \eqref{eq:phiophieq1} is relative to the Lebesgue measure on $[0, T]$. The condition requires that the approximate equality holds for a fraction of time that tends to 1 as $T$ tends to infinity. Therefore, studying time averages appears to be a promising approach to the problem.

Let $\overline{f(t)}$ denote the time average of a time-dependent quantity $f(t)$, defined as
\begin{equation}
\overline{f(t)} = \lim_{T\to\infty} \frac{1}{T} \int_0^T dt  f(t).
\end{equation}
Then  thermalization relative to $\mathcal{A}$ will occur if the following conditions are satisfied:
\begin{align}
&\bullet\quad  \overline{\langle \Phi_t | A | \Phi_t \rangle} \approx \text{Tr}(\rho_E A) \label{eq:condiergo} \\
&\bullet\quad \text{The time variance of } \langle \Phi_t | A | \Phi_t \rangle  \text{ is small.} \label{eq:condivaria}
\end{align}
In fact, if the time average approaches the thermal value and the time variance is small, then the value must closely resemble the thermal value most of the time, leading to \eqref{eq:phiophieq1}. 

Much of the research on thermalization has focused on verifying the validity of these conditions and determining the corresponding requirements on the Hamiltonian and initial conditions. A very weak condition on the Hamiltonian is that it be non-degenerate (which is the generic case); then, $E_\alpha - E_\beta$ vanishes only for $\alpha = \beta$, a condition that will be assumed throughout the following discussion.

 \subsubsection{ETH and Macroscopic Thermalization}\label{sec:smte}
For brevity, let's denote thermalization relative to macroscopic thermal equilibrium as macroscopic thermalization (Sec. \ref{sec:MATEQu}). In this context, it suffices to consider the observable $A= P_\text{eq}$, the projector onto $\Hilbert_\text{eq}$. This specific case has been thoroughly investigated in \cite{GLMTZ10a}, and here we summarize their findings.

In this case, condition \eqref{eq:condivaria} is unnecessary. In fact, given that $\langle \Phi_t|P_\text{eq}| \Phi_t\rangle$ always yields a real number between 0 and 1, if its time average approaches 1, it implies that it must be close to 1 for most of the time. Furthermore, the following result holds: for most $\Phi_0$ (with respect to the uniform measure $\uu$), $\Phi_t$ is mostly in thermal equilibrium. This result is derived from Fubini's theorem on the order of iterated integrals, which indicates that taking the $\uu$-average commutes with taking the time average. Also, it exploits the unitary invariance of $\uu$. In mathematical terms, this can be expressed as:
\begin{align*}
\int \overline{\langle \Phi_t|P_\text{eq}| \Phi_t\rangle} \, \uu(d\Phi_0) &= \overline{\int \langle\Phi_0|e^{-iHt}P_\text{eq}e^{iHt}|\Phi_0\rangle \, \uu(d\Phi_0)} \\
&=\int \langle\Phi_0|P_\text{eq}|\Phi_0\rangle\,\uu(d\Phi_0) \approx 1\,,
\end{align*}
since most $\Phi_0$ are in thermal equilibrium, as expressed by \eqref{eq:mostVectors}. 

The ensemble average of the time average is near 1, so, for most $\Phi_0$, the time average must be near 1, which implies the claim above. Therefore the interesting question is about the behavior of exceptional $\Phi_0$, e.g., of systems which are not in thermal equilibrium at $t=0$. Do they ever go to thermal equilibrium?

From \eqref{eq:basuniphi2}, it follows that 
\begin{align}
\overline{\langle \Phi_t|P_\text{eq}| \Phi_t\rangle} 
&= \sum_{\alpha,\beta} \overline{e^{i(E_\alpha-E_\beta)t}} 
\, c^*_\alpha \, c_\beta
\langle\phi_\alpha|P_\text{eq}|\phi_\beta\rangle\,.
\end{align}
For non-degenerate $H$, the time averaged exponential is $\delta_{\alpha\beta}$, and
\begin{equation}\label{evalave}
\overline{\langle \Phi_t|P_\text{eq}| \Phi_t\rangle} 
= \sum_{\alpha=1} \bigl|c_\alpha\bigr|^2 
\langle\phi_\alpha|P_\text{eq}|\phi_\alpha\rangle\,.
\end{equation}
Thus for thermalization, it is both necessary and sufficient that the right-hand side of \eqref{evalave} closely approaches 1. As discussed in item 3 of Sec.  \ref{sec:TMTE}, the majority of eigenstates of $H$ satisfy the condition $\langle\phi_\alpha|P_\text{eq}|\phi_\alpha\rangle \approx 1$. This ensures macroscopic thermalization for most initial states $\Phi_0$.
However, most states in the system also exhibit macroscopic thermal equilibrium behavior (as noted in item 1 of the same section). Consequently, this result may not offer significant insights into the approach to equilibrium of initial non-equilibrium microstates.

However, it's important to note that if an energy eigenstate $\phi_\alpha$ is not in thermal equilibrium itself, the system remains out of thermal equilibrium when $\Phi_0=\phi_\alpha$ because this state is stationary. Conversely, when the condition
\begin{equation}\label{eq:good}
\langle\phi_\alpha|P_\text{eq}|\phi_\alpha\rangle\approx 1 \quad \text{holds for all }\alpha,
\end{equation}
is satisfied, the system tends to be in thermal equilibrium most of the time, regardless of the initial state $\Phi_0$. 
This observation directly follows from \eqref{evalave}, as the right-hand side represents an average of the quantities $\langle\phi_\alpha|P_\text{eq}|\phi_\alpha\rangle$. Thus we conclude that  
\begin{equation}\label{macrotremaliz}
    \langle \Phi_t|P_\text{eq}| \Phi_t\rangle \approx 1
\end{equation}
for most $t$. Note that Eq. \eqref{eq:good} corresponds to the eigenstate thermalization hypothesis (ETH) within the context of macroscopic equilibrium (see Sec. \eqref{ETH_MATE}). Therefore, if ETH \eqref{eq:good}  holds true, macroscopic thermalization occurs for all initial states $\Phi_0$, highlighting ETH's crucial role as a thermalization condition.

\subsubsection{Macroscopic Thermalization Typicality}\label{sec:smte2}

As expounded in Sec. \ref{sec:ETHT}, macroscopic ETH is typical, meaning it is satisfied for a typical Hamiltonian. Consequently, we can confidently conclude that 
for most Hamiltonians every quantum state tends to spend most of its time in macroscopic thermal equilibrium, regardless of its initial state. This significant observation has been rigorously substantiated by Goldstein et al. in \cite{GLMTZ10a}, making clear the notion that thermalization is a typical property of macroscopic systems.\footnote{This work is heavily influenced by von Neumann's insightful framework \cite{cite66}, but it diverges from his approach by assuming the dominance of thermal equilibrium, a physically motivated assumption.} 

Thus, for a typical system thermalization occurs for all initial states. On the contrary, a contrasting scenario unfolds in the case of atypical systems manifesting  ETH violation. It becomes conceivable that no state beyond the equilibrium set \eqref{ETH_MATE} ever converges towards thermal equilibrium. To exemplify this scenario, one may consider a Hamiltonian $H$ deliberately structured such that every eigenstate can be classified into one of two distinct groups: either it pertains to the equilibrium subspace $\Hilbert_\text{eq}$, representing states inherently in thermal equilibrium, or it remains entirely orthogonal to this subspace.

As underlined in \cite{GHLT17}, this is especially prominent in certain systems that exhibit a phenomenon known as Many-Body Localization (MBL). In MBL systems, the conventional process of thermalization, as described by ETH, is dramatically disrupted. Instead, a unique and non-equilibrium behavior prevails: all states outside the equilibrium set \eqref{ETH_MATE} stay outside of it indefinitely, with no observable fluctuations towards thermal equilibrium. In contrast, all states within the equilibrium set remain in this state indefinitely, exhibiting remarkable stability and an absence of deviations from thermal equilibrium. 

\subsubsection{ETH and Microscopic Thermalization}
Microscopic thermalization is characterized by the fulfillment of the condition \eqref{eq:phiophieq1} for the family of observables   $\mathcal{A}= \mathcal{A}_\text{MITE}$ defined in \eqref{familymite}. According to our discussion in Sec. \ref{sec:ETHT}, microscopic ETH is the condition  that all eigenstates $\phi_\alpha$ of the Hamiltonian $H$ belong to the thermal equilibrium subset MITE as defined in Eq. \eqref{eq:subsetMITE}. 

Several findings \cite{Rig2008, cite48, Lin2009, cite49} rely on microscopic ETH assumptions. They further assume that the Hamiltonian is non-degenerate and features non-degenerate energy gaps, or ``no resonance,'' meaning 
\begin{equation}
\label{energygaps}
E_\alpha-E_\beta \neq E_{\alpha'} - E_{\beta'}\text{ unless }
\begin{cases}\text{either }\alpha=\alpha', \beta=\beta'\\
\text{or }\alpha=\beta, \alpha'=\beta'\,.\end{cases}
\end{equation}
This condition is typically fulfilled in generic scenarios.

In this regard, we shall mention  two results. The first, established in \cite{cite48, Lin2009}, demonstrates that if microscopic ETH holds, then microscopic thermalization occurs for \emph{most} initial states $\Phi_0$.

The second result, demonstrated in \cite{Rig2008}, establishes that microscopic thermalization occurs for all initial states $\Phi_0$, rather than just the majority. However, this hinges on the satisfaction of an additional condition: Srednicki’s extension \cite{Sred96,Sred99} of the ETH to off-diagonal elements, as expressed by
\begin{equation}\label{eq:sredex}
\langle\phi_\alpha|A|\phi_\beta\rangle \approx 0, \text{ for } ,\alpha \neq \beta \text{ and all } A \in \mathcal{A}_\text{MITE}
\end{equation}
(see also \cite{cite50}.)

The proof strategy for the second result revolves around demonstrating the validity of Equations \eqref{eq:condiergo} and \eqref{eq:condivaria}. The fulfillment of Condition \eqref{eq:sredex} is instrumental in establishing the second equation, namely, the smallness of the time variance. However, the computation involved in proving this is rather lengthy, and we shall not present it here. Instead, we refer interested readers to the original literature. For a detailed proof of the validity of Equations \eqref{eq:condiergo} and \eqref{eq:condivaria} using the formalism adopted here, please see \cite{GHLT17}.

\subsubsection{Von Neumann's  Normal Typicality}
\label{sec:vonneumantyp}

In 1929, von Neumann \cite{cite66} established two theorems known as the ``quantum $H$ theorem'' and the ``quantum ergodic theorem'' (QET). However, contrary to their names, these theorems are not quantum analogs of either the $H$ theorem or the ergodic theorem. In fact, the two 1929 theorems bear a strong resemblance to each other, almost serving as reformulations of one another, and they both demonstrate thermalization within von Neumann's equilibrium framework outlined in Sec. \ref{sec:vNNTE}.

In essence, the Quantum Ergodic Theorem (QET) asserts that for a typical Hamiltonian or a typical choice of the decomposition $\mathcal{H}_E=\oplus_M\mathcal{H}_M$, all initial pure states $\Phi_0$ evolve such that for most of time   in the long run,  $\Omega_M/\Omega_E$, i.e., 
\begin{equation}\label{vNtheo12}
\| P_M \Phi_t\|^2 \approx \frac{\Omega_M}{\Omega_E},.
\end{equation}
This assertion, termed ``normal typicality" in \cite{GLMTZ10}, entails that for typical large systems, every initial wave function $\Phi_0$ behaves ``normally." Specifically, it evolves in such a way that the projector $\ket{\Phi_t}\bra{\Phi_t}$, for most times $t$, is macroscopically equivalent to the micro-canonical density matrix.

As discussed in Sec. \ref{sec:vNNTE}, if we assume the dominance of thermal equilibrium, then von Neumann's concept of equilibrium aligns with the notion of macroscopic equilibrium. In this case, when $M=M_\text{eq}$, Eq. \eqref{vNtheo12} reduces to Eq. \eqref{macrotremaliz}, emphasizing macroscopic equilibrium as the more natural definition in this case. This observation likewise extends to the concept of thermalization.

Von Neumann's proof demands significantly more effort and refined methods than the proof of macroscopic thermalization's typicality provided in \cite{GLMTZ10a}. Furthermore, his proof necessitates the ``no resonance" condition \eqref{energygaps}, which becomes unnecessary if the dominance of thermal equilibrium is assumed.

Despite von Neumann laying down the basic framework and tools for subsequent research (including \cite{GLMTZ10a}), arguably,  his conception of thermal equilibrium is not satisfactory from a physical point of view, as noted in Sec. \ref{sec:vNNTE}. After all, the dominance of thermal equilibrium is fundamental in statistical mechanics, and neglecting this aspect is notably peculiar. Nevertheless, his contributions to quantum statistical mechanics are enduring; for further exploration, refer to \cite{GLTZ10} and \cite{GLMTZ10}. In the latter work, a stronger statement of normal typicality is formulated, based on the observation that the bound on deviations from the average specified by von Neumann is unnecessarily coarse. Instead, his proof derives a much tighter and more relevant bound.

\subsubsection{Perspectives on Quantum  vs. Classical Thermalization}

We pause to make some remarks. Equilibrium statistical mechanics relies heavily on the concept of thermalization. However, it's important to note that classical and quantum systems' mechanisms driving thermalization differ. As a consequence, distinct justifications exist for the use of equilibrium ensembles in each scenario.

In classical mechanics, thermalization is expected in realistic systems with a sufficiently large number of constituents: for every macro-state $M$, most initial phase points $X\in\Gamma_M$ will result in the time-evolved phase point $X_t$ spending the majority of its time in the set $\Gamma_{eq}$. While this assertion follows if the system is ergodic, it is actually much weaker than ergodicity. Quantum macroscopic thermalization mirrors this concept by implying, for typical Hamiltonians, that initial states $\Phi_0$ out of thermal equilibrium will spend most of their time in thermal equilibrium. However, it differs in that it applies to \emph{all}, rather than most, initial states $\Phi_0$.

Moreover, quantum macroscopic thermalization entails that, for a typical Hamiltonian $H$, the time evolved state $\Phi_t$ of every initial state $\Phi_0$ satisfies \begin{equation}\label{psitMi}
\scp{\Phi_t}{A|\Phi_t} \approx \Tr(\rho_E A),,
\end{equation} for most times  $t$ in the long run. In this equation, $\rho_{E}$ denotes the standard micro-canonical density matrix, and \eqref{psitMi} is valid for all macroscopic observables $A$ from the family $\mathcal{A}_\text{MATE}$, defined by the decomposition \eqref{eq:orthoDecom} of $\mathcal{H}_E$ into macrostates. This validates the replacement of $\ket{\Phi_t}\bra{\Phi_t}$ with $\rho_E$ for macro-observables in equilibrium.

In the classical case, discussed in Sec. \ref{sec:eesm}, the dominance of the macrostate ${M}_\text{eq}$ implies that all macroscopic variables $M_j = \hat{M}_j (X)$ are nearly constant functions on $\Gamma_E$. In thermal equilibrium, macro variables stabilize around their equilibrium values $ M_j$. Since $\Gamma_{\text{eq}}$ is approximately equal to $\Gamma_E$, the equilibrium value $M_\text {eq}$ of a macrovariable can be accurately calculated using the microcanonical ensemble average $\langle \hat{M}_j\rangle_\text{mc}$.

However, in the quantum case, the substitution of $\ket{\Phi_t}\bra{\Phi_t}$ with $\rho_E$ is not justified for observables $A$ outside the family $\mathcal{A}_\text{MATE}$. For instance, consider a microscopic observable $A$ that does not exhibit nearly constant behavior on the energy shell $\mathcal{H}$. While standard equilibrium statistical mechanics suggests using $\rho_E$ for the expected value of $A$ in equilibrium, similar to the use of $\langle \hat{M}_j\rangle_\text{mc}$ for a variable not nearly constant on the energy shell in the classical case, \textit{quantum macroscopic thermalization, by itself, does not cover this situation}. As discussed earlier, addressing observables of this nature necessitates the fulfillment of another notion of thermal equilibrium, namely microscopic equilibrium. Microscopic thermalization \eqref{psitMi} holds for suitable $A\in \mathcal{A}_\text{MITE}$ and suitable $\Phi_0$, as discussed earlier (indeed, for all $\Phi_0$ under Srednicki’s extension of the ETH to off-diagonal elements). These results are specific to quantum mechanics, whereas the justification for the widespread use of $\rho_E$ in classical statistical mechanics stems from different considerations.

Another noteworthy difference between classical and quantum scenarios involves comparing the subspaces $\mathcal{H}_M$ with the regions $\Gamma_M$. In classical mechanics, each phase point in $\Gamma$ corresponds to one and only one $\Gamma_M$, thereby falling into a single macro-state.  Conversely, a quantum state $\Phi$ may not belong to any specific $\mathcal{H}_M$, often embodying a non-trivial superposition of vectors across various macro-states. This distinction is why we defined equilibrium as being within a neighborhood of $\Hilbert_\text{eq}$ rather than within $\Hilbert_\text{eq}$ itself.

\subsubsection{The Challenge of Thermalization Time Scales}\label{sec:ctts}

Goldstein, Hara, and Tasaki have raised a significant concern regarding the typical behavior of macroscopic thermalization \cite{Har2013,Har2014,GHT14b}. The crux of the issue lies in the disparity between theoretically predicted behavior and real-world observations.

Their studies have revealed significant variability in the relaxation time, ranging from exceedingly long to surprisingly short, depending on the specific system under investigation. Contrary to expectations, they found that the relaxation time is unexpectedly short in scenarios with randomly chosen nonequilibrium subspaces, typically of the order of the Boltzmann time (see below). This suggests that the gradual decay commonly observed in reality is not a prevailing characteristic of randomly chosen nonequilibrium subspaces. It underscores the necessity of considering essential features of realistic nonequilibrium subspaces to fully comprehend macroscopic thermalization in isolated quantum systems.

Their investigation focused on the time scale of thermalization, particularly on escaping from a nonequilibrium subspace $\mathcal{H}_\text{neq}$ across various selections of $\mathcal{H}_\text{neq}$. Following the typicality approach, a common strategy is to opt for a random selection of the nonequilibrium subspace, akin to von Neumann's framework---the framework that we have used to discuss the typicality of macroscopic thermalization in Sec. \ref{sec:smte2}---where discussing a typical Hamiltonian or typical decomposition into macrostates is considered equivalent.

It might be expected that such randomly chosen subspaces would exhibit realistic relaxation times. However, their findings revealed a stark contrast. Regardless of the initial state, even those far from equilibrium, the relaxation time in the described scenario is remarkably short, typically on the order of the Boltzmann time, the average time taken for a typical microscopic fluctuation in a system to occur,  $t_\text{B}= (\hbar/kT)$. Given that $t_\text{B}$ is approximately $10^{-13}$ s at room temperature, this implies that a cup of hot coffee would cool down to room temperature in a mere microsecond. This seemingly paradoxical yet mathematically correct result has significant implications: the gradual temperature decay commonly observed in real-world scenarios is not a prevailing trait for randomly chosen nonequilibrium subspaces $\mathcal{H}_\text{neq}$. This realization underscores the necessity of considering the intrinsic features of realistic nonequilibrium subspaces to gain a comprehensive understanding of thermalization in isolated quantum systems.

Upon closer examination, the atypical slowness of thermalization observed in real-world scenarios might appear less perplexing. Unlike in typical scenarios, where the nonequilibrium subspace is chosen arbitrarily, in reality, it emerges from the macroscopic quantities used to describe the system. This includes standard macroscopic variables, often derived from the aggregation of locally conserved observables---the hydrodynamics fields introduced in Sec. \ref{sec:michydfie}---and the nature of Hamiltonians in realistic systems, which typically involve short-range interactions. Consequently, the resulting subspace and its associated projection operator are expected to exhibit distinctive characteristics. For example, it is conceivable that the commutator between the Hamiltonian and the projection operator is smaller for realistic subspaces compared to those selected randomly. This departure from randomness in determining the nonequilibrium subspace sheds light on why the observed thermalization process in real-world systems may deviate from the behavior expected in typical scenarios.

Realistic systems exhibit unique properties that cannot be revealed through typicality alone. Particularly noteworthy is the phenomenon (emphasized previously at the end of Sec. \ref{SEC:QNMICRO}), where an excess of a locally conserved quantity within a specific region does not dissipate locally but diffuses gradually throughout the entire system, leading to prolonged relaxation times. This highlights the intricate nature of realistic systems and emphasizes the limitations of relying solely on typicality or randomness to comprehend their behavior.

\subsubsection{ETH and Quantum Chaos}
\label{sec:chaosquantum}

The situation surrounding thermalization and the status of the eigenstate thermalization hypothesis (ETH) presents a perplexing conundrum. On one hand, the ETH holds true for a typical macroscopic system, as we have previously underscored. On the other hand, a macroscopic system's thermalization time scale conflicts with real-world systems' observed behavior, which, as we have just seen, exhibit atypical characteristics. This dissonance prompts a deeper inquiry into the status of the ETH and its implications for our understanding of thermalization in quantum systems.

Part of the ongoing research has shifted its focus towards exploring the relationship between the ETH and the chaotic properties of non-integrable quantum systems. This avenue of investigation seeks to uncover potential connections between the behavior of individual quantum eigenstates, as described by the ETH, and the broader chaotic dynamics exhibited by complex quantum systems. By delving into this relationship, researchers aim to gain deeper insights into how quantum chaos influences thermalization processes.

Quantum chaos deals with the study of chaotic behavior within quantum systems. Small perturbations in initial conditions can lead to vastly different outcomes in these systems, akin to chaos in classical dynamics. While classical chaos arises from deterministic nonlinear dynamics, quantum chaos manifests as a complex interplay between the classical and quantum properties of the system. This field of study often involves examining the spectral statistics of quantum Hamiltonians, the level spacing distribution of energy eigenvalues, and the dynamics of systems displaying chaotic behavior.

More precisely, in chaotic systems, the energy level statistics tend to follow the predictions of random matrix theory. The latter describes the statistical properties of matrices representing the Hamiltonians of such systems, and it has been observed that systems with chaotic classical dynamics often have energy spectra that align with the random matrix theory predictions. ETH is more likely to be valid in systems with chaotic spectra because chaotic dynamics lead to the delocalization of energy eigenstates, making them more ``typical'' and thus more likely to exhibit thermalization.

The relationship between ETH and quantum chaos is intriguing. Indeed chaotic quantum systems frequently satisfy the conditions necessary for ETH to hold~\cite{Sred99}. Indeed, the study of quantum chaos may provide valuable insights into the conditions under which ETH is expected to hold.

On the other hand, it may be premature to assert that the ETH does not strictly hold or that thermalization may not occur in integrable non-chaotic quantum systems. Indeed, the recent paper, \textit{Macroscopic Irreversibility in Quantum Systems: ETH and Equilibration in a Free Fermion Chain} by Hal Tasaki \cite{Tasaki24}, suggests a perspective different from the idea that chaoticity is the sole driver of thermalization. Another recent study demonstrating irreversibility, in terms of entropy growth for a quantum  free gas (surely an integrable system!), will be discussed in Section \ref{sec:qentropy}.

In \cite{Tasaki24}, Tasaki investigates the dynamics of a free fermion chain with uniform nearest-neighbor hopping. By examining the system's evolution from an arbitrary initial state with a fixed macroscopic particle number, Tasaki demonstrates that, after a sufficiently long and typical time, the coarse-grained density distribution becomes nearly uniform with an extremely high probability. This result highlights the emergence of irreversible behavior, such as ballistic diffusion, within a quantum system governed by unitary time evolution. Importantly, Tasaki's approach establishes irreversibility without relying on randomness in the initial state or the Hamiltonian, a notable contrast to classical systems where randomness is often a key factor. The breakthrough in his proof is the application of the strong energy eigenstate thermalization hypothesis (ETH) in its large-deviation form, which plays a crucial role in justifying the observed irreversibility.

\quad\\

\quad\\



\subsubsection{Some Observations on Classical and Quantum Chaos }


In the following we briefly discuss quantum chaos and its relation with chaos in classical systems. An analysis of this delicate topic allows us to understand that the classical limit is rather subtle, and the relevance of a coarse grained description for the study of the chaotic properties.
Using the linear structure of the Schr\"odinger equation
we can write  its solution  in the formal way as follows:
\begin{equation}
\label{QM1}
\psi(t)={\cal U}(t) \psi(0)
\end{equation}  
where  the operator ${\cal U}(t)=e^{-i{{\cal H} \over \hbar}t}$ is  unitary. Therefore if we consider
two initial conditions $ \psi(0)$ and $ \psi(0)+\delta \psi(0)$,  one has
\begin{equation}
\label{QM2}
|| \delta \psi(t)||= || \delta \psi(0)||
\end{equation}  
where $|| (\,\,) ||$ denotes the $L_2$ norm.
\\
From the above result, apparently, at variance with classical physics,  it 
seems that one can  conclude  that chaos cannot exist
in quantum world, and this seems to be  a consequence of the linear structure of the equation ruling the evolution law.
Let us note that  although classical mechanics is typically described by nonlinear equations, 
it is formally analogous to quantum mechanics in many respects.
 Indeed, the Liouville equation of classical mechanics affords a linear theory for the evolution of probability density,
 at the cost of switching from a finite dimensional phase space to an infinitely dimensional function space, 
 analogously to the  quantum mechanics  based on the Schr\"odinger equation. 
\\
We briefly discuss such an aspect considering  a deterministic chaotic system
\begin{equation}
\label{QM3}
{d \over dt}{\bf x}={\bf f}({\bf x}) \,\, , \,\, {\bf x} \in R^N
\end{equation}  
and write its evolution in the  form ${\bf x}(t)={\cal S}^t {\bf x}(0)$.
The probability distribution $\rho({\bf x}, t)$ evolves according to
\begin{equation}
\label{QM4}
\partial_t \rho({\bf x}, t)+ \sum_{n=1}^N \partial_{x_n} \Big( f_n({\bf x})  \rho({\bf x}, t) \Big)=0 \, ,
\end{equation}
we  assume that the dynamics is invertible and Liouville theorem holds,  i.e.
$$
\sum_{n=1}^N \partial_{x_n}  f_n({\bf x})  =0
$$
we can write the solution of (\ref{QM4})
\begin{equation}
\label{QM5}
\rho({\bf x}, t)=\rho( {\cal S}^{-t} {\bf x} , 0) \,.
\end{equation}
Even  in the case (\ref{QM3}) is chaotic  the equation for $\rho({\bf x}, t)$ 
can somehow appear, as consequence of  the linear
 structure of the evolution law (\ref{QM4}),  regular,
for instance considering two initial conditions
$\ \rho({\bf x}, 0)$ and $ \rho({\bf x}, 0)+\delta  \rho({\bf x}, 0)$
it is easy to show
$$
\sup_{\bf x} |\delta  \rho({\bf x}, t)|=
\sup_{\bf x} |\delta  \rho({\bf x}, 0)| \,.
$$
In spite of the previous conclusion, one can see that  the presence of  chaos  has  nontrivial   consequences of the  time evolution of $\rho({\bf x}, t)$: 
let us denote  ${\bf y}={\cal S}^{-t} {\bf x}$, from (\ref{QM5}) one has
$$
{\partial  \over \partial x_j } \rho({\bf x}, t)= 
\sum_{n=1}^N  {\partial  \over \partial y_n} \rho({\bf y}, 0)   {\partial y_n \over \partial x_j} \, ,
$$
in presence of  chaos we have that $ {\partial y_n \over \partial x_j}$ increases exponentially with $t$
as $e^{\lambda t}$ where $\lambda$ is the Lyapunov exponent of (\ref{QM3}).
Therefore as a consequence  of the Liouville theorem,
 starting with a smooth  $ \rho({\bf x}, 0)$,  say with a typical spatial scale $l_0$,
 $\rho({\bf x}, t)$ develops   a dendritic structure  whose
typical spatial scale decreases exponentially as  $l_0 e^{- \lambda t}$.

The above result has a  quite  interesting correspondence in  quantum  mechanics~\cite{berry2001chaos}.
Consider a quantum system whose initial condition is practically  classical,
e.g. the   Wigner function  at $t=0$ is smooth and localised on scales 
much larger than $O(\sqrt{\hbar})$.
 Denote by $ \Delta p_0$ and $ \Delta q_0$, respectively, the initial widths 
of the momentum and position distributions (e.g. a Gaussian packet), and let  $ \Delta p_0 \Delta q_0 \sim A_0 \gg \hbar$,
being $A_0$ the action.
Initially, the quantum effects are small, and  basically  the evolution is ruled by classical mechanics,
of course at larger times, say after a certain $t_c$, the quantum nature of the system will prevail.
Let us wonder  about  the dependence of $t_c$ on $A_0$, $\hbar$  and the parameters of the classical Hamiltonian. 

If the classical system is chaotic, the initially smooth packet will be exponentially stretched in the directions with positive 
Lyapunov exponents. Since the phase space volume must be preserved (Liouville theorem) the Wigner function will develop
dendritic structures at small scales, which decrease exponentially in time as $\epsilon_0 e^{-\lambda t}$
 where $\lambda$ is the Lyapunov exponent and $\epsilon_0$ is $O( \Delta p_0)$, or $O( \Delta q_0)$.
  When the smallest scale reaches order $O(\hbar^{1/2})$, i.e. $ \Delta p_0 \Delta q_0  e^{-2 \lambda t} \sim \hbar$
  the quantum effects become relevant. 
  This leads to the following estimate for the crossover time $t_c$:
  \begin{equation}
\label{QM6}
  t_c \sim {1 \over \lambda} \ln {A_0 \over \hbar} \, .
  \end{equation}
  A claim commonly found in many textbooks is that quantum mechanics reduces to classical mechanics, in the classical limit
  ${A_0 / \hbar} \to \infty$. We note that the crossover time $t_c$ diverges, and one may be tempted to conclude 
  that the classical behaviour is then an obvious  consequence of quantum mechanics. 
  On the other hand, in the case of chaotic classical systems, the divergence is so slow that $t_c$ remains definitely small.
  \\
The  above severe limit for $t_c$  is well illustrated by the chaotic tumbling of  Hyperion,  a small satellite of Saturn;
such a minor celestial body has an irregular, potato-like shape which extends for $O(10^2)$ km.
 The irregular (chaotic) motion of Hyperion, due to the interaction with Saturn and the big moon Titan, was quite clearly observed during the Voyager 2 mission.
The classical instability time (the inverse of the Lyapunov exponent) is $\lambda^{-1} \sim 10^2$ days, while the order of magnitude of the classical action is 
$A \sim 10^{58} \hbar$   a quantity that sounds enormous. 
However using  the above numerical values for $A$  and $\lambda$ the estimate (\ref{QM6})  of $t_c$ gives $\simeq 40$ years,
a time interval which is really very small. 
Nobody seriously thinks that in few  decades, astronomers will see quantum effects take over in Hyperion, and its classical chaotic behaviour turn in some quantum quasiperiodic motion~\cite{berry2001chaos}. 
\\
What is the solution of this paradox?
The answer lies in the so-called decoherence effect: Hyperion does not interact with Saturn and Titan only, but also with a myriad of other objects, such as the other moons of Saturn, cosmic dust, photons from the Sun and so on.
If one includes  the environment, in quantum dynamics, one has a smoothing effect that leads to classical behaviour. 
This is the decoherence mechanism:  the quantum suppression of classical chaos involves the interference of waves and their phases, this
happens  after a time  which is much shorter that $t_c$.
We can say that  Hyperion behaves as a classical object over time intervals which are well beyond $40$ years,
 not as a mere consequence of the fact that  $A_0/ \hbar \sim 10^{58} \gg 1$,
  but because of the emergent semiclassical phenomenon, in which the interactions with 
  the external environment play a fundamental role~\cite{zurek2003decoherence}.
\\
Let us now open a parenthesis on 
 the coarse-grained description of systems with a finite number of states,  remarkably 
 such an issue is  very relevant to certain aspects of the semiclassical limit and decoherence~\cite{mantica2000quantum, falcioni2003coarse}.
Let us now consider a deterministic map
\begin{equation}
\label{QM7}
x(t+1)=f(x(t))
\end{equation}
for sake of simplicity we  can consider the $1d$ case with $x \in (0,1)$.
We can introduce spatially discrete approximation of the map, defining:
$$
m(t+1)=F_M(m(t)) \, ,
$$
where $F_M(m(t))= [M f(m(t)/M)]$ and $[\, ]$ indicates the integer part of a real number. 
This dynamics now concerns the integer states $\{0, 1, ...,  M-1 \}$, which represent 
the original ones with accuracy $\eta = 1/M$. 
Whether the original  dynamics (\ref{QM7})  are chaotic or not, those of $F_M$  are 
necessarily  periodic, and  the period is $O(\sqrt{M})$.
Such a fact seems to be in contrast     with the naive idea that the  $M \to \infty$ limit, i.e. the 
$\eta \to 0$  limit recovers the continuous dynamics. 
It is interesting the similarity with  the classical limit of quantum mechanics: because quantum mechanics is governed 
by a linear evolution law, the Schr\"odinger equation, apparently chaos cannot exist in quantum mechanics,
 while it is common in (non-linear) classical mechanics. 
 \\
The above  apparent contradiction can be explained 
in terms of the accuracy of the coarse-graining level, 
 with which both continuous and discrete dynamics are observed,
  and thanks to the introduction of randomness in the quantum mechanical description~\cite{ford1991arnol}.
If the accuracy $\epsilon$ in the random evolution  
 is much larger than the lattice spacing of the discrete dynamics, i.e. $\epsilon \gg \eta$,
 the discrete and the continuous dynamics are practically indistinguishable.
 Analogously, the semiclassical limit applies within a time that grows as  $\hbar$  decreases: 
 the time that a wave-packet takes to spread over a large distance. Therefore, the quantity
$\eta=1/M$   in the $M \to \infty$ limit, the limit of infinitely many states, plays the same role played by $\hbar$ in the $\hbar \to 0$.
\\
This result shows that the decoherence required for quantum mechanics to lead to the 
classical macroscopic behaviour has a simple origin and  classical mechanics can be seen as a sort
 of emergent property of quantum mechanics where the interactions with the environment 
 are modelled using some random variable.
\\
We note that a  rather similar situation is met in the dynamics of purely classical billiards;
we can say that rational polygonal billiards are not chaotic and
play the role of quantum systems, whose ``classical limit'' 
are curved  (convex) billiards, to which they tend as the number of polygonal sides increases indefinitely. 
Indeed, while rational polygonal billiards are nonchaotic systems, 
one can use them to approximate chaotic curved billiard tables to arbitrary precision, 
and ask what happens then to the character of their motion~\cite{VUF93}.
A  class of non-chaotic systems, having zero Lyapunov exponent, but displaying irregular behaviors, for which often, in the literature, 
 the term pseudochaos, initially introduced by Chirikov, is used.
 
Let us conclude this quick overview on quantum chaos by stressing the fact that in the mathematical characterization of chaos in terms of Lyapunov exponents (or equivalently Kolmogorov-Sinai entropy) one has to perform two asymptotic limits, namely infinite time intervals and arbitrary resolution of the state. In many systems, although from a mathematical point of view chaos is absent, one can have a rather ``complex'' behaviour which is highlighted by a non-asymptotic analysis at finite resolution~\cite{Boffetta03}. We can say that the presence, or absence, of chaos in high-dimensional systems is not the main ingredient for the validity of SM. In Section~\ref{sec:conclusionchaos} we will reconsider again this aspect.

\subsection{ Clarifying Misconceptions: Classical and Quantum Entropy Revisited}

Rudolf Clausius introduced the concept of thermodynamic entropy and formulated the second law of thermodynamics in the mid-19th century, asserting that the total entropy of an isolated system can never decrease over time. 
In modern terms, entropy plays a crucial role in resolving what Callen \cite{Callen} defines as the single, overarching problem of thermodynamics: determining the equilibrium state resulting from the removal of internal constraints in a closed, composite system. Callen's approach introduces the entropy function $S=S_\text{C}$ (where the subscript \( C \) stands for Clausius), which depends on the extensive parameters of any composite system. This function is defined for all equilibrium states and possesses the property that the values of extensive parameters, in the absence of internal constraints, maximize entropy over the manifold of constrained equilibrium states.

We will explore the extension of Boltzmann's entropy from classical to quantum realms, emphasizing its significance in characterizing nonequilibrium states. This extension serves as a crucial link between classical Clausius entropy and the quantum domain, shedding light on the behavior of systems that are not in equilibrium. For a deeper exploration of this subject matter, we suggest referring to the work by Goldstein, Lebowitz, Tumulka, and Zanghì in \cite{GLTZ2020}.

\subsubsection{Boltzmann Entropy}

As already recalled in subsection~\ref{sec:bridgelaw}, in the context of classical statistical mechanics Boltzmann's entropy can be written as
\begin{equation}\label{eq:boltzformula}
    S_\text{B} (M) = k_B \log \Omega_M \,, \quad M=M(X)
\end{equation}
where $M$ represents the macrostate, which in turn is a function of the microstate $X$. Thus, $S_\text{B} = S_\text{B}(X)$, and the dependence on $X$—on the time evolution of $X$—and the dominance of thermal equilibrium make the second law understandable in microscopic terms. This was Boltzmann's enduring achievement.

The equation \eqref{eq:boltzformula} maintains its significance in the quantum realm, with $\Omega_M$ denoting the dimensionality of the macrospace $\mathcal{H}_M$, where $X$ is replaced by the quantum state $\Phi$. In the quantum scenario, the expression ``$M=M(X)$'' signifies the macrospace $\mathcal{H}_M$ to which the microstate $\Phi$ belongs. Specifically, for a system with a quantum state $\Phi$ or a density matrix concentrated in $\mathcal{H}_M$, we assign the entropy value $S_\text{B}(\Psi) \equiv S_\text{B}(M)$. 
In fact, as early as 1914, Einstein argued that the entropy of a macrostate should be proportional to the logarithm of the ``number of elementary quantum states" compatible with that macrostate. Schr\"odinger echoed this viewpoint in 1927, within his fully developed wave mechanics framework.

In equilibrium, Boltzmann's entropy $S_\text{B}$ asymptotically coincides with Clausius' entropy $S_\text{C}$, serving as a natural extension to non-equilibrium states in both classical and quantum systems. It's crucial to highlight the significance of incorporating the dimension of the vector space when defining quantum Boltzmann entropy.

Thermodynamic entropy $S_\text{C}$ exhibits an extensive nature, meaning that the total entropy of two systems equals the sum of their individual entropies. This aligns with the dimensionality principle: the dimension of the tensor product of two spaces is the product of the dimensions of each space. Applying the logarithm to the product of these dimensions makes the additivity of thermodynamic entropy evident. Consequently, the properties of $S_\text{B}$ in equilibrium states harmonize with the attributes of $S_\text{C}$ as postulated by Callen. Moreover, the extension provided by $S_\text{B}$ to non-equilibrium states upholds the fundamental characteristics of thermodynamic entropy.

In the classical scenario, the increase in Boltzmann entropy entails the progression of the phase point $X(t)$ towards larger macro sets $\Gamma_M$. Specifically, for most initial phase points $X(0)$ in $\Gamma_M$, $X(t)$ tends towards larger macro sets as $t$ increases until it reaches $\Gamma_{\text{eq}}$. This progression is occasionally interrupted by periods of entropy decrease and temporary returns to previous levels, albeit infrequently, shallowly, and briefly. Once $X(t)$ reaches $\Gamma_{\text{eq}}$, it remains there for an extended duration, except for occasional, shallow, and short-lived entropy fluctuations. A similar narrative applies to the quantum case, as elaborated below.

The clarity of this explanation of entropy increase prompts one to ponder: How could we understand the second law from a microscopic standpoint without grasping entropy as a function of the microstate? However, doubts about this understanding persist even today.

Boltzmann's achievement in demonstrating entropy increase in a dilute gas goes deeper and is closely tied to his development of the integro-differential equation bearing his name. Boltzmann's equation describes the time evolution, taking into account interactions such as collisions, of the distribution function $f$ (see Sec.\ref{eq:Bodiluttoq}), a function of the microstate. In particular, it follows from this equation that 
\begin{equation}\tag{\ref{sec:abmeq1}}
S_\text{B}(M) = k_B \log \Omega_M \sim - k_B \int f \log f \, d\tau.
\end{equation}
increases, in the sense that its time derivative is always greater than or equal to zero, with equality only if $f$ is a (local) Maxwellian. This is known as Boltzmann's H-theorem, which amounts to a derivation of the second law (relative to the partition of the one particle phase space, as recalled in Sec.\ref{eq:Bodiluttoq}).

Over the years, complaints about Boltzmann's explanation of the second law and his notion of entropy have been raised. For instance, Khinchin in 1941 \cite{kh49} criticized individualist accounts of entropy, stating
\begin{quotation}
All existing attempts to give a general proof of this postulate [i.e., $S= k\log \Omega$,] must be considered as an aggregate of logical and mathematical errors superimposed on a general confusion in the definition of the basic quantities.
\end{quotation}

While Khinchin's concerns about the mathematical soundness of Boltzmann's framework may have been valid at the time, they have since been addressed by Lanford, nearly a century after Boltzmann's discovery \cite{Lan75}. Despite this, Khinchin's viewpoint, still echoed today by some, challenges the notion of entropy increase in closed systems. He suggests that the second law should simply assert that when a system in thermal equilibrium has a constraint lifted, the new equilibrium state exhibits higher entropy than the previous one. However, adopting this perspective may limit our ability to explain various phenomena.

The downplaying of Boltzmann's achievement or his potential misunderstanding, which are closely related, has been a consistent theme in the development of statistical mechanics throughout the 20th century and persists today. Investigating the source of this confusion requires examining various factors such as the complexity of Boltzmann's ideas, the mathematical challenges in formalizing his concepts, the philosophical debates surrounding entropy and statistical mechanics, and the historical context in which Boltzmann's work was received and criticized.

Furthermore, the evolution of scientific thought and the emergence of new theoretical frameworks, particularly quantum mechanics with its focus on observables and the role of observers, may have influenced the perception of Boltzmann's contributions over time. Understanding these factors could provide valuable insights into why Boltzmann's ideas have sometimes been undervalued or misunderstood in the development of statistical mechanics. However, exploring these factors in depth is beyond the scope of this review.

\subsubsection{Gibbs-von Neumann Entropy and Shannon Entropy}
Another notion of entropy is Gibbs entropy $ S_G$. Proposed a few decades after Boltzmann's work, Gibbs entropy is widely utilized in equilibrium statistical mechanics. 

In classical mechanics, the Gibbs entropy of a physical system is given by
\begin{equation}\label{SGdef}
S_\text{G}(\rho) = - k_B \int_{\Gamma} \rho(x)  \log \rho(x) \, d\Gamma 
\end{equation}
Here,  $d\Gamma$ denotes the (symmetrized) phase space volume measure, and $\rho$ represents a probability distribution over phase space $\Gamma$.

In the realm of quantum mechanics, we encounter its counterpart, known as the von Neumann entropy, expressed as:
\begin{equation}\label{SGdef2}
S_\text{vN}(\rho) = - k_B \mathrm{Tr} \rho \log \rho
\end{equation}
Here, $\rho$ represents a density matrix defined over the system's Hilbert space. Given the resemblance between these two notions of entropy, we shall collectively refer to them as the Gibbs-von Neumann entropy, denoted by $S_\text{GvN}$.

The immediate challenge with the Gibbs entropy arises from the ambiguity surrounding the choice of distribution or density matrix $\rho$. In the context of a system in thermal equilibrium, $\rho$ typically refers to a Gibbsian equilibrium ensemble, such as the microcanonical, canonical, or grand-canonical ensemble. Consequently, for states in thermal equilibrium, the Boltzmann entropy $S_\text{B}$ and the Gibbs entropy $S_\text{GvN}$ agree to the leading order. Thus, at equilibrium,
\begin{equation}\label{eq:SBSGSC}
    S_\text{B} \approx S_\text{GvN} \approx S_\text{C} \,.
\end{equation}

However, in more general scenarios, determining the appropriate interpretation of $\rho$ becomes less straightforward. While one may consider $\rho$ as representing ignorance, which is meaningful for a probability distribution, this interpretation may be less suitable for a density matrix. Another approach is to associate $\rho$ with a preparation procedure, a concept that holds significance both in classical and quantum contexts. In the quantum case, there is another possibility for the interpretation of $\rho$, namely as a reduced density matrix (see Sec.  \ref{sec:denmatrqm}). 

Regardless of the interpretation of $\rho$, another immediate issue with the Gibbs entropy is its time-invariance:
\[
\frac{d}{dt} S_\text{GvN} (\rho_t) = 0\,.
\]
This property is straightforward to demonstrate. Since the classical and quantum cases are formally identical,  one can undertake this exercise in the quantum case using $\rho_t$ as defined by \eqref{eq:evoldenmat}, i.e., $\rho_t  = e^{-i Ht } \rho \,e^{i Ht}$. Thus, while the increase of $S_\text{B}$ agrees with the increase of $S_\text{C}$ predicted by the second law of thermodynamics, $S_\text{GvN}$ lacks this essential feature of the thermodynamic entropy of an isolated system.\footnote{By replacing $\rho$ with a coarse-grained $\rho_{\text{cg}}$ in the formula for $S_G$, it turns out that ${dS_G(\rho_{\text{cg}})}/{dt} \geq 0$, as suggested by Gibbs and later proven by Ehrenfest. This result immediately gained resonance, notably in the famous book by Tolman, and was extended to the quantum case. However, its proof is almost mathematically trivial, merely following from the convexity property of the logarithm, and is incomparably inferior in depth to Boltzmann's H-theorem.}

The Gibbs-von Neumann entropy $S_\text{GvN}$ shares many similarities with the concept of entropy in information theory and probability. In the late 1940s, Shannon introduced the notion of ``entropy'' for the probability distribution of a random variable. Denoted by $H$ and inspired by Boltzmann's function $H$ in equation \eqref{eq:approxHfun}, i.e., $H(f)=-\sum_j f_j \log f_j$ (for a discrete distribution), it aimed to quantify the level of uncertainty or information content in a message. Shannon regarded the probabilities in his study of optimal data coding for transmission across a noisy channel as objective and did not claim any connection with thermodynamics. Nevertheless, starting with von Neumann, many became fascinated with the idea that thermodynamic entropy is related to the uncertainty of a probability distribution akin to Shannon entropy. 

We won't delve here into the difficulties of this idea, which lead to the fallacy of a subjective interpretation of thermodynamic entropy (see \cite{GLTZ2020}). We simply note that there's a fundamental confusion here between a probability measure on phase space, which formalizes what should be understood by typical behavior, and an objective distribution such as $f$ in kinetic theory.

\subsubsection{Von Neumann Entropy}
The quantum entropy formula \eqref{SGdef2} is commonly associated with von Neumann. However, he expressed dissatisfaction with it within the framework of statistical mechanics, especially when introducing the notion of quantum macrostate alongside the Hilbert space decomposition \eqref{eq:orthoDecom}. In \cite[Sec.~1.3]{cite66}, he lamented that expressions for entropy in the form of \eqref{SGdef2} ``are not applicable here in the way they were intended, as they were computed from the perspective of an observer who can carry out all measurements that are possible in principle--- i.e., regardless of whether they are macroscopic.''

To address this issue, he put forward the following definition of entropy, formulated as a function of the microstate $\Phi$:
\begin{equation}\label{SvN2}
S_\text{vN}(\Phi)=
-k \sum_{M} \| P_M \Phi\|^2 \log \frac{\| P_M \Phi\|^2 }{\Omega_M}\,.
\end{equation}
This proposal aligns more closely with Boltzmann's concept than with Gibbs' by making entropy dependent on the system's microstate.
Similar expressions were advocated recently by Safranek et al.~\cite{SDA17,SDA18}. 

If $\Phi$ is primarily composed of contributions from a limited number of macrostates, and given the substantial size of $\Omega_M$, the term $\| P_M \Phi\|^2 \log | P_M\Phi|^2$ becomes negligible in comparison. As a result, equation \eqref{SvN2} simplifies to the weighted average of the quantum Boltzmann entropies \eqref{eq:boltzformula} associated with different macrostates. Each macrostate is weighted by $\| P_M\Phi\|^2$, indicating the probability of the system being in macrostate $M$. Consequently, if a body is in a quantum superposition of states with significantly different macroscopic characteristics, we would interpret its entropy as an average entropy, rather than considering the system to be in a superposition of states with vastly different entropies.

Maybe von Neumann would have argued against allowing $\Phi$ to involve macroscopic superpositions. As an advocate of the collapse of the quantum state as a consequence of measurements, he would contend that macroscopic systems cannot exist in quantum superposition. This stance would further diminish the attractiveness of the proposal, as uncertainty regarding the actual macroscopic state of the body would contribute to its thermodynamic entropy, leading to nonsensical implications---for example, if we are uncertain about the amount of melted ice in a cup at $0^\circ \text{C}$, we shall not assign to the mixture of water and ice in the cup an entropy depending on our uncertainty (say 1/2 - 1/2, if we know nothing).

As opposed to Gibbs-von Neumann entropy \eqref{SGdef2}, von Neumann entropy \eqref{SvN2} changes with time as bona fide entropy should. However, the ``quantum ergodic theorem" demonstrated by von Neumann, that is, normal typicality \eqref{vNtheo12}, is not sufficient to assess the increase of $S_\text{vN}$ for an isolated system.

\subsubsection{Entanglement Entropy}
\label{sec:enten}
Let's consider the possibility that the density matrix $\rho$ within $S_\text{GvN}$ could represent the system's reduced density matrix as defined by equation \eqref{SGdef2}. 

An immediate drawback of this proposal is that it  has an undesirable implication: the entropy fails to exhibit extensivity. Put differently, the entropy of the combined system, $1+ 2$, does not simply equate to the sum of the entropies of $1$ and $2$. This becomes evident when considering that while the entropies of $1$ and $2$ may each individually increase, the entropy of their union, $1 + 2$, remains constant if it is a closed system, owing to the unitary evolution (apologies for employing the same symbol $S$ for different concepts). Moreover, it's not universally true for all $\rho$ on $\Hilbert_1\otimes \Hilbert_2$ that $S_\text{GvN} (\Tr_2 \rho)$ increases with time, especially when considering time reversal. \new{A clear account on the behaviour of the entanglement entropy with time alongside
the relaxation of a quantum (integrable) system is for instance given in~\cite{AlbaCalabrese17}}.

Another avenue, as discussed in \cite{GLTZ2020}, involves subdividing a macroscopic system, perhaps in a pure state $\Phi$, into small $M$ subsystems. Here, we consider the reduced density matrix of each subsystem $\rho_j$, $j=1, \ldots, M$, and aggregate the Gibbs entropies of $\rho_j$:
\begin{equation}\label{Sentdef}
S_\mathrm{ent}:=-k\sum_{j=1}^M \Tr (\rho_j \log  \rho_j)\,. 
\end{equation}
This ``entanglement" entropy, when the system is in thermal equilibrium, arguably corresponds to thermodynamic entropy under the assumption that the subsystems are sufficiently small and the partition of the macroscopic system is independent. These conditions are linked to the concept of microscopic thermal equilibrium, as discussed in Sec. \ref{sec:mte}. However, for non-equilibrium states, the suitability of \eqref{Sentdef} becomes more debatable, as explored in \cite{GLTZ2020}.

\bigskip
Ultimately,  Boltzmann entropy $S_{B}$ emerges as the most viable choice for the fundamental definition of entropy in quantum mechanics. This is due to its ability to capture quantum systems' essential characteristics while aligning with classical notions of entropy in the appropriate limit. Notably, it provides a robust framework for describing equilibrium states and offers valuable insights into the dynamics of non-equilibrium systems.

\subsubsection{Some Key Insights  on Boltzmann Entropy Growth}

According to the second law of thermodynamics, the total entropy of an isolated system tends to increase over time, embodying the concept of macroscopic irreversibility. The appearance of distinct time-asymmetric behavior in the observed evolution of macroscopic systems, despite the absence of such asymmetry in their microscopic dynamics, stems from the significant difference between microscopic and macroscopic scales. This disparity between scales is at the basis of  Boltzmann's H-theorem.

A simple H-Theorem, as discussed by Goldstein and Lebowitz in a paper dedicated to analyzing various aspects of classical  Boltzmann entropy of non-equilibrium systems \cite{GL04}, smoothly extends to the quantum case as follows. Consider the case where the macrostate $M_t$ satisfies an autonomous deterministic evolution, e.g., the classical or the quantum Boltzmann equation or some hydrodynamic equation possibly involving quantum effects, as for a superfluid. This means that if that evolution carries $M_{t_1} \rightarrow M_{t_2}$, then the unitary microscopic dynamics $U_{t_2-t_1}$ carries $\Hilbert_{M_{t_1}}$ inside $\Hilbert_{M_{t_2}}$, i.e., $U_{t_2-t_1} \Hilbert_{M_{t_1}} \subset \Hilbert_{M_{t_2}}$, with negligible error. Now the fact that Hilbert space dimension is conserved by the unitary time evolution implies that $\Omega_{M_{t_1}} \leq{} \Omega_{M_{t_1}} $, and thus by \eqref{eq:boltzformula} that $S_B(M_{t_2}) \geq S_B(M_{t_1})$ for $t_2 \geq t_1$. The explicit form for the rate of change of $S_B(M_t)$ (including strict positivity) depends on the detailed macroscopic evolution equation. The fact that $\Hilbert_{\text{eq}}$ essentially coincides for large $N$ with the whole energy shell $\Hilbert_E$ (thermal equilibrium dominance) also explains the evolution towards and persistence of equilibrium in an isolated macroscopic system, as we have elaborated in Section \ref{sec:ATTE}.

More generally, not assuming that macro states evolve deterministically, already at this point it is perhaps not unreasonable to expect that in some sense the unitary evolution will carry a pure state $\Phi_t$ from smaller to larger subspaces $\Hilbert_M$. In fact, it was proven in \cite{GLMTZ10} and amply discussed in the previous sections that if one subspace, the equilibrium subspace,  has most of the dimensions in an energy shell, then $\Phi_t$ will sooner or later come very close to that subspace (i.e., reach thermal equilibrium) and stay close to it for an extraordinarily long time (i.e., for all practical purposes, never leave it).

A result in this direction is the quantum counterpart of the following classical statement: Instead of considering the entire history of a system, we focus on two specific times, $t_1$ and $t_2$, along with the corresponding macrostates of the system at these times, denoted as $M_{t_1}$ and $M_{t_2}$, respectively. If the volumes of the corresponding phase space subsets satisfy $\Omega_{M_1} \ll \Omega_{M_2}$, then for any $t \neq 0$, the majority of $X_0 \in \Gamma_{M_2}$ are such that $X_t \notin \Gamma_{M_1}$. 

The quantum analog of this statement is as follows: If the dimensions of subspaces corresponding to the macrostates $M_{t_1}$ and $M_{t_2}$ satisfy $\Omega_{M_1} \ll \Omega_{M_2}$, then for most states $\Phi_0 \in \mathcal{H}_{M_2}$, the state $\Phi_t= U_t\Phi_0$ at time $t \neq 0$ has a negligible component in the $\mathcal{H}_{M_1}$ subspace, i.e., $\| P_{M_1} \Phi_t\| \ll 1$. 

The proof of this proposition is rather simple and does not require special assumptions on the Hamiltonian generating the time evolution $U_t$, as discussed in \cite{GLMTZ10}.  Moreover, the proof only relies on the fact that $\mathcal{H}_{M_1}$ and $\mathcal{H}_{M_2}$ are two mutually orthogonal subspaces of suitable dimensions. As a consequence, the statement remains true if we replace $\mathcal{H}_{M_1}$ by the sum of all macrospaces with dimensions less than that of $\mathcal{H}_{M_2}$. In other words, if $\mathcal{H}_{M_2}$ is much larger than all of the smaller macro spaces combined (a scenario that is not unrealistic), then it is atypical for entropy to be lower at time $t$ than at time $0$.


\label{sec:qentropy}

As already stressed, Lanford's work on the classical dilute gas \cite{Lan75}  has  shed light on the nature of molecular chaos within Boltzmann's analysis, revealing it to be a consequence rather than an assumption, particularly in the context of a dilute gas. By emphasizing factors such as initial conditions, typicality, large numbers, and coarse-graining, Lanford's research underscores the significance of these elements in understanding the emergence of macroscopic irreversibility.

In quantum mechanics, achieving a level of mathematical clarity akin to Lanford's analysis of a dilute gas remains a persistent challenge. The intricate interactions within quantum systems give rise to highly entangled many-body states, presenting formidable obstacles to precise analysis.

Furthermore, a significant conceptual challenge arises in addressing the problem of macroscopic superposition. Determining the entropy of a macroscopic superposition, such as that of cold and hot water, remains elusive. Even assuming the absence of macroscopic superpositions at a given time, the challenge lies in demonstrating the persistence of such a state over time. If the initial state $\Phi_0$ belongs to a single macrostate $M_0$, it is expected that quantum fluctuations of macro variables will remain exceedingly small during the time evolution, resulting in the quantum state $\Phi_t$ at time $t > 0$ also occupying a single macrostate $M_t$. This dynamical hypothesis, as highlighted in \cite{Mori2018}, represents a physically natural yet highly nontrivial proposition.

One might wonder if some form of quantum chaotic behavior is essential to explaining macroscopic irreversibility. The Eigenstate Thermalization Hypothesis (ETH), discussed in previous sections, represents a natural manifestation of chaoticity in both microscopic and macroscopic forms. However, is it really necessary? To answer this question, it is valuable to explore whether irreversibility can arise in integrable systems. This may offer insights into the fundamental mechanisms underlying irreversibility and its relationship with quantum dynamics.

\label{sec:engrqidgas}

A particularly interesting recent contribution in this direction is discussed in \cite{qgas23}, which extends a previous study by some of the same authors on the time evolution of the Boltzmann entropy for a freely expanding classical ideal gas \cite{cgas22}. 


The investigation of the classical ideal gas revealed that Boltzmann entropy may manifest either oscillatory or monotonically increasing behavior contingent upon the selection of macroscopic observables. Consequently, with an appropriate selection of macroscopic variables, it exhibited irreversible behavior within a non-ergodic, non-chaotic, and non-interacting framework. In this context, it is prudent to avoid overinterpreting the relationship between entropy and entropy increase with respect to the choice of macro-observable. This relationship arises solely due to the total disregard of interactions and should be considered as an artifact of the idealization.

In \cite{qgas23}, the authors investigate the temporal evolution of Boltzmann entropy during the non-equilibrium free expansion of a one-dimensional quantum ideal gas. This quantum Boltzmann entropy, denoted as $S_B$, plays a crucial role in quantifying the ``number'' of independent wavefunctions contributing to a specific macroscopic state. Notably, this entropy typically relies on the choice of macro variables, such as the type and extent of coarse-graining, defining the non-equilibrium macroscopic state. However, it's essential to recognize that the extensive part of Boltzmann entropy aligns with thermodynamic entropy under thermal equilibrium conditions.

The study explores two sets of macrovariables: $U$-macrovariables, representing local observables in position space, and $f$-macrovariables, which also encompass momentum space structure. A non-conventional choice of $f$-macrovariables is employed for the quantum gas. For both sets, the corresponding entropies, denoted as $s_f^B$ and $s_U^B$, demonstrate a stable temporal growth pattern. This behavior mirrors observations in classical systems and is influenced by the scale of momentum coarse-graining.



\section{Some additional considerations}
\label{sec:remarks}

\subsection{Probability and dynamics}
In the light of what has been discussed in this Review,
it seems fair to conclude that the success of probability theory statistical ensembles in
describing the properties of macroscopic systems is
due to the large number of degrees of freedom, rather than on properties of the dynamics. One may then be left with the misconception that the microscopic evolution plays no role in this picture. Here we analyze the connections between probabilistic description and microscopic dynamics, by reviewing the different points of view on the subject.

\subsubsection{Probability: the role of dynamics}
It would be impossible to enter here into the details of the endless debate
about the ``true meaning'' of probability. We will limit our discussion to
what is relevant for SM, reminding the two main point of views: 
\begin{enumerate}
 \item probability as a \textit{subjective} degree of belief;
\item probability as the \textit{objective} frequency of occurrence of an event.
\end{enumerate}
The interested reader can find an extensive discussion on this topic in Ref.~\cite{sep-probability-interpret}.
A typical point of view in the SM community has been well summarised by van Kampen~\cite{ka92}: 

\begin{quote}
Objective
  probability deals with the frequency of occurrence of an object or
  events.  Its assertions can therefore be confronted with reality.
  Subjective probability is a degree of belief and cannot be verified
  or falsified. [...]  This point of view is based on the idea that the
  behavior of physical systems is governed by the degree of belief of
  the observer. [...]  In the probability calculus, which deals with the
  objective probability, the main subject is transformation of one
  distribution of probabilities into another.  It should be stressed
  that this way of reasoning can be applied only if an underlying
  probability distribution is given a priori.  This a priori
  probability is determined by the physics of the system one is dealing
  with, not by fairness or lack of information.
  \end{quote}

The ergodic approach can be seen as a natural way to introduce
probabilistic concepts in a deterministic context, thus supporting
the frequentistic interpretation
of probability in the foundations of SM.  An alternative
way (not in contrast with the point of view of Boltzmann) to introduce
probability is by assuming an amount of uncertainty in the initial
conditions. This approach is due to Maxwell~\cite{M79}, who
considers that there are 
\begin{quote}
 a great many systems the properties of which are the same,
  and that each of these is set in motion with a different set values
  for the coordinates and momenta.
\end{quote}

Since in SM one typically deals with a unique system (although with
many degrees of freedom), it is compulsory to find a connection between the time evolution of the system and the probabilistic description. Ensemble theory should be regarded, in this sense, as a practical
mathematical tool, and the ergodic theory (or some of its ``weak''
versions, as that one by Khinchin and Mazur and van der Linden) as an
unavoidable step to justify its validity~\cite{G67}. Of course there is no global
consensus on this: for instance,  Jaynes supported the opposite opinion that
ergodicity is simply irrelevant for Gibbs' method~\cite{J67}.

From the point of view of probability theory, the abstract
problem of ergodicity may appear quite similar in
dynamical systems and statistical
mechanics: in both cases one has to
treat a phase space $\Omega$, an evolution law $U^t$ and an invariant
measure $\mu (A)$. However, while in the study of dynamical systems
one can also deal with only few
degrees of freedom (even a $1D$ map or a $3D$ differential equation
can have non-trivial behaviour), the focus of SM is
for systems with a very large number of variables. With a strong simplification,
one may say that dynamical systems and SM deal with the limit of $t
\to \infty$ and $N \to \infty$, respectively.
 Given the enormous
size of the phase space, it would be meaningless to study
the amount of time spent by the system in a given region of the phase
space: the idea of ergodicity must be considered from a physical
point of view, e.g. following Khinchin's approach. 

\subsubsection{Probability from deterministic dynamics}

Since in deterministic systems chance plays no role, the use of probabilistic concepts and methods in this context may appear, at a first glance, as an unphysical trick, or even a paradox. This was for instance the point of view of Popper~\cite{po02}, who believed that probabilistic concepts are extraneous to a deterministic description of the world. In a letter to him, Einstein wrote: 
\begin{quotation}
    I do not believe that you are right in
  your thesis that it is impossible to derive statistical conclusions
  from a deterministic theory.  Only think of classical statistical
  mechanics (gas theory), or the theory of Brownian movement.
\end{quotation}

Indeed, the use of probability
can become necessary even in a deterministic world. Every time that the state of a system is not perfectly controlled, its probability density must be introduced. The reason for this need typically falls in one of the two following cases: (i) the system has a very large number of degrees of freedom, but only a small fraction of them is accessible or interesting; (ii) the system is chaotic, and therefore  small  uncertainties on the initial conditions  are exponentially amplified.

From an historical point of view, SM was the first example of the use of statistical concepts in physics. It clearly deals with systems of type (i), since
it focuses on a small set of collective variables describing the
thermodynamic properties of a macroscopic system.  Another example is the Brownian motion, which was at the origin of
the modern theory of stochastic processes: in that case one is only interested in the behaviour of the colloidal particle, and disregards the details of the motion of the molecules of the fluid.

 Also chaotic systems typically require probabilistic
descriptions, despite the presence of deterministic dynamics. Having a limited number of
degrees of freedom is not a limitation to the use of a
probabilistic approach.  For instance in
the case of a discrete time map
\begin{align}
{\bf x}_{t+1}= {\bf g}({\bf x}_{t}) \,,
\end{align}
the  time evolution  of the probability density $\rho({\bf x},t)$ is  ruled the Perron-Frobenius operator:
\begin{equation}
\label{d2}
\rho({\bf x},t+1)= {\cal L}_{_{PF}}\rho({\bf x},t)=
\int \rho({\bf y},t) \delta({\bf x}-{\bf g }({\bf y}))  d{\bf y} \,\, .
\end{equation}
In   the one-dimensional case one has
\begin{equation}
\label{d4}
\rho(x,t+1)= \sum_{k}
{\rho(y_k(x),t) \over |g'(y_k(x))| }
\end{equation}
where the $\{y_k(x)\}$ are the pre-images of $x$, i.e.  the points such that
$g(y_k)=x$. Here $g'$ indicates the derivative of $g$.
If the dynamical system is defined on continuous time and it is described by a set of
ordinary differential equations 
\begin{align}
{d \over dt} {\bf x}={\bf f(x)}\,,
\end{align}    
 the density $\rho ({\bf x},t)$ evolves according to 
\begin{equation}
\label{c2}
{ {\partial \rho} \over \partial t} =
{\bf L}\rho= - \nabla \cdot ({\bf f}\rho)  
\end{equation}
where ${\bf L}$ is  sometimes called the Liouville operator.
Let us introduce the invariant probability distribution $\rho^I({\bf x})$, determined by the equation
$$
\rho^I({\bf x})= {\cal L}_{_{PF}}\rho^I({\bf x}) \,\, ,
$$
for the discrete time map, and by
$$
{\bf L}\rho^I({\bf x})=0
$$
for the differential equation.
This distribution has a special status: if at time $t=0$
one has $\rho({\bf x},0)=\rho^I({\bf x})$, then one has $\rho({\bf x},t)=\rho^I({\bf x})$  for any $t>0$.
Particularly relevant is the case of mixing systems, where $\rho^I({\bf x})$ is unique and it is attractive, i.e. for $t \to \infty$ one has
 $\rho({\bf x},t) \to \rho^I({\bf x})$. In this case we can say for sure that the probabilistic approach is not subjective.
 In particular, since $\rho({\bf x},t) \to \rho^I({\bf x})$, the
  invariant distribution does not depend on
  our knowledge of the system. It can be unambiguously reconstructed from measurements over a long sequence ${\bf
  x}_1, {\bf x}_2, ... , {\bf x}_T$.

\subsubsection{Probability and typicality}
In Boltzmann’s approach to SM, probability has no relation to measures of
ignorance or uncertainty.
The use of probability theory was
focused on the macroscopic (thermodynamic) quantities, to obtain the average of
mechanical observables of the microscopic constituents of matter.
He basically adopted frequency of events as the basic notion of
probability~\cite{da18,ce88}.
The interpretation based on the ignorance of the state of the system was introduced, instead, by Gibbs: in his works the probability density functions describe how the microscopic phases of large ensembles of identical
objects are distributed in their phase spaces.
Since computations of time-averages are much harder than ensemble
calculations, one is typically forced to accept the ergodic hypothesis, which amounts
to such equivalence, and proceed as prescribed by Gibbs.
 This approach is particularly meaningful when dealing with some uncertainty on the initial conditions of the system: one
can introduce an initial probability distribution $\rho({\bf x},0)$, which can be viewed as a ``collection of
independent systems'', and then consider its time evolution.  This way of introducing the ensembles has its
practical relevance in fields different from SM. For instance, a
quantitative characterization of the time evolution of the uncertainty
for weather forecasting is obtained by simulating many trajectories originated by
(slightly) different initial conditions (corresponding to an ensemble),
and then comparing the evolution of the different trajectories. In this way one can estimate the reliability degree of the prevision.  The computational limitations due to the huge number of degrees of
freedom of the system limit the number of the trajectories that can be actually performed, typically
$O(10^2)$~\cite{P00, merz2020}. 

A question arises about the link between the probabilistic computations of SM and the results of laboratory experiments, which
are conducted on a single realization of the macroscopic object
(described by an Hamiltonian with a large number of degrees of
freedom) under investigation. In our opinion, the main theoretical issue to be addressed in order
to answer this question is the justification of \textit{typicality},
i.e., of the fact that time averages of macroscopic quantities in the
evolution of a single system are very close to averages of that
quantity over ensembles of microscopically distinct but otherwise
identical replicas of that system~\cite{goldstein2012typicality}.  This fundamental
property can be seen as {\it emergent} in the limit $ N \gg 1$.
To convince ourselves that this is not a hopeless project, we may
refer to one of the best propositions linking  probability and
physics, the Cournot's principle:

\begin{quote}
An event with very small probability will not happen. 
\end{quote}

This statement is strictly related to the one in Jakob
Bernoulli’s celebrated book {\it Ars Conjectandi} (1713), which reads:

\begin{quote}
 Something is morally certain if its probability is so close to
  certainty that shortfall is imperceptible.
\end{quote}

We do not enter the debate about the validity of such a principle: one may refer to~\cite{sh05} for a nice analysis of it. We just recall that
eminent mathematicians, such as P. L\'evy, J. Hadamard, and
A.N. Kolmogorov, considered the Cournot's principle as the only
sensible connection between probability and the empirical world.  That
connection granted, L\'evy stressed the concrete character of
probability, arguing that, at the ontological level~\cite{sh05}:

\begin{quote}
Probability is a physical property just like length and weight. 
\end{quote}

\subsection{SM for non-conservative systems}

At this point it is quite natural to wonder how the concepts presented in this review extend to systems that do not fulfill the usual assumptions of classical SM, for instance non-Hamiltonian models, or systems out of equilibrium.
Of course if one wants to follow the path used in SM, the first thing to do is to find the invariant probability distribution, which is usually a far-from-trivial task. In what follows we discuss an example to make clear that the difficulty is not strictly related to the non-existence of an Hamiltonian. Indeed, provided that a conservative quantity exists, it is usually possible to generalize the derivation adopted for equilibrium systems. Let us consider the Euler equation for
a perfect fluid
\begin{align}
\partial_t {\bf u}+({\bf u} \cdot \nabla ){\bf u}=
 - {1\over \rho_0} \nabla p 
\,\,\, , \,\,\,  \nabla \cdot {\bf u}=0 \,,
\end{align}

where ${\bf u}$ is the velocity field, $\rho_0$ the constant fluid
density, and $p$ the pressure.  Consider a fluid in a box of size $L$
with periodic boundary conditions and a cutoff $\widetilde{K}$

\begin{align}
{\bf u}({\bf x}, t)= { 1 \over L^{3/2}} \sum_{|{\bf k}| < \widetilde{K}} {\hat {\bf u}}({\bf k}, t) e^{i {\bf k} \cdot {\bf x}} \, .
\end{align}

Because of the incompressibility condition ${\hat {\bf u}}({\bf k}, t)
\cdot {\bf k} =0$, it is appropriate to use a set of independent
variables $\{ X_n \}$, evolving according to

\begin{align}
{d X_n \over dt}= \sum_{j,l} M_{njl} X_j X_l \,\,\,\,, \,\,\, n=1, 2, ...., N \sim \widetilde{K}^3 \,,
\end{align}
where $M_{njl}$ is determined by the Navier-Stokes equation.
Once an ultraviolet cutoff is introduced, it is rather simple to build
up an equilibrium SM of the system: it is enough to
use the Liouville theorem and the energy conservation and follow the
usual approach used for the standard SM of
Hamiltonian system~\cite{bo98}.
It is easy to show that
\begin{align}
 \sum_n { \partial \over \partial X_n} {d X_n \over dt}=0
\end{align}
and  
\begin{align}
  {1 \over 2} \sum_n X_n^2=  E= const\,.
\end{align}

Therefore, following  the same reasoning used for the SM of Hamiltonian systems, one obtains
\begin{align}
P(\{ X_n \}) = C \, \delta\Big( {1 \over 2} \sum_n X_n^2 -E \Big)
\end{align}

and equipartition $\langle X_n^2 \rangle =2 E/N$,  in the limit of large $N$  one has a Gaussian distribution

\begin{align}
P(\{ X_n \})  \sim \exp\left(- {\beta \over 2}  \sum_n X_n^2 \right)\,.
\end{align}

The above results for the Euler equation are mathematically correct
and in agreement with numerical computations, but they have a rather
poor relation with the Navier-Stokes equation even in the limit of very
large Reynolds numbers, and therefore the SM for
the inviscid fluids has a very marginal interest for
turbulence~\cite{f95,bo98}.

Let us go back to non conservative systems, in particular to the
chaotic and dissipative case of the Navier-Stokes equation.  In those cases the asymptotic
behaviour of the system will occur on a strange attractor and the
invariant measure cannot be smooth; in particular it is not continuous
with  respect to the Lebesgue measure, singular, and usually
described by a multifractal measure~\cite{BPPV84,PV87}. This unpleasant feature can be prevented introducing a small noise,
in such a way that the problem of the singularity of the invariant measure
is removed.  This procedure is not a mere mathematical trick: it is
rather reasonable to accept the view that the system under
investigation is inherently noisy, e.g., due to the influence of the
external environment.  Therefore one can assume that the ``correct'' measure
is that one obtained adding a (small) noise term (of strength
$\epsilon$) to the dynamical system and then performing the limit
$\epsilon \to 0$.  The measure selected with this procedure is called
``natural'' (or physical) measure and it is, by construction, a
``dynamically robust'' quantity, i.e. it changes in a continuous way when a small perturbation is present. According to Eckmann and
Ruelle~\cite{ER85} this idea dates back to Kolmogorov.

It should be noticed that in any numerical simulation both computers and algorithms are not
``perfect'', i.e. there are unavoidable ``errors'' due to the
truncations, round off and so on. In a similar way in the laboratory
experiments it is not possible to eliminate all the noisy interactions
with the environment. It is therefore evident, at least from a
physical point of view, that numerical simulations and experiments
provide an approximation of the natural measure Once the
technical difficulties about the non singular character of the
invariant measure has been removed one has to found the invariant distribution 
$\rho_{\epsilon}^I({\bf x})$; this is surely a well defined problem
for each value of $\epsilon$.

As already discussed, if the time evolution of the system is ruled by
a deterministic law, one can easily write down the time evolution for
the probability distribution. Due to the presence of randomness in the
dynamics, the evolution law
of $\rho({\bf x},t)$ needs to be modified accordingly.  Let us consider, for instance, the case where each variable $x_i$
is forced by a noisy term $\epsilon_i \eta_i$, where by $\eta_i$ we
denote the components of a white noise vector, i.e.  a Gaussian
process with $\langle \eta_i \rangle=0$ and $\langle \eta_i(t)
\eta_j(t') \rangle=\delta_{ij} \delta(t-t')$. We get for the individual trajectory of $x_i$ the Langevin
equation
\begin{align}
  \frac{dx_n}{dt} = f_n({\bf x}) + \epsilon_n  \eta_n
\end{align}
and for its pdf the Fokker- Planck equation
\begin{equation}
\label{FP}
{ {\partial \rho} \over \partial t} =
 - \nabla \cdot ({\bf f}\rho)  +{ 1 \over 2}\sum_n \epsilon_n^2  { \partial^2 \over \partial x_n   \partial x_n } \rho \, .
\end{equation}
The problem of finding $\rho_{\epsilon}^I({\bf x})$
for a Fokker-Planck evolution is very difficult even in low dimensional
system. \new{Although the invariant
probability distribution can be identified in specific cases -- as a remarkable result we can mention~\cite{matan2014}, where $\rho_{\epsilon}^I({\bf x})$ has been found for Lorenz-like models --,} up to now nobody was able to find it \new{for generic systems} with noisy terms.  This example shows that even for noisy chaotic dissipative systems there is not a definite protocol to determine the invariant probability. This is a fundamental difference with respect to Hamiltonian systems, and in general to dynamics that conserve the phase space volume (e.g. Euler equation), where $\rho_{\epsilon}^I({\bf x})$ is nothing but
the microcanonical distribution.

\subsection{Again about chaos  and statistical laws}
\label{sec:conclusionchaos}

The (re)discovery of chaos in the 1960s had a relevant role in the
redefinition of important  aspects of the scientific  thought, in particular
about topics as determinism,  prediction  and complexity. This is  not the proper place for a  detailed analysis of this subject, and the interested reader can refer to~\cite{bricmont1995chaos, castiglione2008chaos}.
Here we are only interested in the relation between chaos and the validity of SM, for which different points of view have been expressed.

Many  detailed numerical studies of high dimensional Hamiltonian systems  give
 a rather  clear   indication that chaos is neither a necessary nor a sufficient ingredient for justifying, 
 on a dynamical basis, the validity of equilibrium SM. 
Even when chaos is very weak (or absent), one may observe a good agreement between time and ensemble averages. Chaotic behaviors do not necessarily imply such an agreement, as discussed in detail in Section~\ref{sec:chaos}.

The scenario is even more controversial when passing to non-equilibrium SM.
 Several simulations and theoretical works have shown that, in hyperbolic systems, there exists a close relationship between transport coefficients, 
 such as viscosity, thermal and electrical conductivity, and chaos indicators, such as Lyapunov exponents, 
 Kolmogorov-Sinai entropy and escape rates~\cite{gaspard_1998, Dorfman1999}. 
 There is a strong evidence
  that chaos, in the technical sense of positivity of the maximum Lyapunov exponent, is not necessary for 
  non-equilibrium phenomena such as diffusion and conduction, which can be obtained also in systems having vanishing Lyapunov exponent ~\cite{dettmann2000microscopic,dettmann2001note,cecconi2003origin,Cecconi2007}. 
   
The reason  of the  poor  relevance  of chaos for the SM
can be explained in a very simple way as follows.
Important  concepts of chaos, e.g. the Lyapunov exponents, can be defined  in a mathematical rigorous way only
performing  two limits: the perturbation must remain infinitesimal
(in mathematical terms one works with the tangent vectors)  and the observation time must be arbitrarily long.
In the case of low dimensional systems, e.g. the Lorenz's model or the H\'enon's map,  the constraint
on the size of the perturbation, if small enough,  is not really important. On the contrary, it turns out to be relevant when the fluctuations of the degrees of freedom are very different (as in turbulence).
It is now clear that the maximum Lyapunov exponent and the Kolmogorov-Sinai entropy 
 cannot be  completely satisfactory
 for a proper characterization of the many faces of complexity and predictability of systems, such as
 turbulence or, in general, in cases with  many degrees of freedom as in SM~\cite{boffetta2002predictability}.
  At a fist glance it seems natural to suppose that
 there is a relation between quantities of the dynamical systems
realm as  the Kolmogorov-Sinai entropy $h_{KS}$ and the growth of the coarse-grained (Gibbs like) entropy.
However a detailed analysis shows that such a relation is  rather weak, as discussed in Section~\ref{sec:chaos}.
The main reason is the asymptotic nature of $h_{KS}$, i.e. the fact that its value is determined by the
behaviour of the system at very large time intervals. On the contrary
the growth of the coarse-grained entropy involves short time
intervals. Besides, during the early time evolution of the Gibbs entropy one can have entanglement of behaviours at
different characteristic spatial  scales~\cite{PhysRevE.71.016118}.

 Let us briefly  explain the main reason of the irrelevance 
 of chaos for the statistical features of high dimensional systems.
 Consider  a turbulent fluid  with $Re \gg 1$, where $Re= U L/ \nu$ is the Reynolds number ($U$ and $L$ are the characteristic
 velocity and length,  respectively, and $\nu$ is the viscosity). A simple analysis shows that \new{for the Lyapunov exponent one has} $\lambda \sim  (U/L) Re^{\alpha} $,
 where $\alpha \simeq 1/2$, the precise value depending on the intermittent corrections to the Kolmogorov theory~\cite{bo98}.
 This result  is only meaningful if we are interested in the  prediction 
 of perturbations  with a very high resolution, namely if the
 uncertainty on the velocity field is much smaller than the Kolmogorov velocity $U Re^{- 1/4}$.
Instead, the LE cannot have a role for the prediction at large scale, as e.g. in the weather forecasting,
 and it is necessary to generalize the LE for non infinitesimal resolution~\cite{boffetta2002predictability}.
 
 In SM  one has a rather similar scenario. Consider an interacting gas: at a  microscopic level we
 have a chaotic dynamics whose LE is $O(1 / \tau_c)$, where $\tau_c$ is the collision time~\cite{gaspard2022statistical}.
 It is not difficult to understand that  this result cannot be relevant in  the large-scale properties, which do not crucially depend on the microscopic details:  the  characteristic  time of the macroscopic quantities are much larger than $\tau_c$.
 These aspects were discussed  by Dettmann and Cohen, who showed with 
  a clear analysis the practical impossibility
     to observe differences in the diffusive behavior between a genuine deterministic chaotic system (the 2D Lorentz gas with
 circular obstacles) and its non-chaotic variant (the wind-tree Ehrenfest model)~\cite{dettmann2000microscopic,  dettmann2001note}.
  The Ehrenfest model consists of free moving independent particles (wind) that scatter against square obstacles (trees)
   randomly distributed in the plane, with fixed orientation. 
  Due to collisions, particles undergo diffusion; however their motion cannot be chaotic, because a reflection by the flat planes of obstacles does not produce exponential trajectory separation. 
  The divergence is at most algebraic, leading to zero Lyapunov exponent. 
  Of course the above  considerations can be extended to every polygonal scatterer with $N$ sides,
  so there is a whole class of models where diffusion occurs in the absence of chaotic motion.
  
  The analysis  of  the diffusion of free particles against polygonal scatterers 
   clarifies the asymptotic nature of the LE, and its marginal role in SM:
  given   polygonal scatterers with $N$ sides, the true LE (which is zero) can only be observed if a resolution
  much smaller than the length of the side, $O(1/N)$, is available. Therefore in the  limit $N \gg 1$  this effect has no physical relevance.
The necessary ingredient to obtain diffusion in the absence of deterministic chaos is finite-size instability,
i.e. the fact that infinitesimal perturbations are stable, while perturbations of finite size can grow algebraically. This condition typically appears in the presence of   quenched disorder and of quasi-periodic perturbation, see e.g.~\cite{cecconi2003origin,Cecconi2007}.

\subsection{Establishing probabilities without dynamics: the maximum entropy principle}
\label{sec:maxent}

As we mentioned before, part of the scientific community regards SM as a form of statistical inference rather than as a description of objective physical reality. From this point of view, probabilities are interpreted as measures of the degree of ``ignorance'' about a given fact, rather than as quantities which can be physically measured. 
This ``anti-dynamical'' point of view is quite in contrast with the traditional interpretation developed by the founding fathers, and it is in our opinion a bit extreme.

One of the main exponents of this approach was Jaynes, who proposed the maximum entropy principle (MEP) as the rule to find the
probability of a given event in circumstances where only partial
information is available~\cite{J67}.  Let us briefly remind the idea: assuming
the mean values of $m$ independent functions $f_i(\psv{X})$ are known:
\begin{equation}
c_i= \langle f_i \rangle = 
\int f_i(\psv{X})\rho(\psv{X}) d \psv{X} \quad \quad i=1, ... , m \,\,\, ,
\end{equation}
the prescribed rule of the MEP to determine  $\rho(\psv{X})$ is
to maximize the entropy 
\begin{align}
H=  - \int \rho(\psv{X}) \ln \rho(\psv{X}) d \psv{X}
\end{align}
under the constraints $c_i= \langle f_i \rangle$.  
Using the Lagrangian multipliers one easily obtains
\begin{equation}
\rho(\psv{X})= 
{1 \over Z} \exp \sum_{i=1}^m \lambda_i f_i(\psv{X})
\end{equation}
where $\lambda_1, \lambda_2 ...\lambda_m$  are determined by   $c_1, c_2,  ... ,c_m$.
The most frequent, and obvious, objection to this point of view can be
summarized with the Latin motto {\it ``Ex nihilo nihil''}, meaning that it is
not possible to gain insight on a subject from the knowledge of our ignorance.
A nice discussion about an
obvious misuse of the MEP can be found in Ma's  book~\cite{ma85}:
\begin{quotation}
   If I
  ask, ``How many days in a year does it rain in Hsinchu?'' One might
  reply, ``As there are two possibilities, to rain or not to rain, and
  I am completely ignorant about Hsinchu, therefore it rains six
  months in a year.''  This reply relies on the same reason above,
  using ignorance to predict a natural phenomenon, and is equally
  invalid.  Instead, the past record of the rainfall can be obtained
  from the meteorological office. The rainfall in the future can be
  predicted on the basis of the past records plus other
  information. We cannot use ignorance to make predictions. 
\end{quotation}

The MEP approach, when applied to SM, with a fixed
number of particles and the unique constraint on the mean value of the
energy, leads to the usual canonical distribution in a very simple
way.  In an analogous way, imposing also the constraint of the mean
value of the particles, one obtains the grand canonical distribution.
Actually these correct results, which are considered by the supporters
of the MEP as a strong evidence that this approach is relevant for the foundations of SM, are mere coincidences, due to the (lucky) use of the
canonical variables $\psv{X}= ({\bf q}_1, {\bf q}_2,..., {\bf
  q}_N,{\bf p}_1, {\bf p}_2,..., {\bf p}_N)$.

 In a similar way the Maxwell-Boltzmann distribution of dilute gases can be obtained 
 with MEP approach, but, as noted in Sect.~\ref{eq:Bodiluttoq}, also  this is nothing but a coincidence.  
  
Let us clarify the dependence of the result on the used variables:
consider, just for simplicity’s sake, a scalar random variable $x$,
ranging over a continuum, whose probability distribution function is
$p_x(x)$. It is easy to realise that the ``entropy''
\begin{align}
  H_x=-\int  p_x(x) \ln  p_x(x) \, dx
\end{align}
is not an intrinsic quantity of the phenomena concerning $x$.  With a
different parametrisation, i.e. using the coordinate $y = f(x)$ with
an invertible function $f$, rather than $x$, the entropy of the same
phenomenon is given by
\begin{align}
  H_y=-\int  p_y(y) \ln  p_y(y) \, dy
\end{align}
where $p_y(y)$ is determined by the well know formula
\begin{align}
  p_y(y)= {p_x(x) \over |f'(x)| } \Big|_{x=f^{-1}(y)}.
\end{align}
Therefore one has
\begin{align}
  H_y=H_x+\int  p_x(x) \ln | f'(x)|  \, dx\,.
\end{align}
As a consequence, the MEP does not give a unique answer: one obtains
different results if different variables are adopted to describe the
very same phenomenon.  For instance, if we apply the MEP to the
variable $x\in [0, 1]$, without additional information we have
$p_x(x)=1$, while for the variable $y=\cos( \pi x) \in [-1, 1]$ one
has $p_y(y)=1/(\pi \sqrt{1-y^2})$. On the other hand, by applying the MEP to
the variable $y$ one obtains $p_y(y)=1/2$.  Of course the above
conflicting result in the use of MEP is nothing but Bertrand's
paradox~\cite{sh05}: if one tries to compute a probability of a given
event just from the complete ignorance, different results are obtained,
reflecting some hidden assumptions.

In order to avoid this dependence on the choice of variables, Jaynes
later proposed a more sophisticated version of the MEP (REF), in terms of
the relative entropy:
\begin{align}
\tilde H=-\int  p_x(x) \ln \Big(  {p_x(x) \over q(x)} \Big)  \, dx
\end{align}
where $q(x)$ is a known probability density.  It is easy to show
that $\tilde H$ does not depend on the chosen variable $x$, but it
depends on $q(x)$.  On the other hand one must decide how to select
$q(x)$, and this issue is equivalent to the problem of choosing the
``proper variables''.  Therefore, even this more elaborate method is
non-predictive, and we see no reason to pursue the MEP approach
further in the field of SM.

 \subsection{On the Choice of Macroscopic Observables}
\label{sec:otrbmm}
 
We would like clarify  any potential confusion that may have arisen from the approach presented to the problem of thermalization in Section \ref{sec:Ufft} and investigated so far. The distinctions between different senses of thermalization, reflecting various interpretations of thermal equilibrium, hinge on the dependence of the latter on the seemingly arbitrary selection of a family of observables. Should we conclude that the cooling of a cup of coffee, or the increase of entropy to be discussed next, depends on what we choose to observe? Of course not!

These distinctions are fundamental to our conceptualizations of a complex process---thermalization---for which it is beneficial to consider various definitions, each mathematically capturing a facet of the physical complexities inherent in real-world systems. Consider, for instance, the differentiation between microscopic and macroscopic thermalization. Firstly, this contrast underscores a disparity between the classical and quantum understandings of thermalization (there is no classical equivalent of microscopic equilibrium). Secondly, while the former concept strengthens the latter (MICRO implies MACRO), it also enables the characterization of thermalization on microscopic scales beyond the thermodynamic scale, where macroscopic thermalization ensures the validity of thermodynamic descriptions. In thermodynamics, there is nothing contingent on the choice of observables, apart from the trivial sense that each physical system possesses its own thermodynamic description—a fluid's description differs from that of a crystal, for example, but this choice dependence is trivial.

Since our ability to analyze real systems is limited, we often resort to idealized models or the study of typical behavior. The latter approach is particularly illuminating. When we can demonstrate that a property or behavior is typical, we have a reasonable belief that it may also hold true for a specific system, unless compelling reasons suggest otherwise. However, as we have seen, in the case of thermalization, the typical thermalization time scale, in disagreement with observed behavior, indeed suggests otherwise. So we should revisit idealized models to better understand what is happening. 

Consider the case of a free quantum gas, either bosonic or fermionic, as an example. In this system, not all states approach microscopic equilibrium. For instance, as noted in \cite{Tumulka2019}, if all particles possess nearly identical kinetic energy, this equilibrium state persists in the absence of interactions. Even within subregions, all particles maintain essentially the same kinetic energy, a phenomenon starkly contrasting with the canonical density matrix.

Suppose we could show that the ideal gas's macroscopic thermalization holds through an appropriate choice of macroscopic variables, such as hydrodynamic quantities like particle density, but not for kinetic variables. Should we conclude that thermalization depends on the selection of macro-observables?  

Though in highly idealized models, certain macro-observable choices may prove more convenient than others for studying the thermalization process, it would be more prudent to infer that while discrepancies in the choice of macro-observables may arise in idealized models, they are less likely to manifest in more realistic systems, such as a hot cup of coffee. In such cases, we can still rely on fundamental factors like the system's large number of degrees of freedom and the dominance of thermal equilibrium to effectively demonstrate the occurrence of thermalization. This perspective underscores the importance of considering these fundamental factors in understanding thermalization processes across different systems, from idealized models to real-world scenarios. 

It's worth noting that this perspective does not necessarily exclude the role of chaos in the thermalization of real systems. However, chaos might emerge in real systems as a consequence, rather than an assumption, of these two fundamental factors.
Indeed, the macroscopic irreversibility observed in a model such as the fermionic chain studied by Tasaki suggests that the fundamental mechanism behind its emergence does not necessarily rely on assumptions about chaotic behavior.  

Thermalization is a multifaceted phenomenon that encompasses the equilibration of various degrees of freedom within a system. Different degrees of freedom interact and equilibrate at varying rates, so multiple time scales might be involved. For example, in a system consisting of particles with both translational and internal degrees of freedom, the translational motion may equilibrate relatively quickly compared to the internal degrees of freedom. This difference in time scales arises from the varying strengths of interactions between different degrees of freedom. Translational motion, being more directly affected by collisions and interactions with the surroundings, tends to equilibrate faster than internal degrees of freedom, which may involve slower processes such as chemical reactions or changes in molecular configurations. See also the discussion in Sec.~\ref{sec:ergodicitytime}.

Similarly, thermalization may encompass the equilibration of different energy levels or quantum states within a quantum system, each dictated by its characteristic time scale. For instance, in a many-body quantum system, the rapid relaxation of certain local observables to their thermal equilibrium values may contrast with the longer time scales required for the equilibration of locally conserved quantities, as previously emphasized.
Additional insights into the discussed points have been offered in Section \ref{sec:qentropy}.

\section{Conclusions}
\label{sec:conclusions}

Statistical mechanics is an invaluable and irreplaceable tool to
deduce the thermodynamic properties of macroscopic objects from their
microscopic interactions (e.g., to compute their free energy and
correlation functions). While its practical utility is self-evident and 
needs no further discussion, the reasons of its success are still, from 
time to time, the subject of scientific controversies.
In short, the problem of foundations is: why does statistical mechanics work, despite many technical issues being not fully solved and understood?

From its very beginning statistical mechanics has
been characterized by the coexistence of two different souls:
dynamics and probability. A possible way to connect these two realms
is via the ergodic hypothesis, as first realized by
Boltzmann himself.  In the long journey starting from the work and
the reasoning of founding fathers such as Clausius, Maxwell, Boltzmann
and Gibbs, the development of speculation on statistical mechanics
foundations has been characterized by a plethora of interesting and
subtle topics ranging from deep physical intuitions to sophisticated
mathematics, as well as smart numerical simulations and clever
experiments. Among the many breakthroughs of the theory we can mention
the smooth extension of statistical mechanics to quantum systems and
the discovery of deterministic chaos.

Within this scenario, the spectrum of positions to explain the success of the statistical mechanics is quite various. The aim of this review paper
  has been precisely to discriminate between those that, in the
  opinion of the authors, are actually well grounded, and those that can
  be conceptually catching at first sight, but under a more severe
  scrutiny turn out to be less compelling. For instance, after the discovery of deterministic
  chaos most people, under the influence of Prigogine's school, started to believe
  that the presence of positive Lyapunov exponents was the crucial
  ingredient to use probabilistic arguments in physical systems. Instead, robust counterexamples to the claimed relevance of chaos can be found,
  where the presence of many degrees of freedom plays a crucial role, irrespectively of the properties of microscopic
  dynamics. Another approach blessed with remarkable success, but not
  really able to tackle the essence of the foundational
  problem is, in
  the opinion of the authors, the maximum entropy principle. According to it, it is
  not even necessary to look for physical foundations of statistical
  mechanics, since all the statistical properties of a given system can
  be inferred by simply maximizing an information theoretic
  entropy, under the constraints that are known to act on the considered system.

In the previous Sections we tried to present the problem of the foundations
by collecting and discussing some of the most influential works addressing the 
subject, both in the classical and in the quantum domain. The topic is of course very broad, and our selection inevitably suffers from our subjective point of view. 

One of the central take-home messages is that, although ergodicity is not exactly
verified in realistic Hamiltonian models, in systems with many degrees
of freedom  it does hold in a weak sense, which can be sufficient
for the purposes of physics, as showed by Khinchin, Mazur and van der Linden.
In this perspective, as recalled above, chaos is not a fundamental
ingredient for the validity of statistical mechanics.
Once this weak ergodicity is accepted, the law $ S = k_B \ln W$ can be used
to link the microscopic mechanical
quantity $W$ with the macroscopic thermodynamic observable $S$. This fact provides a
conceptual connection between the microscopic dynamics and
the macroscopic world, as well as an operational way to pass from mechanics to thermodynamics.

The above program is still very challenging, in principle. A crucial simplification is due to the fact that macroscopic objects are made of
very large numbers of microscopic constituents, which allows us to use
probabilistic methods, e.g. the statistical ensembles, to study
a single macroscopic system. This possibility is strictly related to the concept of typicality, and is also central in the understanding of macroscopic irreversibility. Once again, the presence of many degrees of freedom is the key aspect.


A key aspect of Boltzmann’s formula, $  S = k_B \ln W  $, is that it applies directly to individual systems, with  $ W $   representing the ``size'' of the macrostate to which the microstate belongs. This formula retains its significance in the quantum realm as well, where the macrostate’s size corresponds to the dimensionality of the macrospace associated with the quantum state, in an appropriate sense. Extending Boltzmann's framework to quantum systems, however, introduces additional complexities, especially in defining macrostates through coarse-grained observables; here, the non-commutativity of quantum operators can add complications. Nevertheless, with suitable techniques, these challenges can be managed, allowing the quantum analysis to closely resemble its classical counterpart.

One of the most notable outcomes of extending Boltzmann’s approach to quantum systems is the concept of quantum equilibrium dominance. This principle suggests that, as in the classical case, the fraction of pure states corresponding to thermal equilibrium is overwhelmingly large compared to the entire energy shell. This underscores the central role of equilibrium thermodynamics and the typicality of thermalization in quantum systems. Furthermore, in the quantum context, this principle is enriched by entanglement—a unique feature absent in classical systems—that enables the exploration of thermal equilibrium at the microscopic level, extending beyond the traditional macroscopic concept.

\newpage
 \section*{Acknowledgements}

MB and AV are grateful for the insightful discussions with E. Aurell, F. Cecconi, M. Cencini, L. Cerino, S. Chibbaro, M. Falcioni, R. Livi, G. Mantica,  U. Marini Bettolo Marconi, L. Peliti, A. Ponno, A. Puglisi, A. Sarracino, L. Rondoni and S. Ruffo. Without them, this work would have never been conceived.
\quad\\
\quad\\
GG wishes to thank as well R. Livi, A. Maritan, A. Ponno and  L. Salasnich for the many inspiring conversations on the subject, and for their long-standing, stimulating collaboration.
\quad\\
\quad\\
NZ extends heartfelt gratitude to the group on Statistical Mechanics led by J. Lebowitz, with whom he has interacted over many decades. This work reflects the knowledge gained from his collaboration with J. Lebowitz, S. Goldstein, and R. Tumulka on the foundations of quantum statistical mechanics. Special thanks to S. Goldstein for his enduring friendship, unwavering support, and impactful mentorship in the foundational aspects of physics. NZ also thanks H. Tasaki for fruitful exchanges.
 \quad\\
 \quad\\
All the authors wish to thank A. Sarracino for reading the manuscript and providing them with useful comments and suggestions.
\quad\\
\quad\\
 MB acknowledges support by ERC Advanced Grant RG.BIO (Contract No. 785932); GG acknowledges partial support from the project MIUR-PRIN2022, “Emergent Dynamical Patterns of Disordered Systems with Applications to Natural Communities”, code 2022WPHMXK; NZ acknowledges the financial support from INFN.
\newpage

\bibliographystyle{abbrvnat}
\bibliography{biblio}

\begin{thebibliography}{248}
\providecommand{\natexlab}[1]{#1}
\providecommand{\url}[1]{\texttt{#1}}
\expandafter\ifx\csname urlstyle\endcsname\relax
  \providecommand{\doi}[1]{doi: #1}\else
  \providecommand{\doi}{doi: \begingroup \urlstyle{rm}\Url}\fi

\bibitem[Aaij et~al.(2013)Aaij, Beteta, Adeva, Adinolfi, Adrover, Affolder, Ajaltouni, Albrecht, Alessio, Alexander, et~al.]{aaij2013first}
R.~Aaij, C.~A. Beteta, B.~Adeva, M.~Adinolfi, C.~Adrover, A.~Affolder, Z.~Ajaltouni, J.~Albrecht, F.~Alessio, M.~Alexander, et~al.
\newblock First observation of c p violation in the decays of b s 0 mesons.
\newblock \emph{Phys. Rev. Lett.}, 110\penalty0 (22):\penalty0 221601, 2013.

\bibitem[Aaij et~al.(2019)Aaij, Beteta, Adeva, Adinolfi, Aidala, Ajaltouni, Akar, Albicocco, Albrecht, Alessio, et~al.]{aaij2019observation}
R.~Aaij, C.~A. Beteta, B.~Adeva, M.~Adinolfi, C.~A. Aidala, Z.~Ajaltouni, S.~Akar, P.~Albicocco, J.~Albrecht, F.~Alessio, et~al.
\newblock Observation of c p violation in charm decays.
\newblock \emph{Phys. Rev. Lett.}, 122\penalty0 (21):\penalty0 211803, 2019.

\bibitem[Abe et~al.(2001)Abe, Abe, Adachi, Ahn, Aihara, Akatsu, Alimonti, Asai, Asai, Asano, et~al.]{abe2001observation}
K.~Abe, R.~Abe, I.~Adachi, B.~S. Ahn, H.~Aihara, M.~Akatsu, G.~Alimonti, K.~Asai, M.~Asai, Y.~Asano, et~al.
\newblock Observation of large cp violation in the neutral b meson system.
\newblock \emph{Phys. Rev. Lett.}, 87\penalty0 (9):\penalty0 091802, 2001.

\bibitem[Alba and Calabrese(2017)]{AlbaCalabrese17}
V.~Alba and P.~Calabrese.
\newblock {Entanglement and thermodynamics after a quantum quench in integrable systems}.
\newblock \emph{Proc. Nat. Acad. Sci.}, 114\penalty0 (30):\penalty0 7947, 2017.
\newblock \doi{10.1073/pnas.1703516114}.

\bibitem[Anderson(1958)]{anderson1958}
P.~W. Anderson.
\newblock Absence of diffusion in certain random lattices.
\newblock \emph{Phys. Rev.}, 109:\penalty0 1492--1505, 1958.

\bibitem[Antenucci et~al.(2015)Antenucci, Crisanti, and Leuzzi]{antenucci2015glassy}
F.~Antenucci, A.~Crisanti, and L.~Leuzzi.
\newblock The glassy random laser: replica symmetry breaking in the intensity fluctuations of emission spectra.
\newblock \emph{Sci. Rep.}, 5\penalty0 (1):\penalty0 16792, 2015.

\bibitem[Arkeryd(1972)]{Arkeryd1972}
L.~Arkeryd.
\newblock On the {Boltzmann} equation.
\newblock \emph{Archive for Rational Mechanics and Analysis}, 45\penalty0 (1):\penalty0 1--16, Jan 1972.
\newblock ISSN 1432-0673.
\newblock \doi{10.1007/BF00253392}.

\bibitem[Arkeryd et~al.(1987)Arkeryd, Esposito, and Pulvirenti]{Arkeryd1987}
L.~Arkeryd, R.~Esposito, and M.~Pulvirenti.
\newblock The boltzmann equation for weakly inhomogeneous data.
\newblock \emph{Comm. Math. Phys.}, 111\penalty0 (3):\penalty0 393–407, Sept. 1987.
\newblock ISSN 1432-0916.
\newblock \doi{10.1007/bf01238905}.
\newblock URL \url{http://dx.doi.org/10.1007/BF01238905}.

\bibitem[Arnold(1963)]{A63}
V.~Arnold.
\newblock Proof of a theorem of an kolmogorov on the preservation of conditionally periodic motions under a small perturbation of the hamiltonian, uspehi mat.
\newblock \emph{Dokl. Akad. Nauk SSSR}, 18\penalty0 (5):\penalty0 113, 1963.

\bibitem[Arnol'd(1964)]{arnold2009instability}
V.~I. Arnol'd.
\newblock Instability of dynamical systems with many degrees of freedom.
\newblock In \emph{Doklady Akademii Nauk}, volume 156, pages 9--12. Russian Academy of Sciences, 1964.

\bibitem[Arnold(1974)]{Arnold}
V.~I. Arnold.
\newblock \emph{Mathematical Methods of Classical Mechanics}.
\newblock Springer-Verlag, New York, 1974.

\bibitem[Aubert et~al.(2001)Aubert, Boutigny, De~Bonis, Gaillard, Jeremie, Karyotakis, Lees, Robbe, Tisserand, Palano, et~al.]{aubert2001measurement}
B.~Aubert, D.~Boutigny, I.~De~Bonis, J.-M. Gaillard, A.~Jeremie, Y.~Karyotakis, J.~Lees, P.~Robbe, V.~Tisserand, A.~Palano, et~al.
\newblock Measurement of cp-violating asymmetries in b 0 decays to cp eigenstates.
\newblock \emph{Phys. Rev. Lett.}, 86\penalty0 (12):\penalty0 2515, 2001.

\bibitem[Ayi(2017)]{Ayi2017}
N.~Ayi.
\newblock From newton's law to the linear boltzmann equation without cut-off.
\newblock \emph{Comm. Math. Phys.}, 350\penalty0 (3):\penalty0 1219--1274, Mar 2017.

\bibitem[Baldovin et~al.(2019)Baldovin, Caprini, and Vulpiani]{baldovin2019irreversibility}
M.~Baldovin, L.~Caprini, and A.~Vulpiani.
\newblock Irreversibility and typicality: A simple analytical result for the ehrenfest model.
\newblock \emph{Physica A}, 524:\penalty0 422--429, 2019.

\bibitem[Baldovin et~al.(2021{\natexlab{a}})Baldovin, Iubini, Livi, and Vulpiani]{baldovin2021}
M.~Baldovin, S.~Iubini, R.~Livi, and A.~Vulpiani.
\newblock Statistical mechanics of systems with negative temperature.
\newblock \emph{Phys. Rep.}, 923:\penalty0 1--50, 2021{\natexlab{a}}.
\newblock ISSN 0370-1573.
\newblock \doi{https://doi.org/10.1016/j.physrep.2021.03.007}.
\newblock URL \url{https://www.sciencedirect.com/science/article/pii/S0370157321001204}.
\newblock Statistical mechanics of systems with negative temperature.

\bibitem[Baldovin et~al.(2021{\natexlab{b}})Baldovin, Vulpiani, and Gradenigo]{BGV21}
M.~Baldovin, A.~Vulpiani, and G.~Gradenigo.
\newblock Statistical mechanics of an integrable system.
\newblock \emph{J. Stat. Phys.}, 183\penalty0 (3):\penalty0 41, 2021{\natexlab{b}}.

\bibitem[Baldovin et~al.(2023)Baldovin, Marino, and Vulpiani]{baldovin2023}
M.~Baldovin, R.~Marino, and A.~Vulpiani.
\newblock Ergodic observables in non-ergodic systems: the example of the harmonic chain.
\newblock \emph{Physica A}, 630:\penalty0 129273, 2023.

\bibitem[Baldovin et~al.(2024)Baldovin, Gradenigo, and Vulpiani]{BVG21}
M.~Baldovin, G.~Gradenigo, and A.~Vulpiani.
\newblock Statistical features of high-dimensional hamiltonian systems.
\newblock In C.~Hidalgo, editor, \emph{{EPS Grand Challenges: Physics for Society in the Horizon 2050}}. IOP Publishing, 2024.

\bibitem[Barnum et~al.(1994)Barnum, Caves, Fuchs, Schack, Driebe, Hoover, Posch, Holian, Peierls, and Lebowitz]{barnum1994boltzmann}
H.~Barnum, C.~M. Caves, C.~Fuchs, R.~Schack, D.~J. Driebe, W.~G. Hoover, H.~Posch, B.~L. Holian, R.~Peierls, and J.~L. Lebowitz.
\newblock Is {Boltzmann} entropy time’s arrow’s archer.
\newblock \emph{Phys. Today}, 47\penalty0 (11):\penalty0 11--15, 1994.

\bibitem[Barr{\'e} et~al.(2001)Barr{\'e}, Mukamel, and Ruffo]{BMR01}
J.~Barr{\'e}, D.~Mukamel, and S.~Ruffo.
\newblock Inequivalence of ensembles in a system with long-range interactions.
\newblock \emph{Phys. Rev. Lett.}, 87\penalty0 (3):\penalty0 030601, 2001.

\bibitem[Basko et~al.(2008)Basko, Aleiner, and Altshuler]{basko2008}
D.~M. Basko, I.~L. Aleiner, and B.~L. Altshuler.
\newblock On the problem of many-body localization.
\newblock In A.~L. Ivanov and S.~G. Tikhodeev, editors, \emph{Problems of Condensed Matter Physics}, pages 50--69. Oxford University Press, 2008.

\bibitem[Batterman(2001)]{ba02}
R.~W. Batterman.
\newblock \emph{The devil in the details: Asymptotic reasoning in explanation, reduction, and emergence}.
\newblock Oxford University Press, 2001.

\bibitem[Benettin and Ponno(2011)]{BP11}
G.~Benettin and A.~Ponno.
\newblock Time-scales to equipartition in the fermi--pasta--ulam problem: finite-size effects and thermodynamic limit.
\newblock \emph{J. Stat. Phys.}, 144:\penalty0 793--812, 2011.

\bibitem[Benettin et~al.(1980)Benettin, Galgani, Giorgilli, and Strelcyn]{benettin1980lyapunov}
G.~Benettin, L.~Galgani, A.~Giorgilli, and J.-M. Strelcyn.
\newblock Lyapunov characteristic exponents for smooth dynamical systems and for hamiltonian systems; a method for computing all of them. part 1: Theory.
\newblock \emph{Meccanica}, 15:\penalty0 9--20, 1980.

\bibitem[Benettin et~al.(2013)Benettin, Christodoulidi, and Ponno]{BCP13}
G.~Benettin, H.~Christodoulidi, and A.~Ponno.
\newblock The fermi-pasta-ulam problem and its underlying integrable dynamics.
\newblock \emph{J. Stat. Phys.}, 152:\penalty0 195--212, 2013.

\bibitem[Benzi et~al.(1984)Benzi, Paladin, Parisi, and Vulpiani]{BPPV84}
R.~Benzi, G.~Paladin, G.~Parisi, and A.~Vulpiani.
\newblock On the multifractal nature of fully developed turbulence and chaotic systems.
\newblock \emph{J. Phys. A}, 17\penalty0 (18):\penalty0 3521, 1984.

\bibitem[Berman and Izrailev(2005)]{BI05}
G.~Berman and F.~Izrailev.
\newblock The fermi--pasta--ulam problem: fifty years of progress.
\newblock \emph{Chaos}, 15\penalty0 (1):\penalty0 015104, 2005.

\bibitem[Berry(1994)]{berry1995asymptotics}
M.~V. Berry.
\newblock Asymptotics, singularities and the reduction of theories.
\newblock In D.~Prawitz, B.~Skyrms, and D.~Westerst{\aa}hl, editors, \emph{{Logic, Methodology and Philosophy of Science IX}}, pages 597--607. Elsevier, 1994.

\bibitem[Berry(2001)]{berry2001chaos}
M.~V. Berry.
\newblock Chaos and the semiclassical limit of quantum mechanics (is the moon there when somebody looks?).
\newblock In R.~J. Russell, P.~Clayton, K.~Wegter-McNelly, and J.~Polkinghorne, editors, \emph{Quantum Mechanics: Scientific perspectives on divine action}, pages 41 -- 54. Vatican Observatory CTNS publications, 2001.

\bibitem[Berthier and Biroli(2011)]{berthier2011theoretical}
L.~Berthier and G.~Biroli.
\newblock Theoretical perspective on the glass transition and amorphous materials.
\newblock \emph{Reviews of modern physics}, 83\penalty0 (2):\penalty0 587, 2011.

\bibitem[Bleher and Sinai(1973)]{bleher1973}
P.~M. Bleher and J.~G. Sinai.
\newblock {Investigation of the critical point in models of the type of Dyson's hierarchical models}.
\newblock \emph{Comm. Math. Phys.}, 33\penalty0 (1):\penalty0 23 -- 42, 1973.

\bibitem[Boffetta et~al.(2002)Boffetta, Cencini, Falcioni, and Vulpiani]{boffetta2002predictability}
G.~Boffetta, M.~Cencini, M.~Falcioni, and A.~Vulpiani.
\newblock Predictability: a way to characterize complexity.
\newblock \emph{Phys. Rep.}, 356\penalty0 (6):\penalty0 367--474, 2002.

\bibitem[Boffetta et~al.(2003)Boffetta, del Castillo-Negrete, L{\'o}pez, Pucacco, and Vulpiani]{Boffetta03}
G.~Boffetta, D.~del Castillo-Negrete, C.~L{\'o}pez, G.~Pucacco, and A.~Vulpiani.
\newblock Diffusive transport and self-consistent dynamics in coupled maps.
\newblock \emph{Phys. Rev. E}, 67\penalty0 (2):\penalty0 026224, 2003.

\bibitem[Bohr et~al.(1998)Bohr, Jensen, Paladin, and Vulpiani]{bo98}
T.~Bohr, M.~H. Jensen, G.~Paladin, and A.~Vulpiani.
\newblock \emph{Dynamical systems approach to turbulence}.
\newblock Cambridge University Press, 1998.

\bibitem[Boltzmann(1877)]{boltzmann1877}
L.~Boltzmann.
\newblock {Bemerkungen \"{u}ber einige Problemeder mechanischen W\"{a}rmetheorie.}
\newblock \emph{Kaiserliche Akademie der Wissenschaften zuWien, mathematisch-naturwissenschaftlicheClasse, Sitzungsberichte}, 7:\penalty0 62--100, 1877.

\bibitem[Boltzmann(1896{\natexlab{a}})]{Bolt96}
L.~Boltzmann.
\newblock Annalen der physik.
\newblock reprinted and translated as Chapter 8 in \cite{Bru66}, 1896{\natexlab{a}}.

\bibitem[Boltzmann(1896{\natexlab{b}})]{boltzmann1896}
L.~Boltzmann.
\newblock {Entgegnung auf die w\"{a}rmetheoretischen Betrachtungen des Hrn. E. Zermelo}.
\newblock \emph{Annalen der Physik}, 293\penalty0 (4):\penalty0 773--784, 1896{\natexlab{b}}.
\newblock \doi{https://doi.org/10.1002/andp.18962930414}.

\bibitem[Boltzmann(1897)]{boltzmann1897}
L.~Boltzmann.
\newblock Zu hrn. zermelo's abhandlung ``ueber die mechanische erkl\"{a}rung irreversibler vorg\"{a}nge''.
\newblock \emph{Annalen der Physik}, 60:\penalty0 392, 1897.

\bibitem[Boltzmann(Part I 1896, Part II 1898)]{Bol1898}
L.~Boltzmann.
\newblock \emph{{Vorlesungen \"uber Gastheorie}}.
\newblock Barth, Leipzig, Part I 1896, Part II 1898.
\newblock English translation by S.G. Brush: {\it Lectures on Gas Theory}. Berkeley: University of California Press (1964).

\bibitem[Boltzmann(1872)]{boltzmann1872}
L.~E. Boltzmann.
\newblock Weitere studien iiber das warmegleichgewicht unter gasmolekiilen.
\newblock \emph{Sitzungsberichte der Akademie der Wissenschafien, Wien}, II\penalty0 (66):\penalty0 275--370, 1872.

\bibitem[Borel(1962)]{Borel}
E.~Borel.
\newblock \emph{Probabilities and life}.
\newblock Dover, London, 1962.

\bibitem[Bricmont(1995)]{bricmont1995chaos}
J.~Bricmont.
\newblock Science of chaos or chaos in science?
\newblock \emph{Annals of the New York Academy of Sciences}, 775\penalty0 (1):\penalty0 131--175, 1995.

\bibitem[Brush(1966)]{Bru66}
S.~G. Brush.
\newblock \emph{Kinetic Theory}.
\newblock Pergamon, Oxford, 1966.

\bibitem[Brush(2003)]{br03}
S.~G. Brush.
\newblock \emph{The Kinetic Theory Of Gases: An Anthology Of Classic Papers With Historical Commentary}, volume~1.
\newblock World Scientific, 2003.

\bibitem[Bunimovich and Sinai(1993)]{BS93}
L.~Bunimovich and Y.~G. Sinai.
\newblock \emph{Statistical mechanics of coupled map lattices}.
\newblock Wiley, San-Francisco, 1993.

\bibitem[Callen(1985)]{Callen}
H.~B. Callen.
\newblock \emph{Thermodynamics and an introduction to thermostatistics}.
\newblock Wiley, New York, NY, 2nd edition, 1985.

\bibitem[Campa et~al.(2009)Campa, Dauxois, and Ruffo]{CDR09}
A.~Campa, T.~Dauxois, and S.~Ruffo.
\newblock Statistical mechanics and dynamics of solvable models with long-range interactions.
\newblock \emph{Phys. Rep.}, 480\penalty0 (3-6):\penalty0 57--159, 2009.

\bibitem[Campa et~al.(2014)Campa, Dauxois, Fanelli, and Ruffo]{CDFR14}
A.~Campa, T.~Dauxois, D.~Fanelli, and S.~Ruffo.
\newblock \emph{Physics of long-range interacting systems}.
\newblock OUP Oxford, 2014.

\bibitem[Campisi(2015)]{campisi2015construction}
M.~Campisi.
\newblock Construction of microcanonical entropy on thermodynamic pillars.
\newblock \emph{Phys. Rev. E}, 91\penalty0 (5):\penalty0 052147, 2015.

\bibitem[Campisi and Kobe(2010)]{CK10}
M.~Campisi and D.~H. Kobe.
\newblock Derivation of the boltzmann principle.
\newblock \emph{American Journal of Physics}, 78\penalty0 (6):\penalty0 608--615, 2010.

\bibitem[Caprara and Vulpiani(2018)]{caprara2018law}
S.~Caprara and A.~Vulpiani.
\newblock Law without law or “just” limit theorems?
\newblock \emph{Foundations of Physics}, 48\penalty0 (9):\penalty0 1112--1127, 2018.

\bibitem[Carati et~al.(2005)Carati, Galgani, and Giorgilli]{CGG05}
A.~Carati, L.~Galgani, and A.~Giorgilli.
\newblock The fermi--pasta--ulam problem as a challenge for the foundations of physics.
\newblock \emph{Chaos}, 15\penalty0 (1):\penalty0 015105, 2005.

\bibitem[Carleman(1933)]{Carleman1933}
T.~Carleman.
\newblock Sur la th{\'e}orie de l'{\'e}quation int{\'e}grodiff{\'e}rentielle de {Boltzmann}.
\newblock \emph{Acta Mathematica}, 60\penalty0 (1):\penalty0 91--146, Mar 1933.
\newblock ISSN 1871-2509.
\newblock \doi{10.1007/BF02398270}.

\bibitem[Castiglione et~al.(2008)Castiglione, Falcioni, Lesne, and Vulpiani]{castiglione2008chaos}
P.~Castiglione, M.~Falcioni, A.~Lesne, and A.~Vulpiani.
\newblock \emph{Chaos and coarse graining in statistical mechanics}.
\newblock Cambridge University Press, 2008.

\bibitem[Cavagna(2009)]{cavagna2009supercooled}
A.~Cavagna.
\newblock Supercooled liquids for pedestrians.
\newblock \emph{Phys. Rep.}, 476\penalty0 (4-6):\penalty0 51--124, 2009.

\bibitem[Cecconi et~al.(2003)Cecconi, del Castillo-Negrete, Falcioni, and Vulpiani]{cecconi2003origin}
F.~Cecconi, D.~del Castillo-Negrete, M.~Falcioni, and A.~Vulpiani.
\newblock The origin of diffusion: the case of non-chaotic systems.
\newblock \emph{Physica D}, 180\penalty0 (3-4):\penalty0 129--139, 2003.

\bibitem[Cecconi et~al.(2007)Cecconi, Cencini, and Vulpiani]{Cecconi2007}
F.~Cecconi, M.~Cencini, and A.~Vulpiani.
\newblock Transport properties of chaotic and non-chaotic many particle systems.
\newblock \emph{J. Stat. Mech.}, 2007\penalty0 (12):\penalty0 P12001–P12001, Dec. 2007.
\newblock ISSN 1742-5468.
\newblock \doi{10.1088/1742-5468/2007/12/p12001}.
\newblock URL \url{http://dx.doi.org/10.1088/1742-5468/2007/12/P12001}.

\bibitem[Cencini et~al.(2009)Cencini, Cecconi, and Vulpiani]{vulpiani2009chaos}
M.~Cencini, F.~Cecconi, and A.~Vulpiani.
\newblock \emph{Chaos: from simple models to complex systems}.
\newblock World Scientific, 2009.

\bibitem[Cercignani(1972)]{cercignani72}
C.~Cercignani.
\newblock On the {B}oltzmann equation for rigid spheres.
\newblock \emph{Transport Theory and Statistical Physics}, 2\penalty0 (3):\penalty0 211--225, 1972.
\newblock \doi{10.1080/00411457208232538}.

\bibitem[Cercignani(1988)]{ce88}
C.~Cercignani.
\newblock \emph{The {Boltzmann} equation and its applications}.
\newblock Springer, 1988.

\bibitem[Cercignani(1998)]{ce98}
C.~Cercignani.
\newblock \emph{{Ludwig Boltzmann: the man who trusted atoms}}.
\newblock Oxford University Press, 1998.

\bibitem[Cercignani et~al.(1994)Cercignani, Illner, and Pulvirenti]{Cercignani1994}
C.~Cercignani, R.~Illner, and M.~Pulvirenti.
\newblock \emph{The Mathematical Theory of Dilute Gases}.
\newblock Springer New York, 1994.
\newblock \doi{10.1007/978-1-4419-8524-8}.
\newblock URL \url{https://doi.org/10.1007/978-1-4419-8524-8}.

\bibitem[Cerino et~al.(2014)Cerino, Gradenigo, Sarracino, Villamaina, and Vulpiani]{CERINO14}
L.~Cerino, G.~Gradenigo, A.~Sarracino, D.~Villamaina, and A.~Vulpiani.
\newblock Fluctuations in partitioning systems with few degrees of freedom.
\newblock \emph{Phys. Rev. E}, 89:\penalty0 042105, 2014.

\bibitem[Cerino et~al.(2016)Cerino, Cecconi, Cencini, and Vulpiani]{cerino2016role}
L.~Cerino, F.~Cecconi, M.~Cencini, and A.~Vulpiani.
\newblock The role of the number of degrees of freedom and chaos in macroscopic irreversibility.
\newblock \emph{Physica A}, 442:\penalty0 486--497, 2016.

\bibitem[Chakraborti et~al.(2022)Chakraborti, Dhar, Goldstein, Kundu, and Lebowitz]{cgas22}
S.~Chakraborti, A.~Dhar, S.~Goldstein, A.~Kundu, and J.~L. Lebowitz.
\newblock Entropy growth during free expansion of an ideal gas.
\newblock \emph{J. Phys. A: Math. Theor.}, 55:\penalty0 394002, 2022.

\bibitem[Chibbaro et~al.(2014)Chibbaro, Rondoni, and Vulpiani]{chibbaro2014reductionism}
S.~Chibbaro, L.~Rondoni, and A.~Vulpiani.
\newblock \emph{Reductionism, emergence and levels of reality}.
\newblock Springer, Berlin, 2014.

\bibitem[Choi(1988)]{Choi}
M.~D. Choi.
\newblock Almost commuting matrices need not be nearly commuting.
\newblock \emph{Proc. Amer. Math. Soc.}, 102:\penalty0 529--533, 1988.

\bibitem[Christenson et~al.(1964)Christenson, Cronin, Fitch, and Turlay]{christenson1964evidence}
J.~H. Christenson, J.~W. Cronin, V.~L. Fitch, and R.~Turlay.
\newblock Evidence for the 2 $\pi$ decay of the k 2 0 meson.
\newblock \emph{Phys. Rev. Lett.}, 13\penalty0 (4):\penalty0 138, 1964.

\bibitem[Cocciaglia et~al.(2022)Cocciaglia, Vulpiani, and Gradenigo]{Cocciaglia2022}
N.~Cocciaglia, A.~Vulpiani, and G.~Gradenigo.
\newblock Thermalization without chaos in harmonic systems.
\newblock \emph{Physica A}, 601:\penalty0 127581, Sept. 2022.
\newblock \doi{10.1016/j.physa.2022.127581}.
\newblock URL \url{https://doi.org/10.1016/j.physa.2022.127581}.

\bibitem[Cornfeld et~al.(2012)Cornfeld, Fomin, and Sinai]{cornfeld2012ergodic}
I.~P. Cornfeld, S.~V. Fomin, and Y.~G. Sinai.
\newblock \emph{Ergodic theory}, volume 245.
\newblock Springer Science \& Business Media, 2012.

\bibitem[Cronin(1981)]{cronin1981cp}
J.~W. Cronin.
\newblock {CP symmetry violation—the search for its origin}.
\newblock \emph{Rev. Mod. Phys.}, 53\penalty0 (3):\penalty0 373, 1981.

\bibitem[Darrigol(2018)]{da18}
O.~Darrigol.
\newblock \emph{{Atoms, Mechanics, and Probability: Ludwig Boltzmann's Statistico-mechanical Writings--an Exegesis}}.
\newblock Oxford University Press, 2018.

\bibitem[Darrigol(2021)]{Darrigol2021BoltzmannsRT}
O.~Darrigol.
\newblock {Boltzmann’s reply to the Loschmidt paradox: a commented translation}.
\newblock \emph{The European Physical Journal H}, 46:\penalty0 1--18, 2021.

\bibitem[de~Pasquale et~al.(1981)de~Pasquale, Tartaglia, and Tombesi]{TOMBESI81}
F.~de~Pasquale, P.~Tartaglia, and P.~Tombesi.
\newblock Stochastic dynamic approach to the decay of an unstable state.
\newblock \emph{Zeitschrift f{\"u}r Physik B Condensed Matter}, 43\penalty0 (4):\penalty0 353--360, Dec 1981.
\newblock ISSN 1431-584X.
\newblock \doi{10.1007/BF01292803}.

\bibitem[Dettmann and Cohen(2000)]{dettmann2000microscopic}
C.~Dettmann and E.~Cohen.
\newblock Microscopic chaos and diffusion.
\newblock \emph{J. Stat. Phys.}, 101:\penalty0 775--817, 2000.

\bibitem[Dettmann and Cohen(2001)]{dettmann2001note}
C.~Dettmann and E.~Cohen.
\newblock Note on chaos and diffusion.
\newblock \emph{J. Stat. Phys.}, 103:\penalty0 589--599, 2001.

\bibitem[Deutsch(1991)]{cite9}
J.~M. Deutsch.
\newblock Quantum statistical mechanics in a closed system.
\newblock \emph{Phys. Rev. A}, 43:\penalty0 2046--2049, 1991.

\bibitem[DiPerna and Lions(1989)]{diperna1989}
R.~J. DiPerna and P.~L. Lions.
\newblock On the cauchy problem for boltzmann equations: Global existence and weak stability.
\newblock \emph{Annals of Mathematics}, 130\penalty0 (2):\penalty0 321--366, 1989.
\newblock ISSN 0003486X.
\newblock URL \url{http://www.jstor.org/stable/1971423}.

\bibitem[Dorfman(1999)]{Dorfman1999}
J.~R. Dorfman.
\newblock \emph{An Introduction to Chaos in Nonequilibrium Statistical Mechanics}.
\newblock Cambridge University Press, Aug. 1999.
\newblock ISBN 9780511628870.
\newblock \doi{10.1017/cbo9780511628870}.
\newblock URL \url{http://dx.doi.org/10.1017/CBO9780511628870}.

\bibitem[Dumas(2014)]{du14}
H.~S. Dumas.
\newblock \emph{{The Kam Story: A Friendly Introduction To The Content, History, And Significance Of Classical Kolmogorov-Arnold-Moser Theory}}.
\newblock World Scientific Publishing Company, 2014.

\bibitem[Eckmann and Ruelle(1985)]{ER85}
J.-P. Eckmann and D.~Ruelle.
\newblock Ergodic theory of chaos and strange attractors.
\newblock \emph{Rev. Mod. Phys.}, 57\penalty0 (3):\penalty0 617, 1985.

\bibitem[Ehrenfest and Ehrenfest(1907)]{ehrenfest1907}
P.~Ehrenfest and T.~Ehrenfest.
\newblock {\"{U}ber zwei bekannte Einw\"{a}nde gegen das Boltzmannsche H-Theorem.}
\newblock \emph{Physikalische Zeitschrift}, 8:\penalty0 311--314, 1907.

\bibitem[Ehrenfest and Ehrenfest(1956, original German version 1912)]{eh56}
P.~Ehrenfest and T.~Ehrenfest.
\newblock \emph{The conceptual foundations of the statistical approach in mechanics}.
\newblock Cornell University Press, New York, 1956, original German version 1912.

\bibitem[Einstein(1949)]{einstein1949}
A.~Einstein.
\newblock Autobiographical notes.
\newblock In P.~A. Schlipp, editor, \emph{{Albert Einstein, Phylosopher-Scientist}}, The Library of Living Philosophers, page~43. sixt printing (1995) edition, 1949.

\bibitem[Ellis(1999)]{ellis}
R.~S. Ellis.
\newblock The theory of large deviations: from boltzmann's 1877 calculation to equilibrium macrostates in 2d turbulence.
\newblock \emph{Physica D}, 133:\penalty0 106--136, 1999.

\bibitem[Emch and Liu(2013)]{emch2013logic}
G.~G. Emch and C.~Liu.
\newblock \emph{The logic of thermostatistical physics}.
\newblock Springer Science \& Business Media, 2013.

\bibitem[Falcioni et~al.(1991)Falcioni, Marconi, and Vulpiani]{FMV91}
M.~Falcioni, U.~M.~B. Marconi, and A.~Vulpiani.
\newblock Ergodic properties of high-dimensional symplectic maps.
\newblock \emph{Phys. Rev. A}, 44\penalty0 (4):\penalty0 2263, 1991.

\bibitem[Falcioni et~al.(2003)Falcioni, Vulpiani, Mantica, and Pigolotti]{falcioni2003coarse}
M.~Falcioni, A.~Vulpiani, G.~Mantica, and S.~Pigolotti.
\newblock Coarse-grained probabilistic automata mimicking chaotic systems.
\newblock \emph{Physical review letters}, 91\penalty0 (4):\penalty0 044101, 2003.

\bibitem[Falcioni et~al.(2005)Falcioni, Palatella, and Vulpiani]{PhysRevE.71.016118}
M.~Falcioni, L.~Palatella, and A.~Vulpiani.
\newblock Production rate of the coarse-grained gibbs entropy and the kolmogorov-sinai entropy: A real connection?
\newblock \emph{Phys. Rev. E}, 71:\penalty0 016118, Jan 2005.
\newblock \doi{10.1103/PhysRevE.71.016118}.
\newblock URL \url{https://link.aps.org/doi/10.1103/PhysRevE.71.016118}.

\bibitem[Falcioni et~al.(2011)Falcioni, Puglisi, Sarracino, Villamaina, and A.Vulpiani]{FALCIONI11}
M.~Falcioni, A.~Puglisi, A.~Sarracino, D.~Villamaina, and A.Vulpiani.
\newblock Estimate of temperature and its uncertainty in small systems.
\newblock \emph{American Journal of Physics}, 79:\penalty0 777--785, 2011.

\bibitem[Fermi(1923)]{F23}
E.~Fermi.
\newblock Dimostrazione che in generale un sistema meccanico normale {\`e} quasi ergodico.
\newblock \emph{Il Nuovo Cimento (1911-1923)}, 25:\penalty0 267--269, 1923.

\bibitem[Fermi(1965)]{fe65}
E.~Fermi.
\newblock \emph{{Collected Papers:(Note E Memorie)}}.
\newblock Accademia Nazionale dei Lincei, Roma and University of Chicago Press, Chicago, 1965.

\bibitem[Fermi et~al.(1955)Fermi, Pasta, Ulam, and Tsingou]{FPU55}
E.~Fermi, P.~Pasta, S.~Ulam, and M.~Tsingou.
\newblock Studies of the nonlinear problems.
\newblock Technical report, Los Alamos National Lab.(LANL), Los Alamos, NM (United States), 1955.

\bibitem[Ford(1983)]{F83}
J.~Ford.
\newblock How random is a coin toss?
\newblock \emph{Physics Today}, 36:\penalty0 40--47, 1983.

\bibitem[Ford et~al.(1991)Ford, Mantica, and Ristow]{ford1991arnol}
J.~Ford, G.~Mantica, and G.~H. Ristow.
\newblock The arnol'd cat: Failure of the correspondence principle.
\newblock \emph{Physica D: Nonlinear Phenomena}, 50\penalty0 (3):\penalty0 493--520, 1991.

\bibitem[Forster(1975)]{Forster}
D.~Forster.
\newblock \emph{Hydrodynamic Fluctuations, Broken Symmetry, and Correlation Functions}.
\newblock Frontiers in Physics 47, XIX, 326 S., London-Amsterdam-Don Mills-Sydney-Tokyo W. A. Benjamin, Inc, 1975.

\bibitem[Frisch(1995)]{f95}
U.~Frisch.
\newblock \emph{{Turbulence: the legacy of A.N. Kolmogorov}}.
\newblock Cambridge University Press, 1995.

\bibitem[Gallagher et~al.(2014)Gallagher, Saint-Raymond, and Texier]{Gallagher2012FromNT}
I.~Gallagher, L.~Saint-Raymond, and B.~Texier.
\newblock \emph{{From Newton to Boltzmann: Hard Spheres and Short-range Potentials}}.
\newblock European Mathematical Society, 2014.

\bibitem[Gallavotti(1972)]{gallavotti1972nota}
G.~Gallavotti.
\newblock Rigorous theory of the {Boltzmann} equation in the {Lorentz} gas.
\newblock Technical report, Istituto di Fisica, Universitá di Roma. Nota interna n. 358, 1972.

\bibitem[Gallavotti(1999)]{ga99}
G.~Gallavotti.
\newblock \emph{Statistical mechanics: A short treatise}.
\newblock Springer Science \& Business Media, 1999.

\bibitem[Gallavotti(2007)]{ga07}
G.~Gallavotti.
\newblock \emph{{The Fermi-Pasta-Ulam problem: a status report}}.
\newblock Springer-Verlag, 2007.

\bibitem[Gallavotti et~al.(2002)Gallavotti, Lebowitz, and Mastropietro]{gallavotti2002large}
G.~Gallavotti, J.~Lebowitz, and V.~Mastropietro.
\newblock Large deviations in rarefied quantum gases.
\newblock \emph{J. Stat. Physics}, 108:\penalty0 831--861, 2002.

\bibitem[Gaspard(1998)]{gaspard_1998}
P.~Gaspard.
\newblock \emph{Chaos, Scattering and Statistical Mechanics}.
\newblock Cambridge Nonlinear Science Series. Cambridge University Press, 1998.
\newblock \doi{10.1017/CBO9780511628856}.

\bibitem[Gaspard(2022)]{gaspard2022statistical}
P.~Gaspard.
\newblock \emph{The Statistical Mechanics of Irreversible Phenomena}.
\newblock Cambridge University Press, 2022.

\bibitem[Gemmer and Mahler(2003)]{gm}
J.~Gemmer and G.~Mahler.
\newblock Distribution of local entropy in the hilbert space of bi-partite quantum systems: Origin of jaynes' principle.
\newblock \emph{Euro. Phys. J. B}, 31:\penalty0 249--257, 2003.

\bibitem[Gemmer et~al.(2004)Gemmer, Mahler, and Michel]{cite11}
J.~Gemmer, G.~Mahler, and M.~Michel.
\newblock \emph{Quantum Thermodynamics: Emergence of Thermodynamic Behavior within Composite Quantum Systems}.
\newblock Lecture Notes in Physics 657. Springer, Berlin, 2004.

\bibitem[Ghofraniha et~al.(2015)Ghofraniha, Viola, Di~Maria, Barbarella, Gigli, Leuzzi, and Conti]{ghofraniha2015experimental}
N.~Ghofraniha, I.~Viola, F.~Di~Maria, G.~Barbarella, G.~Gigli, L.~Leuzzi, and C.~Conti.
\newblock Experimental evidence of replica symmetry breaking in random lasers.
\newblock \emph{Nature communications}, 6\penalty0 (1):\penalty0 6058, 2015.

\bibitem[Gibbs(1902)]{gi02}
J.~W. Gibbs.
\newblock \emph{Elementary principles in statistical mechanics: developed with especial reference to the rational foundations of thermodynamics}.
\newblock C. Scribner's Sons, 1902.

\bibitem[Gogolin and Eisert(2016)]{cite12}
C.~Gogolin and J.~Eisert.
\newblock Equilibration, thermalisation, and the emergence of statistical mechanics in closed quantum systems.
\newblock \emph{Reports on Progress in Physics}, 79:\penalty0 056001, 2016.
\newblock URL \url{http://arxiv.org/abs/1503.07538}.

\bibitem[Goldstein(2001)]{Gol99}
S.~Goldstein.
\newblock Boltzmann's approach to statistical mechanics.
\newblock In J.~Bricmont, D.~D\"urr, M.~C. Galavotti, G.~C. Ghirardi, F.~Petruccione, and N.~Zangh\`{i}, editors, \emph{Chance in Physics: Foundations and Perspectives}, volume 574, pages 39--54. Springer-Verlag, Heidelberg, 2001.

\bibitem[Goldstein(2012)]{goldstein2012typicality}
S.~Goldstein.
\newblock Typicality and notions of probability in physics.
\newblock In Y.~Ben-Menahem and M.~Hemmo, editors, \emph{Probability in physics}, pages 59--71. Springer, 2012.

\bibitem[Goldstein and Lebowitz(2004)]{GL04}
S.~Goldstein and J.~L. Lebowitz.
\newblock On the (boltzmann) entropy of nonequilibrium systems.
\newblock \emph{Physica D}, 193:\penalty0 53--66, 2004.

\bibitem[Goldstein et~al.(2006{\natexlab{a}})Goldstein, Lebowitz, Tumulka, and Zangh\`{i}]{GAP1}
S.~Goldstein, J.~L. Lebowitz, R.~Tumulka, and N.~Zangh\`{i}.
\newblock On the distribution of the wave function for systems in thermal equilibrium.
\newblock \emph{J. Stat. Phys.}, 125:\penalty0 1193--1221, 2006{\natexlab{a}}.
\newblock \doi{10.1007/s10955-006-9210-z}.

\bibitem[Goldstein et~al.(2006{\natexlab{b}})Goldstein, Lebowitz, Tumulka, and Zangh\`{i}]{cite23}
S.~Goldstein, J.~L. Lebowitz, R.~Tumulka, and N.~Zangh\`{i}.
\newblock Canonical typicality.
\newblock \emph{Phys. Rev. Lett.}, 96:\penalty0 050403, 2006{\natexlab{b}}.
\newblock URL \url{http://arxiv.org/abs/cond-mat/0511091}.

\bibitem[Goldstein et~al.(2010{\natexlab{a}})Goldstein, Lebowitz, Mastrodonato, Tumulka, and Zangh\`{i}]{GLMTZ10}
S.~Goldstein, J.~L. Lebowitz, C.~Mastrodonato, R.~Tumulka, and N.~Zangh\`{i}.
\newblock Normal typicality and von neumann’s quantum ergodic theorem.
\newblock \emph{Proceedings of the Royal Society A: Mathematical, Physical and Engineering Sciences}, 466\penalty0 (2123):\penalty0 3203--3224, 2010{\natexlab{a}}.

\bibitem[Goldstein et~al.(2010{\natexlab{b}})Goldstein, Lebowitz, Mastrodonato, Tumulka, and Zangh\`{i}]{GLMTZ10a}
S.~Goldstein, J.~L. Lebowitz, C.~Mastrodonato, R.~Tumulka, and N.~Zangh\`{i}.
\newblock Approach to thermal equilibrium of macroscopic quantum systems.
\newblock \emph{Phys. Rev. E}, 81\penalty0 (1):\penalty0 011109, 2010{\natexlab{b}}.

\bibitem[Goldstein et~al.(2010{\natexlab{c}})Goldstein, Lebowitz, Tumulka, and Zangh\`{i}]{GLTZ10}
S.~Goldstein, J.~L. Lebowitz, R.~Tumulka, and N.~Zangh\`{i}.
\newblock Long-time behavior of macroscopic quantum systems: Commentary accompanying the english translation of john von neumann’s 1929 article on the quantum ergodic theorem.
\newblock \emph{The European Physical Journal H}, 35:\penalty0 173--200, 2010{\natexlab{c}}.

\bibitem[Goldstein et~al.(2013)Goldstein, Hara, and Tasaki]{Har2013}
S.~Goldstein, T.~Hara, and H.~Tasaki.
\newblock Time scales in the approach to equilibrium of macroscopic quantum systems.
\newblock \emph{Phys. Rev. Lett.}, 111:\penalty0 140401, 2013.
\newblock URL \url{http://arxiv.org/abs/1307.0572}.

\bibitem[Goldstein et~al.(2014)Goldstein, Hara, and Tasaki]{Har2014}
S.~Goldstein, T.~Hara, and H.~Tasaki.
\newblock The approach to equilibrium in a macroscopic quantum system for a typical nonequilibrium subspace, 2014.
\newblock URL \url{http://arxiv.org/abs/1402.3380}.

\bibitem[Goldstein et~al.(2015{\natexlab{a}})Goldstein, Hara, and Tasaki]{GHT14b}
S.~Goldstein, T.~Hara, and H.~Tasaki.
\newblock Extremely quick thermalization in a macroscopic quantum system for a typical nonequilibrium subspace.
\newblock \emph{New J. Phys.}, 17:\penalty0 045002, 2015{\natexlab{a}}.

\bibitem[Goldstein et~al.(2015{\natexlab{b}})Goldstein, Huse, Lebowitz, and Tumulka]{GHLT2015}
S.~Goldstein, D.~A. Huse, J.~L. Lebowitz, and R.~Tumulka.
\newblock Thermal equilibrium of a macroscopic quantum system in a pure state.
\newblock \emph{Phys. Rev. Lett.}, 115:\penalty0 100402, 2015{\natexlab{b}}.

\bibitem[Goldstein et~al.(2016)Goldstein, Lebowitz, Mastrodonato, Tumulka, and Zangh\`{i}]{GLMTZ2016}
S.~Goldstein, J.~L. Lebowitz, C.~Mastrodonato, R.~Tumulka, and N.~Zangh\`{i}.
\newblock Universal probability distribution for the wave function of a quantum system entangled with its environment.
\newblock \emph{Commun. Math. Phys.}, 342:\penalty0 965--988, 2016.
\newblock \doi{10.1007/s00220-015-2536-0}.

\bibitem[Goldstein et~al.(2017{\natexlab{a}})Goldstein, Huse, Lebowitz, and Tumulka]{GHLT17}
S.~Goldstein, D.~A. Huse, J.~L. Lebowitz, and R.~Tumulka.
\newblock Macroscopic and microscopic thermal equilibrium.
\newblock \emph{Annalen der Physik}, 529:\penalty0 1600301, 2017{\natexlab{a}}.

\bibitem[Goldstein et~al.(2017{\natexlab{b}})Goldstein, Lebowitz, Tumulka, and Zangh\`{i}]{GLTZ17}
S.~Goldstein, J.~L. Lebowitz, R.~Tumulka, and N.~Zangh\`{i}.
\newblock {Any orthonormal basis in high dimension is uniformly distributed over the sphere}.
\newblock \emph{Annales de l'Institut Henri Poincaré, Probabilités et Statistiques}, 53\penalty0 (2):\penalty0 701 -- 717, 2017{\natexlab{b}}.
\newblock \doi{10.1214/15-AIHP732}.
\newblock URL \url{https://doi.org/10.1214/15-AIHP732}.

\bibitem[Goldstein et~al.(2020)Goldstein, Lebowitz, Tumulka, and Zangh\`{i}]{GLTZ2020}
S.~Goldstein, J.~L. Lebowitz, R.~Tumulka, and N.~Zangh\`{i}.
\newblock Gibbs and boltzmann entropy in classical and quantum mechanics.
\newblock In V.~Allori, editor, \emph{Statistical Mechanics and Scientific Explanation: Determinism, Indeterminism and Laws of Nature}, pages 519--581. 2020.

\bibitem[Gottwald and Oliver(2009)]{gottwald09}
G.~A. Gottwald and M.~Oliver.
\newblock Boltzmann's dilemma: An introduction to statistical mechanics via the kac ring.
\newblock \emph{SIAM Review}, 51\penalty0 (3):\penalty0 613--635, 2009.
\newblock \doi{10.1137/070705799}.

\bibitem[Grad(1967)]{G67}
H.~Grad.
\newblock Levels of description in statistical mechanics and thermodynamics.
\newblock In M.~Bunge, editor, \emph{Delaware Seminar in the Foundations of Physics}, pages 49--76. Springer, 1967.

\bibitem[Gradenigo et~al.(2020)Gradenigo, Antenucci, and Leuzzi]{gradenigo2020glassiness}
G.~Gradenigo, F.~Antenucci, and L.~Leuzzi.
\newblock Glassiness and lack of equipartition in random lasers: The common roots of ergodicity breaking in disordered and nonlinear systems.
\newblock \emph{Phys. Rev. Research}, 2\penalty0 (2):\penalty0 023399, 2020.

\bibitem[Gradenigo et~al.(2021{\natexlab{a}})Gradenigo, Iubini, Livi, and Majumdar]{GILM21}
G.~Gradenigo, S.~Iubini, R.~Livi, and S.~N. Majumdar.
\newblock Localization transition in the discrete nonlinear schr{\"o}dinger equation: ensembles inequivalence and negative temperatures.
\newblock \emph{J. Stat. Mech.}, 2021\penalty0 (2):\penalty0 023201, 2021{\natexlab{a}}.

\bibitem[Gradenigo et~al.(2021{\natexlab{b}})Gradenigo, Iubini, Livi, and Majumdar]{GILM21b}
G.~Gradenigo, S.~Iubini, R.~Livi, and S.~N. Majumdar.
\newblock Condensation transition and ensemble inequivalence in the discrete nonlinear schr{\"o}dinger equation.
\newblock \emph{The European Physical Journal E}, 44:\penalty0 1--6, 2021{\natexlab{b}}.

\bibitem[Griffiths(1994)]{Gri1994}
R.~Griffiths.
\newblock Statistical irreversibility: Classical and quantum.
\newblock In J.~J. Halliwell, J.~P\'rez-Mercader, and W.~H. Zurek, editors, \emph{Physical Origin of Time Asymmetry.}, pages 147--159. Cambridge University Press, 1994.

\bibitem[Hen{\'o}n(1974)]{henon1974integrals}
M.~Hen{\'o}n.
\newblock Integrals of the toda lattice.
\newblock \emph{Phys. Rev. B}, 9\penalty0 (4):\penalty0 1921, 1974.

\bibitem[Henrici and Kappeler(2008)]{henrici2008results}
A.~Henrici and T.~Kappeler.
\newblock Results on normal forms for fpu chains.
\newblock \emph{Communications in mathematical physics}, 278\penalty0 (1):\penalty0 145--177, 2008.

\bibitem[Hurd et~al.(1994)Hurd, Grebogi, and Ott]{HGO94}
L.~Hurd, C.~Grebogi, and E.~Ott.
\newblock On the tendency toward ergodicity with increasing number of degrees of freedom in hamiltonian systems.
\newblock \emph{Hamiltonian Mechanics: Integrability and Chaotic Behavior}, pages 123--129, 1994.

\bibitem[Hájek(2023)]{sep-probability-interpret}
A.~Hájek.
\newblock {Interpretations of Probability}.
\newblock In E.~N. Zalta and U.~Nodelman, editors, \emph{The {Stanford} Encyclopedia of Philosophy}. Metaphysics Research Lab, Stanford University, {W}inter 2023 edition, 2023.

\bibitem[Illner and Pulvirenti(1986)]{Illner1986}
R.~Illner and M.~Pulvirenti.
\newblock Global validity of the {Boltzmann} equation for a two-dimensional rare gas in vacuum.
\newblock \emph{Comm. Math. Phys.}, 105\penalty0 (2):\penalty0 189--203, Jun 1986.
\newblock ISSN 1432-0916.
\newblock \doi{10.1007/BF01211098}.

\bibitem[Illner and Pulvirenti(1989)]{Illner1989}
R.~Illner and M.~Pulvirenti.
\newblock Global validity of the {Boltzmann} equation for two- and three-dimensional rare gas in vacuum: Erratum and improved result.
\newblock \emph{Comm. Math. Phys.}, 121\penalty0 (1):\penalty0 143--146, Mar 1989.
\newblock ISSN 1432-0916.
\newblock \doi{10.1007/BF01218628}.

\bibitem[Izrailev and Chirikov(1966)]{IC66}
F.~Izrailev and B.~Chirikov.
\newblock Statistical properties of a nonlinear string (institute of nuclear physics, novosibirsk, ussr, 1965).
\newblock In \emph{Dokl. Akad. Nauk SSSR}, volume 166, page~57, 1966.

\bibitem[Jancel(2013)]{J69}
R.~Jancel.
\newblock \emph{Foundations of Classical and Quantum Statistical Mechanics: International Series of Monographs in Natural Philosophy}.
\newblock Elsevier, 2013.

\bibitem[Jaynes(1967)]{J67}
E.~T. Jaynes.
\newblock Foundations of probability theory and statistical mechanics.
\newblock In M.~Bunge, editor, \emph{Delaware seminar in the foundations of physics}, pages 77--101. Springer, 1967.

\bibitem[Jensen and Shankar(1985)]{cite29}
R.~V. Jensen and R.~Shankar.
\newblock Statistical behavior in deterministic quantum systems with few degrees of freedom.
\newblock \emph{Phys. Rev. Lett.}, 54:\penalty0 1879--1882, 1985.

\bibitem[Jona-Lasinio(1975)]{J75}
G.~Jona-Lasinio.
\newblock The renormalization group: A probabilistic view.
\newblock \emph{Nuovo Cimento B}, 26:\penalty0 99--119, 1975.

\bibitem[Jona-Lasinio(2001)]{J01}
G.~Jona-Lasinio.
\newblock Renormalization group and probability theory.
\newblock \emph{Phys. Rep.}, 352\penalty0 (4-6):\penalty0 439--458, 2001.

\bibitem[Kac(1947{\natexlab{a}})]{kac1947}
M.~Kac.
\newblock Random walk and the theory of brownian motion.
\newblock \emph{The American Mathematical Monthly}, 54\penalty0 (7):\penalty0 369--391, 1947{\natexlab{a}}.
\newblock ISSN 00029890, 19300972.

\bibitem[Kac(1947{\natexlab{b}})]{kac47}
M.~Kac.
\newblock On the notion of recurrence in discrete stochastic processes.
\newblock \emph{Bullettin of the American Mathematical Society}, 53:\penalty0 1002--1010, 1947{\natexlab{b}}.

\bibitem[Kac(1959)]{ka57}
M.~Kac.
\newblock \emph{Probability and related topics in physical sciences}.
\newblock American Mathematical Soc., 1959.

\bibitem[Kadanoff et~al.(1967)Kadanoff, G{\"o}tze, Hamblen, Hecht, Lewis, Palciauskas, Rayl, Swift, Aspnes, and Kane]{KGHHLPRSAK67}
L.~P. Kadanoff, W.~G{\"o}tze, D.~Hamblen, R.~Hecht, E.~Lewis, V.~V. Palciauskas, M.~Rayl, J.~Swift, D.~Aspnes, and J.~Kane.
\newblock Static phenomena near critical points: theory and experiment.
\newblock \emph{Rev. Mod. Phys.}, 39\penalty0 (2):\penalty0 395, 1967.

\bibitem[Kaneko and Konishi(1987)]{KK87}
K.~Kaneko and T.~Konishi.
\newblock Transition, ergodicity and lyapunov spectra of hamiltonian dynamical systems.
\newblock \emph{Journal of the Physical Society of Japan}, 56\penalty0 (9):\penalty0 2993--2996, 1987.

\bibitem[Khinchin(1949)]{kh49}
A.~Y. Khinchin.
\newblock \emph{Mathematical foundations of statistical mechanics}.
\newblock Courier Corporation, 1949.

\bibitem[Kolmogorov(1954)]{K54}
A.~N. Kolmogorov.
\newblock On conservation of conditionally periodic motions for a small change in hamilton's function.
\newblock 98:\penalty0 527--530, 1954.

\bibitem[Kosterlitz and Thouless(1973)]{Kosterlitz_1973}
J.~M. Kosterlitz and D.~J. Thouless.
\newblock Ordering, metastability and phase transitions in two-dimensional systems.
\newblock \emph{Journal of Physics C: Solid State Physics}, 6\penalty0 (7):\penalty0 1181, apr 1973.
\newblock \doi{10.1088/0022-3719/6/7/010}.

\bibitem[Landau and Lifshitz(1969)]{la69}
L.~D. Landau and E.~M. Lifshitz.
\newblock \emph{{Statistical Physics: Volume 5}}.
\newblock 1969.

\bibitem[Lanford(1973)]{Lan73}
O.~E. Lanford.
\newblock Entropy and equilibrium states in classical statistical mechanics.
\newblock In A.~Lenard, editor, \emph{Statistical Mechanics and Mathematical Problems}, pages 1--113. Springer-Verlag, Berlin, 1973.

\bibitem[Lanford(1975)]{Lan75}
O.~E. Lanford.
\newblock Time evolution of large classical systems.
\newblock In J.~Moser, editor, \emph{Dynamical Systems, Theory and Applications}, pages 1--111. Springer-Verlag, Berlin, 1975.

\bibitem[Lanford(1981)]{LANFORD198170}
O.~E. Lanford.
\newblock The hard sphere gas in the {Boltzmann-Grad} limit.
\newblock \emph{Physica A}, 106\penalty0 (1):\penalty0 70--76, 1981.
\newblock ISSN 0378-4371.
\newblock \doi{https://doi.org/10.1016/0378-4371(81)90207-7}.

\bibitem[Lazarovici and Reichert(2015)]{lazarovici2015typicality}
D.~Lazarovici and P.~Reichert.
\newblock Typicality, irreversibility and the status of macroscopic laws.
\newblock \emph{Erkenntnis}, 80\penalty0 (4):\penalty0 689--716, 2015.

\bibitem[Lebowitz et~al.(1993)]{lebowitz1993boltzmann}
J.~L. Lebowitz et~al.
\newblock Boltzmann's entropy and time's arrow.
\newblock \emph{Physics Today}, 46:\penalty0 32--32, 1993.

\bibitem[Lenci and Rey-Bellet(2005)]{lenci2}
M.~Lenci and L.~Rey-Bellet.
\newblock Large deviations in quantum lattice systems: One-phase region.
\newblock \emph{J. Stat. Phys.}, 119:\penalty0 715--746, 2005.

\bibitem[Linden et~al.(2009)Linden, Popescu, Short, and Winter]{Lin2009}
N.~Linden, S.~Popescu, A.~J. Short, and A.~Winter.
\newblock Quantum mechanical evolution towards thermal equilibrium.
\newblock \emph{Phys. Rev. E}, 79\penalty0 (6):\penalty0 061103, 2009.

\bibitem[Livi et~al.(1986)Livi, Politi, and Ruffo]{LPR86}
R.~Livi, A.~Politi, and S.~Ruffo.
\newblock Distribution of characteristic exponents in the thermodynamic limit.
\newblock \emph{J. Phys. A}, 19\penalty0 (11):\penalty0 2033, 1986.

\bibitem[Livi et~al.(1987)Livi, Pettini, Ruffo, and Vulpiani]{LPRV87}
R.~Livi, M.~Pettini, S.~Ruffo, and A.~Vulpiani.
\newblock Chaotic behavior in nonlinear hamiltonian systems and equilibrium statistical mechanics.
\newblock \emph{Journal of statistical physics}, 48:\penalty0 539--559, 1987.

\bibitem[Loschmidt(1876)]{loschmidt1876}
J.~Loschmidt.
\newblock {\"{U}ber den Zustand des Wärmegleichgewichtes eines Sytsems von Körpern mit Rücksicht auf die Schwerkraft. I.}
\newblock \emph{Kaiserliche Akademie der Wissenschaften zuWien, mathematisch-naturwissenschaftlicheClasse, Sitzungsberichte}, 73:\penalty0 128--142, 1876.

\bibitem[Lychkovskiy(2013)]{Lyc2013}
O.~Lychkovskiy.
\newblock Dependence of decoherence-assisted classicality on the way a system is partitioned into subsystems.
\newblock \emph{Phys. Rev. A}, 87:\penalty0 022112, 2013.
\newblock URL \url{http://arxiv.org/abs/1210.4124}.

\bibitem[Ma(1985)]{ma85}
S.-K. Ma.
\newblock \emph{Statistical mechanics}.
\newblock World Scientific Publishing Company, 1985.

\bibitem[Ma et~al.(2014)Ma, Tan, Yuan, Yuan, and Ao]{matan2014}
Y.~Ma, Q.~Tan, R.~Yuan, B.~Yuan, and P.~Ao.
\newblock Potential function in a continuous dissipative chaotic system: Decomposition scheme and role of strange attractor.
\newblock \emph{International Journal of Bifurcation and Chaos}, 24\penalty0 (02):\penalty0 1450015, 2014.
\newblock \doi{10.1142/S0218127414500151}.
\newblock URL \url{https://doi.org/10.1142/S0218127414500151}.

\bibitem[Mantica(2000{\natexlab{a}})]{M00}
G.~Mantica.
\newblock Quantum algorithmic integrability: the metaphor of classical polygonal billiards.
\newblock \emph{Phys. Rev. E}, 61\penalty0 (6):\penalty0 6434, 2000{\natexlab{a}}.

\bibitem[Mantica(2000{\natexlab{b}})]{mantica2000quantum}
G.~Mantica.
\newblock Quantum algorithmic integrability: the metaphor of classical polygonal billiards.
\newblock \emph{Physical Review E}, 61\penalty0 (6):\penalty0 6434, 2000{\natexlab{b}}.

\bibitem[Maxwell(1879)]{M79}
J.~C. Maxwell.
\newblock On boltzmann’s theorem on the average distribution of energy in a system of material points.
\newblock \emph{Cambridge Philosophical Society’s Transations}, 12:\penalty0 713--741, 1879.

\bibitem[Mazur and Van~der Linden(1963)]{ML63}
P.~Mazur and J.~Van~der Linden.
\newblock Asymptotic form of the structure function for real systems.
\newblock \emph{Journal of Mathematical Physics}, 4\penalty0 (2):\penalty0 271--277, 1963.

\bibitem[Mehra(2001)]{me01}
J.~Mehra.
\newblock \emph{The Golden Age Of Theoretical Physics}.
\newblock World Scientific, 2001.

\bibitem[Merz et~al.(2020)Merz, Kuhlicke, Kunz, Pittore, Babeyko, Bresch, Domeisen, Feser, Koszalka, Kreibich, Pantillon, Parolai, Pinto, Punge, Rivalta, Schroeter, Strehlow, Weisse, and Wurpts]{merz2020}
B.~Merz, C.~Kuhlicke, M.~Kunz, M.~Pittore, A.~Babeyko, D.~N. Bresch, D.~I. Domeisen, V, F.~Feser, I.~Koszalka, H.~Kreibich, F.~Pantillon, S.~Parolai, J.~G. Pinto, H.~J. Punge, E.~Rivalta, K.~Schroeter, K.~Strehlow, R.~Weisse, and A.~Wurpts.
\newblock Impact forecasting to support emergency management of natural hazards.
\newblock \emph{Reviews of Geophysics}, 58\penalty0 (4), 2020.

\bibitem[M{\'e}zard and Parisi(1999)]{mezard1999thermodynamics}
M.~M{\'e}zard and G.~Parisi.
\newblock Thermodynamics of glasses: A first principles computation.
\newblock \emph{Journal of Physics: Condensed Matter}, 11\penalty0 (10A):\penalty0 A157, 1999.

\bibitem[M{\'e}zard et~al.(1987)M{\'e}zard, Parisi, and Virasoro]{mezard1987spin}
M.~M{\'e}zard, G.~Parisi, and M.~A. Virasoro.
\newblock \emph{Spin glass theory and beyond: An Introduction to the Replica Method and Its Applications}, volume~9.
\newblock World Scientific Publishing Company, 1987.

\bibitem[Morgenstern(1954)]{morgenstern54}
D.~Morgenstern.
\newblock General existence and uniqueness proof for spatially homogeneous solutions of the maxwell-boltzmann equation in the case of maxwellian molecules<sup>*</sup>.
\newblock \emph{Proceedings of the National Academy of Sciences}, 40\penalty0 (8):\penalty0 719--721, 1954.
\newblock \doi{10.1073/pnas.40.8.719}.

\bibitem[Mori et~al.(2018)Mori, Ikeda, Kaminishi, and Ueda]{Mori2018}
T.~Mori, T.~N. Ikeda, E.~Kaminishi, and M.~Ueda.
\newblock Thermalization and prethermalization in isolated quantum systems: a theoretical overview.
\newblock \emph{J. Phys. B: At. Mol. Opt. Phys.}, 51:\penalty0 112001, 2018.

\bibitem[M{\"o}ser(1962)]{M62}
J.~M{\"o}ser.
\newblock On invariant curves of area-preserving mappings of an annulus.
\newblock \emph{Nachr. Akad. Wiss. G{\"o}ttingen, II}, pages 1--20, 1962.

\bibitem[M{\"u}ller(2007)]{mu07}
I.~M{\"u}ller.
\newblock \emph{A history of thermodynamics: the doctrine of energy and entropy}.
\newblock Springer Science \& Business Media, 2007.

\bibitem[Nagel(1979)]{nagel1979structure}
E.~Nagel.
\newblock \emph{The structure of science}.
\newblock Hackett Publishing Company Indianapolis, 1979.

\bibitem[Nandkishore and Huse(2015)]{Nand2015}
R.~Nandkishore and D.~A. Huse.
\newblock Many body localization and thermalization in quantum statistical mechanics.
\newblock \emph{Annual Review of Condensed Matter Physics}, 6:\penalty0 15--38, 2015.
\newblock URL \url{http://arxiv.org/abs/1404.0686}.

\bibitem[Netočný and Redig(2004)]{netoc}
K.~Netočný and F.~Redig.
\newblock Large deviations for quantum spin systems.
\newblock \emph{J. Stat. Phys.}, 117:\penalty0 521--547, 2004.

\bibitem[Nicolis and Prigogine(1978)]{nicolis1977self}
G.~Nicolis and I.~Prigogine.
\newblock Self-organization in nonequilibrium systems.
\newblock In J.~Schnakenberg, editor, \emph{From Dissipative Structures to Order through Fluctuations}, number~6, pages 672--672. J. Wiley \& Sons, New York, London, Sydney, Toronto, 1978.
\newblock \doi{https://doi.org/10.1002/bbpc.197800155}.

\bibitem[Niedda et~al.(2023{\natexlab{a}})Niedda, Gradenigo, Leuzzi, and Parisi]{niedda2023universality}
J.~Niedda, G.~Gradenigo, L.~Leuzzi, and G.~Parisi.
\newblock Universality class of the mode-locked glassy random laser.
\newblock \emph{SciPost Physics}, 14\penalty0 (6):\penalty0 144, 2023{\natexlab{a}}.

\bibitem[Niedda et~al.(2023{\natexlab{b}})Niedda, Leuzzi, and Gradenigo]{niedda2023intensity}
J.~Niedda, L.~Leuzzi, and G.~Gradenigo.
\newblock Intensity pseudo-localized phase in the glassy random laser.
\newblock \emph{J. Stat. Mech.}, 2023\penalty0 (5):\penalty0 053302, 2023{\natexlab{b}}.

\bibitem[Oganesyan and Huse(2007)]{huse2007}
V.~Oganesyan and D.~A. Huse.
\newblock Localization of interacting fermions at high temperature.
\newblock \emph{Phys. Rev. B}, 75:\penalty0 155111, 2007.

\bibitem[Ogata(2010)]{ogata2}
Y.~Ogata.
\newblock Large deviations in quantum spin chains.
\newblock \emph{Commun. Math. Phys.}, 296:\penalty0 35, 2010.

\bibitem[Ogata(2013)]{Ogata}
Y.~Ogata.
\newblock Approximating macroscopic observables in quantum spin systems with commuting matrices.
\newblock \emph{Journal of Functional Analysis}, 264:\penalty0 2005--2033, 2013.

\bibitem[Onsager(1949)]{Onsager1949}
L.~Onsager.
\newblock Statistical hydrodynamics.
\newblock \emph{Il Nuovo Cimento}, 6\penalty0 (S2):\penalty0 279--287, Mar. 1949.
\newblock \doi{10.1007/bf02780991}.
\newblock URL \url{https://doi.org/10.1007/bf02780991}.

\bibitem[Orban and Bellemans(1967)]{Orban1967-gq}
J.~Orban and A.~Bellemans.
\newblock Velocity-inversion and irreversibility in a dilute gas of hard disks.
\newblock \emph{Phys. Lett. A}, 24\penalty0 (11):\penalty0 620--621, May 1967.

\bibitem[Paladin and Vulpiani(1987)]{PV87}
G.~Paladin and A.~Vulpiani.
\newblock Anomalous scaling laws in multifractal objects.
\newblock \emph{Phys. Rep.}, 156\penalty0 (4):\penalty0 147--225, 1987.

\bibitem[Palmer(2000)]{P00}
T.~N. Palmer.
\newblock Predicting uncertainty in forecasts of weather and climate.
\newblock \emph{Reports on progress in Physics}, 63\penalty0 (2):\penalty0 71, 2000.

\bibitem[Pandey et~al.(2023)Pandey, Bhat, Dhar, Goldstein, Huse, Kulkarni1, Kundu, and Lebowitz]{qgas23}
S.~Pandey, J.~M. Bhat, A.~Dhar, S.~Goldstein, D.~Huse, M.~Kulkarni1, A.~Kundu, and J.~Lebowitz.
\newblock Boltzmann entropy of a freely expanding quantum ideal gas.
\newblock \emph{J Stat Phys}, 190:\penalty0 142, 2023.

\bibitem[Parisi(1979)]{parisi79}
G.~Parisi.
\newblock Infinite number of order parameters for spin-glasses.
\newblock \emph{Phys. Rev. Lett.}, 43:\penalty0 1754--1756, Dec 1979.
\newblock \doi{10.1103/PhysRevLett.43.1754}.
\newblock URL \url{https://link.aps.org/doi/10.1103/PhysRevLett.43.1754}.

\bibitem[Parisi et~al.(2020)Parisi, Urbani, and Zamponi]{parisi2020theory}
G.~Parisi, P.~Urbani, and F.~Zamponi.
\newblock \emph{Theory of simple glasses: exact solutions in infinite dimensions}.
\newblock Cambridge University Press, 2020.

\bibitem[Peliti(2011)]{peliti2011statistical}
L.~Peliti.
\newblock \emph{Statistical mechanics in a nutshell}.
\newblock Princeton University Press, 2011.

\bibitem[Penrose(2004)]{Pen2004}
R.~Penrose.
\newblock \emph{The Road to Reality}.
\newblock Jonathan Cape, London, 2004.

\bibitem[Pesin(1976)]{MR0410804}
J.~B. Pesin.
\newblock Characteristic {L}japunov exponents, and ergodic properties of smooth dynamical systems with invariant measure.
\newblock \emph{Dokl. Akad. Nauk SSSR}, 226\penalty0 (4):\penalty0 774--777, 1976.
\newblock ISSN 0002-3264.

\bibitem[Pitaevskii and Lifshitz(2012)]{la81}
L.~P. Pitaevskii and E.~Lifshitz.
\newblock \emph{{Physical Kinetics: Volume 10}}.
\newblock Butterworth-Heinemann, 2012.

\bibitem[Poincar{\'e}(1893)]{po92}
H.~Poincar{\'e}.
\newblock \emph{Les m{\'e}thodes nouvelles de la m{\'e}canique c{\'e}leste}, volume~2.
\newblock Gauthier-Villars et fils, imprimeurs-libraires, 1893.

\bibitem[Popescu et~al.(2006)Popescu, Short, and Winter]{PSW06}
S.~Popescu, A.~J. Short, and A.~Winter.
\newblock Entanglement and the foundations of statistical mechanics.
\newblock \emph{Nature Physics}, 2\penalty0 (11):\penalty0 754--758, 2006.

\bibitem[Popper(2002)]{po02}
K.~R. Popper.
\newblock \emph{The Logic of Scientific Discovery}.
\newblock Routledge, 2002.

\bibitem[Prigogine(1984)]{prigogine1984irreversibility}
I.~Prigogine.
\newblock Irreversibility and space-time structure.
\newblock In W.~Horsthemke and D.~Kondepudi, editors, \emph{Fluctuations and sensitivity in nonequilibrium systems}, pages 2--10. Springer, 1984.

\bibitem[Prigogine and Stengers(1979)]{prigogine1979nouvelle}
I.~Prigogine and I.~Stengers.
\newblock \emph{{La nouvelle alliance M{\'e}tamorphose de la science}}.
\newblock 1979.

\bibitem[Pulvirenti and Simonella(2017)]{Pulvirenti2017}
M.~Pulvirenti and S.~Simonella.
\newblock The boltzmann--grad limit of a hard sphere system: analysis of the correlation error.
\newblock \emph{Inventiones mathematicae}, 207\penalty0 (3):\penalty0 1135--1237, Mar 2017.

\bibitem[Pulvirenti et~al.(2014)Pulvirenti, Saffirio, and Simonella]{pulvirenti2014}
M.~Pulvirenti, C.~Saffirio, and S.~Simonella.
\newblock On the validity of the boltzmann equation for short range potentials.
\newblock \emph{Reviews in Mathematical Physics}, 26\penalty0 (02):\penalty0 1450001, 2014.
\newblock \doi{10.1142/S0129055X14500019}.

\bibitem[Ramsey(1956)]{ramsey56}
N.~F. Ramsey.
\newblock Thermodynamics and statistical mechanics at negative absolute temperatures.
\newblock \emph{Phys. Rev.}, 103:\penalty0 20--28, Jul 1956.
\newblock \doi{10.1103/PhysRev.103.20}.
\newblock URL \url{https://link.aps.org/doi/10.1103/PhysRev.103.20}.

\bibitem[Reimann(2007)]{cite46}
P.~Reimann.
\newblock Typicality for generalized microcanonical ensembles.
\newblock \emph{Phys. Rev. Lett.}, 99:\penalty0 160404, 2007.
\newblock URL \url{http://arxiv.org/abs/0710.4214}.

\bibitem[Reimann(2008)]{cite48}
P.~Reimann.
\newblock Foundation of statistical mechanics under experimentally realistic conditions.
\newblock \emph{Phys. Rev. Lett.}, 101:\penalty0 190403, 2008.
\newblock URL \url{http://arxiv.org/abs/0810.3092}.

\bibitem[Reimann(2010)]{cite49}
P.~Reimann.
\newblock Canonical thermalization.
\newblock \emph{New J. Phys.}, 12:\penalty0 055027, 2010.
\newblock URL \url{http://arxiv.org/abs/1005.5625}.

\bibitem[Reimann(2015{\natexlab{a}})]{Reim2015}
P.~Reimann.
\newblock Generalization of von neumann's approach to thermalization.
\newblock \emph{Phys. Rev. Lett.}, 115:\penalty0 010403, 2015{\natexlab{a}}.

\bibitem[Reimann(2015{\natexlab{b}})]{cite50}
P.~Reimann.
\newblock Eigenstate thermalization: Deutsch's approach and beyond.
\newblock \emph{New J. Phys.}, 17:\penalty0 055025, 2015{\natexlab{b}}.
\newblock URL \url{http://arxiv.org/abs/1505.07627}.

\bibitem[Rigol and Srednicki(2012)]{cite53}
M.~Rigol and M.~Srednicki.
\newblock Alternatives to eigenstate thermalization.
\newblock \emph{Phys. Rev. Lett.}, 108:\penalty0 110601, 2012.
\newblock URL \url{http://arxiv.org/abs/1108.0928}.

\bibitem[Rigol et~al.(2008{\natexlab{a}})Rigol, Dunjko, and Olshanii]{cite52}
M.~Rigol, V.~Dunjko, and M.~Olshanii.
\newblock Thermalization and its mechanism for generic isolated quantum systems.
\newblock \emph{Nature}, 452:\penalty0 854--858, 2008{\natexlab{a}}.
\newblock URL \url{http://arxiv.org/abs/0708.1324}.

\bibitem[Rigol et~al.(2008{\natexlab{b}})Rigol, Dunjko, and Olshanii]{Rig2008}
V.~Rigol, M.~Dunjko, and M.~Olshanii.
\newblock Thermalization and its mechanism for generic isolated quantum systems.
\newblock \emph{Nature}, 452:\penalty0 854--858, 2008{\natexlab{b}}.
\newblock URL \url{http://arxiv.org/abs/0708.1324}.

\bibitem[Roberts and Quispel(1992)]{quispel92}
J.~A.~G. Roberts and G.~R.~W. Quispel.
\newblock Chaos and time-reversal symmetry. order and chaos in reversible dynamical systems.
\newblock \emph{Phys. Rep.}, 216:\penalty0 63--177, 1992.

\bibitem[Ruffo(2001)]{R01}
S.~Ruffo.
\newblock Time-scales for the approach to thermal equilibrium.
\newblock In J.~Bricmont, D.~D\"urr, M.~C. Galavotti, G.~C. Ghirardi, F.~Petruccione, and N.~Zangh\`{i}, editors, \emph{Chance in Physics: Foundations and Perspectives}, pages 243--251. Springer-Verlag, Heidelberg, 2001.

\bibitem[Safranek et~al.(2017)Safranek, Deutsch, and Aguirre]{SDA17}
D.~Safranek, J.~M. Deutsch, and A.~Aguirre.
\newblock Quantum coarse-grained entropy and thermodynamics.
\newblock \emph{Phys. Rev. A}, 99:\penalty0 010101, 2017.
\newblock URL \url{http://arxiv.org/abs/1707.09722}.

\bibitem[Safranek et~al.(2018)Safranek, Deutsch, and Aguirre]{SDA18}
D.~Safranek, J.~M. Deutsch, and A.~Aguirre.
\newblock Quantum coarse-grained entropy and thermalization in closed systems.
\newblock \emph{Phys. Rev. A}, 99:\penalty0 012103, 2018.
\newblock URL \url{http://arxiv.org/abs/1803.00665}.

\bibitem[Saito et~al.(1996)Saito, Takesue, and Miyashita]{SaitoEtAl}
K.~Saito, S.~Takesue, and S.~Miyashita.
\newblock System-size dependence of statistical behavior in quantum system.
\newblock \emph{J. Phys. Soc. Japan}, 65:\penalty0 1243, 1996.

\bibitem[Schrödinger(1927)]{Sch1927}
E.~Schrödinger.
\newblock The exchange of energy according to wave mechanics.
\newblock \emph{Annalen der Physik (4)}, 83:\penalty0 956--968, 1927.

\bibitem[Schrödinger(1952)]{Sch1952}
E.~Schrödinger.
\newblock \emph{Statistical Thermodynamics}.
\newblock Cambridge University Press, second edition, 1952.

\bibitem[Shafer and Vovk(2005)]{sh05}
G.~Shafer and V.~Vovk.
\newblock \emph{Probability and finance: it's only a game!}
\newblock John Wiley \& Sons, 2005.

\bibitem[Spohn(2020)]{spohn2020generalized}
H.~Spohn.
\newblock Generalized gibbs ensembles of the classical toda chain.
\newblock \emph{J. Stat. Phys.}, 180\penalty0 (1-6):\penalty0 4--22, 2020.

\bibitem[Srednicki(1994)]{cite57}
M.~Srednicki.
\newblock Chaos and quantum thermalization.
\newblock \emph{Phys. Rev. E}, 50:\penalty0 888--901, 1994.

\bibitem[Srednicki(1996)]{Sred96}
M.~Srednicki.
\newblock Thermal fluctuations in quantized chaotic systems.
\newblock \emph{J. Phys. A}, 29:\penalty0 L75--L79, 1996.

\bibitem[Srednicki(1999)]{Sred99}
M.~Srednicki.
\newblock The approach to thermal equilibrium in quantized chaotic systems.
\newblock \emph{J. Phys. A}, 32:\penalty0 1163–1176, 1999.

\bibitem[Steckline(1983)]{steckline83}
V.~S. Steckline.
\newblock Zermelo, {Boltzmann}, and the recurrence paradox.
\newblock \emph{American Journal of Physics}, 51\penalty0 (10):\penalty0 894--897, 1983.
\newblock \doi{10.1119/1.13373}.

\bibitem[Sugita(2007)]{cite60}
A.~Sugita.
\newblock On the basis of quantum statistical mechanics.
\newblock \emph{Nonlinear Phenomena in Complex Systems}, 10:\penalty0 192--195, 2007.
\newblock URL \url{http://arxiv.org/abs/cond-mat/0602625}.

\bibitem[Sugiura and Shimizu(2013)]{Sugiura13}
S.~Sugiura and A.~Shimizu.
\newblock Canonical thermal pure quantum state.
\newblock \emph{Phys. Rev. Lett.}, 111\penalty0 (1):\penalty0 010401, 2013.

\bibitem[Tasaki(1998)]{Tasaki1}
H.~Tasaki.
\newblock From quantum dynamics to the canonical distribution: General picture and a rigorous example.
\newblock \emph{Phys. Rev. Lett.}, 80:\penalty0 1373--1376, 1998.

\bibitem[Tasaki(2010)]{Tas2010}
H.~Tasaki.
\newblock The approach to thermal equilibrium and "thermodynamic normality", 2010.
\newblock URL \url{http://arxiv.org/abs/1003.5424}.

\bibitem[Tasaki(2016)]{Tas2016}
H.~Tasaki.
\newblock Typicality of thermal equilibrium and thermalization in isolated macroscopic quantum systems.
\newblock \emph{J. Stat. Phys.}, 163:\penalty0 937--997, 2016.

\bibitem[Tasaki(2024)]{Tasaki24}
H.~Tasaki.
\newblock {Macroscopic Irreversibility in Quantum Systems: ETH and Equilibration in a Free Fermion Chain}, 2024.
\newblock URL \url{https://arxiv.org/abs/2401.15263}.

\bibitem[Truesdell(1966)]{T66}
C.~Truesdell.
\newblock \emph{The ergodic problem in classical statistical mechanics}, pages 65--82.
\newblock Springer Berlin Heidelberg, Berlin, Heidelberg, 1966.
\newblock ISBN 978-3-662-29756-8.
\newblock \doi{10.1007/978-3-662-29756-8_5}.
\newblock URL \url{https://doi.org/10.1007/978-3-662-29756-8_5}.

\bibitem[Tumulka(2019)]{Tumulka2019}
R.~Tumulka.
\newblock Lecture notes on mathematical statistical physics, 2019.
\newblock URL \url{www.math.uni-tuebingen.de/de/forschung/maphy/lehre/ss-2019/statisticalphysics/dateien/ lecture-notes.pdf}.

\bibitem[Tumulka and Zangh\`{i}(2005)]{TZ2005}
R.~Tumulka and N.~Zangh\`{i}.
\newblock Smoothness of wave functions in thermal equilibrium.
\newblock \emph{J. Math. Phys.}, 46:\penalty0 112104, 2005.

\bibitem[Uffink(1995)]{U95}
J.~Uffink.
\newblock Can the maximum entropy principle be explained as a consistency requirement?
\newblock \emph{Studies in History and Philosophy of Science Part B: Studies in History and Philosophy of Modern Physics}, 26\penalty0 (3):\penalty0 223--261, 1995.

\bibitem[Van~Kampen(1992)]{ka92}
N.~G. Van~Kampen.
\newblock \emph{Stochastic processes in physics and chemistry}.
\newblock Elsevier, 1992.

\bibitem[Vega et~al.(1993)Vega, Uzer, and Ford]{VUF93}
J.~L. Vega, T.~Uzer, and J.~Ford.
\newblock Chaotic billiards with neutral boundaries.
\newblock \emph{Phys. Rev. E}, 48\penalty0 (5):\penalty0 3414, 1993.

\bibitem[von Neumann(1929)]{cite66}
J.~von Neumann.
\newblock Beweis des ergodensatzes und des h-theorems in der neuen mechanik.
\newblock \emph{Zeitschrift f\"ur Physik}, 57:\penalty0 30--70, 1929.
\newblock URL \url{http://arxiv.org/abs/1003.2133}.
\newblock English translation in European Physical Journal H 35: 201–237 (2010) \url{http://arxiv.org/abs/1003.2133}.

\bibitem[W.~De~Roeck(2006)]{roeck}
K.~N. W.~De~Roeck, C.~Maes.
\newblock Quantum macrostates, equivalence of ensembles, and an h-theorem.
\newblock \emph{J. Math. Phys.}, 47:\penalty0 073303, 2006.
\newblock \doi{10.1063/1.2217810}.

\bibitem[Walecka(1989)]{Walecka89}
J.~D. Walecka.
\newblock \emph{{Fundamentals of Statistical Mechanics. Manuscript and Notes of Felix Bloch}}.
\newblock Stanford University Press, Stanford, CA, 1989.

\bibitem[Wigner(1967)]{wigner67}
E.~P. Wigner.
\newblock Random matrices in physics.
\newblock \emph{SIAM Review}, 9:\penalty0 1--23, 1967.

\bibitem[Wilson(1971)]{PhysRevB.4.3174}
K.~G. Wilson.
\newblock Renormalization group and critical phenomena. i. renormalization group and the kadanoff scaling picture.
\newblock \emph{Phys. Rev. B}, 4:\penalty0 3174--3183, Nov 1971.
\newblock \doi{10.1103/PhysRevB.4.3174}.

\bibitem[Yukalov(2003)]{YUKALOV2003313}
V.~Yukalov.
\newblock Irreversibility of time for quasi-isolated systems.
\newblock \emph{Phys. Lett. A}, 308\penalty0 (5):\penalty0 313--318, 2003.
\newblock ISSN 0375-9601.
\newblock \doi{https://doi.org/10.1016/S0375-9601(03)00056-2}.
\newblock URL \url{https://www.sciencedirect.com/science/article/pii/S0375960103000562}.

\bibitem[Zangh\`{i}(2005)]{zanghi2005fondamenti}
N.~Zangh\`{i}.
\newblock I fondamenti concettuali dell’approccio statistico in fisica.
\newblock In V.~Allori, M.~Dorato, F.~Laudisa, and N.~Zanghì, editors, \emph{La Natura Delle Cose. Introduzione ai Fundamenti e alla Filosofia della Fisica}, pages 139--228. Carocci Roma, 2005.

\bibitem[Zermelo(1896)]{zermelo1896}
E.~Zermelo.
\newblock Ueber einen satz der dynamik und die mechanische wärmetheorie.
\newblock \emph{Annalen der Physik}, 293\penalty0 (3):\penalty0 485--494, 1896.
\newblock \doi{https://doi.org/10.1002/andp.18962930314}.

\bibitem[Zurek(2003)]{zurek2003decoherence}
W.~H. Zurek.
\newblock Decoherence, einselection, and the quantum origins of the classical.
\newblock \emph{Rev. Mod. Phys.}, 75\penalty0 (3):\penalty0 715, 2003.

\bibitem[Zurek(2018)]{zurek2018eliminating}
W.~H. Zurek.
\newblock {Eliminating ensembles from equilibrium statistical physics: Maxwell’s demon, Szilard’s engine, and thermodynamics via entanglement}.
\newblock \emph{Phys. Rep.}, 755:\penalty0 1--21, 2018.

\end{thebibliography}


\end{document}